%% file: thesis.tex
\documentclass[twoside]{iuphd}

\usepackage[utf8]{inputenc}

\usepackage{amsmath}
\usepackage{eurosym}
\usepackage{xcolor}
\usepackage{amssymb}
\usepackage{amsthm}
\usepackage{amscd}
\usepackage{hhline,multirow}  
\usepackage{dcolumn}          
\usepackage[font={sf}]{caption}

\usepackage{bm,curves}
\usepackage{epsfig}
\usepackage{graphicx}
\usepackage{mathrsfs}
\usepackage{slashed}
\usepackage{fancyhdr}

\DeclareGraphicsExtensions{.ps}

\def\gtorder{\mathrel{\raise.3ex\hbox{$>$}\mkern-14mu
        \lower0.6ex\hbox{$\sim$}}}
\def\ltorder{\mathrel{\raise.3ex\hbox{$<$}\mkern-14mu
        \lower0.6ex\hbox{$\sim$}}}
\newcommand{\sW}{\sin^2\theta_W}

\renewcommand{\thesection}{\Roman{section}}
\renewcommand{\thetable}{\Roman{table}}

\newcommand{\nbar}{{\overline n}}
\newcommand{\nslash}{n\hspace*{-0.22cm}\slash\hspace*{0.022cm}}
\newcommand{\nbslash}{\nbar\hspace*{-0.22cm}\slash\hspace*{0.022cm}}
\newcommand{\Tr}{{\rm Tr}}

\newcommand{\nc}{\newcommand}   
\nc{\mb}[1]{\makebox[#1]{}}
\nc{\V}{{\rm v}}
\nc{\W}{{\scriptscriptstyle W}}
\nc{\X}{{\scriptscriptstyle X}}
\nc{\CSV}{{\scriptscriptstyle CSV}}
\nc{\ra}{\rightarrow}
\nc{\alS}{{\alpha_s}}
\nc{\aSpi}{{\frac{\alS}{2\pi}}}
\nc{\api}{{\frac{\alpha}{2\pi}}}
\nc{\dwtilm}{{\delta \widetilde{m}}}
\nc{\ppg}{\pi^+\pi^-\gamma}
\nc{\nubar}{{\overline{\nu}}}
\nc{\nuN}{{\nu N_0}}
\nc{\nubN}{{\overline{\nu} N_0}}
\nc{\snuNC}{{\langle \sigma^{\nuN}_{\NC}\rangle }}
\nc{\snubNC}{{\langle \sigma^{\nubN}_{\NC}\rangle }}
\nc{\snuCC}{{\langle \sigma^{\nuN}_{\CC}\rangle }}
\nc{\snubCC}{{\langle \sigma^{\nubN}_{\CC}\rangle }}
\nc{\snNC}{{\langle \sigma^{\nu p}_{\NC}\rangle }}
\nc{\snbNC}{{\langle \sigma^{\nubar p}_{\NC}\rangle }}
\nc{\snCC}{{\langle \sigma^{\nu p}_{\CC}\rangle }}
\nc{\snbCC}{{\langle \sigma^{\nubar p}_{\CC}\rangle }}
\nc{\Rnu}{{R^{\nu}}}
\nc{\Rnub}{{R^{\overline{\nu}}}}
\nc{\sintW}{{\sin^2 \theta_{\W} }}
\nc{\vp}{{\bf p}}
\nc{\uv}{{u_{\rm v}}}
\nc{\dv}{{d_{\rm v}}}
\nc{\ubar}{{\overline{u}}}
\nc{\dbar}{{\overline{d}}}
\nc{\sbar}{{\overline{s}}}
\nc{\cbar}{{\overline{c}}}
\nc{\Ubar}{{\overline{U}}}
\nc{\Dbar}{{\overline{D}}}
\nc{\Sbar}{{\overline{S}}}
\nc{\Qbar}{{\overline{Q}}}
\nc{\FbWp}{{\overline{F}_2^{Wp}}}
\nc{\FbWD}{{\overline{F}_2^{WD}}}
\nc{\rz}{{1\over \rho_0^2}}
\nc{\gLu} {{g_L^u}}
\nc{\gRu} {{g_R^u}}
\nc{\gLd} {{g_L^d}}
\nc{\gRd} {{g_R^d}}
\nc{\Delu} {{\Delta u^2}}
\nc{\Deld} {{\Delta d^2}}
\nc{\Rnp} {{R^{\nu}_p}}
\nc{\Rnbp} {{R^{\nubar}_p}}
\nc{\Pcs}{{P_{CS}}}
\def\CC{{\scriptscriptstyle CC}}

\def\NC{{\scriptscriptstyle NC}}

\nc{\be}{\begin{equation}}
\nc{\ee}{\end{equation}}
\nc{\bea}{\begin{eqnarray}}
\nc{\eea}{\end{eqnarray}}
\nc{\F}{{\scriptscriptstyle F}}  
\nc{\xF}{{x_{\F}}}
\nc{\Fcc}{F_2^c}

\def\IE{{\it i.e.,}}
\def\EG{{\it e.g.,}}
\def\EA{{\it et al.}}

\def\Psl{\slashed{P} }
\def\psl{\slashed{p} }
\def\ksl{\slashed{k} }
\def\qsl{\slashed{q} }
\def\lsl{\slashed{l} }
\def\Dsl{\slashed{\Delta} }
\def\parsl{\slashed{\partial} }
\def\lan{\langle }
\def\ran{\rangle }

\def\lsim{\mathrel{\rlap{\lower4pt\hbox{\hskip1pt$\sim$}}
    \raise1pt\hbox{$<$}}}
\def\gsim{\mathrel{\rlap{\lower4pt\hbox{\hskip1pt$\sim$}}
    \raise1pt\hbox{$>$}}}

\newcommand*{\defeq}{\mathrel{\vcenter{\baselineskip0.5ex \lineskiplimit0pt
                     \hbox{\scriptsize.}\hbox{\scriptsize.}}}%
                     =}

\def \CQM{{\scriptscriptstyle{CQM}}}
\def \beqn{\begin{eqnarray}}
\def \eeqn{\end{eqnarray}}

\def \bea{\begin{eqnarray}}
\def \beq{\begin{equation}}
\def \eea{\end{eqnarray}}
\def \eeq{\end{equation}}

\def \bwt{\begin{widetext}}
\def \ewt{\end{widetext}}

\graphicspath{{./DIS-Fig/}}
\graphicspath{{./nucl-Fig/}}
\graphicspath{{./Q2-Fig/}}
\graphicspath{{./OC-Fig/}}
\graphicspath{{./TMC-Fig/}}
\graphicspath{{./TMCii-Fig/}}
\graphicspath{{./sidis-Fig/}}
\graphicspath{{./CSV-Fig/}}
\graphicspath{{./IC-Fig/}}
\graphicspath{{./TDIS-Fig/}}

\title{The nonperturbative structure of hadrons}
\author{T.~J.~Hobbs}
\date{August 2014}

\committeechair{J.~T.~Londergan, Ph.D.}
\readertwo{W.~Melnitchouk, Ph.D.}
\readerthree{L.~Kaufman, Ph.D.}
\readerfour{J.~Liao, Ph.D.}
\readerfive{W.~M.~Snow, Ph.D.}
\defensedate{August 11, 2014}

\cryear{2014}

\begin{document}
\maketitle

\acceptancepage

\copyrightpage

\begin{dedication}

\topskip0pt
\vspace*{\fill}

\begin{center}
{\it To my many teachers.}

\vspace*{1.2cm}
\includegraphics[height=0.7cm]{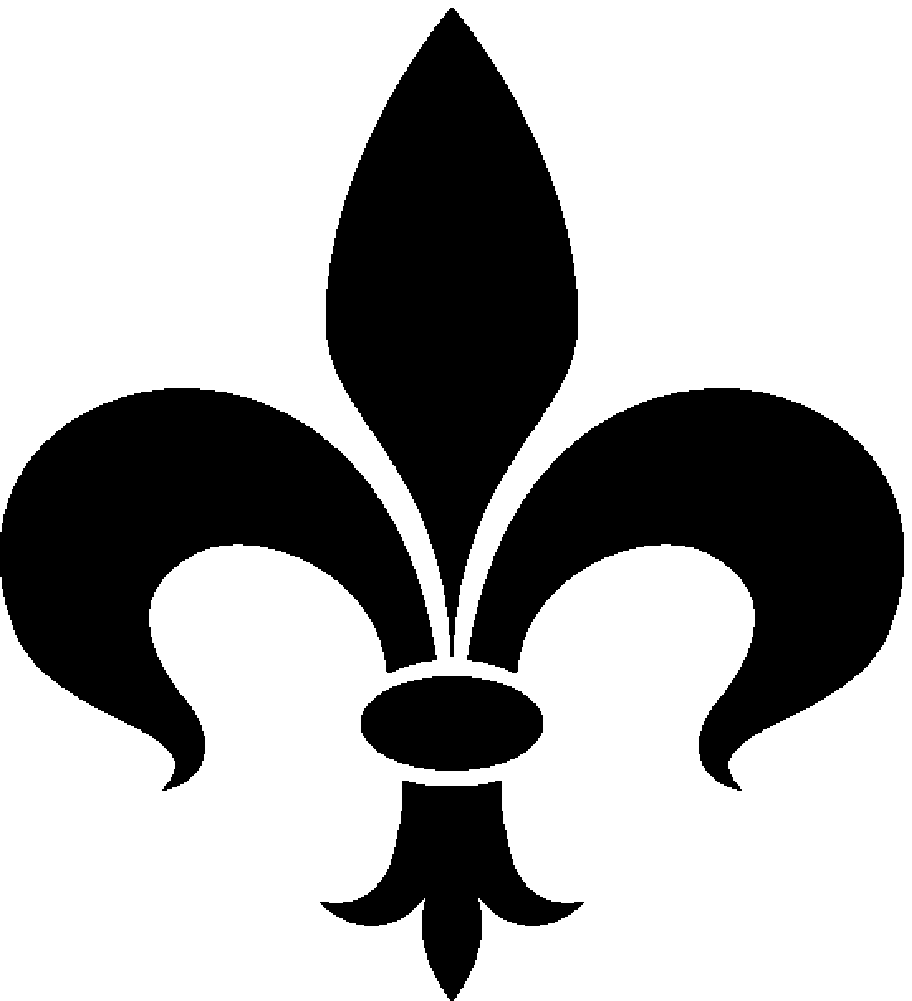}
%
\vspace*{1.2cm}

\begin{quote}
``...what I embody, the principle of life, cannot be destroyed. It is written into the cosmic code,
the order of the universe.'' \\
\hspace*{1cm} --- Heinz R. Pagels, {\it The Cosmic Code} \cite{Heinz}
\end{quote}

\vspace*{0.6cm}

\includegraphics[height=10cm]{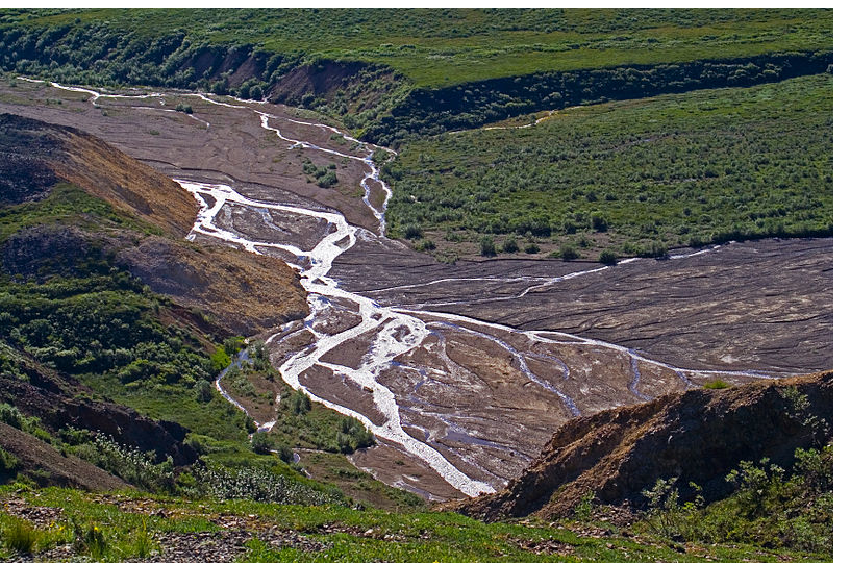}

\vspace*{-.15cm}
\hspace*{1.5cm} Toklat River, AK. \ \ \ \ \ \copyright\ Nic McPhee / flickr.com / CC-BY-SA-2.0
\end{center}

\vspace*{\fill}
\end{dedication}

\begin{acknowledgments}
It has been said, likely with justice, that physicists are something of a peculiar
species. While this may be the case, my ongoing studies in physics have granted me the
considerable fortune of learning from many wonderful and unforgettable people.

At the top of the list must of course be my advisers Tim Londergan at Indiana and
Wally Melnitchouk at JLab. On more occasions than I can conceivably recall, I benefitted
from Tim's ready insights and knowledge, but his unfailingly good humor made the burden of
completing a Ph.D. not only bearable, but enjoyable and timely. Much the same, my collaborations
with Wally began while I was a mere undergraduate, but later success would simply have been impossible
without his vast expertise in the field, pleasant personality, and unhesitating support.

Along the same vein, gratitude is also due my undergraduate faculty mentor at Chicago Jon Rosner
for helpful discussions, collaboration, and support over the years.

In a slightly wider sphere, I owe much to many faculty and personnel at Indiana, among
whom theorists Jinfeng Liao, Brian Serot (who is sorely missed), Adam Szczepaniak, Steve Gottlieb,
and Alan Kosteleck\'y deserve special mention. I benefitted as well from educational interactions with experimentalists
Mike Snow, Lisa Kaufman, Chen-Yu Liu, and Mark Hess; and I will never forget the countless occasions the ever-indefatigable
Moya Wright of IU's CEEM came to my grateful aid.

I would certainly be remiss not to mention my fellow graduate students and postdocs at IU, from whom I
have learned more than could ever be noted here. In particular, Peng Guo, Dan Salvat, Rana Ashkar, Xilin Zhang,
and Dan Bennett especially come to mind and deserve enormous thanks.

I am indebted to many members of the JLab staff and user community, among whom I am happy to count
Alberto Accardi and Pedro Jimenez-Delgado (both terrific collaborators), Cynthia Keppel, Christian Weiss, and
formerly, Mark Paris (my first scientific mentor), Alessandro Bacchetta, and Marc Schlegel. I must also point out
that my education in theoretical physics was enormously facilitated over the years by the support and kindness of
the JLab Theory Center.

I am also grateful to Jerry Miller, Silas Beane, and Huey-Wen Lin at Seattle, as well as
Craig Roberts and Ian Clo\"et at Argonne for years of enlightening discussions and the
hospitality they showed during my recent visits to their institutions.

Within the broader field, for direct collaborations and/or discussions, I wish to thank Tony Thomas (Adelaide),
Dave Murdock (formerly of Tennessee Tech, with whom continued
collaboration was sadly cut short), Chueng Ji (NCSU), Fernanda Steffens (Sao Paulo), Ramona Vogt (UC Davis), Stan Brodsky (SLAC),
Jen-Chieh Peng (Illinois), Jeff Owens (FSU), Michael Ramsey-Musolf and Krishna Kumar (UMass), Kent Paschke and Xiaochao Zheng (Virginia),
Paul Reimer (Argonne), Paul Souder (Syracuse), Fred Olness (SMU), and Simonetta Liuti (UVA).

This work was also enabled by financial support from both the National Science Foundation and the Department of Energy's
Office of Science, for which I am extremely thankful.

I must put down a few words for those closest to me. Throughout my life my immediate family --- father Daniel,
mother Eileen, brother Dan, and sister-in-law Brandi --- has encouraged me at every stage. Your affection, guidance, and
tangible support have made my work possible. To my ``quasi-in-laws'' Ajeeta Khatiwada and Sean Kuvin: our discussions on
physics and life have been a source of relief; I look forward to many years of friendship, and dare I say, collaboration.

And finally, Rakshya: I started graduate school with the goal of achieving my doctorate, but instead found you along
the way. Now after everything the degree itself seems an afterthought --- I simply would never have survived without
you.

\end{acknowledgments}


\begin{abstract}

In this thesis we explore a diverse array of issues that strike at the inherently nonperturbative
structure of hadrons at momenta below the QCD confinement scale. In so doing, we mainly seek a better
control over the partonic substructure of strongly-interacting matter, especially as this relates
to the nonperturbative effects that both motivate and complicate experiments --- particularly DIS; among
others, such considerations entail sub-leading corrections in $Q^2$, dynamical higher twist effects, and hadron mass
corrections. We also present novel calculations of several examples of flavor
symmetry violation, which also originates in the long-distance properties of QCD at low
energy.  Moreover, we outline a recently developed model, framed as a hadronic effective theory amenable
to QCD global analysis, which provides new insights into the possibility of nonperturbative heavy quarks
in the nucleon. This model can be extended to the scale of the lighter mesons, and we assess the accessibility
of the structure function of the interacting pion in the resulting framework.

\vspace*{0.75cm}

\line(1,0){125} \ \ \ \ \ \ \ \ \ \ \line(1,0){125} \ \ \ \ \ \ \ \ \ \ \line(1,0){125}

\vspace*{-0.4cm}

J.~T.~Londergan, Ph.D. \ \ \ \ \ \ \ \ \ \ \ \ \ W.~Melnitchouk, Ph.D. \ \ \ \ \ \ \ \ \ \ \ \ \ L.~Kaufman, Ph.D.

\vspace*{0.75cm}

\line(1,0){125} \ \ \ \ \ \ \ \ \ \ \line(1,0){125}

\vspace*{-0.4cm}

J.~Liao, Ph.D. \ \ \ \ \ \ \ \ \ \ \ \ \ \ \ \ \ \ \ \ \ \ \ \ \ W.~M.~Snow, Ph.D.

\end{abstract}

\tableofcontents

    \pagenumbering{arabic}
    \setcounter{page}{1}

\chapter{Introduction}
\label{chap:ch-intro}
\begin{quote}
``If I were again beginning my studies, I would follow the advice of Plato and start with mathematics.'' \\
\hspace*{0.5cm} --- Galileo Galilei
\end{quote}
\input{the-intro}

\chapter{Invitation: The Handbag Diagram}
\label{chap:ch-DIS}
\begin{quote}
``Lettin' the cat outta the bag is a whole lot easier 'n puttin' it back in.'' \\
\hspace*{0.5cm} --- Will Rogers
\end{quote}
\input{the-DIS}
\input{the-nucl}

\input{the-OPE}

\chapter{Finite-$Q^2$ corrections in electroweak phenomenology}
\label{chap:ch-Q2}
\begin{quote}
``Similarly, many a young man, hearing for the first time of the refraction of stellar light, has thought that doubt was cast
on the whole of astronomy, whereas nothing is required but an easily effected and unimportant correction to put everything
right again.'' \\
\hspace*{0.5cm} --- Ernst Mach
\end{quote}
\input{the-Q2}
\input{the-CSV}

\chapter{Mass corrections to DIS}
\label{chap:ch-TMC}
\begin{quote}
``Principles \\
You can't say A is made of B \\
or vice versa. \\
All mass is interaction.''\\
\hspace*{0.5cm} --- Richard P.~Feynman
\end{quote}
\input{the-TMC}

\input{the-sidis}


\chapter{Nonperturbative charm}
\label{chap:ch-charm}
\begin{quote}
``All the diversity, all the charm, and all the beauty of life are made up of light and shade.'' \\
\hspace*{0.5cm} --- Leo Tolstoy
\end{quote}
\input{the-charm}
\input{the-GA}

\chapter{Epilogue: The pion cloud and final state tagging}
\label{chap:ch-TDIS}
\begin{quote}
``Almost everybody that's well-known gets tagged with a nickname.'' \\
\hspace*{0.5cm} --- Alan Alda
\end{quote}
\input{the-TDISa}

\chapter{Conclusion}
\label{chap:ch-conc}
\begin{quote}
``Reasoning draws a conclusion, but does not make the conclusion certain, unless the mind discovers it by the path of experience.'' \\
\hspace*{0.5cm} --- Roger Bacon
\end{quote}
\input{the-conc}



\include{the-pub}
\pagenumbering{gobble}
\pagestyle{plain}

\addcontentsline{toc}{chapter}{Curriculum Vitae}

\begin{center}
{\normalsize \bf CURRICULUM VITAE}
\end{center}

{\bf Timothy John Hobbs} \ \ \ \ \ \ \ \ \ {\texttt timjhobb@indiana.edu} \ \ \ \ \ \ \ \ \ Citizenship: {\bf USA}

\vspace*{0.4cm}

{\bf \underline{Education}}
\begin{itemize}

  \item Ph.D. Physics (Nuclear/HEP theory), Indiana University, August 2014. \\
        Thesis: {\it The Nonperturbative Structure of Hadrons}; \\
        advisor(s): J. Timothy Londergan and Wally Melnitchouk.

  \item M.S. Physics, Indiana University, December 2010.

  \item B.A. Physics, The University of Chicago, June 2009. \\
        Thesis: {\it Oblique Corrections in Electroweak Physics}; \\
        advisor: Jon Rosner.

  \item B.A. Mathematics, The University of Chicago, June 2009.
\end{itemize}

\vspace*{0.4cm}

{\bf \underline{Teaching Experience}}

Associate Instructor at Indiana University:
\begin{itemize}
\item (Grading), Physics 453 (Quantum Mechanics) \\
       {\it Spring 2012}
\item (Laboratory), Physics 202 (Introductory Electricity \& Magnetism) \\
       {\it Fall/Spring 2011/12}
\item (Recitation), Physics 221/2 (Calculus-based Mechanics/Electricity-Magnetism) \\
       {\it Fall/Spring 2010/11}
\item (Laboratory), Physics 201 (Introductory Mechanics) \\
       {\it Fall/Spring 2009/10}
\end{itemize}

\pagebreak

\vspace*{0.4cm}

{\bf \underline{Journal Articles}}
\begin{enumerate}

\item
  T.~J.~Hobbs, Chueng-Ryong Ji, J.~T.~Londergan and W.~Melnitchouk \\
  in progress: \\
  {\it The role of the Delta in relativistic pion loop corrections to
                                              photon-nucleon vertices}

\item
  T.~J.~Hobbs, P.~Jimenez-Delgado, J.~T.~Londergan and W.~Melnitchouk \\
  Phys.~Rev.~Lett.~({\it submitted}),
  [arXiv:1408.1708 [hep-ph]]: \\
  {\it New limits on intrinsic charm in the nucleon from global analysis of parton distributions}

\item
  T.~J.~Hobbs, J.~T.~Londergan and W.~Melnitchouk \\
  Phys.~Rev.~D {\it 89}, 074008 (2014),
  [arXiv:1311.1578 [hep-ph]]: \\
  {\it Phenomenology of nonperturbative charm in the nucleon}

\item
  M.~Gorchtein, T.~Hobbs, J.~T.~Londergan and A.~P.~Szczepaniak \\
  Phys.\ Rev.\ C {\bf 84}, 065202 (2011),
  [arXiv:1110.5982 [nucl-th]]: \\
  {\it Compton Scattering and Photo-absorption Sum Rules on Nuclei}

\item
  L.~T.~Brady, A.~Accardi, T.~J.~Hobbs and W.~Melnitchouk \\
  Phys.\ Rev.\ D {\bf 84}, 074008 (2011),
  [arXiv:1108.4734 [hep-ph]]: \\
  {\it Next-to leading order analysis of target mass corrections to structure
                                                   functions and asymmetries}

\item
  T.~J.~Hobbs, J.~T.~Londergan, D.~P.~Murdock and A.~W.~Thomas \\
  Phys.\ Lett.\ B {\bf 698}, 123 (2011),
  [arXiv:1101.3923 [hep-ph]]: \\
  {\it Testing Partonic Charge Symmetry at a High-Energy Electron Collider}

\item
  T.~Hobbs and J.~L.~Rosner \\
  Phys.\ Rev.\ D {\bf 82}, 013001 (2010),
  [arXiv:1005.0797 [hep-ph]]: \\
  {\it Electroweak Constraints from Atomic Parity Violation and Neutrino Scattering}

\item
  A.~Accardi, T.~Hobbs and W.~Melnitchouk \\
  JHEP {\bf 0911}, 084 (2009),
  [arXiv:0907.2395 [hep-ph]]: \\
  {\it Hadron mass corrections in semi-inclusive deep inelastic scattering}

\item
  T.~Hobbs and W.~Melnitchouk \\
  Phys.\ Rev.\  D {\bf 77}, 114023 (2008),
  [arXiv:0801.4791 [hep-ph]]: \\
  {\it Finite-Q$^2$ corrections to parity-violating DIS}

\end{enumerate}

\vspace*{0.4cm}

{\bf \underline{Talks and Conferences}}
\begin{itemize}

\item
  Invited talk: ``Phenomenological implications of the nucleon's meson cloud''
  at the Light Cone 2014 International Workshop, Raleigh, NC; May 27, 2014.

\item
  Invited seminar(s): ``Nonperturbative aspects of hadron structure phenomenology''
   at the University of Washington Nuclear Theory Group, Seattle, WA; January 30, 2014
   {\bf AND}
   the Argonne National Lab Physics Division Theory Group, Lemont, IL; February 11, 2014.

\item
  Invited talk: ``Numerical estimates of chiral cloud contributions to tagged proton
  electroproduction'' at the JLab workshop ``Exploring Hadron Structure with Tagged Structure
  Functions,'' Newport News, VA; January 16-18, 2014.

\item
  Contributed talk: ``Phenomenology of nonperturbative heavy quarks in the nucleon''
  at the 2013 Fall Meeting of the APS Prairie Section, Columbia, MO; November 7-9, 2013.

\item
  Contributed talk: ``Comparative study of nonperturbative heavy quarks in the nucleon''
  at the 2013 DNP Meeting, Newport News, Virginia; October 23-26, 2013.

\item
  Contributed talk: ``Nonperturbative heavy quarks in the nucleon'' at
  the 26$^{th}$ Annual Midwest Theory Get-Together, Argonne National Lab;
  September 6 \& 7, 2013.

\item
  Contributed poster: ``Nonperturbative charm content of the nucleon''
  EPJ Web of Conf.~{\bf 66}, 06007 (2014) [INPC 2013];
  at the 2013 International Nuclear Physics Conference,
  Firenze, Italy; June 2-7, 2013.

\item
  Invited talk: ``Non-perturbative (and perturbative) charm contributions
  to various processes'' at the 5th Workshop of the APS Topical Group on
  Hadronic Physics, Denver, Colorado; April 10-12, 2013.

\item
  Contributed talk: ``Intrinsic charm of the proton'' at
  the 25$^{th}$ Annual Midwest Theory Get-Together, Argonne National Lab;
  September 7 \& 8, 2012.

\item
  Invited talk: ``Non-perturbative charm content of the nucleon'' at the
  International Workshop on ``Physics Opportunities at an Electron-Ion
  Collider,'' held at Indiana University, Bloomington; August 20-22, 2012.

\item
  Contributed talk: ``Partonic charge symmetry violation at LHC kinematics'' at
  the 24$^{th}$ Annual Midwest Theory Get-Together, Argonne National Lab;
  September 23 \& 24, 2011.

\item
  Contributed talk/poster: ``The J=0 pole and finite-energy sums rules in real
  Compton scattering'' at the LesNabis School on Amplitude Analysis in Modern
  Physics, Bad Honnef, Germany; August 1-5, 2011.

\item
  Invited talk: ``Target mass corrections to parity-violating DIS,''
  AIP Conf.\ Proc.\  {\bf 1369}, 51 (2011)
  [arXiv:1102.1106 [hep-ph]];
  at the 3rd International Workshop on Nucleon Structure at Large
  Bjorken x, Jefferson Lab; October 13 - 15, 2010.

\item
  Contributed poster: ``Mass corrections in semi-inclusive deep inelastic
  scattering'' at the 2009 JLab Users Group Meeting; June 8-10, 2009.

\item
  Invited talk: ``Finite Q$^2$ corrections to PV DIS'' at the PV DIS
  Workshop, University of Wisconsin, Madison;
  June 3 - 5, 2009.

\item
  Invited talk: ``Parity violation in leptonic DIS'' at the JLab Theory
  Center ``Cake Seminar,'' Newport News, Virginia; September 5, 2007.

\end{itemize}

\vspace*{0.4cm}

{\bf \underline{Awards}}
\begin{itemize}

\item
Indiana University Dept.~of Physics ``Outstanding Graduate Student in Theoretical Research''
(2014), given in Bloomington, IN ($\$$500).

\item
2014 Gary McCartor Fellowship, given by the International Light Cone Advisory Committee (ILCAC)
at the Raleigh, NC Light Cone Meeting ($\$$1000).

\item
Best Young Poster Award: International Nuclear Physics Conference (INPC) 2013
in Firenze, Italy (\EUR{750}).

\item
American Physical Society, FGSA Travel Grant for Excellence in Graduate Research ($\$$500) to
attend the CTEQ-Fermilab School on QCD and Electroweak Phenomenology at PUCP Lima, Peru; July,
30 -- August 9, 2012.

\item
First Place: 2009 JLab Users Group Annual Meeting, Graduate Student Poster Contest ($\$$1000).

\item
Selection for the DOE Office of Science, Science Undergraduate Laboratory Internship Program
(SULI) at Jefferson Lab; 2007 and 2008.

\end{itemize}


\end{document}

%% file: the-intro.tex

In what was arguably the most startling intellectual development of human scientific history, the
early 20$^{\rm th}$ Century heralded the final unification of Planckian Quantum Mechanics and Einsteinian
relativity. Since then, rapid progress has been made in directing the resulting synthesis
-- Quantum Field Theory (QFT) -- toward a total description of microscopic matter.

Though unquestionably novel, the insights that bequeathed
QFT belong to the same historical tradition that in the West originated with the Milesian materialists
Thales and Anaximander, as well as the early atomists of Abdura, among whom the great Democritus \cite{Aristotle}
is likely the best known today.\footnote{Less frequently mentioned, the Jaina tradition and other thinkers within
Hindu civilizations of the Indian subcontinent also postulated the existence of indivisible {\it anu}, most likely
independently of the Greeks as early as the 6$^{\rm th}$ Century BCE.}

Sadly, this enlightened perspective languished in relative obscurity for millenia, not again being
embraced until the chemical analyses of Dalton \cite{Dalton:1808} and Lavoisier; even then, the reality and nature
of the atomic hypothesis remained controversial up to the time of J.~J.~Thomson's discovery
of the electron \cite{Thomson:1897} in 1897, and the complementary discovery of the atomic nucleus \cite{Rutherford:1911}
by Rutherford, Geiger, and Marsden in 1909. With these confirmations as well as observations of the emission
spectra of hydrogen, Bohr's universally recognizable planetary model quickly followed, absorbing
several modifications in response to the wave formalism of Schr\"odinger.

The disunity of special relativity and quantum mechanics was a preventative roadblock to a
theory of (sub-)atomic interactions with electromagnetic fields until the QFT of Dirac \cite{Dirac:1928hu},
Fermi and others was shown to be renormalizable by Bethe; the resulting $U(1)$ theory of QED was
especially facile in describing spectral line shifts of electromagnetic bound states
(\EG~the Lamb shift in hydrogen), as well as the outcomes of various scattering
experiments --- $e^- e^-$ M$\o$ller reactions, the Bhahba process $e^- e^+$, and
electron-muon interactions.

On the other hand, the remarkably short lifetimes involved in nuclear decays indicated that an enormously stronger
force was required to bind a nucleus of net positive electric charge. For this reason, despite successes describing
the electromagnetic interaction, initial doubts regarding the general applicability of field-theoretic methods to the dynamics of
strongly bound matter led to early attempts based on $S$-matrix theory; among these were Regge theory \cite{Regge:1959mz}
and the `bootstrap' scheme \cite{Chew:1968fe} of Chew. Now largely defunct, the latter of these envisioned
a ``nuclear democracy'' of hadronic states, each nested within the other, thereby leading to a situation
in which there simply were no fundamental states. Among its advocates, the bootstrap paradigm was thought
to constrain the S-matrix under the auspices of unitarity, analyticity, and crossing symmetries, but
additionally required a `narrow resonance' approximation to produce scattering amplitudes with mixed success.

As an alternative, Regge theory endeavored to describe amplitudes in strong interaction physics as arising
from exchanges of states of specific angular momentum, themselves belonging to a complex space of
linear `trajectories' $\alpha_J(t)$. This framework leads to a simple prediction for the $(s,\ t)$
dependence of hadronic scattering amplitudes:
\begin{equation}
A(s,\ t)\ \sim\ e^{\alpha_J(t)\ \cdot\ \ln(s/s_0)}\ .
\label{eq:intro-Reg}
\end{equation}
Perhaps ironically, Regge theory remains among the more competitive tools in ongoing efforts
to unite the unresolved details of short- vs.~long-distance physics in QCD as we shall
briefly describe in Chap.~\ref{chap:ch-DIS}.\ref{sec:Compt}.

\begin{figure}[h]
\includegraphics[height=4.3cm]{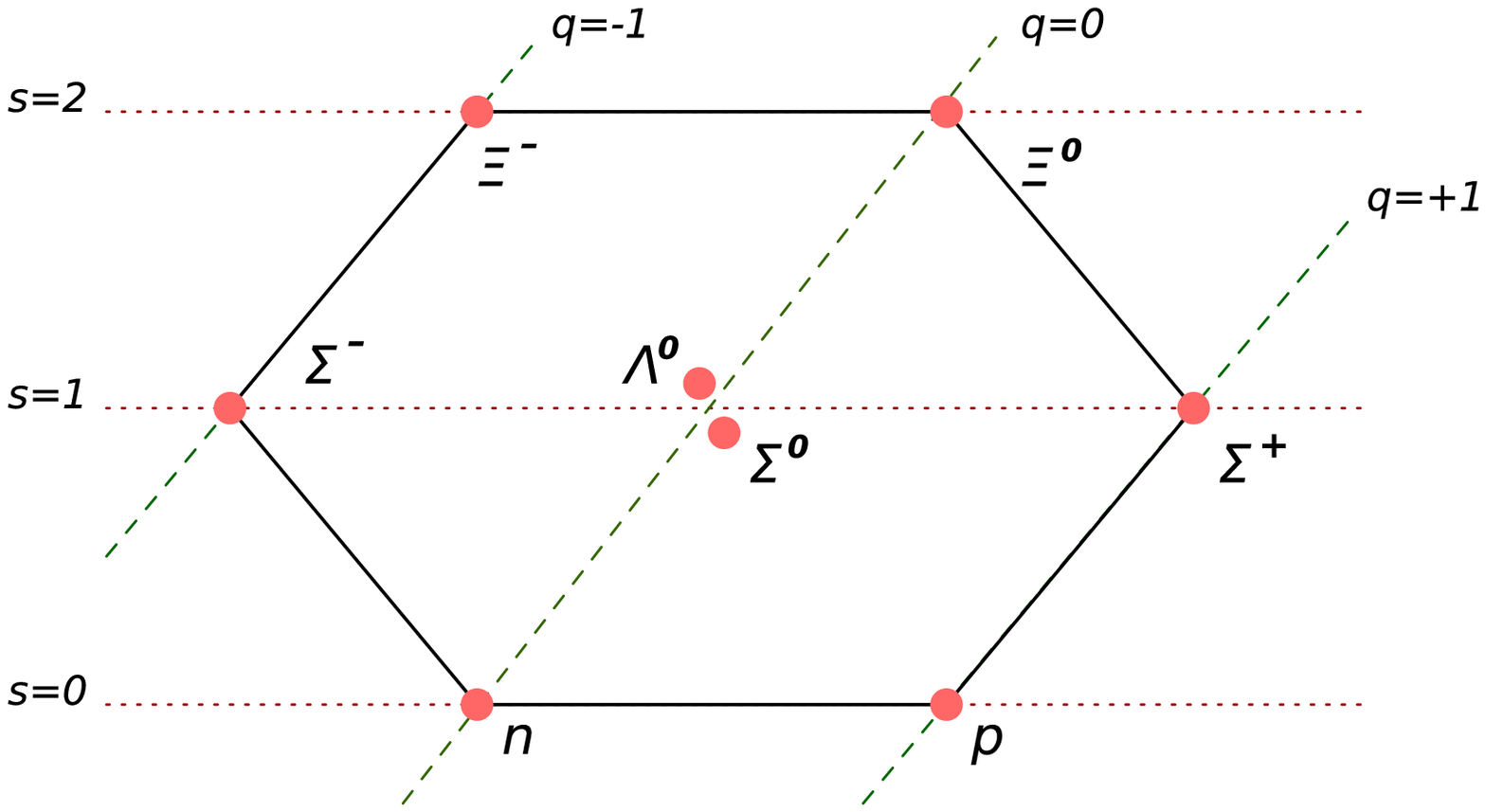} \ \ \ \ \ \
\includegraphics[height=4.3cm]{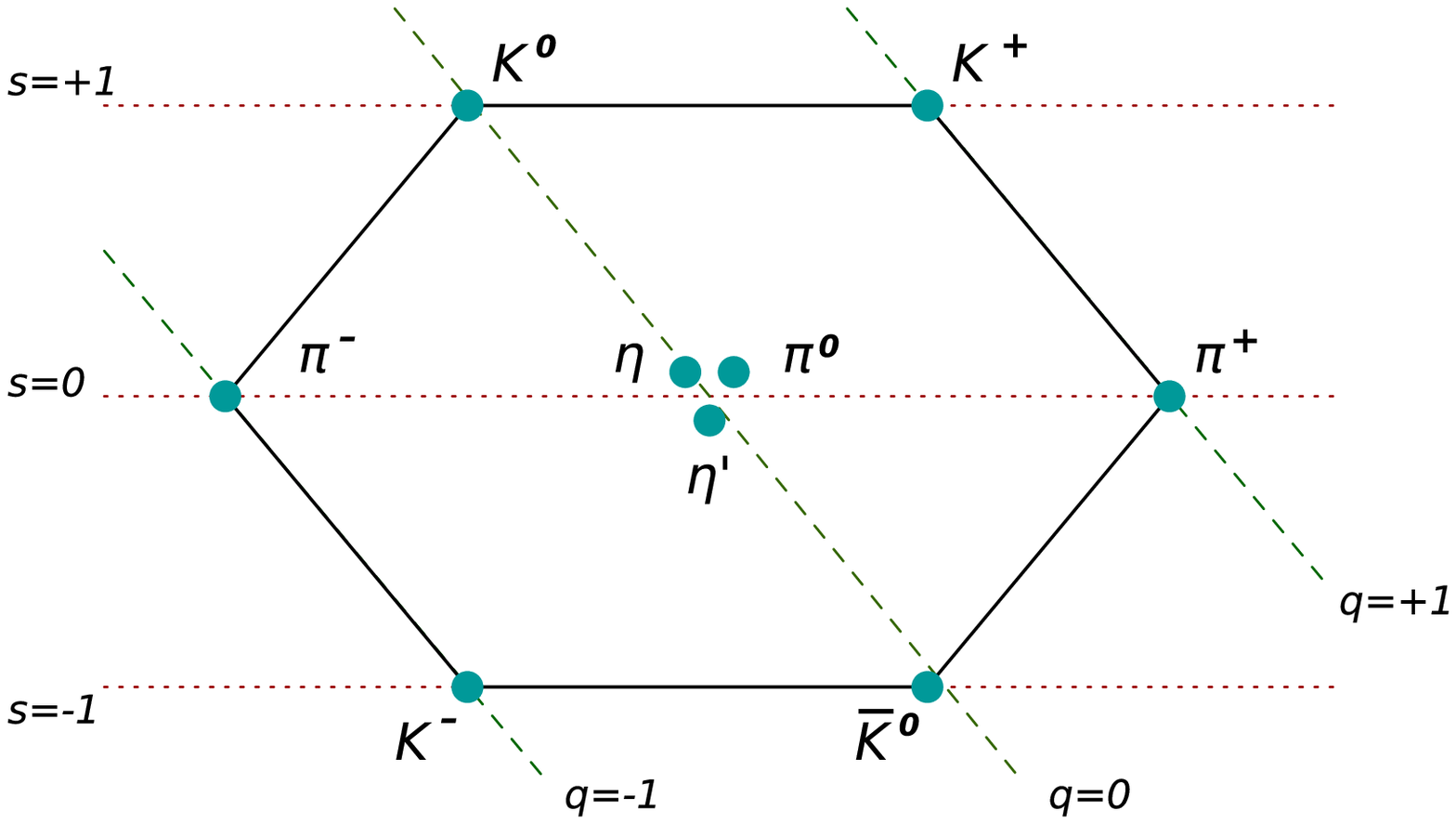} \ \ \ \ \ \
\vspace*{-1cm}
\caption{(Left) The flavor $SU(3)$ octet of spin-1/2 baryons.
(Right) The corresponding pseudoscalar meson nonet.
}
\label{fig:SU3_mult}
\end{figure}
The doubts in more field-theoretic approaches were partially driven by the physical logic embodied in observations
by Landau \cite{Landau:1960} that charge screening phenomena connected to the perturbative calculability of
QED had no clear analogue in the physics of hadrons. While these quandaries stymied theoretical
efforts, advances at the early generation of colliders at Stanford and other experimental facilities revealed
an increasingly rich landscape of mesons and baryons, leading \linebreak Gell-Mann \cite{GellMann:1964nj}
and Zweig \cite{Zweig:1964jf} to the natural suspicion that this proliferation in the hadronic spectrum
was evidence of an underlying flavor symmetry generated by the constituent degrees of freedom --
called `quarks' by Gell-Mann. (Zweig's alternative moniker `aces' never quite caught on.) Hence the
now-famous eight-fold way deduced with Ne'eman presents the lightest spin-$1/2$ baryons as belonging
to a flavor octet generated by an approximate $SU(3)$ flavor symmetry, whereas the higher spin
resonances form a decuplet. Similar flavor multiplets were found to hold for the pseudoscalar and
vector meson nonets. For the sake of illustration, the $SU(3)$ spin-$1/2$ baryons and pseudoscalar mesons are ordered in
typical fashion according to charge and strangeness in the left and right panels of Fig.~\ref{fig:SU3_mult}.

Contemporaneously, the fact that the nucleon was not fundamental and possessed some non-trivial distribution of
electric and magnetic charge was made clear by the characteristic decrease with $Q^2$
of the electromagnetic form factors $G_E(Q^2),\, G_M(Q^2)$ measured in the pioneering elastic electron-proton
experiments conducted by Hofstadter et al.~\cite{Hofstadter:1956qs,Janssens:1965kd}.
At much the same time, some of the first results from the early generation of electron-nucleon
deeply inelastic scattering (DIS) experiments began to emerge; perhaps among the more suggestive
results obtained was the unexpected behavior of the cross section ratio for longitudinally vs.~transversely
polarized photons, namely, that $R = \sigma^L / \sigma^T \sim 0$ for $Q^2 \to \infty$ at fixed $x$.
For reasons that will be explained in greater detail in Chap.~\ref{chap:ch-Q2}, this was a striking
affirmation that the sub-nucleonic constituents of the proton were indeed charge-carrying, spin-1/2 fermions.

All the more, these measurements also presented the first direct experimental
confrontation with Bjorken's current algebra scheme. In another attempt at side-stepping
the problems known to plague formal field theories of the strong interaction, current algebra suggested
that DIS cross sections should depend only upon the single parameter $x$ (rather than the two permitted
by kinematical considerations, $(x,\, Q^2)$, as discussed in Chap.~\ref{chap:ch-DIS}.\ref{sec:DIS}), in a
phenomenon which came to be known as `scaling.' \cite{Bjorken:1968dy}

Physically due to scattering from individual, weakly-interacting partons, scaling was mysterious in the setting of 
generic, perturbative QFTs, in which resummation of corrections to all orders would presumably lead to 
divergences and the irresolvable breaking of scaling. However, the properties of non-Abelian field theories,
together with the dimensional regularization procedures introduced by 't Hooft and Veltman \cite{'tHooft:1972fi} in the end
provide the answer. Such a Yang-Mills theory can be constructed in terms of quarks, with gauge invariance specifying the
interactions. The result is the modern theory of the strong interaction -- quantum chromodynamics
(QCD).

Formally, the degrees of freedom of QCD are quarks and gluons, and their interactions in the asymptotic limit
are governed by a simple\footnote{In practice, a Grassmann algebra must be introduced to invert the
gauge field product $\mathcal{G}^i_{\mu \nu} \mathcal{G}^{i \, \mu \nu}$ and obtain the gluon propagator;
as a side effect, Feddeev-Popov `ghost' terms are thereby generated as well, though they have been suppressed
here for simplicity.} lagrangian:
\begin{align}
{\cal L} &= \sum_q \bar{\psi}_{q,a}(i [\slashed{\partial} + i m_q] \delta_{ab})
         - \alpha_s \gamma^\mu t_{ab}^C A_\mu^C) \psi_q^b
         - \frac{1}{4} \mathcal{G}^i_{\mu \nu} \mathcal{G}^{i\ \mu\nu} \nonumber \\
\mathcal{G}^i_{\mu \nu} &= \partial_\mu A^i_{\nu} - \partial_\nu A^i_{\mu} - \alpha_s f_{ijk} A^j_\mu A^k_\nu,
\label{eq:QCD}
\end{align}
where the QCD $SU(3)$ structure constants are determined from the Gell-Mann algebra by
$[t^i, t^j] = i f_{ijk} \, t^k$.

In the fundamental representation $SU(N)$, the {\it beta function} of
non-Abelian gauge theory describes the dependence of the renormalized strong coupling $\alpha_s$ on the
regularization scale $\mu$ to an arbitrary order in perturbation theory. At leading order, the famous result as
first isolated from the lagrangian of Eq.~(\ref{eq:QCD}) by Gross, Wilczek, and Politzer \cite{Gross:1973id} was
found to be
\begin{align}
\beta (\alpha_s)\ =\ &\partial \alpha_s \Large/ \partial \log \mu \nonumber\\
=\ &- \left( {11 \over 3}\ T_A\ -\ {4 \over 3} n_F\ T_R \right)\ {\alpha_s^2 \over 2\pi}\ +\ \mathcal{O}(\alpha_s^3)\ , \nonumber\\
T_A = 3,\ \,\,\, T_R = 1/2\ \implies\ &- \left( 11\ -\ {2 \over 3} n_F \right)\ {\alpha_s^2 \over 2\pi}\ +\ \mathcal{O}(\alpha_s^3)\ ,
\label{eq:QCDbeta}
\end{align}
which is clearly negative for any choice of flavor number $n_F \le 16$. This profound result, possible only in the
context of non-Abelian gauge theories, is in fact the finding that unifies the disparate problems just described and
renders them solvable.

\begin{figure}[h]
\includegraphics[height=5.7cm]{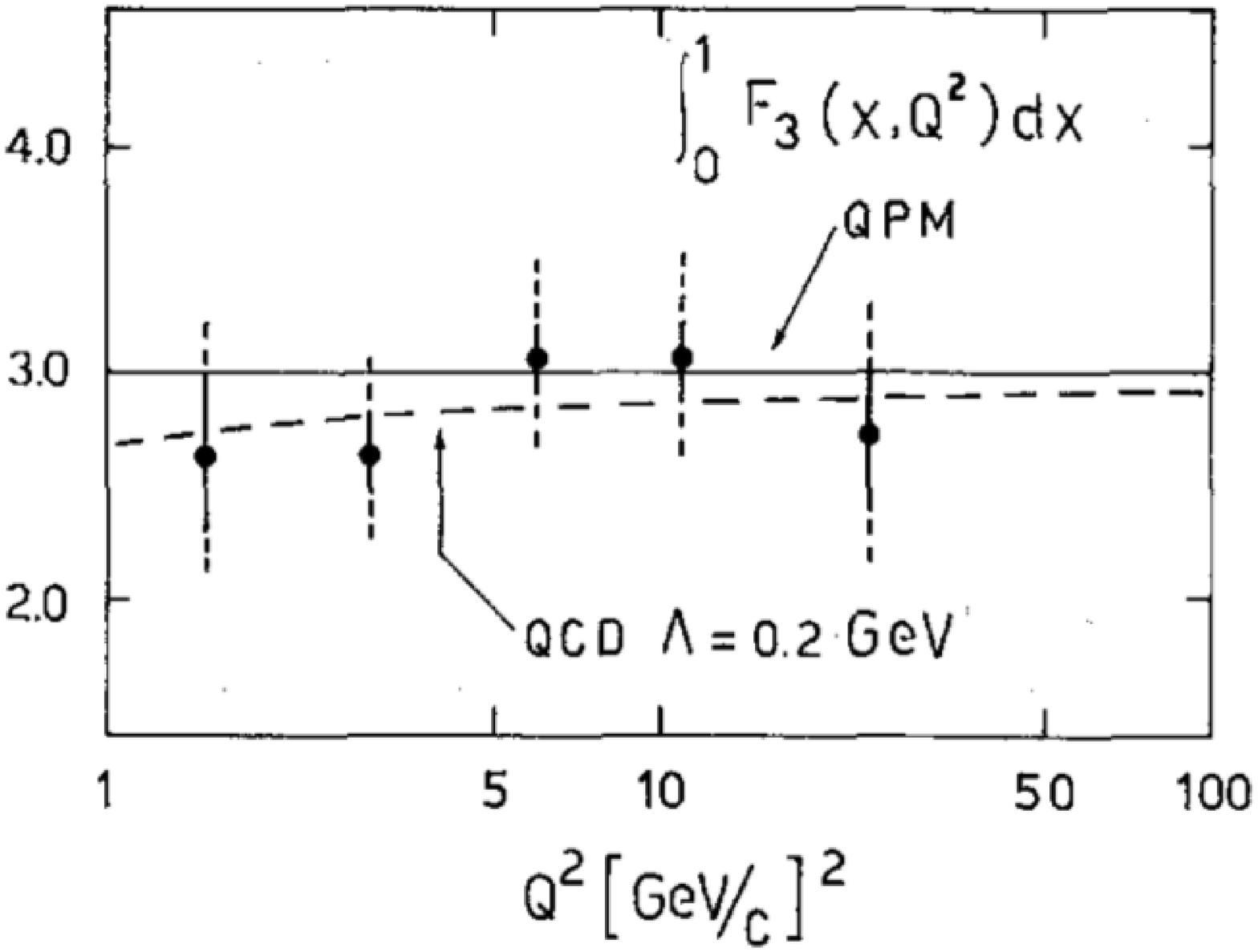} \ \ \
\raisebox{0.48cm}{\includegraphics[height=5.4cm]{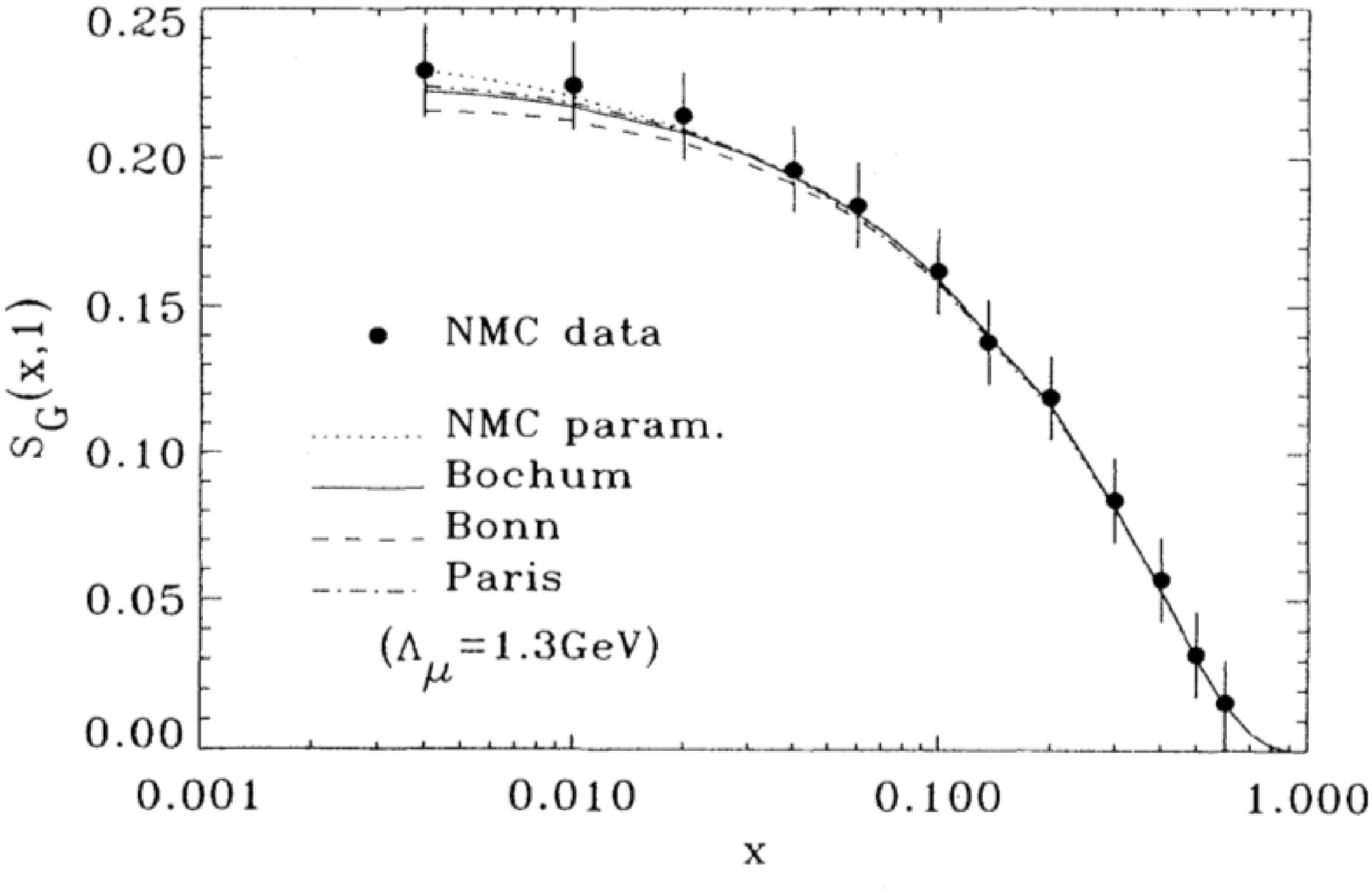}}
\vspace*{-1cm}
\caption{(Left) The WA25 test \cite{Allasia:1985hw} of the GLS sum rule defined by Eq.~(\ref{eq:GLS}).
(Right) The analogous NMC experimental test \cite{Allasia:1990nt} of the Gottfried sum rule compared with
various model predictions of \cite{Melnitchouk:1992eu}. The data are plotted in bins of a lower integration bound $x$,
such that the relation in Eq.~(\ref{eq:Gott}) is recovered in the limit $x \rightarrow 0$.
}
\label{fig:SR_test}
\end{figure}

When the predictions of QCD are married to the quark-parton model (QPM) formulated by Feynman with the impulse
approximation \cite{Feynman:1969ej}, the basic framework for hadronic phenomenology emerges.
Apropos, a crucial confirmation of the basic contours of the QPM came in
the form of various {\it sum rules} devised with current algebra under the assumption that the partons responsible for
Bjorken scaling were in fact the quark-level degrees of freedom of QCD. Actually, there are several such relations, all of
which emanate from number conservation arguments applied to the electroweak structure functions to be introduced in detail in
Chap.~\ref{chap:ch-Q2}.

In particular, the QPM treats the {\it nonperturbative} portion of the spin-independent nucleon wavefunction
as being dominated by its {\it valence} quark content, which is itself represented by the $C$-odd combinations of quark and antiquark
distributions $q_v(x) = q(x) - \bar{q}(x)$. These distributions are inherently probabilistic and therefore satisfy normalization
conditions in the proton (in accordance with the quarks' fractional charges):

\begin{equation}
\int_0^1\ dx\ u_v(x, Q^2)\ =\ 2\ , \,\,\,\,\,\,\, \int_0^1\ dx\ d_v(x, Q^2)\ =\ 1\ .
\label{eq:val_sum}
\end{equation}

Of course these quantities reside in the structure functions moments, and impose certain behaviors that may be readily
derived; of especial relevance to this thesis are the weak interaction Gross-Llewellyn-Smith (GLS) and Gottfried sum rules,
which we list up to first-order corrections in $\alpha_s$ as\footnote{Though we shall discuss them only in passing in
Chap.~\ref{chap:ch-Q2}, similar relations exist for spin-polarized observables -- e.g., $g_i(x, Q^2) \sim \Delta q \pm \Delta \bar{q}$.}
\begin{subequations}
\begin{align}
\label{eq:GLS}
(\mathrm{GLS})\ \,\,\,\,\, &\rightarrow\ \,\,\,\,\, {1 \over 2}\ \int_0^1\ dx\
\Big( F^{W^-}_3(x, Q^2) + F^{W^+}_3(x, Q^2) \Big)\ =\ 3\ , \\
(\mathrm{Gottfried})\ \,\,\,\,\, &\rightarrow\ \,\,\,\,\,\,\,\,\,\, \int_0^1\ {dx \over x}\
\Big( F^p_2(x, Q^2) - F^n_2(x, Q^2) \Big)\ =\ 1/3\ ,
\label{eq:Gott}
\end{align}
\end{subequations}
where the latter result of $1/3$ for the Gottfried sum rule assumes a flavor-symmetric light quark sea.
Thus, in connecting the partonic constituents of the nucleon to a conserved baryon number and other global properties
of hadrons, the `na\"ive' QPM is impressively accurate as the comparisons of Eqs.~(\ref{eq:GLS} - \ref{eq:Gott}) with data
from WA25 and NMC confirm in Fig.~\ref{fig:SR_test}.

In this and other respects, QCD and the parton model have been vindicated as remarkably successful descriptions of an enormous
range of hadronic physics; this is particularly true at scales larger than a characteristic mass $\Lambda_{QCD} \sim 1$ GeV
determined from the running of $\alpha_s (Q^2)$ as well as global analyses of hadronic data. Despite this triumph, perturbative
QCD (pQCD) as formulated in Eq.~(\ref{eq:QCD}) does not determine the infrared, long-distance dynamics that must be responsible
for hadron structure --- in this sense, many of the remaining difficulties in strong interaction physics might be described
as nonperturbative.

For instance, while the careful measurement and analysis of sum rules was a key verification of the parton model
and QCD, they still receive potentially important contributions from nonperturbative corrections and other effects beyond
those stipulated by pQCD as well. Accessing and describing such sources of nonperturbative physics is therefore a principal goal
in the ongoing quest to connect the UV behavior of QCD to physics of confined systems and understand how hadronic
structure arises from the basic features of QCD. Various effective field theories (such as will be described in part
in this thesis) have been an obvious device for carrying such investigations forward on the theoretical side.

Experimentally, DIS is uniquely disposed to probe the intermediate regions where the onset of perturbative scaling occurs, and
to better control nonperturbative physics. As such, it is the aim of this thesis to describe a number of recent theoretical advances
in better understanding specific sources of nonperturbative physics, with a special focus on the phenomenology of DIS.

After a brief introduction in Chap.~\ref{chap:ch-DIS} of some of the more important properties of the DIS handbag diagram and analytical 
tools required for many of our calculations, we turn to the electroweak phenomenology of DIS in Chap.~\ref{chap:ch-Q2}. Specifically,
various parity-violating experiments promise unprecedented sensitivity in the continuing effort to uncover possible physics beyond the
Standard Model (SM). Here we shall review newly found sources of phenomenology, and assess their potential impact.
Beyond this, parity violation may also prove a means of directly accessing parton-level breaking of charge symmetry -- a nonperturbative
effect of importance to analyses of sum rules of the type given in Eqs.~(\ref{eq:GLS} - \ref{eq:Gott}), for example.

Inspired by these issues, we present in Chap.~\ref{chap:ch-TMC} a comprehensive analysis of target mass corrections -- so called
``kinematical'' higher twist effects.  Hadronic masses are themselves inherently nonperturbative, and we present various calculations and
schemes for their evaluation in both inclusive and semi-inclusive DIS.

In Chap.~\ref{chap:ch-charm}, we present a novel model calculation of nonperturbative or {\it intrinsic} charm in the nucleon. We formulate
our model in terms of effective hadronic degrees of freedom in a study of deep relevance to the important transition from confinement to
asymptotically free quarks, which can be thought to occur at momenta comparable to heavy quark masses.

In the penultimate Chap.~\ref{chap:ch-TDIS} we present a low mass analogue of the two-step model of Chap.~\ref{chap:ch-charm}
which originates in chiral perturbation theory ($\chi$PT). The resulting framework permits an analysis of the nucleon's pion
cloud and in this light we consider possible extractions of the pion structure function $F^\pi_2$, as well as related dynamics.

Lastly, we survey possible extensions of this work and conclude in Chap.~\ref{chap:ch-conc}.


%% file: the-DIS.tex

Chief among the aims of modern QCD and its low-momentum effective formulations is a rigorous
description of the partonic substructure of hadronic matter. The putative constituent particles of
baryons and mesons --- the quarks and gluons --- interact both strongly and electromagnetically (in
the case of the quarks). As such, arguably the `cleanest' means of accessing information about the
multiplicity and momentum distributions (among other things) of these constituent particles is through
the interaction of an external probe in events like deeply inelastic scattering (DIS) [$l N \to l' X$],
whereby a leptonic scatterer $l$ (\EG~an electron or neutrino) interacts with the target nucleon $N$
via photon exchange (in electromagnetic processes), hence leaving an {\it inclusive} final state in which only
the energy and angle of the scattered lepton $l'$ are directly measured.

The properties of DIS are such that it is uniquely enabled to probe the inner landscape and dynamics of
hadrons: the inelasticity of DIS events implies some absorption of energy by the target, with
consequent excitations of its internal degrees of freedom (partons). On the other hand, the `deepness' of
the process results from the kinematics we now discuss, which facilitate a high level of spatial resolution
relative to the $\sim 1$ fm length scale of the nucleon.

\section{The DIS Process}
\label{sec:DIS}

\begin{figure}[h]
\includegraphics[height=6.3cm]{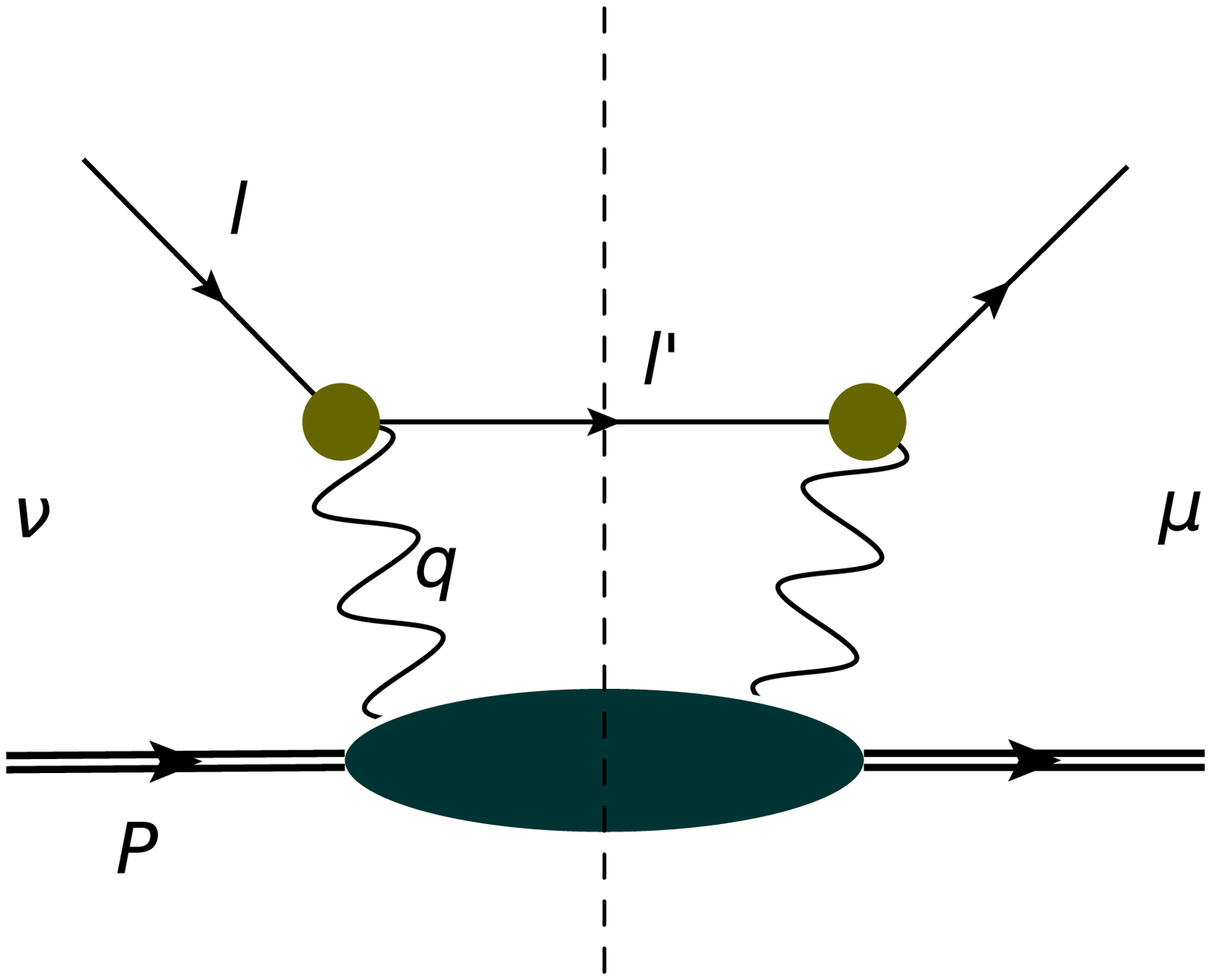} \ \ \ \ \ \
\raisebox{0.44cm}{\includegraphics[height=6cm]{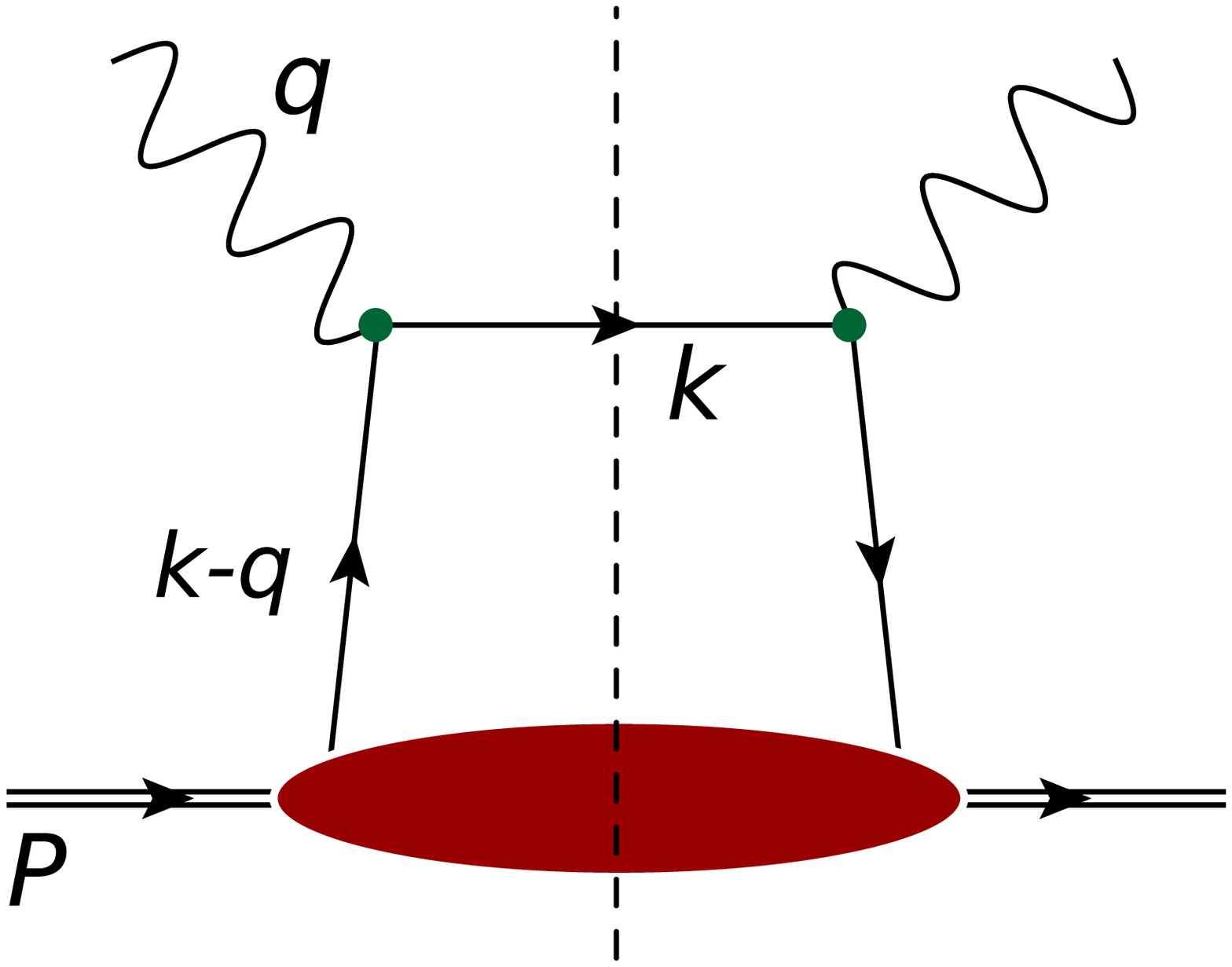}}
\caption{(Left) A representation of the DIS handbag diagram, illustrating
the separate origins of the $L^{\mu\nu}$ and $W_{\mu\nu}$ tensors.
(Right) The interaction of a nucleon with the virtual
photon, as represented by the blue ``blob'' at left. The diagram
is strictly leading twist, and follows from the photon scattering from an
individual constituent quark, thereby giving the correlation
functions $\Phi_{q,\bar{q}}$ of Eq.~(\ref{eq:quark_cors}).
}
\label{fig:DIS}
\end{figure}

The kinematics of the DIS process that represents a main focus of this thesis are deceptively
simple. The process sketched in Fig.~\ref{fig:DIS} entails the scattering of leptons
[$l^\mu = (E; \boldsymbol{l})$] from an on-shell nucleon target (predominantly for this thesis) of
4-momentum $P$ and mass $M$ via a virtual exchange boson of momentum $q$ and virtuality $Q^2 \equiv -q^2 \gg M^2$.
Hence, the conserved energy of the photon-nucleon system is necessarily
\begin{align}
W^2\ &=\ \Big( P + q \Big)^2\ =\ \Big( P^2 + 2 P\cdot q + q^2 \Big) \nonumber\\ 
&=\ M^2 + Q^2\ \left( {1 \over x} - 1 \right)\ ,
\label{eq:W-sq}
\end{align}
where we have identified the invariant Bjorken limit scaling parameter $x = Q^2 / (2P\cdot q)$.

The Bjorken limit ensures the validity of the DIS description, and, after a suitable boost
to a large target momentum frame, of the impulse approximation as well. The latter compels a
picture of the photon-nucleon interaction in which time dilation ensures that the incident
photon scatters incoherently from the nucleon's constituent quarks. Formally, the Bjorken
limit implies $Q^2,\ P \cdot q\ \rightarrow \infty$ for fixed $x$. Of course, actual
experimental measurements are typically performed in a target rest frame in which
\begin{equation}
x\ =\ {Q^2 \over 2 P \cdot q}\ \rightarrow\ {Q^2 \over 2 M \nu}\ ,
\end{equation}
and $Q^2 \sim 4 E E'\ \sin^2{(\theta_e /2)}$, where $\theta_e$ is the lab frame angle of the
scattered lepton and $\nu = E - E'$ the inelastic energy transfer to the target. As such,
it is also useful to define an inelasticity fraction $y = \Big( P \cdot q \Big) \Big/ \Big( P \cdot l \Big)$.

With these definitions, we deduce that the DIS differential cross section must inhabit a parameter space
spanned by $x, y, Q^2$, and is then given by the contraction of a {\it leptonic} tensor $L^{\mu\nu}$ and
a corresponding {\it hadronic} tensor $W_{\mu\nu}$:
\begin{equation}
{d^3 \sigma \over dx dy d\phi_S} = {M \over x s} \, \left({\alpha \over Q}\right)^2
L^{\mu\nu}(l,l',\lambda) \,W_{\mu\nu}(q,P,S)\ ;
\label{eq:DIS-CSI}
\end{equation}
after averaging over the nucleon spin $S$ and lepton helicity $\lambda$, this can be reduced to the simpler
form
\begin{equation}
{d^2 \sigma \over dQ^2 dx} = {\pi \over M} \, \left({\alpha \over E Q x}\right)^2
L^{\mu\nu}(l,l') \,W_{\mu\nu}(q,P)\ .
\label{eq:DIS-CSII}
\end{equation}
Thus, amplitudes for electron-nucleon scattering are typically separable into independent components
representing the harder leptonic and softer hadronic interactions. The former encodes the coupling
of the scattered lepton with the exchange photon. From the `handle' of the diagram shown in
Fig.~\ref{fig:DIS} the simplest lepton tensor, following spin averages in both the initial and
intermediate electron states, is found to be
\begin{align} 
L^{\mu\nu} &= {1 \over 2} \, \sum_{s,s'} \,
\mathrm{Tr} \big[\bar{\chi}_e^{s}(l) \gamma^\mu \chi_e^{s'}(l') \,
\bar{\chi}_e^{s'}(l') \gamma^\nu \chi_e^s(l) \big] \nonumber \\
&= {1 \over 2} \mathrm{Tr} \, \big[ \lsl \gamma^\mu \lsl' \gamma^\nu \big]
\nonumber \\
&= 2 \left( l^\mu l'^\nu + l^\nu l'^\mu - l \cdot l' g^{\mu\nu} \right)\ .
\label{eq:lep}
\end{align} 
We have used the cyclicity of the trace, as well as our convention for (approximately) massless
leptons ---
\begin{equation}
\sum_{s=\pm 1/2} \chi_e^s(l) \,\, \bar{\chi}_e^s(l) = \lsl \defeq \gamma \cdot l\ .
\label{eq:lep_avg}
\end{equation}

For processes in which the lepton helicity is explicitly retained (for instance, the parity-violating
physics described in Chap.~\ref{chap:ch-Q2}), one recovers
\begin{equation}
L^\gamma_{\mu\nu}
= 2 \left( l_\mu l'_\nu + l'_\mu l_\nu - l \cdot l'g_{\mu \nu}
	 + i \lambda\varepsilon_{\mu\nu\alpha\beta}\ l^\alpha l'^\beta
    \right)
\label{eq:PV_Ltens}
\end{equation}
for the $L^\gamma_{\mu\nu}$ tensor of Eq.~(\ref{eq:DIS-CSI}).

Whereas the analytic behavior of the lepton-boson vertex is generally under control, the dynamics
involved in the corresponding hadronic structure $W_{\mu\nu}$ are inherently nonperturbative in
the context of QCD and must therefore be constrained by experimental inputs. In spite of this
indeterminacy, a considerable amount of information can be deduced from consideration of the
analytic properties of $W_{\mu\nu}$.

\begin{figure}
\includegraphics[height=5.8cm]{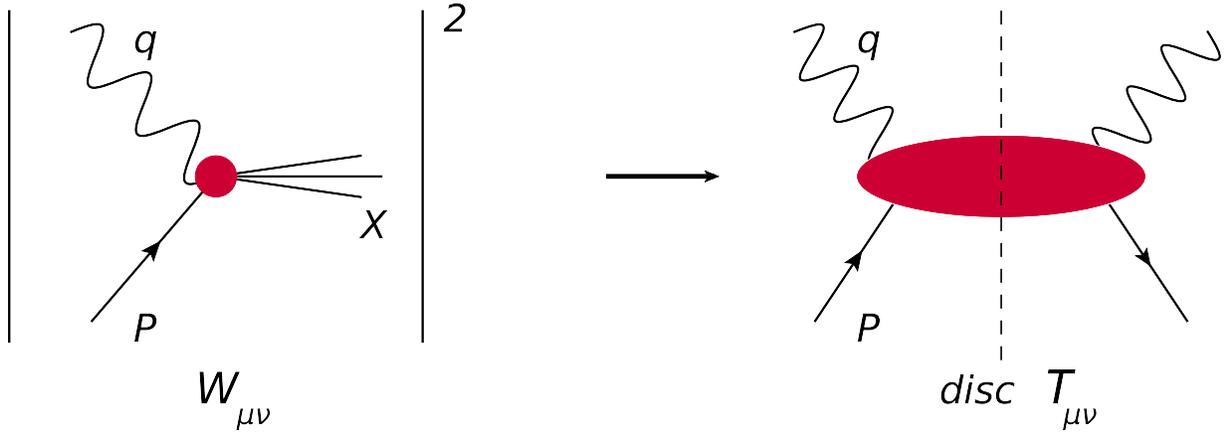}
\caption{An illustration of the Cutkosky procedure, and the relation
between the inclusive DIS process (left) and the cut Compton amplitude
$\mathrm{disc} \, T_{\mu\nu}$ (right).
}
\label{fig:Compton}
\end{figure}

On general grounds, the hadronic tensor of Eq.~(\ref{eq:DIS-CSII}) can be expanded explicitly in terms
of hadronic current operators $J_\mu$ as suggested by the left-hand diagram of Fig.~\ref{fig:Compton}:
\begin{equation}
W_{\mu\nu}^\alpha\ (P, q)
= {1 \over 2M}
 (2\pi)^3\ \sum_X \,
  \langle X | J_\mu^{\alpha}(0)    | N \rangle^*
  \langle X | J_\nu^{\alpha}(0) | N \rangle \,
  \delta^{(4)}(P + q - k_X)\ .
\label{eq:W-EW}
\end{equation}
(We require the indices $\alpha = \{\gamma, \gamma Z, Z\}$ to specialize to specific neutral
exchanges, as will become relevant for the treatment of electroweak processes in Chap.~\ref{chap:ch-Q2}.)

We now more thoroughly explore the connection between the hadronic tensor $W_{\mu\nu}$ that enters
the cross section for various QCD and electroweak processes and the more fundamental Compton
amplitude $T_{\mu\nu}$. Writing Eq.~(\ref{eq:W-EW}) more carefully, one can obtain
\begin{equation}
2M W_{\mu\nu}(P, q) = \frac{1}{2\pi} \sum_X \int \frac{d^3 k_X}{(2\pi)^3 2k_X^0}
\, (2\pi)^4 \, \delta^{(4)}(q+P-k_X) \, \langle P|J_\mu(0)|k_X \rangle
\langle k_X|J_\nu(0)|P \rangle\ ,
\label{eq:HadcI}
\end{equation}
such that the Fourier transformation of the 4-dimensional $\delta$-function given by
\begin{equation}
\delta^{(4)}(q+P-k_X) = \frac{1}{(2\pi)^4} \int d^4\xi \, \exp
\left( i[q+P-k_X] \cdot \xi \right)
\end{equation}
permits a translation of the current operators in Eq.~(\ref{eq:HadcI}).
This is allowed by the gauge invariance of the lagrangian from which the
Compton amplitude is derived, which in turn implies a current conservation
$\partial^\mu J_\mu \equiv 0$. Provided that the currents of Eq.~(\ref{eq:HadcI})
possess a {\it leading twist}\footnote{Formally, by `twist' we refer to a property
of operators determined by the difference of their $\tau =$ ``spin'' $-$ dimension.
In Sec.~\ref{sec:OPE} we shall see that the leading contribution in operator product
expansions must enter for $\tau = 2$, or {\it twist-$2$}.} bilinear form of
$J_\mu(0) = \bar{\psi}(0) \, \Gamma_\mu \, \psi(0)$ for some Dirac structure
$\Gamma_\mu$ (\EG~$\Gamma_\mu = \gamma_\mu$), we may make use of
\begin{align}
e^{-i k_X \cdot \xi} \, \psi(0) \, | k_X \rangle &= \psi(\xi) \, | k_X \rangle\ ,
\ \ \ \ \ \ \ \ 
\langle P | \, \bar{\psi}(0) \, e^{+i P \cdot \xi} = \langle P | \, \bar{\psi}(\xi)\ ,
\end{align}
to rewrite Eq.~(\ref{eq:HadcI}) as
\begin{equation}
2M W_{\mu\nu}(P, q) = \frac{1}{2\pi} \int d^4 \xi e^{iq \cdot \xi}
\, \langle P|J_\mu(\xi) \, J_\nu(0)|P \rangle\ .
\label{eq:HadcII}
\end{equation}
We have exploited the completeness of intermediate states in the cut ``blob'' of the Compton amplitude
--- i.e.,
\begin{equation}
{1 \over (2\pi)^3}\ \sum_X\ \frac{d^3 k_X}{2k_X^0} \, |k_X \rangle \langle k_X|\ \equiv\ \mathbf{1}\ .
\end{equation}

As a general consequence of the QFT optical theorem,
this connects immediately to the matrix elements of the forward virtual Compton amplitude via
\begin{equation}
W_{\mu \nu} = \frac{1}{\pi}\ \mathrm{Im}\ T_{\mu\nu}\ \equiv\ \frac{1}{\pi}\ \mathrm{disc}\ T_{\mu\nu}\ ,
\label{eq:comp-had}
\end{equation}
{\it viz.}
\begin{align}
T_{\mu \nu} &= i\, \int d^4 \xi \, \exp(iq \cdot \xi)
\langle N(p)|T\,\, J_\mu(\xi)\ J_\nu(0)|N(p) \rangle\ , \nonumber\\
&= i\, \int d^4 \xi \, \exp(iq \cdot \xi)
\langle N(p)|\ [J_\mu(\xi), J_\nu(0)]|\ N(p) \rangle\ .
\label{eq:Comp}
\end{align}
This can be seen through the canonical procedure: application of the Cutkosky rules
\cite{Cutkosky:1960sp} to the cut Compton amplitude, as illustrated in Fig.~\ref{fig:Compton}.
Schematically, the unitarity of the $S$-matrix $S^\dagger S \equiv 1$ also constrains the
associated $T$-matrix due to the definition $S \defeq 1 + iT$. The field-theoretic analogue of
this relation is precisely what is shown diagrammatically in Fig.~\ref{fig:Compton}. The
gist of the fundamental procedure of \cite{Cutkosky:1960sp} is that the discontinuity
induced by the real axis branch cut in the exchange momenta $k_X$ of the Compton `blob' in
Fig.~\ref{fig:Compton} can be accessed by replacing the internal integrations as
\begin{align}
\int\ {d^4k_X \over (2\pi)^4}\ &\rightarrow\ \int\ {d^4k_X \over (2\pi)^4}\ (2\pi)^4\ \delta^{(4)}
\left( P + q -k_X \right)\ {1 \over k^2_X - m^2 + i\epsilon}\ ;
\label{eq:Cut1}
\end{align}
without loss of generality, we have assumed the exchanged momentum of the Compton `blob'
to be carried by scalar constituents --- hence the explicit propagator of Eq.~(\ref{eq:Cut1}).
By the residue theorem applied in the complex $k_X$ plane, we may thus isolate the discontinuity across
the real axis $k_X > 2m$ branch cut by placing the $k_X$ propagator on its mass-shell:
\begin{align}
{1 \over k^2_X - m^2 + i\epsilon}\ \rightarrow\ -2\pi i\ \delta(k^2_X - m^2)\ ,
\end{align}
with the result of this scheme being the desired integration measure; that is,
\begin{align}
\int\ dk^0_X\ {d^3k_X \over (2\pi)^3}\ (2\pi)^4\ \delta^{(4)} &\left( P + q -k_X \right)\ \delta( k^2_X - m^2)\
\nonumber\\ 
&=\ {1 \over (2\pi)^3}\ \int\ {d^3k_X \over 2 k^0_X}\ (2\pi)^4\ \delta^{(4)} (P + q - k_X)\ .
\end{align}
In fact, when wedded to the hadronic transition matrix elements
$\langle P|J_\mu(0)|k_X \rangle \langle k_X|J_\nu(0)|P \rangle$ and incoherently summed over $X$ this result is consistent
with Eq.~(\ref{eq:HadcI}), thus formalizing the connection between $W_{\mu\nu}$ and the operator structure of $T_{\mu\nu}$.

We shall proceed further by deconstructing $T_{\mu\nu}$ in terms of local current operators in
Sec.~\ref{sec:OPE}. At the same time, in anticipation of subsequent developments within
this thesis we note that a straightforward extension of the preceding formalism connects
the hadronic tensor $W_{\mu\nu}$ to the internal quark-quark correlation functions that
will be the subject of later modeling attempts, particularly those of Chap.~\ref{chap:ch-charm}.

We can follow the same Cutkosky paradigm as just outlined for the single nucleon example, but
modify the handbag diagram to incorporate simple scattering from constituent quarks \cite{Bacchetta:2002xd}
of $4$-momentum $k$ as indicated in the right diagram of Fig.~\ref{fig:DIS}; the quark-level $W_{\mu\nu}$ tensor then becomes
\begin{align}
2MW_{\mu\nu}\ &=\ e^2_q\ \sum_X\ \int {d^3k_X \over (2\pi)^3\ 2k^0_X}\ \int d^4k\ \delta(k^2 - m^2)\
\theta(k^0 - m)\ \delta^{(4)}(P + q - k_X - k) \nonumber\\
&\times \bigg\{ \lan P|\ \bar{\psi}(0)\ |k_X \ran \lan k_X|\ \psi(0)\ |P \ran\ \gamma_\mu\ (\ksl +m) \gamma_\nu\ \nonumber\\
&\,\,\,\,\, -\ \lan P|\ \psi(0)\ |k_X \ran \lan k_X|\ \bar{\psi}(0)\ |P \ran\ \gamma_\nu\ (-\ksl +m) \gamma_\mu \bigg\}\ .
\end{align}
If we repeat some of the same manipulations used to rewrite Eq.~(\ref{eq:HadcI}), this can be rendered as
\begin{align}
2MW_{\mu\nu}\ &=\ e^2_q\ \int d^4k\ \delta(k^2 - m^2)\ \theta(k^0 - m)\ \int {d^4 \xi \over (2\pi)^4}\ e^{-i(q-k)\cdot \xi} \nonumber\\
&\times \bigg\{ \lan P|\ \bar{\psi}(\xi)\ \psi(0)\ |P \ran \gamma_\mu\ (\ksl +m) \gamma_\nu\ \nonumber\\
&\,\,\,\,\, -\ \lan P|\ \psi(\xi)\ \bar{\psi}(0)\ |P \ran \gamma_\nu\ (-\ksl +m) \gamma_\mu \bigg\}\ ;
\end{align}
moreover, a simple inspection of the last equation suggests a compact form for the quark-level correlation functions. Evidently,
they are
\begin{subequations}
\begin{align}
\Phi_q(P,k,q)\ &=\ \int {d^4 \xi \over (2\pi)^4}\ e^{-i(q-k)\cdot \xi}\,\,
\lan P|\ \bar{\psi}(\xi)\ \psi(0)\ |P \ran\ , \\
\Phi_{\bar{q}}(P,k,q)\ &=\ \int {d^4 \xi \over (2\pi)^4}\ e^{-i(q-k)\cdot \xi}\,\,
\lan P|\ \psi(\xi)\ \bar{\psi}(0)\ |P \ran\ .
\end{align}
\label{eq:quark_cors}
\end{subequations}
The existence of the correlators $\Phi_{q, \bar{q}}$ implies the validity the quark-parton model expressions
we shall develop later in this thesis for the {\it structure functions} that emerge from $W_{\mu\nu}$. In
essence, it is the objects $\Phi^{q, \bar{q}}$ that encode the nonperturbative long-distance
correlations of quarks in the nucleon. It is these that remain beyond first principles computations
in QCD, and hence are among the main goals of DIS and other experiments.

Next, in Sec.~\ref{sec:Compt} we turn to the Compton amplitude in the real $Q^2 = 0$ limit. This will
be explored in the context of dispersive Kramers-Kr\"onig relations that enable one to extract
surprising electromagnetic properties of the nucleon from the asymptotic (i.e., $\nu \to \infty$) behavior
of the $\gamma N \rightarrow \gamma N$ reaction.
Following this, in Sec.~\ref{sec:OPE} we discuss basic features of DIS amplitudes in the operator product expansion
(OPE) framework, which permits the twist decomposition of the hadronic observables that are
a principal focus of this thesis. This treatment, with its scale-dependent factorization of short- from long-distance
physics, will be in sharp contrast to the $Q^2 = 0$ techniques of Sec.~\ref{sec:Compt}. 

%% file: the-nucl.tex

\section{Compton Scattering in the $Q^2 = 0$ limit}
\label{sec:Compt}

\begin{figure}
\includegraphics[height=4.4cm]{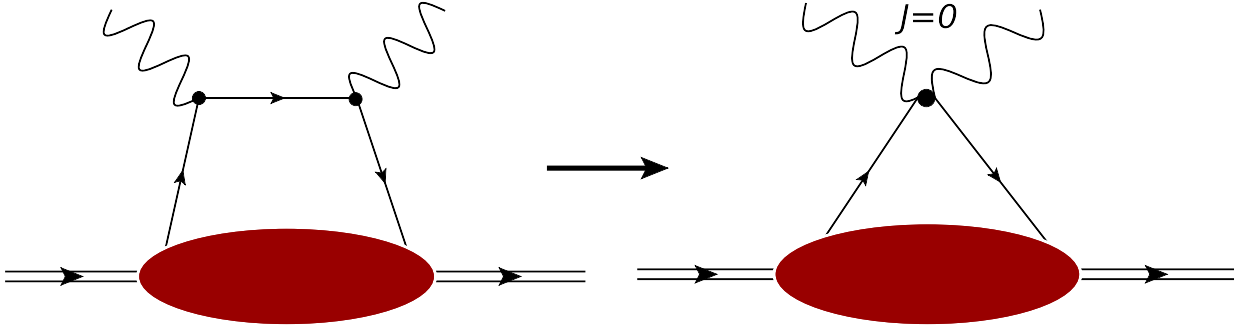}
\caption{The fixed-pole contribution to the Compton amplitude may
  arise due to an effective, local two-photon coupling to elementary
  constituents within the proton.}
\label{fig:diag}
\end{figure}

In the last section, we examined the relation between the DIS amplitude and analytic
properties of the virtual Compton diagram. We continue that analysis in the present
section by considering the Compton amplitude in the different setting of exclusive photoproduction reactions
of the form $\gamma N \rightarrow \gamma N$, outlining the results of a recent publication \cite{Hobbs:2011}.
In such situations both the initial and final state photons are purely real, and the associated forward
(\IE~$q=q'$) amplitude consequently selects uniquely constrained hadronic matrix elements. In particular,
such physics is especially amenable at higher energies to Regge theory as mentioned in Chap.~\ref{chap:ch-intro},
hence offering a means of identifying possible dualities connecting short-distance physics to
long-range dynamics. Moreover, for asymptotic photon energies ($\nu \to \infty$) the photon
can couple locally to the constituent quark currents of the nucleon as depicted in Fig.~\ref{fig:diag},
resulting in a universal (\IE~energy-independent) contribution to the scattering amplitude that
has historically been thought to originate with a $J=0$ Regge pole \cite{Damashek:1969xj,Brodsky:2008qu}.
This observation is driven by the logic that the pointlike quark-photon vertex of Fig.~\ref{fig:diag} is dominated 
by spin-$0$ behavior, and would therefore be incorporated using the Regge language of Eq.~(\ref{eq:intro-Reg}) as an
$\alpha_{J=0}(t=0) \equiv 0$ contribution to the forward Compton amplitude --- an energy-independent constant.
Making a precise measurement of the $J=0$ pole has thus been a strong object of interest for some decades, as
doing so amounts to a basic test of the properties of QCD; this is because the $J=0$ pole contains information about the basic,
energy-independent structure of the coupling of photons to the fundamental sources of electromagnetic charge
within the theory --- the constituent quarks \cite{Brodsky:2008qu}.

There have in fact been a number of studies which have attempted to extract the
$J=0$ pole; these have arrived at various numerical results more-or-less consistent with
the Thomson term, $\mathrm{Re}\, T_1(0) = -3 \, \mu$b$\cdot$GeV. These include the
pioneering work of Damashek and Gilman \cite{Damashek:1969xj}, as well as
results found in \cite{Dominguez:1970wu} and \cite{Shibasaki:1971}
($\mathrm{Re}\, T_1^{J=0}=(-3\pm2) \mu{\rm b \cdot GeV}$) and a slightly later
study \cite{Tait:1972}
($\mathrm{Re}\, T_1^{J=0}=(-3^{+4}_{-5}) \mu{\rm b \cdot GeV}$).

The advent of higher energy data at $\nu \gtrsim 10$ TeV, however, has made a re-analysis timely.
In our recent calculation \cite{Hobbs:2011}, a new attempt was made to carefully extract the fixed $J=0$ contribution in the spirit
of \cite{Damashek:1969xj}, as well as to construct a series of consistent finite-energy sum rules (FESRs) on the basis of energy
{\it scale separation}, by which we refer to the qualitatively distinct behaviors of the total photoproduction cross section
that dominate within well-separated energy regimes. After a brief discussion of the basic theory, we therefore present a novel
determination of the $J=0$ contribution, which indeed suggests a difference between the nucleon Thomson term and the quark-level
fixed pole.
%
\subsection{The Real Compton Amplitude}
Real photon scattering is in fact simply a limit of the virtual photon case discussed in Sec.~\ref{sec:DIS},
corresponding to $Q^2 \rightarrow 0$. More formally, in this special circumstance
we make the following kinematical definitions: for real photons we take the $4$-momentum to be
$q^\mu = (\nu, {\bf q})$ such that $q^2 = \nu^2 - {\bf q}^2 = -Q^2 \equiv 0$; the
corresponding polarization vectors are then $\epsilon^\mu = (0,\boldsymbol{\epsilon})$. As before,
the photon energy is $\nu = P \cdot q / M$. For our purposes, the object of dispersion
relations at finite energy are the Compton amplitudes --- especially for the nucleon
spin-averaged process. These amplitudes reside within the Compton T-matrix, which may be
taken from the Compton tensor of Eq.~(\ref{eq:Comp}) after the appropriate contractions:
\begin{equation}
\mathscr{T} = e^2 \, \epsilon'^{* \mu} \epsilon^\nu \, T_{\mu\nu}\ .
\label{eq:T-mat}
\end{equation}
In the forward limit, we have $q = q'$ and may expand the RHS of the last expression as
\begin{align}
e^2 \, T_{\mu\nu} &= \bar{u}_N(P') \,
\big( T_1(\nu) \, g_{\mu\nu} + iT_2(\nu) \, \sigma_{\mu\nu} \big)
\, u_N(P)\ , \nonumber \\
\mathscr{T} &= 8\pi M \, \bar{u}_N(P') \,
\big( \boldsymbol{\epsilon}'^* \cdot \boldsymbol{\epsilon} \, T_1(\nu) \, +
\, i \sigma \cdot [\boldsymbol{\epsilon}'^* \times \boldsymbol{\epsilon}] \, iT_2(\nu) \big)
\, u_N(P)\ .
\label{eq:Camp}
\end{align}
Within this last expression, the simple spin-averaged Compton amplitude $T_1(\nu)$ contains
much information bearing upon nucleon substructure, some of which has been newly extracted in our
analysis \cite{Hobbs:2011}. The fact that $T_1(\nu)$ is an analytic function of the complex
parameter $\nu$ implies this information can be accessed via the well-known Kramers-Kr\"onig
relations.

Generically, for a function $f(z) = f_1(z) +if_2(z)$ analytic in the upper-half complex plane,
one can show \cite{toll56} by the residue theorem that
\begin{equation}
f_1(z) = {1 \over \pi} \mathcal{P}\!\!\!\int \limits_{-\infty}^\infty dz'\, {f_2(z') \over z' - z}\ .
\end{equation}
If an odd behavior $f_2(-z) = -f_2(z)$ is ascribed to the imaginary part, several straightforward manipulations
yield
\begin{align}
f_1(z) &= {1 \over \pi} \mathcal{P}\!\!\! \int \limits_{-\infty}^\infty dz'\, {z' f_2(z') \over z'^2 - z^2}
+ {z \over \pi} \mathcal{P}\!\!\! \int \limits_{-\infty}^\infty dz'\, {f_2(z') \over z'^2 - z^2} \nonumber \\
       &= {1 \over \pi} \mathcal{P}\!\! \int \limits_{0}^{\infty} dz'^2\, {f_2(z') \over z'^2 - z^2}\ .
\end{align}
The finite behavior of the last integral in the $z \to \infty$ may be ensured by implementing a {\it subtraction}
of the form $[f_1(z) - f_1(0)]/z^2$, which provides the once-subtracted dispersion relation we require ---
\begin{equation}
f_1(z) = f_1(0)\, +\, {z^2 \over \pi} \mathcal{P}\!\! \int \limits_{0}^{\infty} {dz'^2 \over z'^2}\, {f_2(z') \over z'^2 - z^2}\ ,
\end{equation}
where the oddness of $f_2$ has ensured $f_2(0) = 0$.

%
Depending essentially only upon its analyticity and unitarity, the spin-averaged
forward Compton scattering amplitude $T_1(\nu)$ of Eq.~(\ref{eq:Camp}) can be cast
into such a dispersive relation. Identifying the subtraction constant as the
standard low-energy Thomson limit, the `master formula' of this analysis follows:
\begin{align} 
\label{eq:DR}
{\rm Re} \, T_1(\nu) &= -\frac{Z^2}{A^2} \frac{\alpha}{M}
+\frac{\nu^2}{\pi} \int_{0}^\infty\frac{d\nu'^2
}{\nu'^2(\nu'^2-\nu^2)} {\rm Im} \, T_1(\nu') \nonumber \\
&= - \frac{Z^2}{A^2} \frac{\alpha}{M}
 +\frac{\nu^2}{2\pi^2} \int_0^\infty\frac{d\nu'}{\nu'^2-\nu^2}\sigma(\nu')\ . 
\end{align} 
where for convenience we have suppressed the explicit principal value $\mathcal{P}$ notation
in Eq.~(\ref{eq:DR}) and the following. Also, for generality, we normalize to the mass
number $A$. The second line of Eq.~(\ref{eq:DR}) emerges after a simple application of the
optical theorem for real scattering
\bea
{\rm Im} \, T(\nu)&=&\frac{\nu}{4\pi}\sigma(\nu)\ , \label{opt}
\eea
which is in clear analogy with Eq.~(\ref{eq:comp-had}). That the subtraction constant
in the dispersion relations of Eq.~(\ref{eq:DR}) should go as
$T_1(0) = -(Z^2 \alpha) \big/ (A^2 M)$ is required by the $\nu \to 0$ behavior of $T_1(\nu)$:
namely, in the infrared limit the electromagnetic probe can only be sensitive to `global'
properties of the target, \EG~its charge and mass. This is precisely the Thomson term.

With this we pause momentarily to take stock of the significance of the last few deductions. For
functions which are analytic in the upper-half of the complex plane, dispersion relations can
be written down which connect their real and imaginary parts. That this can be done in the context
of scattering theory for complex amplitudes implies their causal structure, which by the optical
theorem permits observable cross sections to be related to the real part of the underlying amplitude
as we have done in Eq.~(\ref{eq:DR}). This property allows us to analyze the contributions to
$\mathrm{Re}\, T_1 (\nu)$ using photoproduction measurements, and specifically, constrain the $J=0$
pole $\mathrm{Re}\, T_1^{J=0}(\nu)$.
%
\subsection{Finite-Energy Sum Rules}
\begin{figure}[ht]
\vspace*{1.3cm}
\includegraphics[height=8cm]{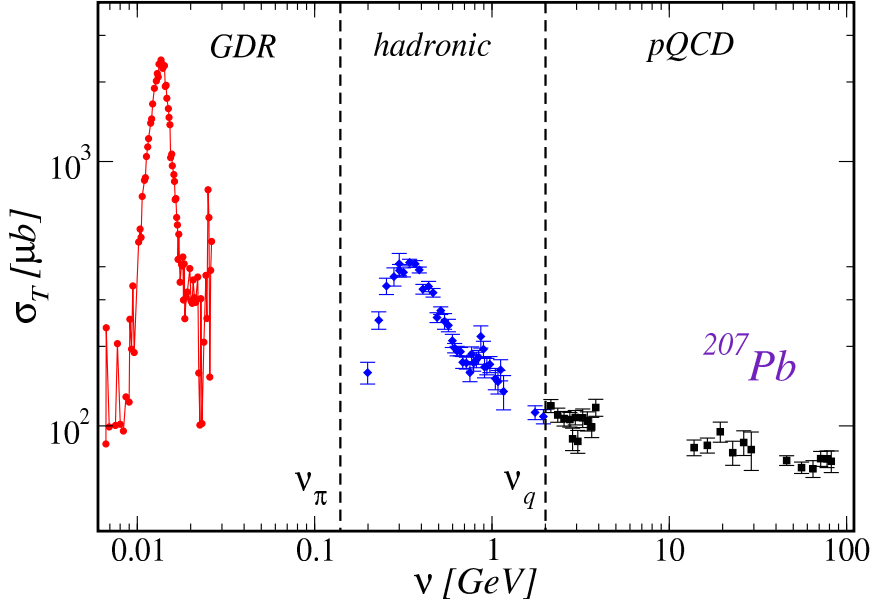}
\caption{
Scale separation is evident in the photo-absorption cross section for $^{207}$Pb. 
Data are taken from \cite{Harvey:1964,Hesse:1970cy,Caldwell:1973,Caldwell:1979,Bianchi} 
and show the distinct regimes in $\nu$ for which scattering is predominantly from collective
nuclei (giving rise to the GDR phenomenon), constituent nucleons, and pQCD quarks/gluons.
} 
\label{fig:regions}
\end{figure}

By construction, the dispersive integral of Eq.~(\ref{eq:DR}) strictly includes photo-absorption cross
sections all the way up to infinite energy; however, the approximate scale separation evident in Fig.~\ref{fig:regions}
for the $^{207}$Pb cross section $\sigma_T(\nu)$ between the nuclear ($\nu \lesssim m_\pi$)
and hadronic ($\nu \lesssim 2$ GeV) domains allows us to approximate the integral with a more restricted range of
photo-absorption data. That is to say that the $\nu \to \infty$ contribution we aim to extract for perturbatively
free partons has a corresponding analogue at lower energies; in this case, at energies sufficiently large relative to
the giant dipole resonances (GDR) of Fig.~\ref{fig:regions}, the nuclear Compton amplitude is dominated by local,
pointlike interactions with individual constituent nucleons.
%
%
As we now demonstrate, this observation leads to the famous nuclear photo-absorption FESR due to Thomas, Reiche, and
Kuhn (TRK) \cite{Thomas:1925a,Levinger:1960}. Again, as shown in Fig.~\ref{fig:regions}, nuclear deformation resonances
(\IE~GDRs) saturate the photo-absorption cross section for $\nu \lesssim E_{\max} \approx 30$ MeV, such that we may compute
the dispersive relation of Eq.~(\ref{eq:DR}) up to $\nu_{max}  \lesssim  m_\pi$, which approximately demarcates the purely
nuclear physics from the hadronic scale beyond which single-nucleon resonances primarily contribute to the cross section,  
\begin{align}
{\rm Re} \, T(\nu_{max}) &\approx\ -\frac{Z^2}{A^2} \frac{\alpha
  }{M}-\frac{1}{2\pi^2}\int_{0}^{E_{max}}d\nu'\sigma(\nu') \nonumber\\
&\approx\ -\frac{Z}{A}\frac{\alpha}{M}\,\,\,\,\,\,\, \iff \,\,\,\,\,\,\, \nu_{max} \lesssim m_\pi\ .
\label{eq:s1}
\end{align}
Taking note that $\alpha/M \approx 3.03\mbox{ mb~MeV}$, as well as the assumption in the previous relation that the low-energy scattering
is controlled by coherent interactions with individual nucleons, the TRK sum rule appears following some trivial reconfigurations ---
\begin{eqnarray}
\label{eq:TRK}
&& \int_{0}^{E_{max}} d\nu \sigma (\nu) = 2 \pi^2 \frac{NZ}{A^2} 
\frac{\alpha}{M} 
 \approx 60 \frac{NZ}{A^2} \mathrm{ mb~MeV}\ ,
\end{eqnarray}
which has been found to hold at the $\sim30$\% level for an array of nuclei.
Actually, the integration on the LHS of Eq.~(\ref{eq:s1}) is only approximately
consistent with the dispersion relation of Eq.~(\ref{eq:DR}), and is obtained using
an expansion of the form
\begin{equation}
\nu^2\int_0^{E_{max}}\!\!\!\!\!\!d\nu'\frac{\sigma(\nu')}{\nu^2-\nu'^2}
= \left[1 + \frac{ \langle \nu^2 \rangle}{\nu^2} +\dots\right] 
\int_{0}^{E_{max}}\!\!\!\!\!\!d\nu'\sigma(\nu')\ ,  
\label{eq:cor_nucl}
 \end{equation} 
which is a sound approximation at first order so long as the spectrum-averaged mean squared energy satisfies 
\begin{equation} 
\langle \nu^2 \rangle\ =\ \left( \int_{0}^{E_{max} }d\nu'\ \nu'^2\sigma(\nu') \right)
\Bigg/
\left( \int_{0}^{E_{max} }d\nu'\ \sigma(\nu') \right)\ \ll\ 1 \ . 
    \end{equation} 
For typical nuclei the first-order correction term of Eq.~(\ref{eq:cor_nucl}) represents a $\sim$10\%
effect, and neglecting it is therefore acceptable for the illustrative purposes here.

%
The arguments leading to the TRK sum rule are by no means unique, and readily
generalize to the higher energies of the nucleon excitation (\IE~resonance) region,
which roughly corresponds to $m_\pi \lesssim \nu \lesssim 2$ GeV.
Beyond this kinematical regime, resonance excitations vanish as seen in Fig.~\ref{fig:regions},
and the photo-production cross section is describable with a slow-varying background. Physically,
this qualitative behavior is attributable to incoherent scattering off the constituent quarks of
the nucleon. Building upon the analogy with Eq.~(\ref{eq:s1}), we therefore write down a FESR
driven by scattering from constituent quarks:
\bea
{\rm Re} \, T^{\CQM}_1 \equiv \mbox{Re}\, T_1(\nu_{max})  &\approx&- \frac{1}{A} 
\sum_{q\in A} \frac{\alpha}{m_q}   e_q^2 \nonumber \\ & = &  
- \frac{3Z + 2N}{A} \frac{\alpha}{M}\ . 
\label{eq:TRK_rhs1}
\eea

Though tempting, it is not enough to proceed exactly as before with the TRK relation by
simply associating ${\rm Re}\, T^{\CQM}_1$ with the nuclear Thomson term of Eq.~(\ref{eq:s1})
added to the photo-absorption cross section integrated up to a large energy $E_{max} \gtrsim 2$ GeV
relative to the nucleon resonance region.
The simple reason for this is the high energy behavior of $\sigma_T(\nu)$, which, rather than vanishing,
instead monotonically approaches the analytic Froissart\footnote{Relying only on the analyticity and unitary
of forward scattering amplitudes like $T_1(\nu)$, the Froissart bound restricts the energy dependence of
the associated total cross section to $\sigma \le A \ln^2 s$, where $A$ is a constant, and for our purposes
the Mandelstam variable evaluates to $s\ =\ M \cdot (2\nu + M)$.} bound \cite{Froissart:1961} as seen in Fig.~\ref{fig:HE}. 
Such an increase in the cross section can be understood as a feature of scattering from the asymptotically free
partons of pQCD; as it happens, the $\nu$ dependence of these dynamics is ideally suited to Regge theory as already
mentioned in Eq.~(\ref{eq:intro-Reg}). We explicitly model the contributions of reggeon and pomeron exchange to the cross section
as
\begin{equation} 
\sigma^{R+\mathbb{P}}(\nu)  = \sigma^R_T  + \sigma^{\mathbb{P}}_T = \sum_{i=R,\mathbb{P}} c_i \left( \frac{\nu}
{\mbox{1 GeV}} \right)^{\alpha_i(0)-1} \ ,  
\label{regge} 
\end{equation} 
where for dimensional reasons we include explicit factors of 1 GeV, and the pomeron and reggeon contributions are fixed by
standard numerical choices for the trajectory intercepts 
$\alpha_R(0) = 0.5$ and $\alpha_{\mathbb{P}}(0) = 1.097$, respectively; we also again normalize to nucleon number for generality. These
cross sections may be related to complex amplitudes of the form 
\begin{align}
T_1^{R+\mathbb{P}}(\nu)  = T_1^R + T_1^{\mathbb{P}}   &=  - \sum_{i=R,\mathbb{P}} \frac{c_i}{4\pi}\frac{1 + e^{-i \pi \alpha_i (0)}}
  {\sin \pi \alpha_i (0)} \nu^{\alpha_i (0)} \nonumber \\
  &= \frac{\nu^2}{2\pi^2} \int_0^\infty \frac{d\nu'}{\nu'^2 - \nu^2} \sigma^{R+\mathbb{P}}(\nu')\ . 
\label{eq:Regge}
\end{align}
We may therefore rewrite Eq.~(\ref{eq:DR}) by simply adding and subtracting the
high energy contributions of Eqs.~(\ref{regge}) and (\ref{eq:Regge}), hence leading to
\begin{align}  
\label{DR-sub} 
  \mbox{Re}\, T_1(\nu)  &=  - \frac{Z^2}{A^2} \frac{\alpha}{M}
     + \frac{\nu^2}{2\pi^2}  \int_0^\infty d\nu' 
  \frac{ \sigma(\nu')  - \sigma^{R+\mathbb{P}}(\nu') }{\nu'^2 - \nu^2}  
  +  \mbox{Re} \, T_1^{R+\mathbb{P}} (\nu)  \\
&= - \frac{Z^2}{A^2} \frac{\alpha}{M}
   -  \frac{1}{2\pi^2}  \int_0^{E}  d\nu'   \sigma(\nu')  
+   \sum_{i=R,\mathbb{P}} \frac{c_i  \mbox{GeV} }{2\pi^2\alpha_i(0) } 
\left( \frac{E}{\mbox{1 GeV}} \right)^{\alpha_i(0)} 
  + \, \mbox{Re} \, T_1^{R+\mathbb{P}} (\nu)\ , \nonumber
\end{align} 
where we obtain the second line by taking the asymptotic limit $\nu \to \infty$ to
evaluate the subtraction in the integrand of the first line.
In Eq.~(\ref{DR-sub}) $E \sim 2$ GeV is the energy beyond which the 
difference between the data $\sigma(\nu)$ and the high-energy asymptotic form 
$\sigma^{R+\mathbb{P}}(\nu)$ is negligible.

We proceed by identifying the constituent quark analog of the TRK constant computed in Eq.~(\ref{eq:TRK_rhs1})
as the contribution from scattering off bound quarks; doing so, we may then recast the LHS of Eq.~(\ref{DR-sub}) as
\begin{equation}
\mbox{Re}\, T_1(\nu) = \mbox{Re} \, T_1^{\CQM} \,
   +\,   \frac{c_{\mathbb{P}} \mbox{GeV}}{2\pi^2\alpha_P(0) } 
 \left( \frac{E}{\mbox{1 GeV}} \right)^{\alpha_{\mathbb{P}}(0)} \,
   +\, \mbox{Re} \, T_1^{R+\mathbb{P}} (\nu)\ ,
\label{eq:CQDR}
\end{equation}
thereby leading to a new phenomenological FESR at quark-level after several rearrangements:
\begin{align}
\mbox{Re} \, T_1^{\CQM}  &= - \frac{Z^2}{A^2} \frac{\alpha}{M} 
   -   \frac{1}{2\pi^2}  \int_0^{E}  d\nu' \sigma(\nu')
   +   \frac{c_R \mbox{GeV}}{2\pi^2\alpha_R(0) } 
 \left( \frac{E}{\mbox{1 GeV}} \right)^{\alpha_R(0)} \,\,\, \implies \nonumber \\
  - \left( 2 +\frac{ZN}{A^2} \right)\frac{\alpha}{M}    & =
   -  \frac{1}{2\pi^2}  \int_0^{E}  d\nu'   \sigma(\nu')
    +   \frac{c_R  \mbox{GeV}}{2\pi^2\alpha_R(0) } \left( \frac{E}{\mbox{1 GeV}} \right)^{\alpha_R(0)}\ .
\label{eq:CQM-FESR}
\end{align} 
We note that the pomeron term must be included explicitly in Eq.~(\ref{eq:CQDR}) to account
for the fact that $\mathbb{P}$-exchanges possess quantum numbers consistent with the vacuum (putatively arising
from gluonic pQCD mechanisms), and therefore cannot directly contribute to $\mbox{Re} \, T_1^{\CQM}$.

Aside from this novel FESR, we may also exploit Eq.~(\ref{DR-sub}) to determine the scale independent $J = 0$ pole
described at the start of Sec.~\ref{sec:Compt}; namely, this is the difference between the complete forward Compton amplitude
up through the excitation region and the smooth Regge theory background:
\begin{align}  
  \mbox{Re} \, T_1^{J=0} &\defeq \lim_{\nu \to \infty}  [\mbox{Re} \, T_1(\nu) - \mbox{Re} \, T_1^{R+\mathbb{P}}(\nu)] \nonumber \\ 
  & = - \frac{Z^2}{A^2} \frac{\alpha}{M}   
   -  \frac{1}{2\pi^2}  \int_0^{E}  d\nu'   \sigma(\nu')  
   + \sum_{i=R,\mathbb{P}} \frac{c_i  \mbox{GeV} }  {2\pi^2\alpha_i(0) } 
  \left( \frac{E}{\mbox{1 GeV}} \right)^{\alpha_i(0)}\ . 
\label{j0} 
\end{align} 
%
%
\subsection{The $J=0$ Pole and CQM FESR} 
To extract a numerical determination of the $J=0$ fixed pole from the FESR given in Eq.~(\ref{j0}), it is first
necessary to parametrize the $\nu$ dependence of the hadronic cross-section --- we do so by fitting a sum of Breit-Wigner
resonances over a smooth background \cite{Hobbs:2011}.
Moreover, the Regge theory background is chosen so that it explicitly matches onto the Regge
cross section, 
\begin{equation}
\sigma^{R+\mathbb{P}}(\nu) = \left( 1-e^{-\frac{2(\nu-\nu_\pi)}{M}}\right)\, \cdot\, \left[c_R
  \left(\frac{\nu}{1\,\mathrm{GeV}}\right)^{0.097}\, +\, c_{\mathbb{P}} \left(\frac{\nu}{1\,\mathrm{GeV}}\right)^{-0.5}\right]\ ,
\label{eq:backg}
\end{equation}
where the prefactor ensures that the background vanishes at pion threshold.

\begin{figure}[h]
\vspace*{1.4cm}
\includegraphics[height=8cm]{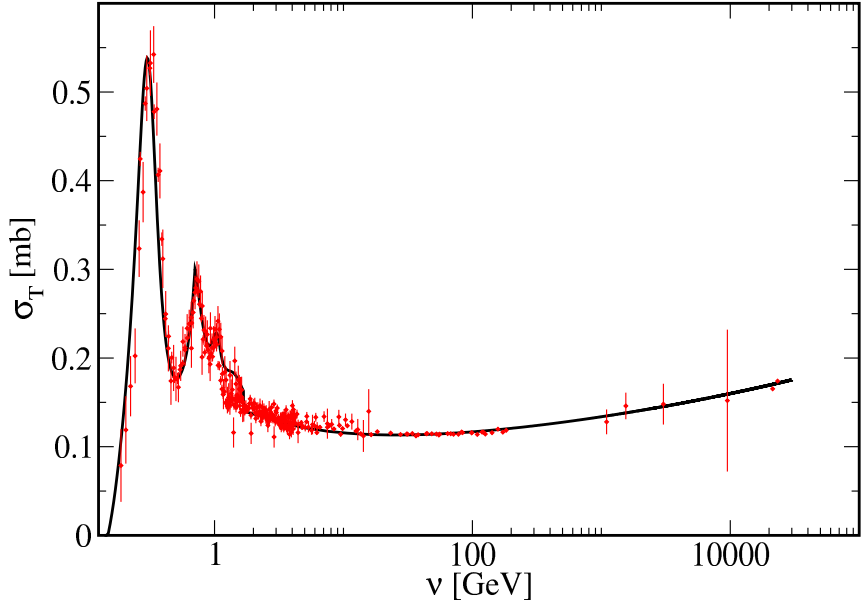}
\caption{The fit to the total photoproduction cross section off the proton (red dots) using
a sum of Breit-Wigner resonances in the low-energy region, and a continuous Regge background
extended from higher energies (solid curve).}
\label{fig:HE}
\end{figure}

As mentioned, we take the Regge intercepts from previous fits to photo-absorption data on the proton \cite{Breitweg:1999},
with the consequent high-fidelity description of the global proton photo-production data shown in Fig.~\ref{fig:HE}.
Constraining the model in Eq.~(\ref{eq:backg}) to modern data, which now extend to larger values of $\nu$, leads to
a significant enhancement in the contributions from reggeon exhange, and we find the parameters $c_R = 68.0\mu$b, and
$c_{\mathbb{P}}=99.0\mu$b most adequately adapt the Regge background to the $\nu \gtrsim 2$ GeV data. This contrasts
significantly compared to fits from the benchmark work by Damashek and Gilman \cite{Damashek:1969xj}, which at the time lacked the
high-energy rise due to pomeron exchange:
\begin{equation}
\sigma^{R+\mathbb{P}}(\nu) = \left( 96.6\ +\ 70.2\ \left[ {\nu \over 1\ \mathrm{GeV}} \right]^{-1/2} \right)\ \mu \mathrm{b}\ ;
\label{eq:DGbackg}
\end{equation}
improvements in data at high $\nu$ thus lead to a rather different numerical description of photoproduction.
Assembling these various elements, we are able to estimate the $J=0$ fixed pole, now finding \cite{Hobbs:2011}
\begin{equation}
{\rm Re}\, T^{J=0}_1\ =\ (-0.72\pm0.35)\ \mu{\rm b\,GeV}\ ;
\end{equation}
when connected to modern high-energy data, the dispersive approach therefore produces a fixed pole contribution which
is markedly different from previous estimates \cite{Damashek:1969xj,Dominguez:1970wu,Shibasaki:1971,Tait:1972}.
Given that these had been consistent with the standard Thomson term result Re $T_1(0)=-3.03 \mu$bGeV as summarized
at the start of Sec.~\ref{sec:Compt}, this new discrepancy is for the first time confirmational that the main
contribution to Eq.~(\ref{j0}) does not come from the coherent nucleonic Thomson term alone, but from local
interactions of the type shown in Fig.~\ref{fig:diag}. Tracing the separate origin of these contributions can
in principle be accomplished by means of the CQM FESR developed in Eq.~(\ref{eq:CQM-FESR}), for example, and urges additional
analytic and experimental investigations of exclusive processes like $\gamma N \rightarrow \gamma N$.

%% file: the-OPE.tex
\section{The Operator Product Expansion}
\label{sec:OPE}

We conclude this chapter with a pedagogical overview of a computational technique
of considerable utility in analyzing hadronic matrix elements. As it is these
that encode the details of nonperturbative structure extracted in DIS, a brief
description will help to contextualize the arguments of
Chap.~\ref{chap:ch-Q2} -- \ref{chap:ch-TMC}. 

The hadronic tensor $W_{\mu\nu}$ of Eq.~(\ref{eq:HadcII}) is the Fourier transform
of a time-ordered product of currents that possess a definite operator structure at the level
of the bilinears $J_\mu(\xi) = \bar{\psi}(\xi) \, \Gamma_\mu \, \psi(\xi)$. Without much
loss of generality, one might concisely state the object of QFT applied to hadronic
structure as a program for understanding the inherently {\it non-local} correlations of
constituent fields within the nucleon (for instance) in terms of specific {\it local}
operators of definite dimension, spin, parity, etc. Of course, the nucleon is
nonperturbative by nature, and the array of operators that contribute to the product
$J_\mu(\xi) \, J_\nu(0)$ is in principle unbounded. To rectify this impasse, an
expansion is required that provides a natural decomposition of the non-local
product $J_\mu(\xi) \, J_\nu(0)$ in terms of local operators $\hat{\mathcal{O}}^i(0)$
in a fashion that gives a specific ordering. This is precisely the description of
Wilson's Operator Product Expansion (OPE) \cite{Wilson:1969zs}.

In broadest terms, the OPE implies the following separation: for two generic operators
$\Omega_1(\xi_1)$, $\Omega_2(\xi_2)$, their product can be expanded as the sum
\begin{align}
\Omega_1(\xi_1)\, \Omega_2(\xi_2) \, &= \, \sum_i \, C_i(\xi_1 - \xi_2) \,
\hat{\mathcal{O}}^i \left( \xi_2 \right) \,\,\,\, \implies \nonumber \\
J(\xi)\, J(0) \, &= \, \sum_i \, C_i(\xi) \, \hat{\mathcal{O}}^i (0)\ ,
\label{eq:OPE1}
\end{align}
where the the sum over $i$ counts various operator structures, and the coefficients $C_i(\xi)$
are generally singular in the $\xi \rightarrow 0$ limit (in fact, the leading contributions to
the OPE originate in the $\hat{\mathcal{O}}^i$ associated with coefficients $C_i(\xi)$ which
are most singular for $\xi^2 \sim 0$). Moreover, the scale dependence $Q^2$ relative to the renormalization
parameter $\mu$ is fully contained within the coefficient functions, which satisfy the Callan-Symanzik renormalization
group equation \cite{Symanzik:1970rt}
\begin{equation}
\left( \mu {\partial \over \partial \mu} + \beta(g) {\partial \over \partial g} - \gamma^N \right)
C^N \left( {Q^2 \over \mu^2}, g \right) \, \equiv \, 0\ ,
\label{eq:Call-Sym}
\end{equation}
which requires knowledge of the QCD Beta function $\beta(g)$ [defined in Eq.~(\ref{eq:QCDbeta})] and anomalous
dimension $\gamma^N$.

Fundamentally, the OPE posits that the Fourier-transformed product of hadronic currents defined by
Eq.~(\ref{eq:Comp}) can be expanded in a closed set of Lorentz structures; these may be compiled as \cite{TMC}
\begin{align}
\label{eq:OPE_gen}
T_{\mu\nu} &= \sum_{i, N}\ \bigg[ \left(g_{\mu\nu} - {q_\mu q_\nu \over q^2}\right)\ q_{\mu_1} \dots q_{\mu_N}
C^N_{L,i}\left({Q^2 \over m^2}, g^2\right) \\
&+\ \left(g_{\mu \mu_1} q_\nu q_{\mu_2} + g_{\nu \mu_2} q_\mu q_{\mu_1} - g_{\mu\nu}q_{\mu_1}q_{\mu_2}
+ g_{\mu\mu_1} g_{\nu\mu_2} Q^2 \right) q_{\mu_3} \dots q_{\mu_N}\ C^N_{2,i}\left({Q^2 \over m^2}, g^2\right) \nonumber\\
&-\ i\epsilon_{\mu\nu\alpha\beta}\ g_{\alpha \mu_1} q_\beta q_{\mu_2}\ \dots\ q_{\mu_N}\ C^N_{3,i}\left({Q^2 \over m^2}, g^2\right) \bigg]\,\,
\times\,\, \bigg( {2 \over Q^2} \bigg)^N\ \lan N(P)|\ \mathcal{O}^{\mu_1 \dots \mu_N}_i(0)\ |N(P) \ran\ . \nonumber
\end{align}

Lorentz covariance allows the nucleon spin-averaged matrix elements of local operators to be
expanded in the general form
\begin{equation}
\lan N(P)|\ \mathcal{O}^{\mu_1 \dots \mu_N}_i(0)\ |N(P) \ran\ =\ A^{(N)}_i\ 
\cdot\ \bigg[P^{\mu_1}\ P^{\mu_2}\ \dots\ P^{\mu_N} \bigg]\ - \{\ \mathrm{trace}\,\, \mathrm{terms}\ \}\ ,
\label{eq:OPE_mat}
\end{equation}
in which the hadronic matrix elements of the operators $\mathcal{O}_i$ have been factorized from the momenta
$P^{\mu_1}\ P^{\mu_2}\ \dots\ P^{\mu_N}$; also, in Eq.~(\ref{eq:OPE_mat}) `trace terms' refers to combinations
involving $g^{\mu_a \mu_b}$ that ensure the tracelessness of the final expansion. This full form is only
necessary for a thorough treatment at leading twist which includes fully summed power corrections. Ignoring these
for the sake of illustration at present (in fact they will be needed later in Chap.~\ref{chap:ch-TMC} to evaluate target
mass effects), we insert the un-symmetrized expression of Eq.~(\ref{eq:OPE_mat}) [\IE~excluding the trace terms]
into Eq.~(\ref{eq:OPE_gen}), leading to the desired combination of coefficient functions and hadronic
matrix elements. A few simple tensor contractions are sufficient to yield
\begin{equation}
T_{\mu\nu} = \sum_{i, N}\ \bigg[ \kappa^L_{\mu\nu}\ C^N_{L,i}(Q^2)\ +\ \kappa^2_{\mu\nu}\ C^N_{2,i}(Q^2)\ 
+\ \kappa^3_{\mu\nu}\ C^N_{3,i}(Q^2) \bigg]\,\,
\times\,\, \bigg( {1 \over x} \bigg)^N\ A_i^{(N)}\ ,
\label{eq:OPE_genII}
\end{equation}
where the rank-2 objects
\begin{subequations}
\begin{align}
\kappa^L_{\mu\nu}\ &=\ g_{\mu\nu} + {q_\mu\ q_\nu \over Q^2}\ , \\
\kappa^2_{\mu\nu}\ &=\ \left( {P_\mu P_\nu \over \nu^2} \right) {1 \over Q^2}\ +\ \Big( P_\mu\ q_\nu + P_\nu\ q_\mu \Big) \Big/ \nu\
-\ g_{\mu\nu}\ , \\
\kappa^3_{\mu\nu}\ &=\ i\epsilon_{\mu\nu\alpha\beta} {P_\alpha\ q_\beta \over q}\ ,
\end{align}
\label{eq:kappa}
\end{subequations}
follow from the expression in Eq.~(\ref{eq:OPE_mat}) up to contributions from the trace terms, which are only relevant for higher power
corrections in $(1/Q^2)$. As mentioned, the role of these additional terms in generating target mass corrections will be discussed subsequently in
Chap.~\ref{chap:ch-TMC}.\ref{sec:TMC}.

\begin{figure}[t]
\includegraphics[height=3.25cm]{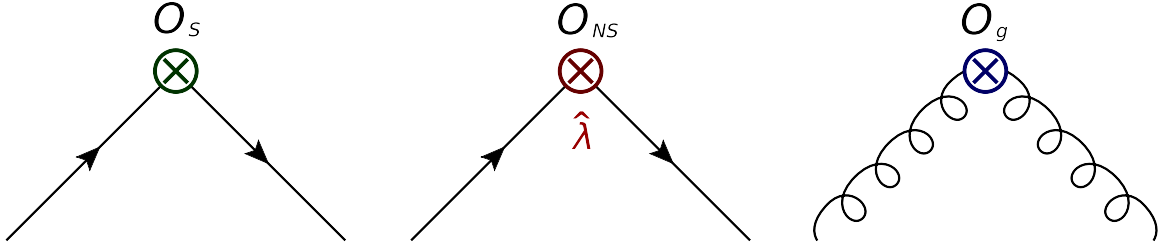}
\caption{
Leading twist operator structures $\mathcal{O}_i$ that contribute to the OPE of Eq.~(\ref{eq:OPE_gen}).
}
\label{fig:dia-OPE}
\end{figure}

At leading twist, however, basic symmetries require that only three primary operator structures can
contribute to the parton-level product of currents $J(\xi)\, J(0)$ \cite{Greiner}. These are the singlet
current operator $i=S$
\begin{equation}
\mathcal{O}^{\mu_1 \dots \mu_N}_S(0)\ = {i^{n-1} \over 2n!}\ \left( \gamma^{\mu_1}\
D^{\mu_2}\ \dots\ D^{\mu_N}\,\, + \mu_a \mu_b\,\,\, \mathit{permutations} \right)\ ,
\end{equation}
non-singlet ($i=NS$)
\begin{equation}
\mathcal{O}^{\mu_1 \dots \mu_N}_{NS}(0)\ = {i^{n-1} \over 2n!}\ \left( \gamma^{\mu_1}\
D^{\mu_2}\ \dots\ D^{\mu_N}\ \hat{\lambda}\,\, + \mu_a \mu_b\,\,\, \mathit{permutations} \right)\ ,
\end{equation}
and tree-level couplings of the glue ($i=g$)
\begin{equation}
\mathcal{O}^{\mu_1 \dots \mu_N}_g(0)\ = {i^{n-2} \over n!}\ \left( \hat{G}^{\mu_1 \mu}\
D^{\mu_2}\ \dots\ D^{\mu_{N-1}}\ \hat{G}^{\mu_N}_\mu\,\, + \mu_a \mu_b\,\,\, \mathit{permutations} \right)\ ,
\end{equation}
all of which are represented pictorially in Fig.~\ref{fig:dia-OPE}.

The operator decomposition of hadronic currents is most easily studied in terms of the moments of DIS
structure functions, which can be unfolded from Eq.~(\ref{eq:OPE_gen}) by again invoking the analyticity of
$T_{\mu\nu}$. Employing the same logic we exploited in Sec.~\ref{sec:Compt} to project the desired helicity
amplitudes from the Compton amplitude, the residue theorem in the parameter $\omega = (1/x) \in \mathbb{C}$
implies
\begin{equation}
{1 \over 2\pi i}\ \oint_C\ \omega^{-N}\ T_{\mu\nu}
\equiv\ {2 \over \pi}\ \int_1^\infty\ d\omega\ \omega^{-N}\ \mathrm{Im}\, T_{\mu\nu}\ ;
\label{eq:cauch}
\end{equation}
the latter quantity on the RHS, $\mathrm{Im}\, T_{\mu\nu}$, can be related explicitly to the DIS structure functions
we require on the grounds of Eq.~(\ref{eq:comp-had}). Again the Cauchy theorem allows the LHS of Eq.~(\ref{eq:cauch})
to be evaluated term-wise, and a simple change of variables on the RHS gives
\begin{equation}
{2 \over \pi}\ \int_1^\infty\ d\omega\ \omega^{-N}\ \mathrm{Im}\, T_{\mu\nu}\
\rightarrow\ 2\ \int_0^1\ dx \cdot x^{N-2}\ W_{\mu\nu}\ .
\label{eq:cauchII}
\end{equation}
With this, it is then enough to go term-by-term in the $\kappa_{\mu\nu}$ tensors of Eq.~(\ref{eq:OPE_genII}) and
equate the various moments within Eq.~(\ref{eq:cauchII}); we thus get the leading twist moments $M^N(Q^2)$:
\begin{subequations}
\begin{align}
M^N_{L/2}(Q^2)\ &=\ \int_0^1\ dx \cdot x^{N-2}\ F_{L/2}(x, Q^2)\ =\ \sum_i\ A^{(N)}_i \cdot C^N_{L/2,\ i}(Q^2)\ , \\
M^N_3(Q^2)\ &=\ \int_0^1\ dx \cdot x^{N-1}\ F_3(x, Q^2)\ =\ A^{(N)}_{NS} \cdot C^N_{3,\ NS}(Q^2)\ .
\end{align}
\label{eq:massless_mom}
\end{subequations}
The OPE therefore provides relations between the moments $M^N(Q^2)$ of structure functions $F(x,Q^2)$
and matrix elements $A^{(N)}_i$ of the operators $\mathcal{O}_i$ depicted in Fig.~\ref{fig:dia-OPE} at leading twist.
The structure functions themselves may be obtained from Eq.~(\ref{eq:massless_mom}) by applying an analytic
transform, as we shall demonstrate for a specific example related to target mass corrections in Eq.~(\ref{eq:Mellin})
of Chap.~\ref{chap:ch-TMC}.

All the more, the scale dependence in the formalism resides entirely within the calculable functions $C^N(Q^2, g)$, which
run according to Eq.~(\ref{eq:Call-Sym}). Thus, if the perturbatively calculable beta function and anomalous dimension
are known, the dependence on $Q^2$ of the moments $M^N(Q^2)$ can also be determined, and from that, the evolution of
$F(x,Q^2)$. To leading order in $\alpha_s$ the running of, \EG~moments of non-singlet structure functions like $F_3(x,Q^2)$
can be written simply as \cite{Ross:1978xk}
\begin{equation}
M^{N,(0)}_3(Q^2)\ =\ \left\{ {\alpha_s(Q^2) \over \alpha_s(\mu^2)} \right\}^{\gamma^{N,(0)}_{NS} \Big/ 2\beta_0}
\cdot M^N_3(\mu^2)\ ,
\label{eq:RSach}
\end{equation}
where the parameter $\beta_0$ of the leading order beta function may be taken from Eq.~(\ref{eq:QCDbeta}) as\linebreak
$\beta_0 \defeq 11 - 2n_F/3$, and $\mu$ is an initial renormalization scale at which the OPE is applied. In particular,
$\gamma^{N=1,(0)}_{NS} = 0$ --- a useful fact which guarantees the leading order scale invariance of non-singlet moments,
including the valence quark distributions to be introduced in later chapters.


%% file: the-Q2.tex

As emphasized in the preceding chapters, energetic lepton-nucleon
scattering has been the primary source of knowledge regarding the nucleon's quark
and gluon (\IE~parton) substructure. In keeping with the chronology we introduced
in Chap.~\ref{chap:ch-intro}, the preponderance of this information has come from
DIS of electrons (or muons), while neutrino DIS has yielded complementary constraints
on valence and sea parton distribution functions (PDFs) via the weak current.

Though recent decades have seen increased activity, a less thoroughly explored
method involves the interference of electromagnetic and weak currents, which is
capable of selecting unique partonic flavor combinations --- a quality attributable
to the parity-violating operator structures involved.
The interference approach consists of measuring the small $\gamma$--$Z^0$
amplitude in the neutral current DIS of a polarized electron
from a hadron $h$, $\vec{e}\ h \to e\ X$.
Because the axial current is sensitive to the polarization of the
incident electron, measurement of the asymmetry between left- and
right-hand polarized electrons is thus proportional to the $\gamma$--$Z^0$
interference amplitude.

As we alluded in the introductory chapter, 1970s parity-violating DIS (PVDIS)
measurements were responsible for key early confirmations of
the Standard Model (SM) of particle physics \cite{Prescott,Cahn}. All the more, after four decades
experimental techniques are now sophisticated enough to enable the measurement of left-right asymmetries
as small as a few parts-per-billion, while current and next-generation
facilities will be able to upgrade the statistics of earlier experiments by an order
of magnitude \cite{JLab6,JLab12}.

Aside from these issues, precision studies in the electroweak sector have garnered special interest
in recent years as the assault on the long-standing dark matter (DM) problem has grown more elaborate
(as have other general tests of the SM). The SM inputs of the {\it electroweak} (EW)
sector are conceivably sensitive to various potential non-SM dynamics. Often, this novel physics is
imagined as arising from undiscovered processes that might predominate at some scale
$\Lambda > \Lambda_{EW}$ beyond that of the EW sector, as hypothetically occurs (for example) in certain
predictions of technicolor or other supersymmetric extensions of the SM, leptoquarks, composite
fermions, etc; the resulting physics might then be encapsulated in new four-fermion contact interactions
of the form \cite{Kumar:2013yoa}
\begin{equation}
\mathcal{L}^{BSM}_{eff}\ =\ {g^2 \over (1 + \delta)\, \Lambda} \sum_{i,j = L,R}
\eta^\psi_{ij}\ \bar{e}_i \gamma_\mu e_i\, \cdot\, \bar{\psi}_j \gamma^\mu \psi_j\ ,
\label{eq:BSM}
\end{equation}
where $\psi$ can represent the Dirac fields of quarks or leptons, $\delta = 0, 1$ for $\psi = e,\, \neq e$,
and the matrix elements $\eta^\psi_{ij} = \pm 1, 0$. Concisely stated, the principal aim of
modern searches for physics beyond the SM is to better constrain the scale $\Lambda$ and
interaction strength $g$ that enter effective interactions of the type $\mathcal{L}^{BSM}_{eff}$
defined in Eq.~(\ref{eq:BSM}), the existence of which would produce observable differences from
the effective four-fermion interactions that can be computed within the standard EW theory as we illustrate
shortly.

Experience teaches that the hypothetical signals for these novel dynamics would almost certainly require
an unprecedented level of precision to access, and to provide this it is first necessary to understand
the predictions of the SM electroweak theory for electron-nucleon DIS, as well as the potential
power corrections and other nonperturbative effects that could complicate such experimental
determinations.

\section{The Electroweak Lagrangian}
\label{sec:EW_intro}

The $SU(2) \times U(1)$ GWS theory \cite{Glashow:1961tr,Salam:1968rm} prescribes a self-contained form for the propagator
structure and interactions within the EW sector. Using
standard notation, the lagrangian for the left-handed ($\Psi$) and
right-handed ($\psi$) fermion fields can be put down as
\begin{align}
\mathcal{L}^{EW} &= \sum_i \,\, \bar{\psi}_i \, \left( i \parsl - m_i - {g \, m_i \mathcal{H} \over 2 m_W} \right) \, \psi_i
\nonumber\\
&- \sum_i \,\, \hat{q_i} \, \bar{\psi}_i \, \gamma^\mu \, \psi_i A_\mu
\nonumber\\
&- {g \over 2\sqrt{2}} \sum_i \,\, \bar{\Psi}_i \, \gamma^\mu \mathcal{P}^- \left(T^- W^-_\mu + T^+ W^+_\mu \right) \, \Psi_i
\nonumber\\
&- {g \over 2 \cos{\theta_W}} \sum_i \,\, \bar{\psi}_i \gamma^\mu \left(g_V^i - g_A^i \gamma^5 \right) \psi_i \, Z_\mu\ ,
\label{eq:L_EW}
\end{align}
with $\hat{q}_i = e \cdot q_i$, where $q_i$ is the fractional charge of the $i^{th}$ fermion species. The projection
operator we have taken to be $\mathcal{P}^- \defeq 1 - \gamma^5$, $\mathcal{H}$ is the Higgs field, $\theta_W$ the famous
weak mixing angle first introduced by Weinberg \cite{Weinberg:1967tq}, and $g$ represents the gauge coupling constant.

Unlike the singlet right-handed mass eigenstates $\psi$, the left-handed fields are doublets
\begin{equation}
\Psi = \left( \begin{array}{c} \nu_i \\ l^-_i \end{array} \right)\ , \,\,\,\,\,
\left( \begin{array}{c} u_i \\ d'_i \end{array} \right)\ ,
\label{eq:Psi}
\end{equation}
which mix under the CKM matrix \cite{Cabibbo:1963yz} in the latter case of the quark fields; this
induces the flavor-mixing of the three quark generations, such that, for example, $d' \approx V_{ud} d + V_{us} s$
after neglecting the heaviest generation.

With this, we can interpret the SM interactions provided by Eq.~(\ref{eq:L_EW}): the
first, kinetic term contains the fermion propagators and Yukawa-type coupling following
the Higgs' spontaneous acquisition of a non-zero vacuum expectation value (VEV), whereas the
second term codifies the electromagnetic photon-fermion coupling. The third and fourth terms provide
the quark and lepton couplings to the gauge fields, which are of particular relevance for parity
violation studies.
For instance, using the definition of Eq.~(\ref{eq:Psi}) for light quarks in Eq.~(\ref{eq:L_EW})
results in the charged current (CC) coupling 
\begin{align}
\mathcal{L}^{CC} &= {-g \over 2\sqrt{2}} \, \sum_i \, \left( \bar{u}_i, \bar{d}'_i \right)
\gamma^\mu \mathcal{P}^- \, \left[ T^- W^-_\mu + T^+ W^+_\mu \right] \,
\left( \begin{array}{c} u_i \\ d'_i \end{array} \right)  \nonumber \\
&= {-g \over 2\sqrt{2}} \, \Big( \bar{u} \gamma^\mu (1-\gamma^5) d' \, W^+_\mu \,\,
+ \,\, \bar{d}' \gamma^\mu (1-\gamma^5) u \, W^-_\mu \Big)\ ,
\label{eq:CC}
\end{align}
where again the mixed state $d'$ is determined from CKM matrix elements as
$d'_i = \sum_j \, \mathcal{V}_{ij} d_j$. These considerations figure importantly in direct searches
for parton-level violation of charge symmetry as discussed in Sec.~\ref{sec:deut-pCSV}.

The source of parity violation in neutral current (NC) electron scattering from hadrons is the $Z^0$ exchange
mechanism; to describe this process, we require the quark and lepton couplings that can be taken from
the fourth term of Eq.~(\ref{eq:L_EW}):
\begin{subequations}
\begin{align}
\mathcal{L}^e_Z &=\ {-g \over 2 \cos{\theta_W}} \, \bar{e} \, \gamma^\mu \left(g^e_V - g^e_A \gamma^5 \right) e \, Z_\mu\ , \\
\mathcal{L}^u_Z &=\ {-g \over 2 \cos{\theta_W}} \, \bar{u} \, \gamma^\mu \left(g^u_V - g^u_A \gamma^5 \right) u \, Z_\mu\ , \\
\mathcal{L}^d_Z &=\ {-g \over 2 \cos{\theta_W}} \, \bar{d} \, \gamma^\mu \left(g^d_V - g^d_A \gamma^5 \right) d \, Z_\mu\ .
\label{eq:L_Z}
\end{align}
\end{subequations}
\begin{figure}[h]
\includegraphics[height=6.5cm]{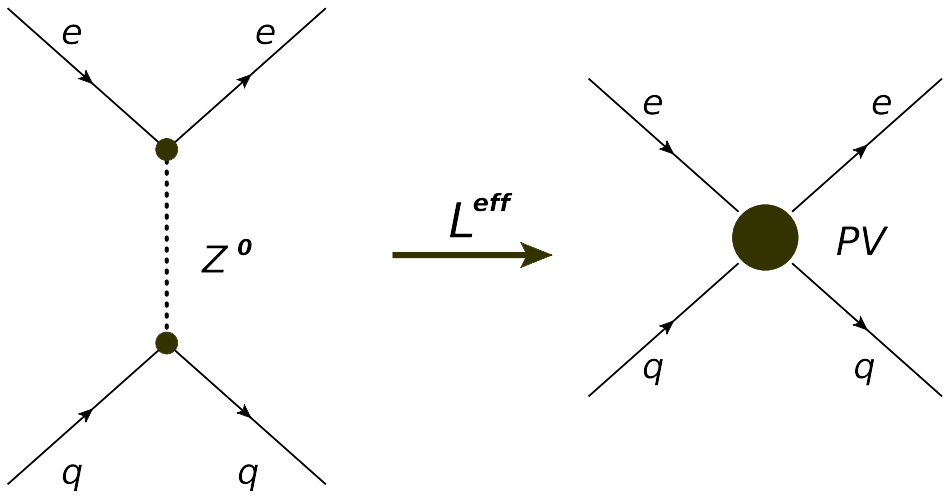}
\caption{
For $Q^2 \ll m^2_Z$, the exchange mechanism at left can be used to compute the effective
parity-violating interaction given in Eq.~(\ref{eq:LPV_eff}).
}
\label{fig:LZ_EW}
\end{figure}
These interactions supply the vertex structure in the $Z^0$-exchange diagram in the LHS of Fig.~\ref{fig:LZ_EW}. With
them, we can compute an amplitude, which we call $A_Z$:
\begin{eqnarray}
A_Z &=& \left({g^2 \over 4 \cos^2{\theta_W}} \right) \, \bar{e} \, \gamma^\mu \left(g^e_V - g^e_A \gamma^5 \right) e \, Z_\mu\,
Z_\nu\, \bar{u} \, \gamma^\nu \left(g^u_V - g^u_A \gamma^5 \right) u \,\, + \,\,  \{d\,\, \mathit{terms}\} \nonumber \\
&=& \left({g^2 \over 4 \cos^2{\theta_W}} \right) \, \bar{e} \, \gamma^\mu \left(g^e_V - g^e_A \gamma^5 \right) e \,
\left[ {g_{\mu\nu} \over k^2 - m^2_Z + i\epsilon} \right]
\, \bar{u} \, \gamma^\nu \left(g^u_V - g^u_A \gamma^5 \right) u \nonumber \\
&& \,\,\,\,\,\,\,\,\,\,\,\,\,\,\,\,\,\,\, \hspace{1.25cm} + \,\, \{d \ \ \mathrm{terms} \}\ ,
\label{eq:Amp_Z}
\end{eqnarray}
where we have used a Feynman gauge expression for the $Z^0$ propagator. The behavior of the amplitude in Eq.~(\ref{eq:Amp_Z})
under the parity operator ensures that only terms linear in $\gamma^5$ contribute to parity violation; taken together with a 
limit in which the exchange momentum of the $Z^0$ is small (\IE~, $k^2 \ll m^2_Z$), this results in
\begin{eqnarray}
A^{PV}_Z &=& \left({g^2 \over 4 \cos^2{\theta_W}} \right) \, \bar{e} \, \gamma^\mu \left(g^e_V - g^e_A \gamma^5 \right) e \,
\left[ {-g_{\mu\nu} \over m^2_Z } \right]
\, \bar{u} \, \gamma^\nu \left(g^u_V - g^u_A \gamma^5 \right) u \,\, + \,\,  \{d\,\, \mathit{terms}\} \,\,\, \implies \nonumber \\
&=& \left({g^2 \over 4 m^2_Z \cos^2{\theta_W}} \right) \, \left[ g^e_V \cdot g^u_A \, \bar{e} \, \gamma^\mu e \,
\bar{u} \, \gamma_\mu \gamma^5 \, u \,\, 
+ g^e_A \cdot g^u_V \, \bar{e} \, \gamma^\mu \gamma^5 e \, \bar{u} \, \gamma_\mu \, u \right] \nonumber \\
&& \,\,\,\,\,\,\,\,\,\,\,\,\,\,\,\,\,\,\, \hspace{1.25cm} + \,\, \{d \ \ \mathrm{terms} \}\ .
\label{eq:AmpII_Z}
\end{eqnarray}
This last expression amounts to an effective, parity-violating four-fermion lepton-quark interaction. In fact, after imposing the
so-called `custodial' symmetry $m^2_W = m^2_Z \cdot \cos^2{\theta_W}$, and using the canonical definition of the effective Fermi coupling
\begin{equation}
{G_F \over \sqrt{2}} = {g^2 \over 8 m^2_W} \approx {g^2 \over 8 m^2_Z \cos^2{\theta_W}}\ ,
\end{equation}
a convenient form for the flavor $SU(2)$ parity-violating lagrangian presents itself:
\begin{equation}
{\cal L}^{\rm PV}
= \frac{G_F}{\sqrt{2}}
  \left[
    \bar{e} \gamma^\mu \gamma_5e
    \left( C_{1u} \bar{u} \gamma_\mu u
	 + C_{1d} \bar{d} \gamma_\mu d
    \right)
  + \bar{e} \gamma^\mu e
    \left( C_{2u} \bar{u} \gamma_\mu \gamma_5 u
	 + C_{2d} \bar{d} \gamma_\mu \gamma_5 d
    \right)
 \right]\ ,
\label{eq:LPV_eff}
\end{equation}
in which the tree-level electroweak couplings are:
\begin{subequations}
\label{eq:C12}
\begin{align}
C_{1u}
&=\ 2\, g^e_A \cdot g^u_V\ 
 =\ -\frac{1}{2}\ +\ \frac{4}{3}\sW\ , \\
C_{1d}
&=\ 2\, g^e_A \cdot g^d_V\
 =\ \ \ \, \frac{1}{2}\ -\ \frac{2}{3}\sW\ ,\\
C_{2u}
&=\ 2\, g^e_V \cdot g^u_A\
 =\ -\frac{1}{2}\ +\ 2\sW\ ,	\\
C_{2d}
&=\ 2\, g^e_V \cdot g^d_A\
 =\ \ \ \, \frac{1}{2}\ -\ 2\sW\ .
\end{align}
\end{subequations}
An overall factor of $2$ has been absorbed into the parity-violating coupling constants
$C_{1q}$ and $C_{2q}$, and we have used the general definition
\begin{subequations}
\begin{eqnarray}
g^i_V &=& \tau^i_z - 2q_i \cdot \sin^2{\theta_W}\ , \\
g^i_A &=& \tau^i_z\ ,
\end{eqnarray}
\end{subequations}
where $\tau^i_z$ is the isospin projection and $q_i$ the fractional charge in
units of $e$ of the $i^{th}$ fermion species.

Having these conventions in hand, we can finally write the relevant vector and axial-vector couplings
of the EW lagrangian in Eq.~(\ref{eq:L_EW}) as
\begin{subequations}
\label{eq:gAgV}
\begin{align}
g^u_V\  &= \ +\frac{1}{2}\ -\ \frac{4}{3}\sW\ , \\
g^d_V\  &= \ -\frac{1}{2}\ +\ \frac{2}{3}\sW\ , \\
g^e_V\  &= \ -\frac{1}{2}\ +\ 2\sW\ ,
\end{align}
\end{subequations}
with $g^e_A = -1/2$ and $g^{u,d}_A = \pm 1/2$. It is precisely these, \IE~the combinations of coupling
constants in Eqs.~(\ref{eq:C12}) and (\ref{eq:gAgV}), that precision DIS measurements aspire to extract
with enough sensitivity to challenge SM predictions. This logic depends on the fact that the effective
couplings $C_{1q}$ and $C_{2q}$ really only depend upon $\sin^2 \theta_W$; hence, by independently
measuring $C_{1q},\ C_{2q}$ in PVDIS experiments, a fundamental test of the EW theory itself may be
constructed. Namely, such precision measurements should have the capacity to observe or exclude interactions
of the type represented in Eq.~(\ref{eq:BSM}), which might otherwise interfere with the mechanism
of Fig.~\ref{fig:LZ_EW}.
\begin{figure}[h]
\includegraphics[height=11cm]{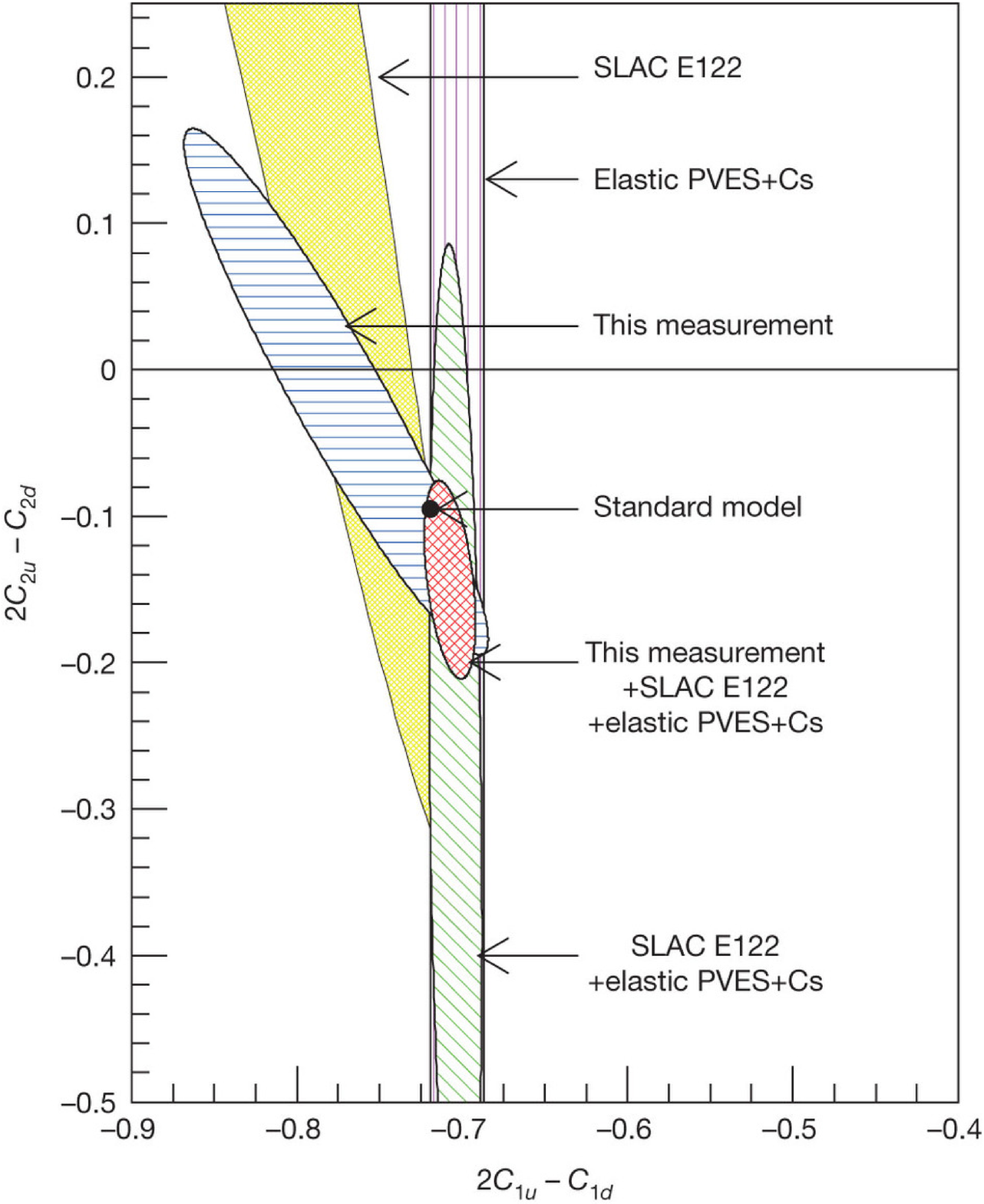}
\caption{
Taken from \cite{Wang:2014bba}, we see that the parameter space of EW couplings is already constrained by
a number of measurements from both atomic physics and electron scattering experiments.
}
\label{fig:C1C2}
\end{figure}
The progress made in such SM tests can be conveniently parametrized in a space spanned by the linear combinations
$2C_{1u}-C_{1d}$ vs.~$2C_{2u}-C_{2d}$, as illustrated in Fig.~\ref{fig:C1C2}. The excluded regions of that plot
emphasize that much progress has indeed already been made in constraining the parameter space, with the strongest
constraints coming from atomic parity violation measurements (specifically, of the weak charge $Q_W$ in Cesium
\cite{Hob10}) and electron scattering experiments --- especially the recent results obtained by the JLab PVDIS
Collaboration \cite{Wang:2014bba}. For the former, the weak charge\footnote{The weak analogue of
electromagnetic charge, direct calculation yields $Q^p_W = 1-4\sin^2 \theta_W$ for the proton and $Q^n_W = -1$
for the neutron, again, at tree-level.} of an arbitrary nucleus of $Z$ protons and
$N$ neutrons can be determined at tree-level from Eq.~(\ref{eq:LPV_eff}) to be \cite{Deandrea:1997wk}
\begin{equation}
Q^{Z,N}_W\ =\ -2 \Big( C_{1u} \cdot [2Z +N] + C_{1d} \cdot [Z+2N] \Big)\ ;
\end{equation}
thus, parity-violating transitions within electron clouds surrounding heavy nuclei (for example, $6S \rightarrow 7S$
in $^{133}$Cs) may serve as an alternative means of constraining the parameter space of Fig.~\ref{fig:C1C2}.

SM tests aside, it has also been suggested more recently that PVDIS can be used to probe
parton distribution functions in the largely unmeasured region of
high Bjorken-$x$ \cite{Souder,SLAC3}.
In particular, the PVDIS asymmetry for a proton is proportional
to the ratio of $d$ to $u$ quark distributions at large $x$.
Current determinations of the $d/u$ ratio rely heavily on inclusive
proton and deuteron DIS data, and there are large uncertainties in
the nuclear corrections in the deuteron at high $x$ \cite{NP}.
While novel new methods have been suggested to minimize the nuclear
uncertainties \cite{A3,BONUS,Comment}, the use of a proton target alone
would avoid the problem altogether.

In this chapter we shall follow the arguments of \cite{Hobbs08} in order to
examine the accuracy of the parton model predictions for the PVDIS
asymmetries in realistic experimental kinematics at finite $Q^2$.
In particular, in Sec.~\ref{sec:PVDIS} we provide a complete set of
formulas for cross sections and asymmetries for scattering polarized
leptons from unpolarized targets, including finite-$Q^2$ effects.
PVDIS from the proton is discussed in Sec.~\ref{sec:pro}, where we test
the sensitivity of the extraction of the $d/u$ ratio at large $x$ to
finite-$Q^2$ corrections.
One of the main uncertainties in the calculation is the ratio of
longitudinal to transverse cross sections for the $\gamma$--$Z^0$
interference, for which no empirical information currently exists, and
we provide some numerical estimates of the possible dependence of the
left-right asymmetry on this ratio.
We also briefly explore in Sec.~\ref{sec:pol} the possibility of using
PVDIS with polarized targets to constrain quark helicity distributions
at large $x$.
As a comprehensive discussion of polarized PVDIS in the parton model was
previously given by Anselmino {\em et al.} \cite{Ans}, we here perform
a numerical survey of the sensitivity of polarized PVDIS asymmetries
to spin-dependent PDFs.

Finally, for deuteron targets, we examine in Sec.~\ref{sec:deut} how the asymmetry
is modified in the presence of finite-$Q^2$ corrections, and where
these can pose significant backgrounds for extracting standard model
signals. In addition, we highlight in Sec.~\ref{sec:deut-pCSV} a novel high-energy
process on the deuteron based on \cite{Hobbs:2011vy} that may offer a novel degree of
sensitivity to quark-level charge symmetry breaking (CSV).
%
%
\section{Nucleon Structure from Parity-Violation}
\label{sec:PVDIS}

Kinematically, we again work in the framework of DIS as established in
Chap.~\ref{chap:ch-DIS}. In this setting, we discuss the general decomposition
of the hadronic tensor, and provide formulas for the PV asymmetry in terms of
structure functions, and in the parton model in terms of PDFs.
\begin{figure}[ht]
\includegraphics[height=7.5cm]{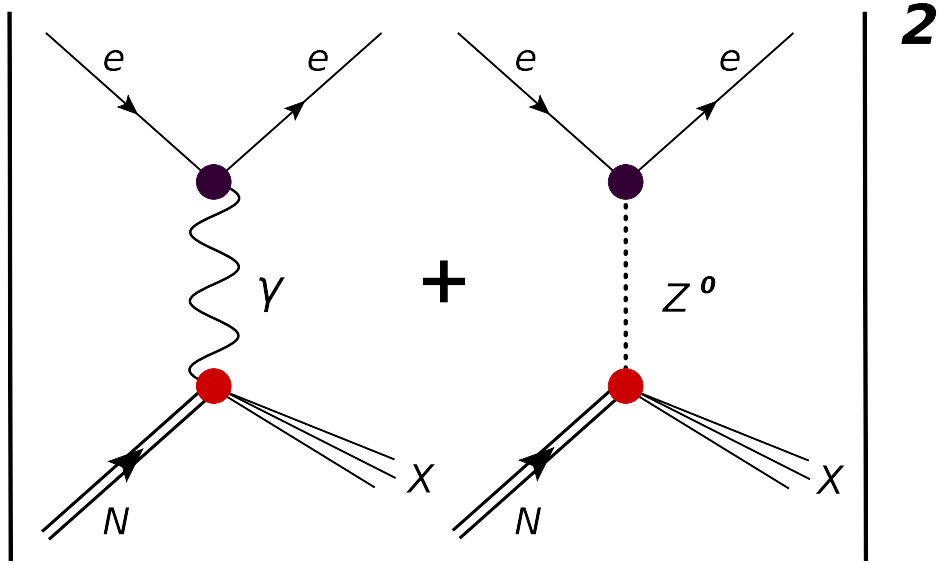}
\caption{
The dominant parity-violating contribution to neutral current DIS
derives from the interaction of purely electromagnetic (left diagram)
and purely weak amplitudes (right diagram).
}
\label{fig:gZ-int}
\end{figure}
%

\vspace*{0.15cm}
{\it Hadronic Tensor.}
\vspace*{0.15cm}

We begin with the differential cross section for inclusive
electron--nucleon scattering, which in general can be written as
the squared sum of the $\gamma$- and $Z^0$-exchange amplitudes.
We will consider contributions to the cross section from the pure
$\gamma$ exchange amplitude $A_\gamma$ and the $\gamma$--$Z$
interference $A_{\gamma Z}$ as depicted in Fig.~\ref{fig:gZ-int};
the purely weak $Z^0$ exchange contribution to the cross section
is strongly suppressed relative to these by the weak coupling $G_F$
and is therefore not considered in the numerical analysis following
these derivations.

Formally, the cross section can be written in terms of products
of leptonic and hadronic tensors as \cite{Ans,TWbook}:
\begin{align}
\label{eq:dsig}
\frac{d^2\sigma}{d\Omega dE'}
&= {\alpha^2 \over Q^4}
  {E' \over E}\,\, \bigg|\, A_\gamma + A_Z\, \bigg|^2\ , \nonumber\\
&= {\alpha^2 \over Q^4}
  {E' \over E}
  \left( L^\gamma_{\mu \nu} W^{\mu \nu}_\gamma\
     +\ {G_F \over 4 \sqrt{2} \pi \alpha}
         L^{\gamma Z}_{\mu \nu} W^{\mu\nu}_{\gamma Z}
     +\ {G^2_F \over 32 \pi^2 \alpha^2}
         L^Z_{\mu \nu} W^{\mu\nu}_Z
  \right)\ ,
\end{align}
where $E$ and $E'$ are the (rest frame) electron energies,
$Q^2$ is (minus) the $4$-momentum transfer squared, and
$\alpha$ is the electromagnetic fine structure constant, keeping
with the conventions laid out in Chap.~\ref{chap:ch-DIS}.\ref{sec:DIS}.

Following Eq.~(\ref{eq:lep}), the lepton tensor $L^i_{\mu\nu}$ encodes the coupling of the scattered electron/neutrino
to the exchange boson(s) [\IE~($i = \gamma,\ \gamma Z,\ Z$)] as \cite{Beringer:1900zz}
\begin{eqnarray}
\label{eq:LgZ}
L^\gamma_{\mu\nu}
&=& 2 \left( l_\mu l'_\nu + l'_\mu l_\nu - l \cdot l'g_{\mu \nu}
	 + i \lambda\varepsilon_{\mu\nu\alpha\beta}\ l^\alpha l'^\beta
    \right)\ , \nonumber \\
L^{\gamma Z}_{\mu\nu} &=& (g^e_V + \lambda g^e_A)\ L^\gamma_{\mu\nu}\ , \nonumber \\
L^{Z}_{\mu\nu} &=& (g^e_V + \lambda g^e_A)^2\ L^\gamma_{\mu\nu}\ , \nonumber \\
L^{W}_{\mu\nu} &=& (1 + \lambda \cdot e)^2\ L^\gamma_{\mu\nu}\ ,
\end{eqnarray}
for leptons of charge $e = \pm 1$ and helicity $\lambda = \pm 1$. Note that the structure
of $L^{W}_{\mu\nu}$ prohibits charged current exchanges for positive-helicity electrons
or negative-helicity positrons.

The complementary hadronic tensors for the electromagnetic, interference, and weak
contributions are then given by:
\begin{eqnarray}
W_{\mu\nu}^\gamma
&=& {1 \over 2M}
  \sum_X\,
    \langle X | J_\mu^\gamma | N \rangle^*
    \langle X | J_\nu^\gamma | N \rangle
    \times (2\pi)^3\ \delta(k_X - p - q)\ , \cr
W_{\mu\nu}^{\gamma Z}
&=& {1 \over 2M}
  \sum_X 
  \left\{
    \langle X | J_\mu^{Z} | N \rangle^*
    \langle X | J_\nu^{\gamma}    | N \rangle
  + \langle X | J_\mu^{\gamma}    | N \rangle^*
    \langle X | J_\nu^{Z} | N \rangle
  \right\}  \times (2\pi)^3\ \delta(k_X - p - q)\ , \cr
W_{\mu\nu}^Z
&=& {1 \over 2M}
  \sum_X\, 
    \langle X | J_\mu^Z | N \rangle^*
    \langle X | J_\nu^Z | N \rangle
    \times (2\pi)^3\ \delta(k_X - p - q)\ ,
\end{eqnarray}
where $M$ is again the nucleon mass, and $J_\mu^{\gamma, \gamma Z, Z}$ correspond
to the electromagnetic, interference, and weak hadronic current, respectively,
cf.~Eq.~(\ref{eq:W-EW}).
In general, the hadronic tensor for a nucleon with spin $4$-vector
$S^\mu$ can be written in terms of three spin-independent and five
spin-dependent structure functions \cite{Ans}:
\begin{eqnarray}
\label{eq:DIS_SFs}
W_{\mu\nu}^i
&=& - \frac{g_{\mu\nu}}{M}\ F_1^i\
 +\ \frac{p_\mu p_\nu}{M\ p\cdot q}\ F_2^i\
 +\ \frac{i \varepsilon_{\mu\nu\alpha\beta} p^\alpha q^\beta}
	 {2M p\cdot q}\ F_3^i				  \\
&+& \frac{i \varepsilon_{\mu\nu\alpha\beta}}{p\cdot q}
    \left( q^\alpha S^\beta\ g_1^i
	 + 2x p^\alpha S^\beta\ g_2^i
    \right)\
 -\ \frac{p_\mu S_\nu + S_\mu p_\nu}{2 p\cdot q}\ g_3^i	 
 + \frac{S \cdot q\ p_\mu p_\nu}{(p \cdot q)^2}\ g_4^i
 +\ \frac{S \cdot q\ g_{\mu\nu}}{p \cdot q}\ g_5^i\ , \nonumber
\end{eqnarray}
for both the electromagnetic ($i=\gamma$) and interference
($i=\gamma Z$) currents.
Each of the structure functions generally depend on two variables,
usually taken to be $Q^2$ and the Bjorken scaling variable
$x = Q^2/2M\nu$, where $\nu$ is the DIS energy transfer.

Performing the appropriate contractions of $W_{\mu\nu}$ with its
leptonic counterpart, we can write a general expression in terms of
the structure functions of Eq.~(\ref{eq:DIS_SFs}). Up to kinematical
prefactors independent of lepton helicity and target spin (and therefore
of no consequence for left-right helicity or spin asymmetries), we have
\begin{align}
\sigma_\lambda = {d^2\sigma_\lambda \over d\Omega dE'} &= \sum_{\{i=\gamma,\ \gamma Z,\ Z\}} \eta_i\ g^e_i\,
\biggl\{ 2xy\,F_1^i +
{2\over y} \left( 1-y-{xyM \over 2E}\right) (F_2^i + g_3^i)
-2\lambda x\left( 1-{y\over2} \right)F_3^i \nonumber\\
&-2\lambda x\left(2-y-{xyM \over E}\right)g_1^i
+4\lambda {x^2 M\over E}\, g_2^i + {2\over y} \left( 1-y-{xyM \over 2E}\right)\, g_3^i \nonumber\\
&-{2\over y}\left( 1+{xM \over E}\right)\left(1-y-{xyM\over 2E}
\right)g_4^i +2xy\left(1+{xM\over E}\right) g_5^i \biggl\}\ ,\ \mathrm{and} \nonumber\\
& \hspace*{-0.7cm} \eta_\gamma = 1\ , \ \ \ \eta_{\gamma Z} = {G_F \over 4\sqrt{2}\pi\alpha}\ ,\ \ \ \eta_Z = {G^2_F \over 32\pi^2 \alpha^2}\ ; 
\label{eq:CrSec}
\end{align}
for the sake of computing the asymmetry, we let $g^e_\gamma = 1$, $g^e_{\gamma Z} = g^e_V + \lambda g^e_A$,
and $g^e_Z = (g^e_V + \lambda g^e_A)^2$, after choosing a convenient set of coordinates to treat a longitudinally
spin-polarized nucleon target.

Below we will consider scattering of a polarized electron from an
unpolarized hadron target, in which only the spin-independent
structure functions $F_{1-3}^{\gamma Z}$ enter.
Asymmetries resulting from scattering of an unpolarized electron
beam from a polarized target, which are sensitive to the
spin-dependent structure functions $g_{1-5}^{\gamma Z}$, will be
discussed in Sec.~\ref{sec:pol}.

\vspace*{0.15cm}
{\it Beam Asymmetries.}
\vspace*{0.15cm}

The PV interference structure functions can be isolated by constructing
an asymmetry between cross sections for right- ($\sigma_{\lambda=+1}$) and left-hand
($\sigma_{\lambda=-1}$) polarized electrons:
\begin{equation}
\label{eq:APVdef}
A^{\rm PV} = \frac{\sigma_{\lambda=+1} - \sigma_{\lambda=-1}}{\sigma_{\lambda=+1} + \sigma_{\lambda=-1}}\ ,
\end{equation}
in which $\sigma_\lambda \equiv d^2\sigma_\lambda/d\Omega dE'$ as defined in Eq.~(\ref{eq:CrSec}).
Since the purely electromagnetic contribution to the
cross section is independent of electron helicity, it cancels in the
numerator, essentially leaving only the $\gamma$--$Z$ interference term due
to the strong suppression of the purely weak process by the squared coupling $G^2_F$.
The denominator, on the other hand, contains all contributions, but is
dominated by the purely electromagnetic component.
In terms of structure functions, the PVDIS asymmetry may thus be written as
\begin{equation}
A^{\rm PV}
= - \left( {G_F Q^2 \over 4 \sqrt{2} \pi \alpha} \right)
  { g^e_A
    \left( 2xy F_1^{\gamma Z} - 2 [1 - 1/y + xM/E] F_2^{\gamma Z} \right)
  + g^e_V\
    x (2-y) F_3^{\gamma Z}
  \over
    2xy F_1^\gamma - 2 [1 - 1/y + xM/E] F_2^\gamma
  }\ ,
\end{equation}
where $y=\nu/E$ is the lepton fractional energy loss.

In the Bjorken limit ($Q^2, \nu \to \infty$, $x$ fixed),
the interference structure functions $F_1^{\gamma Z}$ and
$F_2^{\gamma Z}$ are related by the Callan-Gross relation,
$F_2^{\gamma Z} = 2x F_1^{\gamma Z}$, similar to the electromagnetic
$F_{1,2}^\gamma$ structure functions \cite{Ans}. This behavior may
be encapsulated equivalently by the vanishing of a longitudinal
structure function $F^i_L = F^i_2 - 2xF^i_1$ corresponding to the
associated component of Eq.~(\ref{eq:OPE_gen}).
At finite $Q^2$, however, corrections to the Callan-Gross relation are usually
parametrized in terms of the ratio of the longitudinal to transverse
virtual photon cross sections:
\begin{equation}
R^{\gamma (\gamma Z)}\
\equiv\ \frac{\sigma_L^{\gamma (\gamma Z)}}{\sigma_T^{\gamma (\gamma Z)}}\ 
=\ r^2 \frac{F_2^{\gamma (\gamma Z)}}{2x F_1^{\gamma (\gamma Z)}} - 1\ ,
\end{equation}
for both the electromagnetic ($\gamma$) and interference ($\gamma Z$)
contributions, with
\begin{equation}
r^2 = 1 + {Q^2 \over \nu^2} = 1 + {4 M^2 x^2 \over Q^2}\ .
\end{equation}

In terms of this ratio, the PVDIS asymmetry can be written more
compactly as:
\begin{equation}
A^{\rm PV}
= - \left( {G_F Q^2 \over 4 \sqrt{2} \pi \alpha} \right)
  \left[ g^e_A\ Y_1\ \frac{F_1^{\gamma Z}}{F_1^{\gamma}}\
     +\ {g^e_V \over 2}\ Y_3\ \frac{F_3^{\gamma Z}}{F_1^{\gamma}} 
  \right]\ ,
\label{eq:APV}
\end{equation}
where the functions $Y_{1,3}$ parametrize the dependence on $y$
and on the $R$ ratios:
\begin{subequations}
\label{eq:Y}
\begin{align}
\label{eq:Y1}
Y_1
&= \frac{ 1+(1-y)^2-y^2 (1-r^2/(1+R^{\gamma Z})) - 2xyM/E }
	{ 1+(1-y)^2-y^2 (1-r^2/(1+R^\gamma)) - 2xyM/E } 
   \left( \frac{1+R^{\gamma Z}}{1+R^\gamma} \right)\ ,	\\
\label{eq:Y3}
Y_3
&= \frac{ 1-(1-y)^2 }
	{ 1+(1-y)^2-y^2 (1-r^2/(1+R^\gamma)) - 2xyM/E }
   \left( \frac{r^2}{1+R^\gamma} \right)\ .
\end{align}
\end{subequations}
In the Bjorken limit, the kinematical ratio $r^2 \to 1$,
while the longitudinal cross section vanishes relative to the
transverse, $R^i \to 0$, for both $i = \gamma$ and $\gamma Z$ as we
described in Chap.~\ref{chap:ch-intro}. Physically, we are now in a
position to correctly interpret this behavior as a feature of asymptotic
scattering from pointlike constituent quarks, for which a purely leading
twist calculation is an accurate treatment.
On the other hand, for kinematics relevant to future experiments ($Q^2 \sim$~few GeV$^2$,
$\nu \sim$~few GeV), the factor $2xyM/E$ provides a small correction,
and can for practical purposes be dropped.
In this case the functions $Y_1$ and $Y_3$ have the familiar limits
\cite{Prescott}:
\begin{subequations}
\label{eq:Ybj}
\begin{align}
\label{eq:Y1bj}
Y_1 &\to 1\ ,	\\
\label{eq:Y3bj}
Y_3 &\to \frac{1-(1-y)^2}{1+(1-y)^2}\ \equiv\ f(y)\ .
\end{align}   
\end{subequations}
Typically the contribution from the $Y_3$ term is much smaller
than from the $Y_1$ term because $g_V^e \ll g_A^e$, although for
quantitative comparisons it must be included.

\vspace*{0.15cm}
{\it Electroweak Structure Functions.}
\vspace*{0.15cm}

The PVDIS asymmetry $A^{\rm PV}$ can be evaluated from knowledge
of the electromagnetic and interference structure functions.
At leading twist, the electroweak structure functions may be
expressed in terms of PDFs;
for reference these are listed as follows (at leading order in $\alpha_s$):
\begin{equation}
\begin{aligned}[c]
F_1^\gamma(x) &= \frac{1}{2} \sum_q e_q^2\ (q(x) + \bar{q}(x))\ ; \\
F_2^\gamma(x) &= 2x F_1^\gamma(x)\ ; \\
F_3^\gamma(x) &= 0\ ;
\end{aligned}
\hspace*{0.8cm}
\vspace*{0.2cm}
\begin{aligned}[c]
\vspace*{0.2cm} F_1^{\gamma Z}(x) &= \sum_q e_q\ g_V^q\ (q(x) + \bar{q}(x))\ ; 	\\
F_2^{\gamma Z}(x) &= 2x F_1^{\gamma Z}(x)\ ;			\\
F_3^{\gamma Z}(x) &= 2 \sum_q e_q\ g_A^q\ (q(x) - \bar{q}(x))\ ;
\end{aligned}
\label{eq:EW-SFs}
\end{equation}
in Eq.~(\ref{eq:EW-SFs}) the quark $q$ and antiquark $\bar q$ distributions are defined
with respect to the proton.

For the sake of completeness, we note the
weak neutral collection using our conventions to be
\begin{subequations}
\label{eq:FgZ}
\begin{align}
F_1^Z(x) &= \frac{1}{2} \sum_q (g^{q\, 2}_A + g^{q\, 2}_V )\ (q(x) + \bar{q}(x))\ , 	\\
F_2^Z(x) &= 2x F_1^Z(x)\ ,			\\
F_3^Z(x) &= 2 \sum_q g_A^q\ g_V^q\ (q(x) - \bar{q}(x))\ .
\end{align}
\end{subequations}

In terms of PDFs, the PV asymmetry in Eq.~(\ref{eq:APV}) can be neatly
written as:
\begin{equation}
A^{\rm PV}
= - \left( {G_F Q^2 \over 4 \sqrt{2} \pi \alpha} \right)
    \left( Y_1\ a_1\ +\ Y_3\ a_3 \right)\ ,
\label{eq:a1a3ex}
\end{equation}
with the hadronic vector and axial-vector terms being respectively given by
\begin{equation}
a_1 = \frac{2 \sum_q  e_q\ C_{1q}\ (q+\bar q)}
	   {  \sum_q  e_q^2\       (q+\bar q)}\ ;
\hspace*{1cm}
a_3 = \frac{2 \sum_q  e_q\ C_{2q}\ (q-\bar q)}
	   {  \sum_q  e_q^2\       (q+\bar q)}\ .
\label{eq:a13}
\end{equation}
In this analysis we will focus on the large-$x$ region dominated by
valence quarks, so that the effects of sea quark will be negligible.

At finite $Q^2$, corrections to the parton model expressions appear in
the form of perturbatively generated $\alpha_s$ corrections, target mass
corrections \cite{TMC}, as well as higher twist ($1/Q^2$ suppressed)
effects.
Some of these effects have been tentatively investigated in the
literature \cite{HT}, and in Chap.~\ref{chap:ch-TMC} we consider the
issue of TMCs, but in the present chapter we focus on the
finite-$Q^2$ effects on the asymmetry arising from non-zero values
of $R^{\gamma(\gamma Z)}$, which to date have not been systematically considered.
While data and phenomenological parameterizations are available for
$R^\gamma$ \cite{R1990,R1998,RJLab}, currently no empirical
information exists on $R^{\gamma Z}$.
In our numerical estimates below, we shall consider a range of
possible behaviors for $R^{\gamma Z}$ and examine its effect on
$A^{\rm PV}$.

\subsection{PVDIS on the Proton}
\label{sec:pro}

Parity-violating DIS on a proton target has recently been discussed
as a means of constraining the ratio of $d$ to $u$ quark distributions
at large $x$ \cite{Souder} --- a quantity that has the means of differentiating
among various quark models of the nucleon's valence structure.
At present the $d/u$ ratio is essentially unknown beyond $x \sim 0.6$
due to large uncertainties in the nuclear corrections in the deuteron,
which is the main source of information on the $d$ quark distribution
\cite{NP,Comment}.
Several new approaches to determining $d/u$ at large $x$ have been
proposed, for example using spectator proton tagging in semi-inclusive
DIS from the deuteron \cite{BONUS} (similar to the process analyzed in
Chap.~\ref{chap:ch-TDIS}), or through a ratio of $^3$He and
$^3$H targets to cancel the nuclear corrections \cite{A3}.
The virtue of the PVDIS method is that, rather than using different
hadrons (or nuclei) to select different flavors, here one uses [the
interference of] different gauge bosons to act as a flavor ``filter,''
thereby avoiding nuclear uncertainties altogether.

\begin{figure}[h]
\vspace*{1.4cm}
\includegraphics[height=8cm]{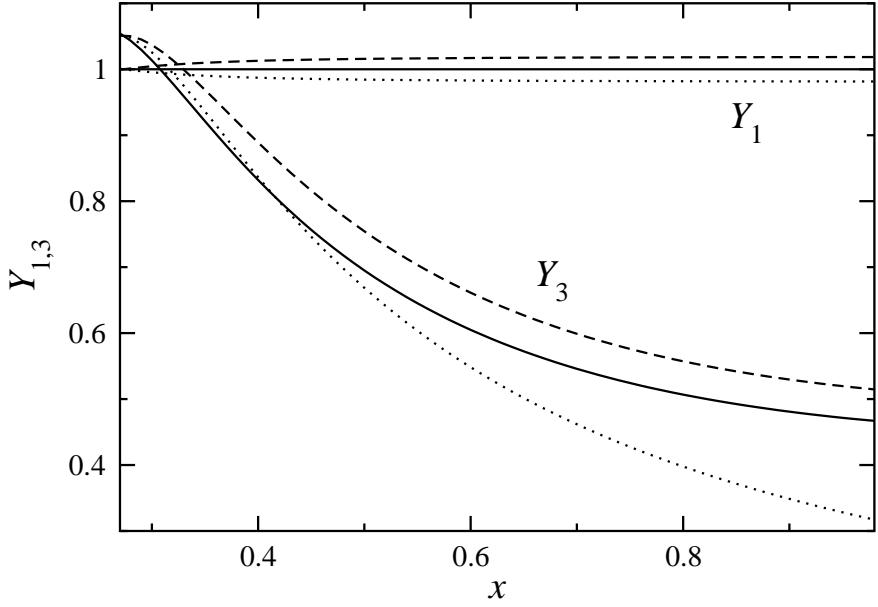}
\caption{$Y_1$ and $Y_3$ as a function of $x$, for $Q^2=5$~GeV$^2$
	and $E=10$~GeV.
	For $Y_1$, the solid line (at $Y_1=1$) corresponds to
	$R^{\gamma Z} = R^\gamma$ \cite{R1990}, while the dotted
	lines correspond to $\pm 20\%$ deviations of $R^{\gamma Z}$
	from $R^\gamma$.
	For $Y_3$, the Bjorken limit result ($R^\gamma=0, r^2=1$)
	is given by the dotted curve, the dashed has $R^\gamma=0$
	but $r^2 \neq 1$, while the solid represents the full result.}
\label{fig:Y13}
\end{figure}

In the valence region at large $x$, the PV asymmetry is sensitive
to the valence $u$ and $d$ quark distributions in the proton.
Here the functions $a_1$ and $a_3$ in Eqs.~(\ref{eq:a13}) for the
proton can be simplified to:
\begin{subequations}
\begin{equation}
\label{eq:a1p}
a_1^p = \frac{12 C_{1u} - 6 C_{1d}\ d/u}{4 + d/u}\ ,
\end{equation}
and
\begin{equation}
\label{eq:a3p}
a_3^p = \frac{12 C_{2u} - 6 C_{2d}\ d/u}{4 + d/u}\ .
\end{equation}  
\end{subequations}
This reveals that both $a_1^p$ and $a_3^p$ depend directly on the $d/u$ quark
distribution ratio.


To explore the relative sensitivity of the proton asymmetry
$A^{\rm PV}_p$ to the vector and axial vector terms, in
Fig.~\ref{fig:Y13} we show the functions $Y_1$ and $Y_3$ for
the proton as a function of $x$, evaluated at $Q^2 = 5$~GeV$^2$,
for a beam energy $E = 10$~GeV (which we will assume throughout).
For $Y_1$, the solid line (at $Y_1=1$) corresponds to
$R^{\gamma Z} = R^\gamma$, while the dashed (dotted) curves around it
represent $+(-) 20\%$ deviations of $R^{\gamma Z}$ from $R^\gamma$.
For $Y_3$, the Bjorken limit result ($R^\gamma=0, r^2=1$) is given by
the dotted curve, the dashed curve has $R^\gamma=0$ but $r^2 \neq 1$,
while the solid represents the full result with $R^\gamma \neq 0$
and $r^2 \neq 1$.
In all cases we use $R^\gamma$ from the parameterization of
Ref.~\cite{R1990}.
The results with the parameterization of Ref.~\cite{R1998} are very
similar, and are consistent within the quoted uncertainties.

Note that at fixed $Q^2$, the large-$x$ region also corresponds to
low hadronic final state masses $W$, so that with increasing $x$ one
eventually encounters the resonance region at $W \lesssim 2$~GeV (akin
to the behavior illustrated for exclusive photoproduction in Fig.~\ref{fig:HE}
of Chap.~\ref{chap:ch-DIS}.\ref{sec:Compt}).
For $Q^2 = 5$~GeV$^2$ this occurs at $x \approx 0.62$, and for
$Q^2 = 10$~GeV$^2$ at $x \approx 0.76$.
This may introduce an additional source of uncertainty in the
extraction of the PV asymmetry at large $x$, arising from possible
higher twist corrections to structure functions.
In actual experimental conditions, the value of $Q^2$ can be varied
with $x$ to ensure that the resonance region is excluded from the
data analysis.
For the purposes of illustrating the finite-$Q^2$ effects in our
analysis, we shall fix $Q^2$ at the low end of values attainable
with an energy of $E = 10$~GeV, namely $Q^2 = 5$~GeV$^2$.

\begin{figure}[h]
\vspace*{1.5cm}
\includegraphics[height=9cm]{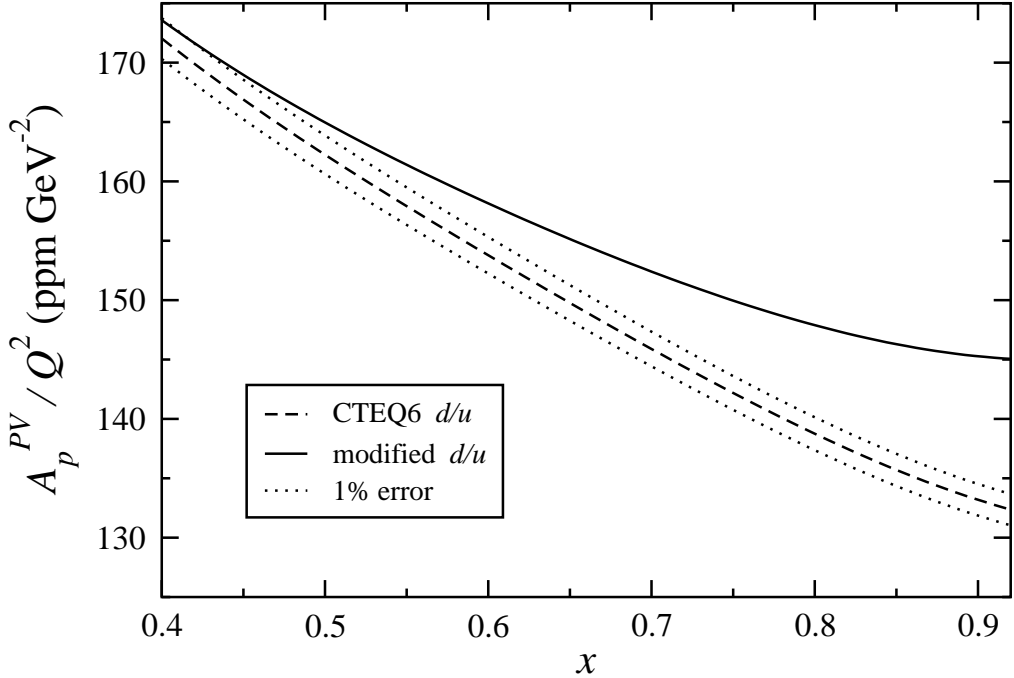}
\caption{Proton PV asymmetry $A^{\rm PV}_p/Q^2$ as a function of $x$,
	for $Q^2 = 5$~GeV$^2$, in parts per million (ppm)
	$\cdot$~GeV$^{-2}$.
	The prediction with the standard CTEQ6 PDFs (dashed) is
	compared with that using a modified $d/u$ ratio at large $x$
	(solid).
	A $\pm 1\%$ uncertainty band (dotted) is shown around the
	standard CTEQ6 prediction.}
\label{fig:APV_du}
\end{figure}

The sensitivity of the proton asymmetry $A^{\rm PV}_p$, measured in
parts per million (ppm), to the $d/u$ ratio is illustrated in
Fig.~\ref{fig:APV_du} as a function of $x$, for $Q^2 = 5$~GeV$^2$,
where $A^{\rm PV}_p/Q^2$ is shown.
Here we assume that $R^{\gamma Z} = R^\gamma$, so that the coefficient
$Y_1$ in the vector term is unity.
For the $u$ and $d$ distributions we use the CTEQ6 PDF set \cite{CTEQ},
in which the $d/u$ ratio vanishes as $x \to 1$, along with a modified
$d/u$ ratio which has a finite $x \to 1$ limit of 0.2 \cite{NP},
$d/u \to d/u + 0.2\ x^2\ \exp(-(1-x)^2)$ \cite{Peng}, motivated by
theoretical counting rule arguments \cite{FJ}.
Also shown (dotted band around the CTEQ6 prediction) is a $\pm 1\%$
uncertainty, which is a conservative estimate of what may be expected
experimentally at JLab with 12~GeV \cite{Souder,JLab12}.
The results indicate that a signal for a larger $d/u$ ratio would
be clearly visible above the experimental errors.


At finite $Q^2$ the asymmetry $A^{\rm PV}_p$ depends not only on the
PDFs, but also on the longitudinal to transverse cross sections ratios
$R^\gamma$ and $R^{\gamma Z}$ for the electromagnetic and $\gamma Z$
interference contributions, respectively.
A number of measurements of the former have been taken at SLAC and
JLab \cite{R1990,R1998,RJLab}, and parameterizations of $R^\gamma$
in the DIS region exist. As such, the contribution from $R^\gamma$
is under comparatively better control, both theoretically and
experimentally.



These effects are to be compared with the relative change in
$A^{\rm PV}_p$ arising from different large-$x$ behaviors of the
$d/u$ ratio (dashed curve), expressed as a difference of the
asymmetries with the standard CTEQ6 \cite{CTEQ} PDFs and ones
with a modified $d/u$ ratio \cite{NP,Peng},
$\delta^{(d/u)} A_p^{\rm PV}/A_p^{\rm PV (0)}$.
where $A_p^{\rm PV (0)}$ is computed in terms of the standard
(unmodified) PDFs.
This is of the order 2\% for $x \sim 0.5$, but rises rapidly to
$\sim 10\%$ for $x \sim 0.9$.
While the kinematical and $R^\gamma$ corrections are smaller than
the (maximal) $d/u$ effect on the asymmetry, these must be included
in the data analysis in order to minimize the uncertainties on the
extracted $d/u$ ratio.

\begin{figure}[h]
\vspace*{1.5cm}
\includegraphics[height=9cm]{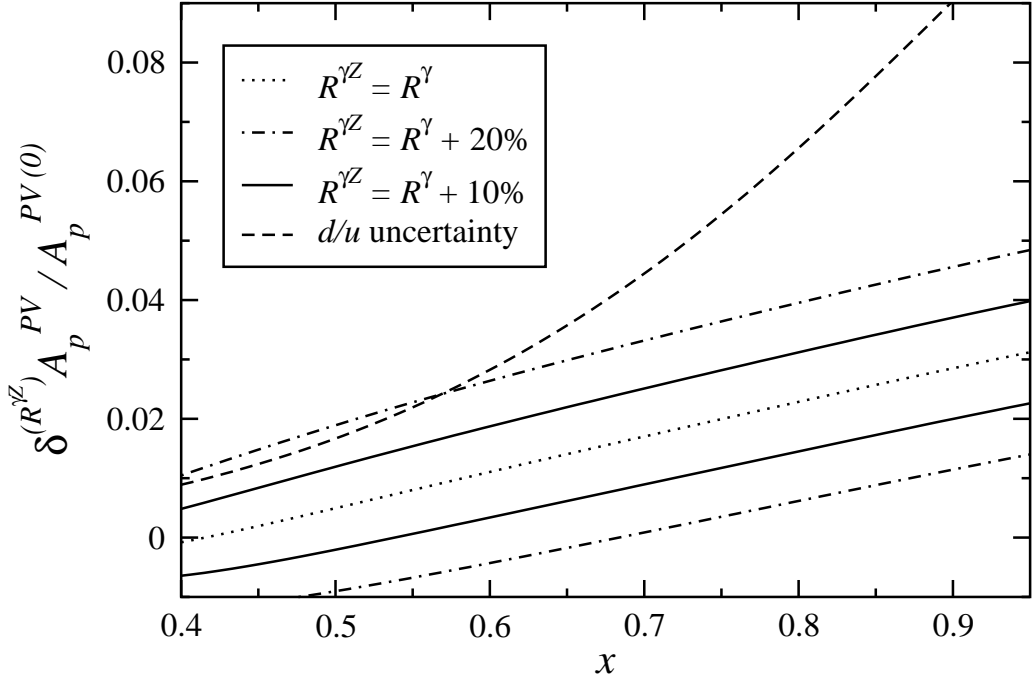}
\caption{Relative effects on the proton PV asymmetry $A^{\rm PV}_p$
	from the $\gamma Z$ interference ratio $R^{\gamma Z}$
	compared with the Bjorken limit asymmetry $A^{\rm PV\ (0)}_p$.
	The baseline result for $R^{\gamma Z} = R^\gamma$ (dotted)
	is compared with the effects of modifying $R^{\gamma Z}$ by
	$\pm 10\%$ (solid) and $\pm 20\%$ (dot-dashed),
	for $Q^2=5$~GeV$^2$.
	For reference the relative uncertainty
	$\delta^{(d/u)}A_p^{\rm PV}/A_p^{\rm PV (0)}$ from the
	$d/u$ ratio is also shown (dashed).}
\label{fig:dRgZ_Ap}
\end{figure}

In contrast to $R^\gamma$, no experimental information currently
exists on the interference ratio $R^{\gamma Z}$.
Since $R^{\gamma Z}$ enters in the relatively large $Y_1$ contribution
to $A^{\rm PV}_p$, any differences between $R^{\gamma Z}$ and
$R^\gamma$ could have important consequences for the asymmetry.
At high $Q^2$ one expects that $R^{\gamma Z} \approx R^\gamma$ at
leading twist, if the PVDIS process is dominated by single quark
scattering.
At low $Q^2$, however, since the current conservation constraints
are different for weak and electromagnetic probes, there may be
significant differences between these \cite{RlowQ}.

To explore the potential effects of $R^{\gamma Z}$ on $A^{\rm PV}_p$
we therefore consider several possible scenarios for the ratios.
These are illustrated in Fig.~\ref{fig:dRgZ_Ap}, where we plot the
ratio
$\delta^{(R^{\gamma Z})} A^{\rm PV}_p / A^{\rm PV (0)}_p$,
where
$\delta^{(R^{\gamma Z})} A^{\rm PV}_p$ is the difference between
the full asymmetry and that calculated in Bjorken limit kinematics,
$A^{\rm PV (0)}_p$.
The baseline correction with $R^{\gamma Z}=R^\gamma$ (dotted curve), with
$R^\gamma$ from Ref.~\cite{R1990}, is compared with the effects
of modifying $R^{\gamma Z}$ by $\pm 10\%$ (solid) and $\pm 20\%$
(dot-dashed).
The result of such a modification, which comes through the $Y_1$ term
in the asymmetry, is an $\approx$ 1\% (2\%) shift of $A^{\rm PV}_p$
relative to the $R^{\gamma Z}$-independent asymmetry.
For $x \lesssim 0.6$, a 20\% difference between $R^{\gamma Z}$ and
$R^\gamma$ would be comparable to, or exceed, the maximal $d/u$
uncertainty considered here (dashed curve), although at larger $x$ the
sensitivity of $A^{\rm PV}_p$ to $d/u$ becomes increasingly stronger.
As with the $R^\gamma$ corrections discussed at length in \cite{Hobbs08}, the
possible effects on the asymmetry due to $R^{\gamma Z}$ are potentially
significant, which partially motivates work in subsequent chapters (esp.,
Chap.~\ref{chap:ch-TMC}) to estimate possible differences with $R^\gamma$.

\subsection{Spin-Polarized PVDIS}
\label{sec:pol}

In this section we explore the possibility of extracting
{\em spin-dependent} PDFs in parity-violating unpolarized-electron 
scattering from a {\em polarized} hadron.
In particular, we examine the sensitivity of the polarized proton,
neutron and deuteron PVDIS asymmetries to the polarized $\Delta u$
and $\Delta d$ distributions at large $x$, where these are poorly
known.
The $\Delta d$ distribution in particular remains essentially
unknown beyond $x \approx 0.6$.

Using Eq.~(\ref{eq:CrSec}), the PV differential cross-section (with respect to the variables
$x$ and $y$) for unpolarized electrons on longitudinally polarized
nucleons can generally be written in terms of five spin-dependent
structure functions \cite{Ans}:
\begin{eqnarray}
\frac{d^2 \sigma^{\rm PV}}{dxdy}(\bar{\lambda}, S_L)
&=& 2x \left( 2-y-\frac{xyM}{E} \right) g_1^{\gamma Z}\
 -\ \frac{4x^2M}{E}\ g_2^{\gamma Z}\
 +\ \frac{2}{y} \left( 1-y-\frac{xyM}{2E} \right) g_3^{\gamma Z}
	\nonumber\\
& &
 -\ \frac{2}{y} \left( 1+\frac{xM}{E} \right)
	        \left( 1-y-\frac{xyM}{2E} \right) g_4^{\gamma Z}\
 +\ 2xy \left( 1+\frac{xM}{E} \right) g_5^{\gamma Z}\ ,
\end{eqnarray}
where the nucleon (longitudinal) spin vector $S_L$ is given by 
$S^{\mu}_L = (0;0,0,1)$, and $\bar{\lambda}$ is the average over 
$\lambda = +1$ and $\lambda = -1$ [see Eq.~(\ref{eq:CrSec})].
The analogue of the PV asymmetry in Eq.~(\ref{eq:APVdef}) for a
polarized target can be defined as:
\begin{equation}
\Delta A^{\rm PV}
= \frac{\sigma^{\rm PV} (\bar{\lambda}, S_L)
      - \sigma^{\rm PV} (\bar{\lambda}, -S_L)}
       {\sigma^{\rm PV} (\bar{\lambda}, S_L)
      + \sigma^{\rm PV} (\bar{\lambda}, -S_L)}\ ,
\end{equation}
where $\sigma^{\rm PV} (\bar{\lambda}, S_L)
\equiv d^2 \sigma^{\rm PV}/dxdy$.
Some of the structure functions $g_{1-5}^{\gamma Z}$ have simple
parton model interpretations, whereas others do not; either way,
at present there is essentially no phenomenological information about them.
In order to proceed, we shall therefore consider the asymmetry in
the high energy limit, $M/E \to 0$, which eliminates the structure
function $g_2^{\gamma Z}$.
In this limit, the operator product expansion gives rise to the
relation $g_3^{\gamma Z} - g_4^{\gamma Z} = 2xg_5^{\gamma Z}$,
which further eliminates one of the functions.
Furthermore, in the parton model the $g_4^{\gamma Z}$ structure
function vanishes, leaving the Callan-Gross-like relation
$g_3^{\gamma Z} = 2x g_5^{\gamma Z}$.
In terms of the remaining two structure functions, the spin-dependent
PV asymmetry may be written:
\begin{equation} 
\Delta A^{\rm PV}
= \frac{G_F Q^2}{4 \sqrt{2}\pi\alpha}
  \left( g_A^e\ f(y)\ \frac{g_1^{\gamma Z}}{F_1^{\gamma}}\
      +\ g_V^e\ \frac{g_5^{\gamma Z}}{F_1^{\gamma}}
  \right)\ ,
\end{equation}  
where the kinematical factor $f(y)$ is given in Eq.~(\ref{eq:Y3bj}).

As an aside we note that the clean isolation of such spin-polarized observables
could present yet another opportunity to test the predictions and
structure of QCD and the SM; this is evident by the form of spin-polarized
sum rules analogous to the relations mentioned in Eqs.~(\ref{eq:GLS} \& \ref{eq:Gott}),
which may be derivable in the QPM in terms of quark-level quantities and electroweak
parameters. For example, this is the case with the electromagnetic Bjorken sum rule
\cite{Bjorken:1966jh},
\begin{align}
S_B\ =\ \int_0^1\ dx \left\{ g^\gamma_{1p}(x,Q^2)\ -\ g^\gamma_{1n}(x,Q^2) \right\}\ = {g_A \over 6} + \mathrm{O}(\alpha_s)\ ,
\end{align}
where $g_A \sim 1.26$ is the axial charge of the nucleon predictable by the Adler-Weisberger relation.

\begin{figure}[h]
\vspace*{1.4cm}
\includegraphics[height=9cm]{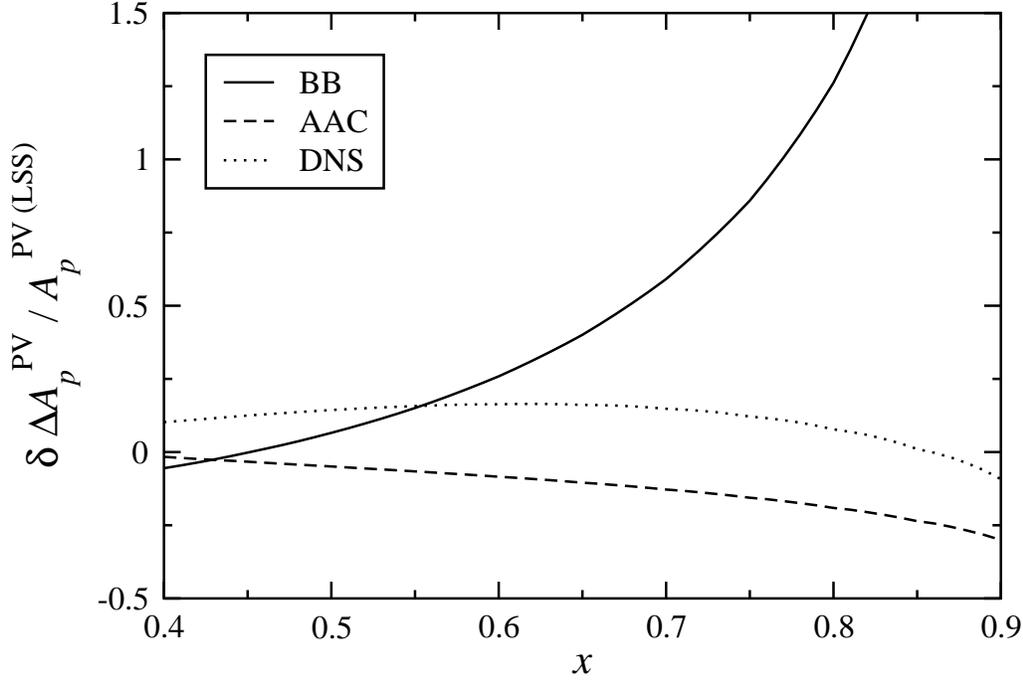}
\caption{Sensitivity of the polarized proton PV asymmetry
	$\Delta A^{\rm PV}_p$ on the spin-dependent
	$\Delta u$ and $\Delta d$ distributions.
	The asymmetries for the BB \cite{BB} (solid), AAC \cite{AAC}
	(dashed) and DNS \cite{DNS} (dotted) distributions are
	evaluated relative to the baseline asymmetry for the LSS
	PDFs \cite{LSS}.}
\label{fig:dDA}
\end{figure}

In the QCD parton model the $g_1^{\gamma Z}$ and $g_5^{\gamma Z}$
structure functions can be expressed in terms of helicity dependent
PDFs $\Delta q$ as \cite{Ans}:
\begin{subequations}
\begin{eqnarray}
g_1^{\gamma Z}
&=& \sum_q e_q\ g_V^q
    \left( \Delta q + \Delta \bar{q} \right)\ , \\
g_5^{\gamma Z}
&=& \sum_q e_q\ g_A^q\
    \left( \Delta q - \Delta \bar{q} \right)\ ,
\end{eqnarray}
\end{subequations}
where $\Delta q$ is a function of $x$ and $Q^2$.
Using these expressions, the PV asymmetries for proton, neutron and 
deuteron (which in this analysis we take to be a sum of proton and 
neutron) targets can then be written \cite{KK}:
\begin{subequations}
\begin{eqnarray}
\Delta A^{\rm PV}_p
&=& \frac{6\ G_F Q^2}{4 \sqrt{2}\pi\alpha}
    \left[ (2 C_{1u} \Delta u - C_{1d} \Delta d) f(y)
	 + (2 C_{2u} \Delta u - C_{2d} \Delta d)
    \right]
    \left( {1 \over 4 u + d} \right)\ ,			\\
\Delta A^{\rm PV}_n
&=& \frac{6\ G_F Q^2}{4 \sqrt{2}\pi\alpha}
    \left[ (2 C_{1u} \Delta d - C_{1d} \Delta u) f(y)
	 + (2 C_{2u} \Delta d - C_{2d} \Delta u)
    \right]
    \left( {1 \over u + 4 d} \right)\ ,			\\
\Delta A^{\rm PV}_d
&=& \frac{3\ G_F Q^2}{10 \sqrt{2}\pi\alpha}
    \left[ (2 C_{1u} - C_{1d}) f(y)
         +  2 C_{2u} - C_{2d}
    \right]
    \left( {\Delta u + \Delta d \over u + d} \right)\ .
\end{eqnarray}
\end{subequations}

In Fig.~\ref{fig:dDA} we illustrate the sensitivity of the proton
asymmetry $\Delta A^{\rm PV}_p$ to the $\Delta u$ and $\Delta d$ PDFs,
by comparing the difference $\delta \Delta A^{\rm PV}_p$ in the
asymmetry arising from different parameterizations \cite{BB,AAC,DNS},
relative to the LSS parameterization \cite{LSS}. The spread among these
various schemes reflects varying model assumptions that determine
the initial input parametrizations, which are in turn constrained by the limited
available data.
The resulting effects in $\Delta A^{\rm PV}_p$ at intermediate
$x$, $x \sim 0.5$--0.6, are of order 20\%; however, these increase rapidly with $x$.
At $x \approx 0.7$--0.8 the AAC \cite{AAC}, DNS \cite{DNS} and LSS
\cite{LSS} parameterizations give asymmetries that are within
$\sim 20\%$ of each other, whereas the BB fit \cite{BB} deviates
by 50--100\% in this range --- a consequence of the lack
of precise experimental constraints, especially at high $x$.
The results for neutron and deuteron targets are found to be very
similar to those in Fig.~\ref{fig:dDA}.
While this does not constitute a systematic error on the uncertainty
in $\Delta A^{\rm PV}_p$ due to PDFs, it does indicate the sensitivity
of polarized PVDIS to helicity distributions at large $x$, and suggests
that a measurement of $\Delta A^{\rm PV}_p$ at the 10--20\% level could
discriminate between different PDF behaviors.

Finally, for completeness we indicate that in principle it should be possible to extract 
data on the PDF quantities $\Delta u / u$ and $\Delta d / d$ 
from polarized asymmetries given knowledge of the unpolarized
PDF behavior and experimental values for the DIS couplings.

Making the appropriate combinations of single-nucleon polarized asymmetries,
we obtain:
\begin{subequations}
\begin{eqnarray}
\frac{\Delta u}{u} 
&=& \frac{\sqrt{8} \pi \alpha}{3G_F Q^2} \cdot
    { \left( 2 (C_{1u} f(y) + C_{2u}) (4 + d/u) \Delta A^{PV}_p 
        + (C_{1d}+C_{2d})(4 d/u + 1) \Delta A^{PV}_n
    \right)
  \over 4 [C_{1u}f(y)+C_{2u}]^2 + [C_{1d}f(y)+C_{2d}]^2}\ ,  \nonumber\\
\\
\frac{\Delta d}{d} 
&=& \frac{\sqrt{8} \pi \alpha}{3G_F Q^2} \cdot 
    \frac{[2(C_{1u}f(y) + C_{2u})(4+u/d) \Delta A^{PV}_n 
        + (C_{1d}+C_{2d})(4 u/d+1)\Delta A^{PV}_p]}
       {4[C_{1u}f(y)+C_{2u}]^2 + [C_{1d}f(y)+C_{2d}]^2}\ . \nonumber\\
\end{eqnarray}
\end{subequations}
Alternatively, we may write these in terms of the proton and deuteron
asymmetries:
\begin{subequations}
\begin{eqnarray}
\frac{\Delta u}{u} 
&=& \frac{\sqrt{8}\pi\alpha
  [(4+d/u)C_1(y)\Delta A^{PV}_p + 5(1+d/u)[C_{1d}f(y)+C_{2d}]\Delta A^{PV}_d]}
  {3G_F Q^2 (C_1(y) \cdot C_2(y)}\ ,    \\
\frac{\Delta d}{d} 
&=& \left(\sqrt{8}\pi\alpha [\frac{5(u/d+1)\Delta A^{PV}_d}{C_1(y)} 
        - \frac{(4u/d+1)\Delta A^{PV}_p}{C_{1u}f(y)+C_{2u}} \right) \Bigg/      
       \left( 1+\frac{C_{1d}f(y)+C_{2d}}{2(C_{1u}f(y)+C_{2u})} \right)\ . \nonumber\\
\end{eqnarray}
\end{subequations}
Thus, with precise measurements of $\Delta A^{PV}_p,\, \Delta A^{PV}_n,$ and $\Delta A^{PV}_d$
the light quark polarizations might be better constrained --- a fact that urges
further acquisition of spin-polarized DIS asymmetry measurements, particularly
at high $x$.
%
\section{Scattering from the Isoscalar Deuteron}
\label{sec:deut}

For parity-violating scattering from an isoscalar deuteron, the
dependence of the left-right asymmetry on PDFs cancels in the
parton model, so that the asymmetry is determined entirely by
the weak mixing angle, $\theta_W$.
The deuteron asymmetry is therefore a sensitive test of effects beyond
the parton model, such as higher twist contributions, or of more exotic
effects such as charge symmetry violation in PDFs or new physics beyond
the SM.

In fact as early as the late 1970s parity-violating DIS on the deuteron provided
important early tests of the SM \cite{Prescott,Cahn}.
In the parton model, the asymmetry for an isoscalar deuteron becomes
independent of hadronic structure, and is therefore given entirely by electroweak 
coupling constants.
At finite $Q^2$, however, contributions from longitudinal structure
functions, or from higher twist effects, may play a role.
The higher twists have been estimated in several phenomenological 
model studies \cite{HT}, while
more recently, it has been suggested that PVDIS on a deuteron could
also be sensitive to charge symmetry violation (CSV) effects in PDFs
(see Ref.~\cite{CSVreview} for a review of CSV in PDFs).
In this section we explore the contributions from kinematical
finite-$Q^2$ effects and the longitudinal structure functions $F^i_L$ on the
PV asymmetry, and assess their impact on the extraction of CSV effects.
Having done so, we outline a possible measurement at higher $Q^2$ which
would be ideally suited to an envisioned electron-ion collider and may well
yield an unambiguous CSV signal.

\subsection{Electroweak Structure}
\label{sec:deut_EW}
Assuming the deuteron is composed of a proton and a neutron, and 
neglecting possible differences between free and bound nucleon PDFs,
the functions $a_1$ and $a_3$ in Eq.~(\ref{eq:a13}) for a deuteron target become:
\begin{subequations}
\begin{eqnarray}
\label{eq:a1d}
a_1^d &=& \frac{6}{5} \left( 2 C_{1u} - C_{1d} \right)\ ,	\\
\label{eq:a3d}
a_3^d &=& \frac{6}{5} \left( 2 C_{2u} - C_{2d} \right)\ .
\end{eqnarray}  
\end{subequations}
If in addition $R^\gamma_d \approx R^\gamma_p$ and
$R^{\gamma Z}_d \approx R^{\gamma Z}_p$, as is observed experimentally
\cite{R1990}, then the $y$-dependent terms in the deuteron asymmetry
become
$Y_1^d \approx Y_1^p \equiv Y_1$ and $Y_3^d \approx Y_3^p \equiv Y_3$.
The PV asymmetry can then be written as:
\begin{equation}
A^{\rm PV}
= -\left( \frac{3 G_F Q^2}{10 \sqrt{2}\pi\alpha} \right)
  \left[ Y_1 \left( 2 C_{1u} - C_{1d} \right)\
      +\ Y_3 \left( 2 C_{2u} - C_{2d} \right)\
  \right]\ ,
\label{eq:deut_asym}
\end{equation}
which in the Bjorken limit ($Y_1 \to 1$, $Y_3 \to f(y)$) becomes
independent of hadron structure, and is a direct measure of the
electroweak coefficients $C_{iq}$.

\begin{figure}[h]
\vspace*{1.5cm}
\includegraphics[height=9cm]{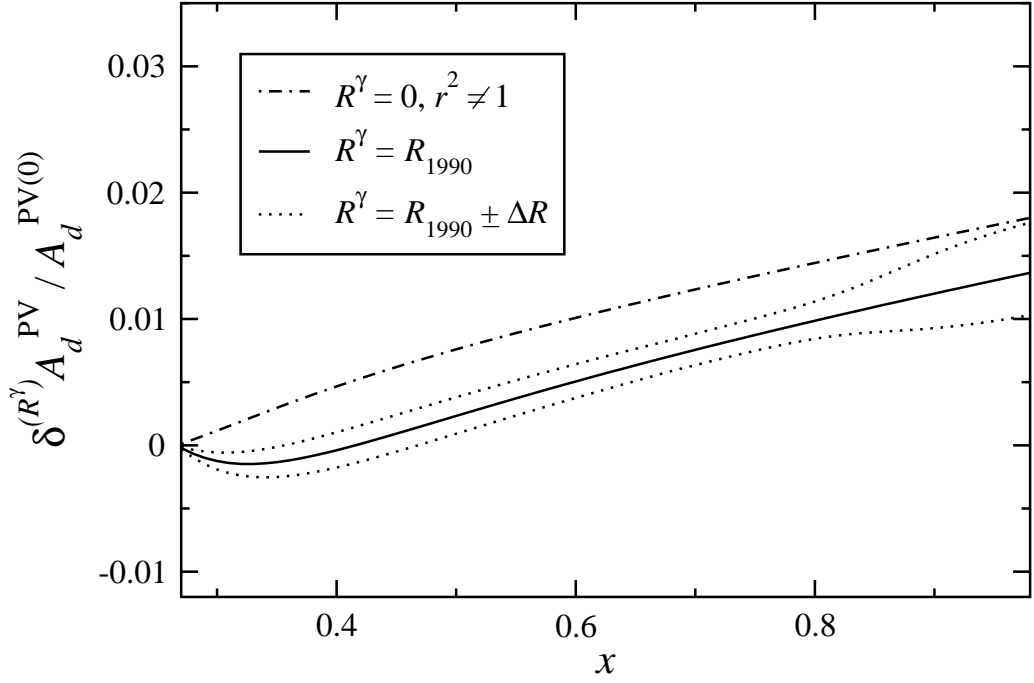}
\caption{Relative effects on the deuteron PV asymmetry
	$A^{\rm PV}_d$ from the electromagnetic ratio $R^\gamma$
	(with $R^{\gamma Z}=R^\gamma$), compared with the Bjorken
	limit asymmetry $A^{\rm PV\ (0)}_d$.
	The full results (solid), for $Q^2=5$~GeV$^2$, are compared
	with those for $R^\gamma=0$ (but $r^2 \neq 1$) (dot-dashed),
	with the dotted curves representing the uncertainty on
	$R^\gamma$ from Ref.~\cite{R1990}.}
\label{fig:dRg_Ad}
\end{figure}

\begin{figure}[h]
\vspace*{1.5cm}
\includegraphics[height=9cm]{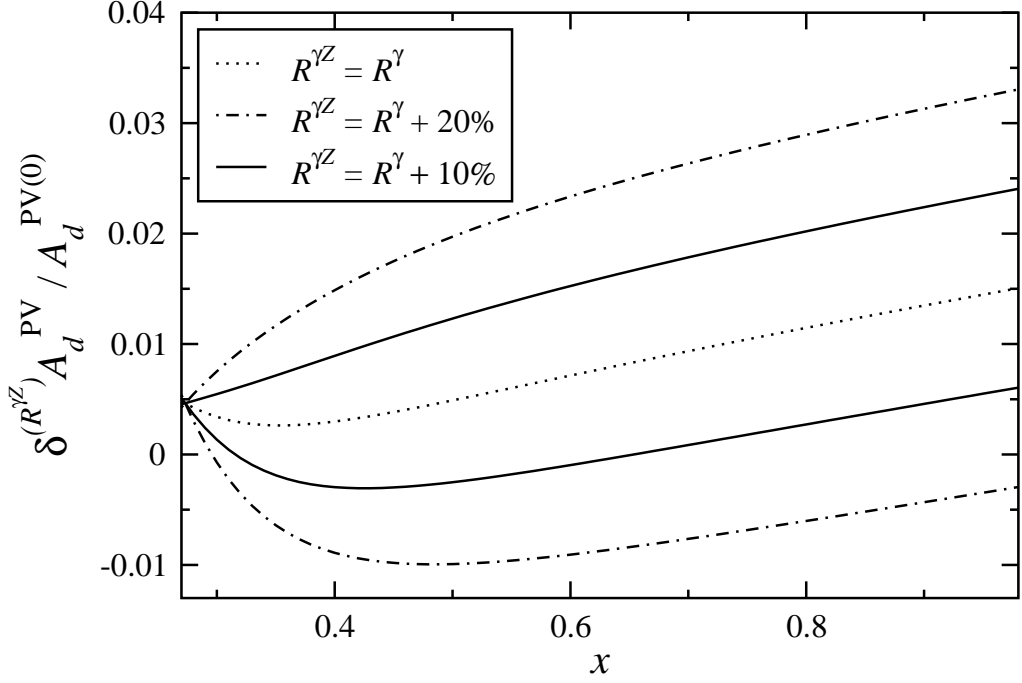}
\caption{Relative effects on the deuteron PV asymmetry $A^{\rm PV}_d$
	from the $\gamma Z$ interference ratio $R^{\gamma Z}$
	compared with the Bjorken limit asymmetry $A^{\rm PV\ (0)}_d$.
	The baseline result for $R^{\gamma Z} = R^\gamma$ (dotted)
	is compared with the effects of modifying $R^{\gamma Z}$ by
	$\pm 10\%$ (solid) and $\pm 20\%$ (dot-dashed),
	for $Q^2=5$~GeV$^2$.}
\label{fig:dRgZ_Ad}
\end{figure}

In Fig.~\ref{fig:dRg_Ad} the relative effect on $A^{\rm PV}_d$
from $R^\gamma$ is shown via the ratio
$\delta^{(R^\gamma)} A^{\rm PV}_d / A^{\rm PV (0)}_d$,
where
$\delta^{(R^\gamma)} A^{\rm PV}_d$ is the difference between the
full asymmetry and that calculated in Bjorken limit kinematics, 
$A^{\rm PV (0)}_d$.
The correction due to $R^\gamma$ is comparatively smaller in the deuteron.
The effect on $A^{\rm PV}_d$ from the purely kinematical $r^2$ correction
in the $Y_3$ term (with $R^\gamma=0$) is an increase of order $1\%$ over
the Bjorken limit asymmetry in the range $0.5 \lesssim x \lesssim 0.9$.
Inclusion of the $R^\gamma$ ratio cancels the correction somewhat,
reducing it to $\lesssim 0$--0.5\% for $x \lesssim 0.6$, and to
$\lesssim 0.5$--1\% for $x > 0.6$.

The effects of a possible difference between $R^{\gamma Z}$ and
$R^\gamma$ are illustrated in Fig.~\ref{fig:dRgZ_Ad} through the
ratio 
$\delta^{(R^{\gamma Z})} A^{\rm PV}_d / A^{\rm PV (0)}_d$,
where
$\delta^{(R^{\gamma Z})} A^{\rm PV}_p$ is the difference between   
the full and Bjorken limit asymmetries.
As for the proton in Fig.~\ref{fig:dRgZ_Ap}, the baseline correction
with $R^{\gamma Z}=R^\gamma$ (dotted curve, equivalent to the solid
curve in Fig.~\ref{fig:dRg_Ad}) is compared with the effects of
modifying $R^{\gamma Z}$ by a constant $\pm 10\%$ (solid) and
$\pm 20\%$ (dot-dashed).
This results in an additional $\approx$ 1\% (2\%) shift of
$A^{\rm PV}_d$ for a 10\% (20\%) modification relative to the
baseline asymmetry for $x > 0.5$.
Such effects will need to be accounted for if one wishes to compare
with SM predictions, or when extracting CSV effects
in PDFs, which we discuss in the next subsection.

\subsection{Charge Symmetry Violation}
\label{sec:deut-CSV}
Aside from these general considerations regarding sub-leading $Q^2$ corrections
to helicity asymmetry measurements from them, scattering from deuteron targets
represents a fertile testing ground for the violation of partonic charge symmetry. 
By definition, charge symmetry is a precise operation involving isospin, which governs
the interchange of protons and neutrons, or equivalently, of up and down quarks. The
charge symmetry operator $\Pcs$ corresponds to a rotation of $180^{\circ}$ about the $2$
axis in isospin space, such that 
\begin{align}
\Pcs &= e^{i\pi T_2} \ , \nonumber \\ 
\Pcs |u\rangle &= -|d\rangle; \hspace{0.6cm} \Pcs |d\rangle = |u\rangle\ .
\label{eq:PCSdef}
\end{align}
It is of particular importance because at low energies, where it has been 
studied extensively, charge symmetry is a far better symmetry than isospin 
in general, typically being respected to better than 
1\%~\cite{Henley:1979,Miller:1990iz,Miller:2006tv}. It is therefore natural to assume 
that charge symmetry is also valid at the partonic level and, indeed, almost 
all analyses of parton distribution functions (PDFs) assume charge symmetry, 
whether the assumption is stated or not. The importance of charge symmetry 
violation in PDFs within the context of tests of the Standard Model has 
recently been of considerable interest~\cite{Bentz:2009yy,Londergan:2003ij}.
In the discussion above an implicit assumption has been made
that charge symmetry is exact --- namely, that the quark distributions
in the proton and neutron are related by $u^p = d^n$ and $u^n = d^p$.
Quark mass differences and electromagnetic effects are expected,
however, to give rise to small violations of charge symmetry in
PDFs, which may be parametrized by:
\begin{subequations}
\begin{eqnarray}
\delta u &=& u^p - d^n\ , \\
\delta d &=& d^p - u^n\ .
\end{eqnarray}
\end{subequations}
Non-zero values of $\delta u$ and $\delta d$ have been predicted in
nonperturbative models of the nucleon \cite{CSVmodels}, and can in
addition arise from radiative QED effects in $Q^2$ evolution 
\cite{MRSTCSV,MRSTQED,GJRQED}.

\begin{figure}[h]
\vspace*{1.5cm}
\includegraphics[height=9cm]{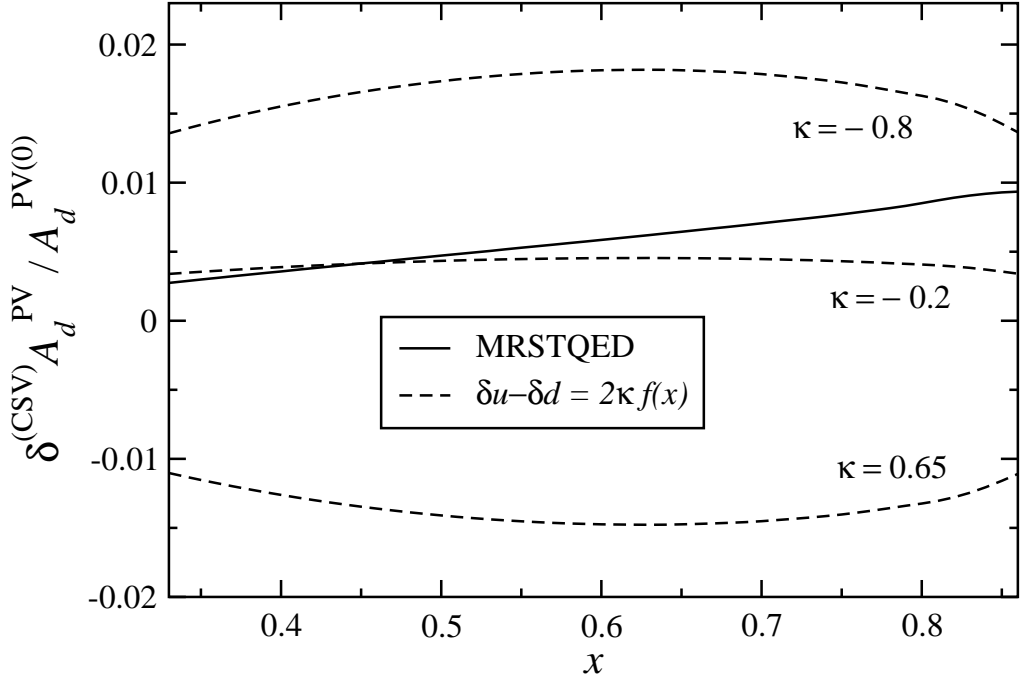}
\caption{Relative effects on the deuteron PV asymmetry $A^{\rm PV}_d$
	of CSV in PDFs, compared with the charge symmetric asymmetry.
	The CSV distributions $\delta u - \delta d$ are from the
	MRSTQED fit \cite{MRSTQED} (solid) and from the parameterization
	$\delta u - \delta d = 2 \kappa f(x)$ (dashed, see text),
	with $\kappa = -0.2$ (best fit), and the two 90\% confidence
	levels, $\kappa = -0.8$ and $\kappa = +0.65$ \cite{MRSTCSV}.}
\label{fig:dCSV}
\end{figure}

It is convenient to define the $u$ and $d$ quark distributions
in the presence of CSV according to \cite{KK}:
\begin{subequations}
\begin{eqnarray}
u\ \equiv\ u^p - \frac{\delta u}{2} &=& d^n + \frac{\delta u}{2}\ , \\
d\ \equiv\ d^p - \frac{\delta d}{2} &=& u^n + \frac{\delta d}{2}\ . 
\end{eqnarray}
\end{subequations}
With these definitions, the deuteron functions $a_1^d$ and $a_3^d$
in the $A_d^{\rm PV}$ asymmetry can be written:
\begin{subequations}
\begin{eqnarray}
a_1^d &=& a_1^{d(0)} + \delta^{\rm (CSV)} a_1^d\ ,	\\
a_3^d &=& a_3^{d(0)} + \delta^{\rm (CSV)} a_3^d\ ,
\end{eqnarray}
\end{subequations}
where $a_1^{d(0)}$ and $a_3^{d(0)}$ are given by Eqs.~(\ref{eq:a1d})
and (\ref{eq:a3d}), respectively.
The fractional CSV corrections are then given by:
\begin{subequations}
\begin{eqnarray}
\label{eq:csva1}
\frac{\delta^{\rm (CSV)} a_1^d}{a_1^{d(0)}}
&=& \left( -\frac{3}{10} + \frac{2 C_{1u} + C_{1d}}{2 (2 C_{1u} - C_{1d})}
    \right)
    \left( \frac{\delta u - \delta d}{u + d} \right)\ ,	\\
\frac{\delta^{\rm (CSV)} a_3^d}{a_3^{d(0)}}
&=& \left( -\frac{3}{10} + \frac{2 C_{2u} + C_{2d}}{2 (2 C_{2u} - C_{2d})}
    \right)
    \left( \frac{\delta u - \delta d}{u + d} \right)\ .
\label{eq:csva3}
\end{eqnarray}
\end{subequations}

\begin{figure}[h]
\vspace*{1.5cm}
\includegraphics[height=9cm]{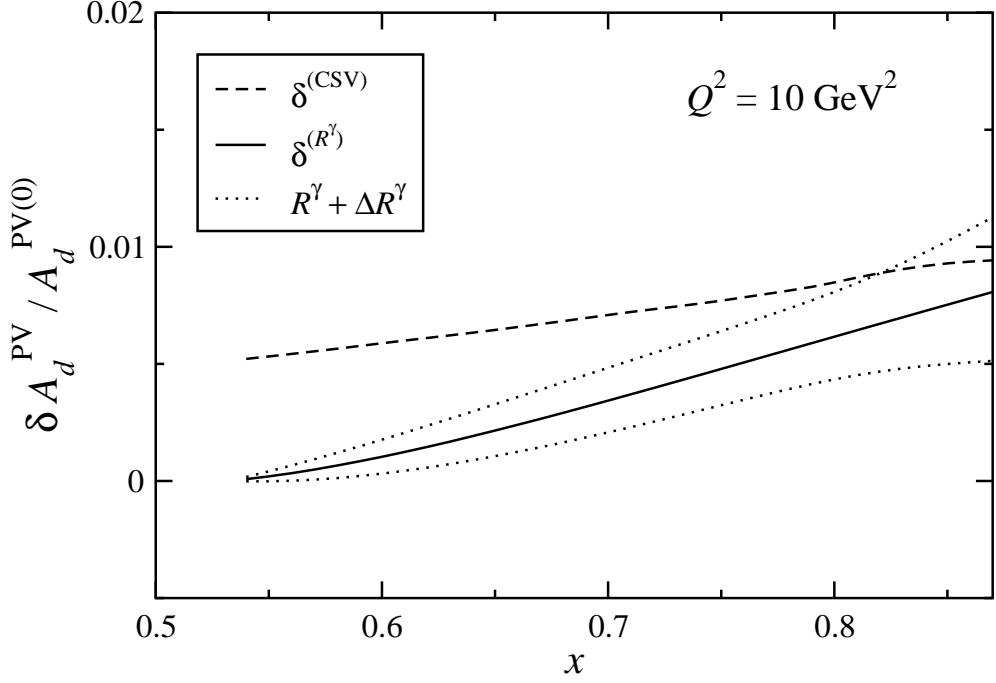}
\caption{Relative effects on the deuteron PV asymmetry $A^{\rm PV}_d$
	from CSV in PDFs \cite{MRSTCSV} (dashed, see Fig.~\ref{fig:dCSV})
	and from $R^\gamma$ \cite{R1990} (with $R^{\gamma Z}=R^\gamma$,
	solid) at $Q^2 = 10$~GeV$^2$, compared with the charge symmetric
	asymmetry in Bjorken limit kinematics.
	The shaded area represents the uncertainty in $R^\gamma$.}
\label{fig:dRg_AdQ10}
\end{figure}

Using these definitions, we plot in Fig.~\ref{fig:dCSV} the effect of
CSV in valence PDFs on the deuteron asymmetry $A_d^{\rm PV}$.
The asymmetry using the MRSTQED paramterization \cite{MRSTQED} of
$\delta u - \delta d$ (solid curve) gives an $\approx 0.5$--1\% effect
for $0.5 \lesssim x \lesssim 0.9$, similar to the effect predicted
from nonperturbative (\EG~bag model) calculations \cite{CSVmodels}. In this
case, the constraint to $\delta u - \delta d$ follows from a QCD global
fit that explicitly incorporates QED corrections due to finite photon distribution functions
$\gamma_p(x,Q^2) \neq \gamma_n(x,Q^2)$. Alternatively, and more directly,
MRST also performed a comparable estimate by means of the phenomenological fit \cite{MRSTCSV}
\begin{align}
\delta u - \delta d\ =\ 2 \kappa \cdot &f(x)\ , \nonumber\\
&f(x)\ =\ x^{-1/2} (1-x)^4 (x-0.0909)\ ,
\label{eq:CSVfit}
\end{align}
where $\kappa$ is a free parameter; this resulted in a similar correction to the 
asymmetry, $\sim 0.5\%$ for most of the $x$ range considered.
The best fit gives $\kappa = -0.2$, although the constraints on
$\kappa$ are relatively weak, with values of $\kappa = -0.8$ and
$+0.65$ giving $\sim 1.5$--2\% effect for $0.5 \lesssim x \lesssim 0.8$
at the 90\% confidence level.

For the central values (best fit parameters), the magnitude of the CSV
effect on the asymmetry at $Q^2 = 5$~GeV$^2$ is similar to that due to
the finite-$Q^2$ kinematics ($r^2 \neq 1$, $R^\gamma \neq 0$) seen in
Fig.~\ref{fig:dRg_Ad}, and may be smaller than that due to possible
differences between $R^{\gamma Z}$ and $R^\gamma$ in 
Fig.~\ref{fig:dRgZ_Ad}.
Unless the finite-$Q^2$ corrections are known to greater accuracy than
at present, they may impede the unambiguous extraction of CSV effects
from the asymmetry.

On the other hand, since the finite-$Q^2$ corrections are expected to
decrease with $Q^2$, a cleaner separation should be possible at larger
$Q^2$, insofar as the CSV effects appear at leading twist.
In Fig.~\ref{fig:dRg_AdQ10} the effect of $R^\gamma$ on $A^{\rm PV}_d$
(solid) is compared with the CSV results \cite{MRSTCSV} for different
$\kappa$ values (dashed) at $Q^2 = 10$~GeV$^2$.
The deviation from the Bjorken limit kinematics of the
$\delta^{(R^\gamma)}$ curve is clearly less than the corresponding
result at $Q^2 = 5$~GeV$^2$ in Fig.~\ref{fig:dRg_Ad} (the shaded region
here indicates the uncertainty in $R^\gamma$), whereas the CSV results
are similar to those at the lower $Q^2$.
The contrast is especially striking at $x \sim 0.6$, where the CSV
effects are several times larger than the correction to $A^{\rm PV}_d$
due to $R^\gamma$.
At larger $x$ the CSV effects for the central $\kappa$ value become
comparable to the $R^\gamma$ uncertainty, however, and the 90\% confidence
level corrections ($\kappa = -0.8$ and +0.65) are of the order 2\%
and are still several times larger than the $R^\gamma$ uncertainty.

These results suggest that if the CSV effects in PVDIS from the
deuteron are of the order $\sim 0.5\%$, the optimal value of $x$
to observe them would be $x \sim 0.6$ at $Q^2=10$~GeV$^2$.
If the CSV effects are of order $\sim 2\%$, they should be clearly
visible over a larger $x$ range, even up to $x \approx 0.8$.
Note that the minimum value of $x$ attainable at the $Q^2=10$~GeV$^2$
kinematics ($x \approx 0.53$) is somewhat smaller than at the lower
$Q^2$ vales because at fixed incident energy and $Q^2$, the fractional
lepton energy loss exceeds unity at higher $x$.
%

%% file: the-CSV.tex
%

\subsection{Partonic CSV at high $Q^2$}
\label{sec:deut-pCSV}

Of course, although it has been observed in hadronic processes \cite{Stephenson:2003dv},
no unambiguous violation of charge symmetry has been seen at the partonic level
thus far; this is aside from the fact that the one QCD global analysis that allowed for
CSV converged upon a non-zero effect --- albeit with very large errors~\cite{MRST03}.
The current upper limits are consistent with the validity of partonic charge 
symmetry in the range 5-10\%~\cite{Londergan:2009kj}. Theoretical models as we have
seen tend to produce estimations of charge symmetry violation (CSV) in PDFs which 
for many observables give effects at roughly 
the 1\% level~\cite{Londergan:2009kj,Lo98a}. As just argued, this reality presents
experimentalists with a considerable challenge, first to observe effects of this magnitude, and 
then to isolate the signal from competing effects of similar size.

In marked contrast to the situation at moderate $Q^2$ outlined in Sec.~\ref{sec:deut-CSV},
a new facility has recently been proposed that would collide electrons 
or positrons from an electron accelerator with protons or deuterons from 
the LHC~\cite{LHeC}. In this paper we will show that such a 
facility (given the name LHeC) has the potential to produce charge symmetry 
violating effects which are considerably larger than those expected with 
other facilities. We will review the effect in question, show the results 
of theoretical calculations for the proposed CSV effects, and discuss why 
they ought to be expected to be relatively large at energies 
accessible to an electron-ion collider. It should be noted we make no
presumptions regarding the likelihood or possible timetable for the
construction of an LHeC [or similar electron-ion collider (EIC)]; in fact
a wide range of proposals \cite{Accardi:2012qut} and suggestions have
been made. Rather, we aim only to demonstrate the physics capabilities
of an archetypal EIC, and thereby provide some motivation for the experimental
searches that might be undertaken at such a hypothetical facility.

\begin{figure}[ht]
\includegraphics[width=5.5in]{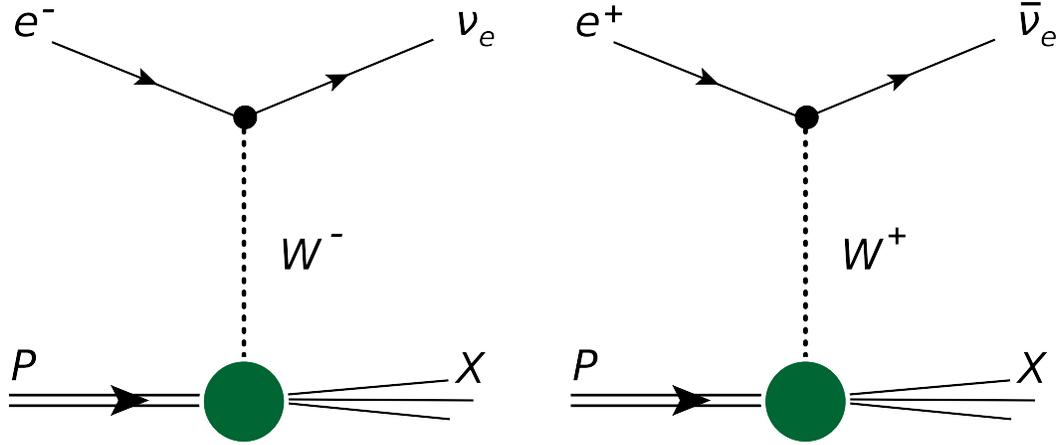}
\caption{Schematic picture of charged-current neutrino production in DIS 
induced by an electron or positron on a proton.} 
\label{Fig:fig1}
\end{figure}

The reactions of interest are the charged current (CC) cross sections for 
electron and positron deep inelastic scattering at energies in the range 50-100 GeV 
on protons and deuterons at LHC energies, \IE~several TeV. These are important 
because they directly and unambiguously probe the flavor structure of the 
proton PDFs in the valence region. Consider the 
deep inelastic reaction $(e^-,\nu_e)$, whereby a high-energy electron incident 
on a proton produces a neutrino. The process results from a $W^-$ which 
is absorbed on quarks from the proton, as shown schematically in 
Fig.~\ref{Fig:fig1}, while the final hadronic state is characteristically unobserved. The signature 
for this process is disappearance of the electron, together with very large 
deposition of energy in the hadronic sector. 

The $F_2$ structure function for the CC reaction on a proton has the form   
\be
F_2^{W^- p}(x) = 2x[ u(x) + c(x) + \dbar(x) + \sbar(x)] .  
\label{eq:F2ep}
\ee
Of course, these reactions will occur at extremely high energies 
and very large $Q^2$, such that we assume corrections to the $F_2$ structure 
functions in Eq.~(\ref{eq:F2ep}) [and others]
arising from quark mixing matrices, quark masses, or higher twist effects 
to be completely negligible. 

We can also consider the corresponding reaction for positrons on protons, 
$(e^+,\nubar_e)$. This reaction involves the absorption of a $W^+$ on the 
proton, with the resulting $F_2$ structure function  
\be
F_2^{W^+ p}(x) = 2x[ \ubar(x) + \bar{c}(x) + d(x) + s(x)] .  
\label{eq:F2ePp}
\ee
Using Eqs.~(\ref{eq:F2ep} \& \ref{eq:F2ePp}), we can straightforwardly calculate
the $F_2$ structure functions (per nucleon) on the deuteron, finding  
\bea
F_2^{W^- D}(x) &=& x[ u^+(x) + d^+(x) + 2c(x) + 
2\sbar(x)- \delta d(x) - \delta\ubar(x)]\ ; \nonumber \\  
F_2^{W^+ D}(x) &=& x[ u^+(x) + d^+(x) + 2\bar{c}(x) + 2s(x)- 
\delta\dbar(x) - \delta u(x)]\ .  
\label{eq:F2eD}
\eea

Note that in Eq.~(\ref{eq:F2eD}) we have introduced combinations of quark parton distribution 
functions (PDFs) that are even or odd under charge conjugation, as well as the CSV 
PDFs 
\bea
 q^{\pm}(x) &=&  q(x) \pm \bar{q}(x) \ ; \nonumber \\ 
\delta u(x) &=& u^p(x) - d^n(x) \ ; \nonumber \\ 
\delta d(x) &=& d^p(x) - u^n(x) \ .  
\label{qpmdef}
\eea
There are analogous relations to Eq.~(\ref{qpmdef}) for the antiquark CSV 
PDFs, and for the remainder of this section we assume that $\cbar(x) = c(x)$
[actually, deviations away from this hypothesis will be presented systematically
in Chap.~\ref{chap:ch-charm}]. 
The distributions $q^-(x)$, which involve the differences between quark 
and antiquark PDFs (alternatively, they are the $C$-odd combinations of 
quark distributions), are the \textit{valence} parton distributions for a 
given quark flavor.   

We now define the following quantity (the master-expression of this section), 
\be
R^-(x) \equiv \frac{2(F_2^{W^- D}(x) - F_2^{W^+ D}(x))}{F_2^{W^- p}(x) + 
F_2^{W^+ p}(x)} .
\label{eq:Rmin}
\ee
The quantity $R^-(x)$ is given by the difference in the $F_2$ structure 
functions per nucleon for electron-deuteron and positron-deuteron CC 
reactions, divided by the average $F_2$ structure function for CC 
reactions on protons initiated by electrons and by positrons.  

Using Eqs.~(\ref{eq:F2ep}), (\ref{eq:F2ePp}) and (\ref{eq:F2eD}) we can 
straightforwardly show that the quantity $R^-(x)$ in (\ref{eq:Rmin}) has the 
form 
\be
 R^-(x) = \frac{x[ -2s^-(x) + 
  \delta u^-(x) - \delta d^-(x)]}{x[u^+(x) + d^+(x) + s^+(x) + 2c(x)]} \ .
\label{eq:F2rat}
\ee 
Thus $R^-(x)$ is proportional to the valence quark CSV parton distributions 
plus the strange quark asymmetry (the difference between the strange and 
antistrange PDFs). Insofar as the strange quark asymmetry exists, it should 
be large only at quite small Bjorken $x < 0.1$, while theoretical estimates 
of the valence CSV parton distributions 
\cite{Sather:1991je,Rodionov:1994cg} suggest that for $Q^2 \sim 10$ 
GeV$^2$ they peak at values $x \sim 0.4$.     

\begin{figure}[ht]
\includegraphics[height=6.5cm]{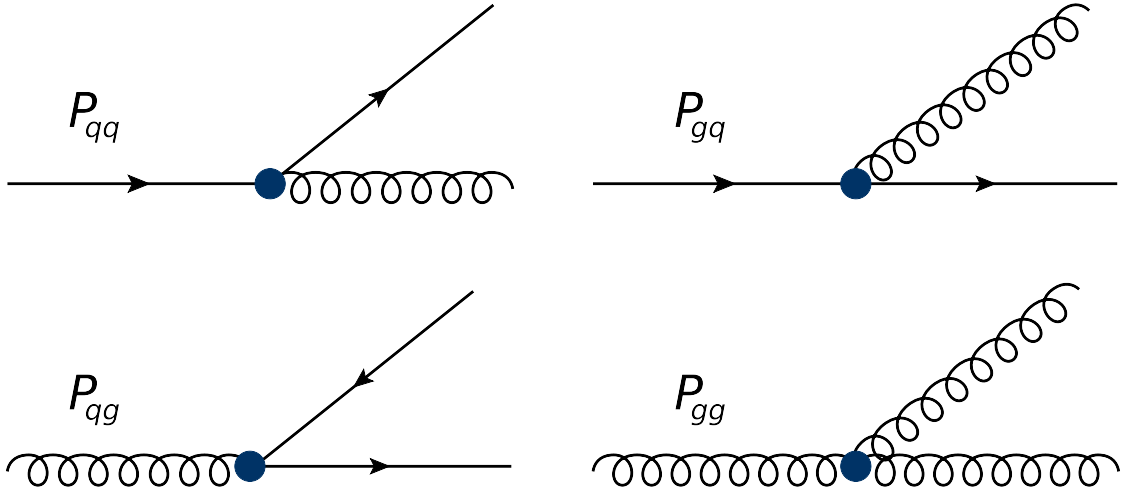}
\caption{
Schematic picture of quarks coupling to QCD gluons. This gives the 
origin of QCD splitting relevant for $C$-even parton 
distributions and gluons.
}
\label{fig:QCDsplt}
\end{figure}

\begin{figure}[ht]
\includegraphics[height=6cm]{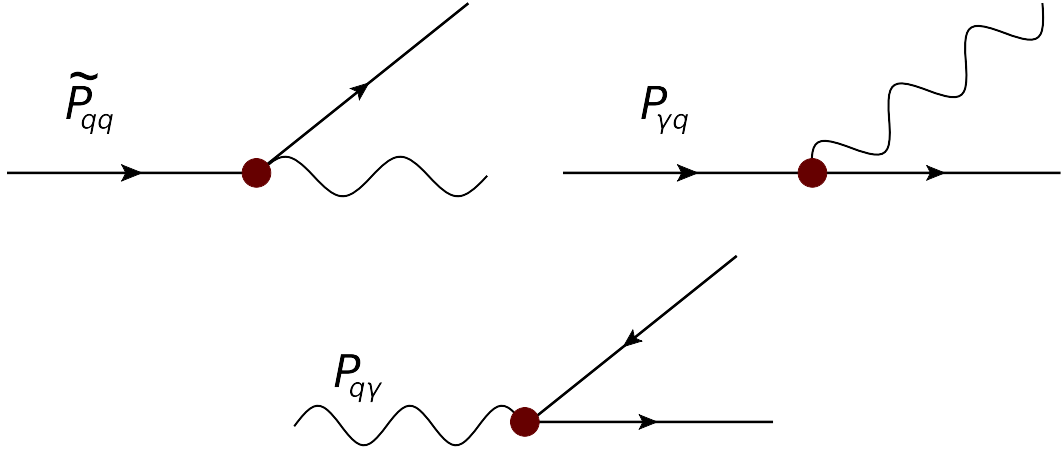}
\caption{Schematic picture of quarks coupling to photons. This gives the 
origin of the QED splitting which produces CSV effects in parton 
distribution functions.
\label{fig:QEDsplt}}
\end{figure}

We thus consider a hypothetical collider with 50 GeV electrons or positrons colliding 
on protons and deuterons of roughly 7 TeV energy ---
similar to what might be achievable at a putative `LHeC-type' machine.  
More specifically, we consider charged-current reactions at such a facility with $Q^2 = 10^5$ 
GeV$^2$, and simulate the two dominant mechanisms for charge symmetry violation 
in parton distribution functions. The first of these arises from radiation 
of a photon by a quark, such as is shown schematically in 
Fig.~\ref{fig:QEDsplt}; such contributions were first calculated systematically by MRST
\cite{Martin:2004dh} and Gl\"uck \EA~\cite{Gluck:2005xh}. QED corrections of this kind are 
analogous to the familiar couplings of gluons to quarks portrayed in Fig.~\ref{fig:QCDsplt}, 
aside from the fact that photons do not exhibit self-coupling as do gluons. 
Inclusion of these `QED splitting' terms will necessarily induce charge symmetry 
violation in parton distribution functions due to the differing electromagnetic (EM) 
couplings of the photon to up and down quarks.  

The behavior of parton distributions with increasing $Q^2$ is given by the 
DGLAP evolution equations 
\cite{Dokshitzer:1977sg,Gribov:1972ri,Altarelli:1977zs}. We 
expand the DGLAP evolution equations to lowest order in both the 
strong coupling $\alS$ and the electromagnetic coupling $\alpha$,  
\bea
\frac{\partial q_i(x,\mu^2)}{\partial \ln \mu^2} &=& \aSpi \left[ 
  P_{qq}\otimes q_i + P_{qg}\otimes g \right] + 
  \api e_i^2 \widetilde{P}_{qq}\otimes q_i \ ; \nonumber \\ 
 \frac{\partial g(x,\mu^2)}{\partial \ln \mu^2} &=& \aSpi \left[ 
  \sum_j P_{gq}\otimes q_j + P_{gg}\otimes g \right] + 
  \api e_i^2 \widetilde{P}_{qq}\otimes q_i \ ; \nonumber \\ 
\frac{\partial \gamma(x,\mu^2)}{\partial \ln \mu^2} &=& 
  \api \sum_j e_j^2 P_{\gamma q}\otimes q_j \ . 
 \label{eq:DGLAP}
\eea
In Eq.~(\ref{eq:DGLAP}), $q_i(x,\mu^2)$ is the parton distribution 
for a given flavor $i$, $g(x,\mu^2)$ is the gluon distribution, and 
$\gamma(x,\mu^2)$ the photon parton distribution mentioned earlier \cite{Martin:2004dh};
these probability distributions are collectively bounded at some starting
scale $Q^2=\mu^2$ by a global momentum conservation rule of the form
\begin{equation}
\int_0^1 dx\ x \left\{ \sum_q q(x,\mu^2)\ +\ g(x,\mu^2)\ +\ \gamma(x,\mu^2) \right\}\ \equiv\ 1\ ,
\end{equation}
where the convolutions in Eq.~(\ref{eq:DGLAP}) we take to be
\be
P\otimes q = \int_x^1 \, \frac{dz}{z} P(z)\ q(\frac{x}{z}, \mu^2) \ .
\label{eq:convol}
\ee
The splitting functions themselves are given by  
\bea
 \widetilde{P}_{qq}(z) &=& \frac{P_{qq}(z)}{C_F}\ ; \hspace{0.3 cm} 
  P_{\gamma q}(z)\ =\ \frac{P_{gq}(z)}{C_F}\ ; \nonumber \\    
  P_{q\gamma}(z) &=& \frac{P_{qg}(z)}{T_R}\ ; \hspace{0.3 cm} 
  P_{\gamma \gamma}(z)\ =\ \sum_j \frac{-2e_j^2}{3} \delta (1-z)\ .
\label{eq:Pdef}
\eea

Fortunately, the QCD probability distributions above are directly calculable from the amplitudes given in
Fig.~\ref{fig:QCDsplt}; at leading order we have
\begin{subequations}
\begin{equation}
P_{qq}(z) = 2\delta(1-z) - {4 \over 3}(1+z)
          + {8 \over 3} \left( {1 \over 1-z} - \delta(1-z) \cdot \int_0^1 dz' {1 \over 1-z'} \right)\ ,
\end{equation}
\begin{equation}
P_{gq}(z) = {4 \over 3}\left\{1 + (1-z)^2 \right\} \big/ z\ ,
\end{equation}
\begin{equation}
P_{qg}(z) = {1 \over 2}\left\{z^2 + (1-z)^2 \right\} \big/ z\ ,
\end{equation}
\begin{align}
P_{gg}(z) = \left( {11 \over 2} - {N_F \over 3} \right)\delta(1-z) &+ 6\ \bigg\{ z(1-z) + {1 \over z} - z \nonumber\\
          &+\ \left( {1 \over 1-z} - \delta(1-z) \cdot \int_0^1 dz' {1 \over 1-z'} \right) \bigg\}\ ,
\end{align}
\label{eq:QCD_DGLAP}
\end{subequations}
while relations analogous to Eqs.~(\ref{eq:DGLAP} \& \ref{eq:QCD_DGLAP}) hold for antiquarks. Taking 
the valence combinations from Eq.~(\ref{qpmdef}), we obtain scaling relations for up 
and down valence quarks  
\bea
\frac{\partial u^-(x,\mu^2)}{\partial \ln \mu^2} &=&  
 \aSpi P_{qq}\otimes u^- + \frac{2\alpha}{9\pi}\widetilde{P}_{qq}\otimes u^- \ ; 
  \nonumber \\ 
\frac{\partial d^-(x,\mu^2)}{\partial \ln \mu^2} &=&  
 \aSpi P_{qq}\otimes d^- + \frac{\alpha}{18\pi}\widetilde{P}_{qq}\otimes d^- 
 \label{eq:valevol}
\eea
For the valence CSV parton distributions we desire on the other hand, since $\delta u^-(x) = u_p^-(x) - 
d_n^-(x)$, from Eq.~(\ref{eq:valevol}) we may obtain the evolution equations to lowest order in $\alS$
and $\alpha$,       
\bea
\frac{\partial[\delta u^-(x,\mu^2)]}{\partial \ln \mu^2} &\approx& 
  \api (e_u^2 - e_d^2)\widetilde{P}_{qq}\otimes u^- \ ; 
  \nonumber \\ 
 \frac{\partial[\delta d^-(x,\mu^2)]}{\partial \ln \mu^2} &\approx& 
  -  \api (e_u^2 - e_d^2)\widetilde{P}_{qq}\otimes d^- \ . 
 \label{eq:CSVtrunc}
\eea

Eqs.~(\ref{eq:valevol} \& \ref{eq:CSVtrunc}) describe how the valence and CSV quark
distributions respectively evolve with $Q^2$, to lowest order in both $\alS$ and $\alpha$.
Note that the total strength of the CSV effect as embodied by the first moments $\int_0^1dx\ \delta q^-$ is
preserved (approximately) under $Q^2$ evolution by Eq.~(\ref{eq:RSach}). Physically,
the above formalism establishes that, with increasing $Q^2$,  
partons radiate gluons and photons which carry off momentum; since the total 
momentum fraction carried by quarks is given by the second moment of the parton 
distributions, as $Q^2$ increases, the parton distribution functions will shift 
towards progressively smaller $x$.

Comparison of Eqs.~(\ref{eq:valevol}) and (\ref{eq:CSVtrunc}) makes clear that the 
radiation from valence quarks will be greater than that from the valence CSV 
distributions. This occurs because to lowest order in $\alS$ and $\alpha$, 
valence quark evolution contains contributions from both gluon and photon radiation, 
whereas the valence CSV distribution has only a single term from photon radiation. 
This suggests that with increasing $Q^2$, the valence parton  
distributions would experience a larger shift to low $x$ than would the valence 
CSV distributions. We note that the quantity $R^-(x)$ defined in 
Eq.~(\ref{eq:F2rat}) is proportional to the ratio of valence CSV distributions to 
valence PDFs, at a given $x$ value. Thus, if the CSV valence distributions are becoming 
large relative to the valence PDFs at high $Q^2$, we expect the quantity 
$R^-(x)$ to grow as $Q^2$ increases; specifically one would expect the ratio to 
increase logarithmically with $Q^2$. 

Eq.~(\ref{eq:CSVtrunc}), the QCD evolution equations for the valence CSV parton 
distributions, have been solved by Gl\"uck \EA~\cite{Gluck:2005xh} and also by
MRST \cite{Martin:2004dh}; both made slightly different 
approximations for the initial conditions, and we point out that while the effect of 
photon radiation is clear, it is far less obvious that the boundary conditions 
imposed on the calculations are appropriate. That is to say, it is not unambiguous
{\it a priori} that low scales typical of quark models necessarily represent
the appropriate place to set the CSV effect to zero. Therefore, in the absence of a compelling
theoretical derivation it would extremely helpful to test the idea experimentally.

As stated, MRST \cite{Martin:2004dh} confronted this method with experimental data by
explicitly including QED effects in the DGLAP equations via the photon 
distribution $\gamma(x,Q^2)$ introduced in Sec.~\ref{sec:deut} with a scale dependence
determined by Eq.~(\ref{eq:DGLAP}). 
Specifically, MRST aimed to identify this quantity in the direct photon production
process $ep \rightarrow e\gamma X$ where the final state $e$ and $\gamma$ are produced with equal and 
opposite large transverse momentum. This mechanism has been observed by the ZEUS 
Collaboration in $ep$ collisions at $\sqrt{s}= 300$ and 318 GeV 
\cite{Chekanov:2004wr}. The observed cross sections were in reasonable 
agreement with the MRST calculations but disagreed with calculations done using
the Monte Carlo simulations PYTHIA \cite{Sjostrand:2000wi} and HERWIG 
\cite{Marchesini:1991ch}. Again, it would hence be useful to have other experimental tests 
of this method for including radiation of photons by partons, and the 
experiment suggested here could provide additional confirmation of this method. 
 
The second source of valence parton CSV we consider arises naturally from the 
mass difference between the $u$ and $d$ quarks and may be calculated within
light cone quark models.
Formally, models~\cite{TWbook,Jaffe:1983hp,Signal:1989yc} generally specify the quark distributions in a
fashion similar to the derivation of $\Phi_{q,\bar{q}}$ in Chap.~\ref{chap:ch-DIS}.\ref{sec:DIS}: 
\begin{subequations}
\begin{align}
q (x, \mu^2)\ &=\ {1 \over 4\pi} \ \int d\xi e^{-iMx\xi}\ \langle N | \gamma^+ \psi^\dagger (\xi^-)
 \gamma^+ \psi(0) | N\rangle\ \hspace*{1cm} \implies \\
&=\ {1 \over 2} \sum_X \, \int {d^3 k \over 2E_k (2\pi)^3}\ |\langle X |\frac{1+ \gamma^0\gamma^3}{2} 
 \psi(0) | N\rangle |^2\, \times\, \delta (M(1-x) - p_X^+)\ , \ \ \mathrm{and} \\
\bar{q} (x, \mu^2)\ &=\ {1 \over 2} \sum_X \, \int {d^3 k \over 2E_k (2\pi)^3}\ |\langle X |\frac{1+ \gamma^0\gamma^3}{2} 
 \psi^\dagger(0) | N\rangle |^2\, \times\, \delta (M(1-x) - p_X^+)\ ,
\end{align}
\label{eq:LF-qx}
\end{subequations}
by which we conclude the light-front result for the $q_v(x) = q(x) - \bar{q}(x)$ valence-type distributions to be
\begin{align}
q_v(x, \mu^2)\ &= M \sum_X \, |\langle X |\frac{1+ \gamma^0\gamma^3}{2} 
 \psi_v(0) | N\rangle |^2\, \times\, \delta (M(1-x) - p_X^+)\ ,
\label{eq:qvx}
\end{align}
where we have defined the field of the valence quark as $\psi_v \defeq \psi - \psi^\dagger$, and an explicit linear
dependence on the nucleon mass $M$ has issued due to the fact that $dk^+ = Mdx$ in the integration measures of Eq.~(\ref{eq:LF-qx}).

Technically, Eq.~(\ref{eq:qvx}) denotes the process where a valence quark is removed 
from a nucleon $|N\rangle$, and the result is summed over all final states 
$|X\rangle$. The quantity $p_X^+$ is the energy of the state following removal 
of a valence quark with momentum $k$, while $\mu^2$ represents the 
starting value for the $Q^2$ evolution of the 
parton distributions. Eq.~(\ref{eq:qvx}) is formally exact and provides a natural 
starting point for calculations which provide the correct support to the 
PDFs. 

Model quark wavefunctions are found to be nearly invariant under the small 
mass changes typical of CSV~\cite{Rodionov:1994cg}, so we concentrate on the 
breaking of partonic charge symmetry associated with energy shifts resulting 
from $u$ and $d$ quark  
mass differences. In particular, we consider the effect of 
the $n-p$ mass difference $\delta M \equiv 
M_n - M_p = 1.3$ MeV, as well as the difference in diquark masses arising from the
current quark mass difference between up and down quarks. We define the 
quantity 
\be 
\dwtilm = m_{dd} - m_{uu} \, ,
\label{eq:mtilde}
\ee
for which we have a robust estimate $\dwtilm \sim 4$ MeV \cite{Bickerstaff:1989ch}. 
We determine CSV valence PDFs by calculating the variation of quark model 
parton distributions in Eq.~(\ref{eq:qvx}) 
with respect to these quantities, \IE 
\be 
  \delta q_v \approx \frac{\partial q_v}{\partial (\dwtilm)}\dwtilm 
  + \frac{\partial q_v}{\partial (\delta M)}\delta M \ .
\label{eq:iso}
\ee
Using the last relation, the valence charge symmetry violating parton 
distributions may be obtained by evaluating variations with respect to diquark 
and nucleon masses of quark model valence parton distributions. 
The resulting PDFs then account for the quark and nucleon mass differences 
that generate CSV effects. 
  
 
In practice, we compute using the approach first developed by Sather \cite{Sather:1991je}, 
which is founded upon a static quark picture that ignores partonic transverse momentum $k_\perp$.
On the down side, such models do not necessarily ensure
the desire support in the quark distributions, but for our purposes the Sather prescription gives numerically
similar results when compared with alternative models \cite{Rodionov:1994cg} that do not possess this limitation.
By applying Eq.~(\ref{eq:iso}) to Eq.~(\ref{eq:qvx}) within this scheme, an analytic approximation relating 
valence quark CSV to derivatives of the valence PDFs emerges. For instance, the Sather prescription approximates
the CSV effect in the $u_v(x)$ distribution due to the nucleon mass difference $\delta M$ as
\begin{equation}
\delta u^M(x,Q^2_0)\ \approx\ -{|\delta M| \over M}\ \left( u_v(x,Q^2_0) + \{x-1\} \cdot {du_v(x,Q^2_0) \over dx} \right)\ .
\end{equation}
As a rule, the analytic approximation of Sather is appropriate only at $Q^2$ values appropriate 
for quark model calculations, \IE~$Q^2 \sim 0.25-0.5$ GeV$^2$.    

We hence proceed with the Sather prescription, differentiating valence parton distribution 
functions to obtain valence CSV PDFs. For this purpose we use the MRST2001 
parton distributions \cite{Martin:2001es} at the starting scale, $Q_0^2 = 1$ 
GeV$^2$. This is slightly too large a value of $Q^2$ for the validity of 
Sather's analytic approximation, but 
the resulting errors should be small. We then insert the resulting 
CSV PDFs into the DGLAP evolution equations and evolve to the $Q^2$ 
appropriate for the electron collider experiments, finding results similar 
to those obtained using the CSV distributions of Rodionov \EA. 

Since the CSV effects arising from QED splitting effects and from quark 
mass differences are essentially independent, we simply add the two effects 
directly to produce an overall CSV effect.

Finally, there is one additional term that enters into the quantity $R^-(x)$ of 
Eq.~(\ref{eq:F2rat}), namely,
the strange quark momentum asymmetry~\cite{Signal:1987gz,Thomas:2000ny}
\be 
xs^-(x) \equiv x[s(x) - \overline{s}(x)]\ .
\label{eq:sminus}
\ee
Strange (antistrange) parton distributions can be measured through 
opposite-sign dimuon production initiated by neutrinos (antineutrinos). 
A neutrino undergoes a charged-current reaction, producing a $\mu^-$ and 
a $W^+$, which is then absorbed on an $s$ quark producing a charm quark. The 
charm quark subsequently undergoes a semileptonic decay producing a 
$\mu^+$ and an $s$ quark. The cross section for this process is 
proportional to the strange quark distribution. The corresponding reaction 
initiated by an antineutrino measures the antistrange PDF. 

Dimuon cross sections have been measured by the CCFR \cite{Bazarko:1994tt} 
and NuTeV \cite{Goncharov:2001qe} experiments, by which one 
might extract the quantity $xs^-(x)$. These analyses have been undertaken 
by five groups: CTEQ \cite{Lai:2007dq}; Mason \EA~\cite{Mason:2007zz}; 
the NNPDF Collaboration \cite{Ball:2009mk}; MSTW \cite{Martin:2009iq}; 
and Alekhin, Kulagin and Petti \cite{Alekhin:2009mb}. Hence, to control
the potential contributions to $R^-$ from strangeness asymmetries, we
exploited the NuTeV neutrino analysis of Mason \EA~by
constructing an analytic fit to the results of Ref.~\cite{Mason:2007zz}
at $Q^2 = 16$ GeV$^2$ with the form 
\be 
xs^-(x) = A x^b \exp{(-cx)} \cdot (x - 0.004) \ .
\label{eq:sfit}
\ee
The resulting strange quark asymmetry was then inserted into the DGLAP evolution 
equation and evolved to the putative LHeC scale. 

 \begin{figure}[ht]
\hspace*{-0.2cm}
\includegraphics[width=8.6cm]{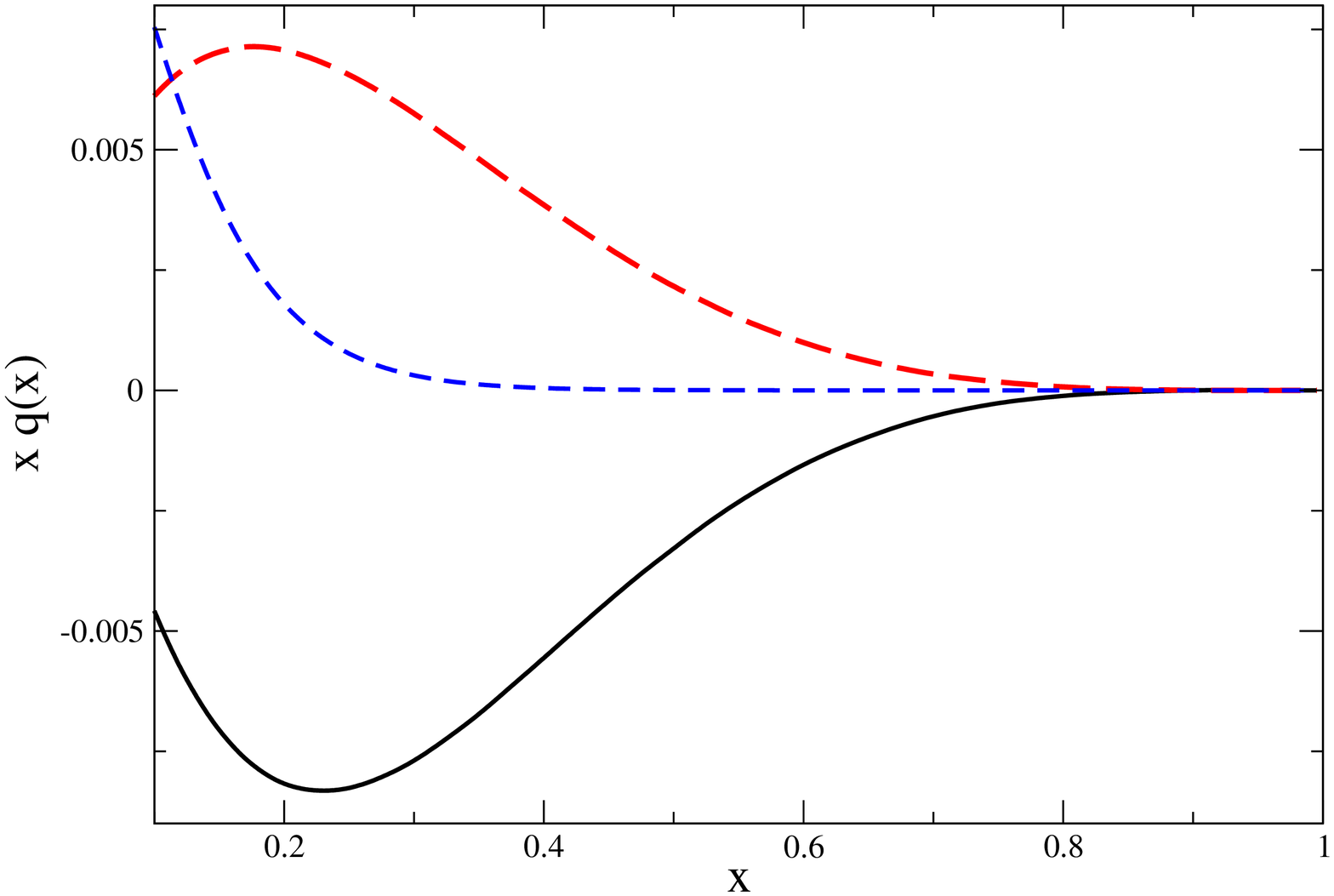}
\includegraphics[width=8.6cm]{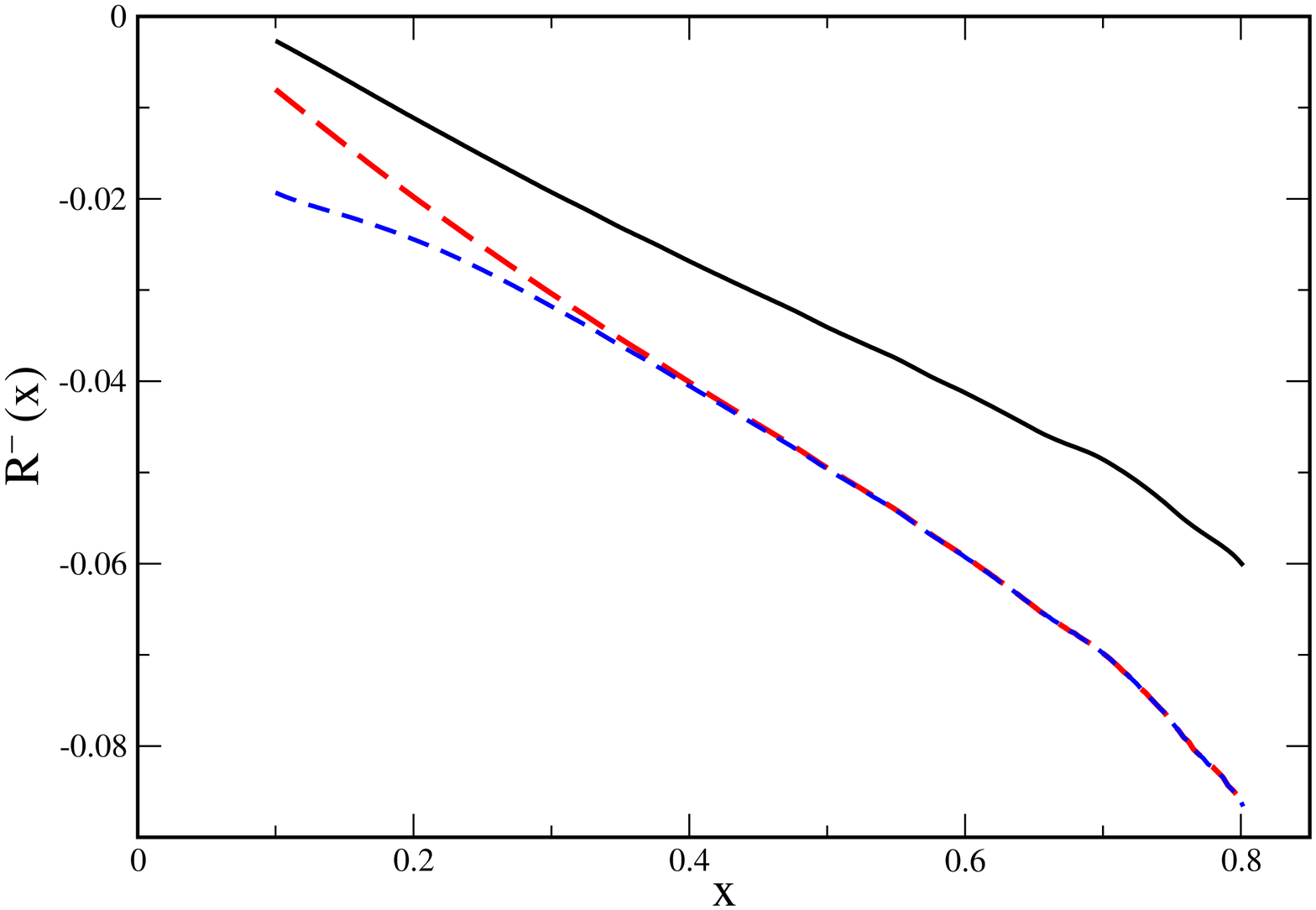}
\caption{(Left) Parton distributions that occur in the numerator 
of Eq.~(\protect\ref{eq:F2rat}). Solid curve: $x\delta u^- (x)$; long-dashed 
curve: $x\delta d^- (x)$; short-dashed curve: $xs^-(x)$. The PDFs have been evolved 
to $Q^2 = 10^5$ GeV$^2$. (Right) Contributions to the quantity $R^-(x)$ vs.~$x$ from 
Eq.~(\protect\ref{eq:F2rat}), where the PDFs are evolved to $Q^2 = 10^5$ GeV$^2$. 
Solid curve: contribution from QED splitting parton CSV term only; long-dashed 
curve: includes contribution also from quark mass CSV term; short-dashed curve: 
contribution from all terms including strange quark asymmetry.  
\label{Fig:numer}}
\end{figure}

Having gathered the requisite pieces, the parton distribution functions that occur in the numerator of 
Eq.~(\ref{eq:F2rat}) are plotted in the left panel of Fig.~\ref{Fig:numer}. The solid black curve 
is $x\delta u^- (x)$, the red long-dashed curve is $x\delta d^- (x)$, and the 
blue short-dashed curve is $xs^-(x)$. As one might expect, 
the valence CSV distributions peak at a relatively large 
value $x \sim 0.2$, while the strange quark asymmetry peaks at an extremely 
small $x$ value. Note that due to valence quark normalization [as defined in
Eq.~(\ref{eq:val_sum})], all of 
these quantities must have zero first moment, \IE~$\langle q(x) \rangle 
= 0$, where $q = [\delta u^-, \delta d^-, s^-]$. The strange quark 
asymmetry has zero first moment because the proton has no net strangeness; 
the valence CSV distributions must have zero first moment because otherwise 
this would change the total number of valence quarks in the neutron. As a result,
each of these curves must cross zero at a small value of $x$ (not shown in 
Fig.~\ref{Fig:numer}).

Another notable point is that the signs of these quantities are such that (for 
values of $x$ above the crossover point for all of the parton 
distributions)   
all three contributions should add together in the numerator of 
Eq.~(\ref{eq:F2rat}). The right panel of Fig.~\ref{Fig:numer} shows the expected value of 
$R^-(x)$ vs.~Bjorken $x$. The solid curve on the right of Fig.~\ref{Fig:numer} includes 
only the QED splitting contribution to partonic CSV. The long-dashed curve 
includes both QED splitting and quark mass contributions to valence quark 
CSV, while the short-dashed curve includes all three terms in the 
numerator of Eq.~(\ref{eq:F2rat}), including also the contribution from 
strange quark asymmetry.

\begin{figure}[h]
\vspace*{1.4cm}
\includegraphics[height=9cm]{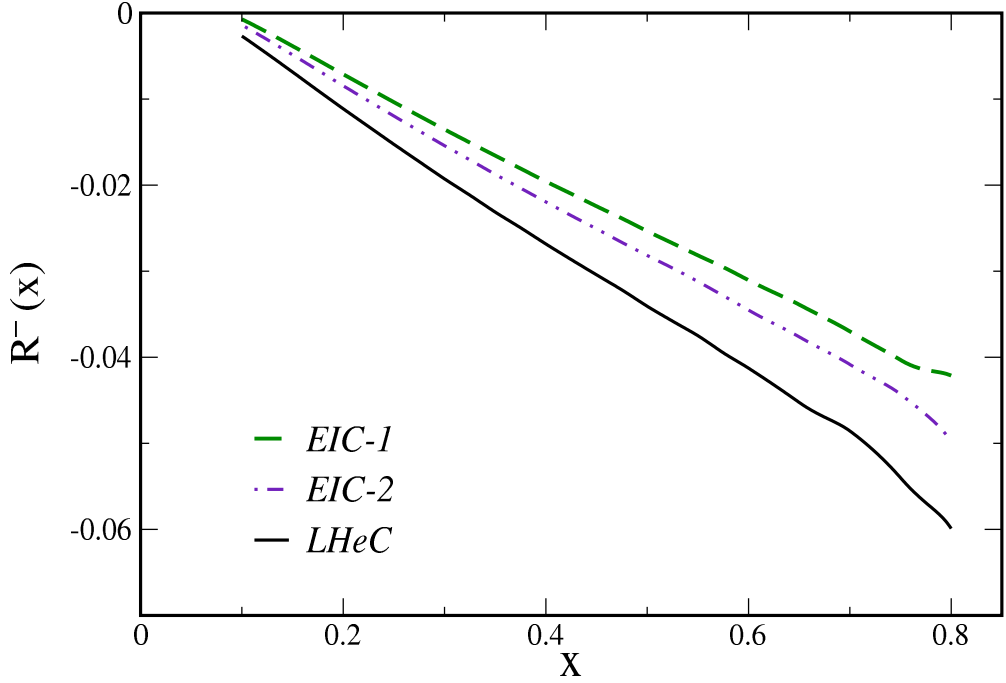}
\caption{
A comparison of the CSV effects generated only by the MRSTQED breaking evolved via Eq.~(\ref{eq:valevol})
to the putative LHeC scale with the more restrained kinematics of a first- and second-stage medium
energy electron-ion colliders described in-text. 
}
\label{fig:MEIC}
\end{figure}

We see that for large $x > 0.2$ the strange quark contribution is essentially 
negligible,but the predicted values of $R^-(x)$ are considerable: for $x = 0.6$ the 
calculated ratio is greater than 6\%. This is quite a sizable result for 
partonic CSV terms, which for most observables yield effects at the 1\% level or 
smaller~\cite{Londergan:2009kj}. This confirms our argument that, while both 
the valence quark and valence CSV distributions shift to lower $x$ values with 
increasing $Q^2$, the CSV distributions experience a smaller shift 
(because to lowest order the CSV valence distributions only radiate 
photons while the valence parton PDFs radiate both gluons and photons), 
and thus 
for a given $x$ the ratio of valence CSV distributions to valence PDFs should 
increase slowly with $Q^2$. Our best theoretical estimate of the ratio 
$R^-(x)$ from 
Eq.~(\ref{eq:F2rat}) at large $x$ values is predicted to be rather large, of 
the order of several percent. For reasonably large values $x > 0.1$, the 
ratio $R^-(x)$ is composed of relatively equal contributions from valence 
parton CSV effects arising from quark mass differences and from QED radiation. 
Thus the quantitative values obtained for the ratio $R^-(x)$ can provide a 
further check on the assumptions made in determining charge symmetry violation 
arising from QED radiation.

We finish up with one additional observation: significant partonic CSV signals embodied by $R^-(x,Q^2)$
may in fact not even require the vaunted kinematics of a hypothetical LHeC-type machine. This is suggested by the
$x$ dependence plotted in Fig.~\ref{fig:MEIC}, wherein we compare the QED-generated contributions to
$R^-$ at LHeC with the same computation at the more restricted kinematics of lower-energy machines \cite{Accardi:2012qut},
labeled `EIC-1' and `EIC-2.' For each, we assume the collision of $5,\,20$ GeV electrons on $20,\,250$ GeV deuterons
at $Q^2 = 300,\, 2000$ GeV$^2$ and center-of-mass energy $\sqrt{s} = 45,\, 140$ GeV, respectively. The fact that the lower
energy curves quite closely tract the LHeC prediction signifies that the evolution enhancements to $R^-$ may not require
such large intervals in $Q^2$ to become apparent, bolstering the accessibility of these effects at more modest kinematics;
at the same time, the event rates corresponding to the $W$ exchange mechanisms of Eq.~(\ref{eq:F2eD}) remain to be estimated,
and must be carefully simulated in the future.

It would therefore seem that a hypothetical high energy electron/positron collider whose beams interact 
with deuteron beams from the LHC may produce the most promising observable 
with which to search for partonic charge symmetry violating effects, although less energetic reactions
may also be useful. 

%% file: the-TMC.tex

Independent of the inherent interest in extending and strengthening
the preceding formalism to analyze hadronic interactions, traditional
DIS continues to build upon its earlier successes by probing new regions
of experimental phase space. While the perturbative domain of large
$Q^2$ and small parton momentum fraction $x$ has received considerable
attention both experimentally and theoretically, the previous chapter
demonstrated that the region of large $x$ and low $Q^2$ ($\sim 1 - 2$~GeV$^2$),
involves various nonperturbative effects which remain comparatively
unexplored. This is perhaps unsurprising given the difficulty
in reliably computing the various corrections that are needed to
describe data in this region. Among the more vexing of these
nonperturbative dynamics are the effects due to target mass corrections
(TMCs) associated with finite values of $M^2/Q^2$, where $M$ is the
nucleon mass, as well as higher twist terms arising from long-range
nonperturbative multi-parton correlations.

This is particularly relevant for imminent proposed measurements at
Jefferson Lab \cite{BONUS12, MARATHON, SOLID} and elsewhere, which seek to probe
very large $x$ ($x \sim 0.85$) and provide improved access to many
of the observables described in Chap.~\ref{chap:ch-Q2}: the ratio of
$d$ to $u$ quark distributions, partonic CSV, and SM tests in
parity-violating asymmetries.
Aside from this, better knowledge of PDFs at large $x$ is imperative
in searches for new physics signals at collider experiments such as
those of the Tevatron or LHC at large rapidities\footnote{The parton
momentum fractions $x_1, x_2$ probed in generating particles of
rapidity $y \defeq (1/2) \ln \Big(E + p^3 \Big/ E - p^3 \Big)$ are
given by a simple relation:
\begin{equation}
x_{1,2}\ =\ {Q \over \sqrt{s}}\ \exp(\pm y)\ . \nonumber
\end{equation}
} or for heavy mass particles \cite{Brady11}, as well as at more central
rapidities where uncertainties in large-$x$ PDFs at low $Q^2$ can, through
$Q^2$ evolution, affect cross sections at small $x$ and large $Q^2$
\cite{Kuhlmann00}.

\section{Kinematical Higher Twist Effects}
\label{sec:kin-HT}

Among these effects, the most amenable to direct computation are,
in principle, the TMCs, as well as their more general counterpart in
semi-inclusive reactions, hadron mass corrections (HMCs). Therefore,
drawing upon the findings of \cite{Brady:2011uy}, we present in this chapter a
systematic analysis of several frameworks for the computation of TMCs/HMCs,
and assess the phenomenological impact on DIS observables of especial
importance.

In Sec.~\ref{sec:TMC} we summarize the main results for TMCs in the
OPE and collinear factorization (CF) formulations for the $F_1$, $F_2$, $F_3$ and $F_L$
structure functions at NLO, and exhibit some numerical examples.
Implications for various observables are discussed in Sec.~\ref{sec:obs},
with special attention given to the longitudinal to transverse cross section
ratios $R$, and parity-violating (PV) DIS asymmetries on the proton and
deuteron, which are sensitive to $\gamma Z$ interference structure functions.
We also quantify the effects of perturbative NLO corrections on the $R^{\gamma Z}$ ratio
for the $\gamma Z$ interference, about which essentially nothing is known
empirically. We conclude the chapter by extending our mass correction prescriptions
to the more general setting of semi-inclusive DIS (SIDIS). As a slightly more exclusive
counterpart to fully inclusive DIS, SIDIS by definition implies the production of specific
final state hadrons, which necessitates a more subtle formalism for simultaneously incorporating
both the initial and final state hadron masses. In Sec.~\ref{sec:semi-intro} we present the contours
of just such a model, developed in \cite{AHM09}, which exploits the properties of CF as formulated in
momentum space.
%
\section{Target Mass Corrections}
\label{sec:TMC}

In this section we review the kinematic corrections to structure
functions arising from scattering at finite values of $Q^2/\nu^2$.
We consider several frameworks for the TMCs, including the
conventional one based on the operator product expansion of Chap.~\ref{chap:ch-DIS}.\ref{sec:OPE},
and various approximations to it, as well as a number of prescriptions
using collinear factorization at leading and next-to-leading order
in $\alpha_s$.

As discussed by Nachtmann \cite{Nachtmann73}, these target mass effects,
sometimes called ``kinematical higher twists'' are in fact associated with
leading twist operators (hence contain no additional information
on the nonperturbative parton correlations), even though they give rise to
$Q^2/\nu^2 = 4 x^2 M^2/Q^2$ corrections, where $\nu = Q^2/2Mx$
is the energy transfer.  Nachtmann further showed that one could
generalize the standard operator product expansion (OPE) of
structure function moments to finite $Q^2$ such that only operators
of a specific twist would appear at a given order in $1/Q^2$.
The resulting target mass corrected structure functions can then
be derived through an inverse Mellin transformation, as shown by
Georgi and Politzer \cite{GP76} and demonstrated below (for a thorough
review of TMCs in the OPE approach see Ref.~\cite{TMC}).

Later, an alternative formulation in terms of collinear factorization
(CF) was used by Ellis, Furmanski and Petronzio \cite{EFP83} to derive
TMCs including the effects of off-shell partons and parton transverse
motion.  While the OPE and CF formulations yield identical results
for leading twist PDFs, they differ in the details of how the target
mass corrections are manifested at finite $Q^2$.  Other versions of
TMCs were subsequently derived \cite{AOT94,KR02,AQ08} within the CF
formalism using various assumptions about the intrinsic properties
of partons and higher twist contributions, leading to rather large
differences in some cases \cite{AQ08}.

In any case, the structure functions for the scattering of an unpolarized
lepton from an unpolarized nucleon are defined in terms of the nucleon
hadron tensor introduced in Eq.~(\ref{eq:HadcI}),
\begin{subequations}
\label{eq:WmunuC}
\begin{eqnarray}
W_{\mu\nu}
&=& {1 \over 4 \pi} \int d^4 z\,e^{i q \cdot z}
    \langle P \left|
	\left[ J_\mu^\dagger(z), J_\nu(0) \right]
    \right| P \rangle
\label{eq:Wmunu_def}					\\
&=& \left( -g_{\mu\nu} + \frac{q_\mu q_\nu}{q^2} \right) F_1(x,Q^2)\
 +\ \left( p_\mu - \frac{p\cdot q}{q^2} q_\mu \right)
    \left( p_\nu - \frac{p\cdot q}{q^2} q_\nu \right)
    {F_2(x,Q^2) \over p \cdot q}			\nonumber\\
& &
 -\ i\epsilon_{\mu\nu\alpha\beta} q^{\alpha} p^{\beta}
    \frac{F_3(x,Q^2)}{2 p\cdot q}\, ,
\end{eqnarray}
\end{subequations}
where again $J_\mu$ is the electromagnetic or weak current operator
for a given virtual boson ($\gamma$, $Z$ or $W^\pm$).
Here $P$ and $q$ are again the nucleon and exchanged boson $4$-momenta,
respectively, with $q^2 = -Q^2$ as before.

The structure functions $F_{1,2}$ are related to the product of
two vector or two axial-vector currents, while $F_3$ arises from the
interference of vector and axial-vector currents.  The $F_1$ structure
function is proportional to the transverse virtual boson cross section,
and $F_2$ is given by a combination of transverse and longitudinal
cross sections.  It is convenient to also study the longitudinal
structure function introduced in Chap.~\ref{chap:ch-Q2},
\begin{equation}
\label{eq:FLdef}
F_L(x,Q^2)\ =\ \rho^2 F_2(x,Q^2) - 2x F_1(x,Q^2)\, ,
\end{equation}
where
\begin{equation}
\rho^2\ =\ 1 + \frac{4 x^2 M^2}{Q^2}\ \equiv\ r^2\ ,
\end{equation}
with $r^2 = 1 + Q^2 / \nu^2$ having been used in Chap.~\ref{chap:ch-Q2}.
In the following we will summarize target mass corrections for each
of these structure functions computed within the various approaches
outlined above.

\subsection{Operator product expansion}
\label{ssec:OPE}

Target mass corrections to structure functions were first systematically
considered by Georgi and Politzer \cite{GP76} in the framework of the
operator product expansion, and in the current section we reprise the essential
details of their formalism. Actually, the basic foundation has already been
introduced in Chap.~\ref{chap:ch-intro}.\ref{sec:OPE} in which we outlined
the main idea involving the expansion of the current operators of
Eq.~(\ref{eq:WmunuC}) into the general form presented in Eq.~(\ref{eq:OPE_gen}).

Formally, the OPE generates mass-dependent $1/Q^2$ corrections by 
introducing covariant derivatives into the twist-two quark bilinears
of $J_\mu J_\nu$ in Eq.~(\ref{eq:Wmunu_def}):
$\bar{\psi}\gamma^{\mu} D^{\mu_1} \cdots D^{\mu_N} \psi$; since each
derivative $D^{\mu_i}$ increases both the dimension and spin of the
operator by one unit, the twist (dimension minus spin) remains
unchanged. Rather than the simple expression of Eq.~(\ref{eq:OPE_mat}),\linebreak
this practice demands a more complicated expansion of
$\lan N(P)|\ \mathcal{O}^{\mu_1 \dots \mu_N}_i(0)\ |N(P) \ran$
--- one that explicitly incorporates the $g^{\mu_a \mu_b}$ trace
terms responsible for $1/Q^2$ corrections.

Making the replacement $N \rightarrow 2k$ in the language of \cite{GP76},
we can rearrange Eq.~(\ref{eq:OPE_mat}) as
\begin{align}
\Pi^{\mu_1\ \dots\ \mu_{2k}} &\defeq
\lan N(P)|\ \mathcal{O}^{\mu_1 \dots \mu_{2k}}_i(0)\ |N(P) \ran \nonumber\\
&=\ \sum_{j=0}^k\ (-1)^j\ {(2k -j)! \over 2^j\ (2k)!}\
g\ \dots\ g\,\, P\ \dots\ P\ ,
\label{eq:OPE_TMC-ex}
\end{align}
where $g\ \dots\ g$ refers to a symmetric combination of $j$ metric terms $\sim g^{\mu_a \mu_b}$,
while the $P\ \dots\ P$ is a product of $2k - 2j$ factors of $P^{\mu_a}$. Note that the `trace terms'
mentioned in, \EG~Eq.~(\ref{eq:OPE_mat}) are thus accounted for by the $j > 0$ contributions inside
the summation of Eq.~(\ref{eq:OPE_TMC-ex}). By the same logic on the other hand, taking only the $j=0$
term recovers the leading contribution to the OPE found in Chap.~\ref{chap:ch-DIS}.\ref{sec:OPE} ---
\begin{equation}
\Pi^{\mu_1\ \dots\ \mu_{2k}} = P^{\mu_1}\ \dots\ P^{\mu_{2k}}\ .
\end{equation}
The calculation is then carried forward by inserting the expansion of Eq.~(\ref{eq:OPE_TMC-ex}) into
the RHS for the OPE of $T_{\mu\nu}$ given in Eq.~(\ref{eq:OPE_gen}). The contributions for specific
structure functions moments may be isolated from Eq.~(\ref{eq:OPE_TMC-ex}) by projecting the
coefficients of the tensor structures $\kappa^{L,2,3}_{\mu\nu}$ defined in Eq.~(\ref{eq:kappa}); for
instance, projecting the coefficient of $\kappa^2_{\mu\nu}$ and keeping track of combinatoric
factors as shown in \cite{GP76} results in
\begin{align}
\int_0^1 dx\ x^{N-2}\ F^{\rm OPE}_2(x,Q^2)\ &=\ \sum_{j=0}^\infty \left({P^2 \over Q^2}\right)^j
{(N+j)! \over j!(N-2)!}\ {A^{(N+2j)} \cdot C^{N+2j}_2 \over (N+2j)!\ (N+2j-1)!}\ , \nonumber\\
&=\ \sum_{j=0}^\infty \left({M^2 \over Q^2}\right)^j
{(N+j)! \over j!(N-2)!}\ {M^{(0),\ N+2j}_2 \over (N+2j)!\ (N+2j-1)!}\ ,
\label{eq:OPE_F2TMC}
\end{align}
in which on-shellness of the nucleon grants $P^2 = M^2$. In Eq.~(\ref{eq:OPE_F2TMC}), we have recognized the fact
that the matrix elements $A^{(N+2j)}$ are due to the correlated quark fields inside the nucleon;
for our purposes here, we can relate them back to the ``massless'' structure functions of
Eq.~(\ref{eq:massless_mom}):
\begin{equation}
A^{(N+2j)} \cdot C^{N+2j}_2\ =\ M^{(0),\ N+2j}_2\ =\ \int_0^1 dz\ z^{N+2j-2}\ F^{(0)}_2(z)\ ,
\end{equation}
whence the second line of Eq.~(\ref{eq:OPE_F2TMC}) follows.

It remains only to unfold the resulting target mass corrected structure functions. This may
be accomplished by means of an inverse Mellin transformation; \IE~generically, for a suitably
analytic function $f(x)$,
\begin{equation}
M^n\ =\ \int_0^1 dx\ x^{n-1}\ f(x)\ \implies\ f(x)\
=\ {1 \over 2\pi i} \int_{-i \infty}^{+i \infty} dn\ x^{-n}\ M^n\ .
\label{eq:Mellin}
\end{equation}
Applying this transform to extract $F^{\rm OPE}_2(x,Q^2)$ from Eq.~(\ref{eq:OPE_F2TMC}) at last yields
the desired result:
\begin{equation}
\label{eq:OPE_F2}
F_2^{\rm OPE}(x,Q^2)
= {(1+\rho)^2 \over 4 \rho^3} F_2^{(0)}(\xi,Q^2)\
 +\ {3x(\rho^2-1) \over 2\rho^4}
    \left[ h_2(\xi,Q^2) + {\rho^2-1 \over 2 x \rho} g_2(\xi,Q^2)
    \right]\ .
\end{equation}
A very similar procedure suffices for the other $F^{\rm OPE}_i(x,Q^2)$ \cite{GP76},
which we tabulate for completeness:
\begin{eqnarray}
F_1^{\rm OPE}(x,Q^2)
&=& {1+\rho \over 2\rho} F_1^{(0)}(\xi,Q^2)\
 +\ {\rho^2-1 \over 4\rho^2}
    \left[ h_2(\xi,Q^2) + {\rho^2-1 \over 2 x \rho} g_2(\xi,Q^2)
    \right],					\nonumber\\
F_L^{\rm OPE}(x,Q^2)
&=& {(1+\rho)^2\over 4 \rho} F_L^{(0)}(\xi,Q^2)
 +\ {x(\rho^2-1) \over \rho^2}
    \left[ h_2(\xi,Q^2) + {\rho^2-1 \over 2 x \rho} g_2(\xi,Q^2)
    \right],                                     \nonumber\\
\label{eq:OPE_F3}
F_3^{\rm OPE}(x,Q^2)
&=& {(1+\rho) \over 2 \rho^2} F_3^{(0)}(\xi,Q^2)
 +\ {(\rho^2-1) \over 2\rho^3} h_3(\xi,Q^2).
\label{eq:OPE}
\end{eqnarray}
Again, $F_i^{(0)}$ are the structure functions in the ``massless'' $M^2/Q^2 \to 0$
limit, evaluated at the modified scaling variable $\xi$
\cite{Nachtmann73,Greenberg71},
\begin{equation}
\xi = {2x \over 1 + \rho}\, ,
\label{eq:Nacht}
\end{equation}
which approaches $x$ as $M^2/Q^2 \to 0$.
The functions $h_2$, $g_2$ and $h_3$ are associated with higher order
terms in $M^2/Q^2$ and are given by \cite{GP76,TMC}
\begin{subequations}
\label{eq:hg}
\begin{eqnarray}
h_2(\xi,Q^2)
&=& \int_\xi^1 du\, {F_2^{(0)}(u,Q^2) \over u^2}\, ,\\
g_2(\xi,Q^2)
&=& \int_\xi^1 du\, \int_u^1 dv {F_2^{(0)}(v,Q^2) \over v^2}\
 =\ \int_\xi^1 du\, (u-\xi) {F_2^{(0)}(u,Q^2) \over u^2}\, ,\\
h_3(\xi,Q^2)
&=& \int_\xi^1 du\, {F_3^{(0)}(u,Q^2) \over u}\, .
\end{eqnarray}
\end{subequations}
(Note that the function $g_2$ here should not be confused with the
spin-dependent $g_2$ structure function measured in polarized
lepton--nucleon scattering.)

The expressions in Eqs.~(\ref{eq:OPE}) are known to suffer from the
``threshold problem'', in which the target mass corrected (leading
twist) structure functions do not vanish as $x \to 1$, and are in
fact nonzero in the kinematically forbidden $x > 1$ region, where
for a proton target baryon number conservation would be violated.
This is clear from the ${\cal O}(1)$ terms in Eqs.~(\ref{eq:OPE})
in which the massless functions $F_i^{(0)}$ are evaluated at $\xi$.
Because at any finite $Q^2$ value one has
	$\xi < \xi_0 \equiv \xi(x=1) < 1$,
for any input function $F_i^{(0)}$ which is nonzero for $0 < x < 1$,
the target mass corrected function at $x=1$ will not vanish,
$F_i^{\rm OPE}(x=1,Q^2<\infty) > 0$.
A number of attempts have been made to ameliorate the threshold
problem \cite{Tung79,Steffens06} using various prescriptions and
{\it ans\"atze}, although none of these is unique and without
additional complications \cite{TMC}.

Recently, Kulagin and Petti \cite{KP06} showed that by expanding the
target mass corrected structure functions to leading order in $1/Q^2$,
the resulting functions have the correct $x \to 1$ limits,
\begin{eqnarray}
F_1^{1/Q^2}(x,Q^2)
&=&
\frac{1}{4}
\left(5-\rho^2\right)F_1^{(0)}(x,Q^2)\
-\ \frac{1}{4} \left(\rho^2-1\right)
\left[ x F_1^{(0)\, \prime}(x,Q^2)
     -\ h_2(x,Q^2)
\right],					\nonumber\\
\label{eq:KP1_F2}
F_2^{1/Q^2}(x,Q^2)
&=&
\left(2-\rho^2\right)F_2^{(0)}(x,Q^2)\
-\ \frac{1}{4} \left(\rho^2-1\right)
\left[ x F_2^{(0)\, \prime}(x,Q^2)
     -\ 6 x h_2(x,Q^2)
\right],					\nonumber\\
F_L^{1/Q^2}(x,Q^2)
&=&
F_L^{(0)}(x,Q^2)\
-\ \frac{1}{4} \left(\rho^2-1\right)
\left[ x F_L^{(0)\, \prime}(x,Q^2)
     -\ 4 x h_2(x,Q^2)
\right],					\nonumber\\
F_3^{1/Q^2}(x,Q^2)
&=&
\frac{1}{4}
\left(7-3\rho^2\right)F_3^{(0)}(x,Q^2)\
-\ \frac{1}{4} \left(\rho^2-1\right)
\left[ x F_3^{(0)\, \prime}(x,Q^2)
     -\ 2 h_3(x,Q^2)
\right].						\nonumber\\
& &
\label{eq:KP1}
\end{eqnarray}
While avoiding the threshold problem, this prescription, however, raises
the question of whether the $1/Q^2$ approximation is sufficiently
accurate for structure functions near $x \approx 1$ at moderate $Q^2$.
To test the convergence of the $1/Q^2$ expansion at large $x$,
we further expand the OPE results (\ref{eq:OPE}) to include
${\cal O}(1/Q^4)$ corrections,
\begin{eqnarray}
F_1^{1/Q^4}(x,Q^2)
&=& F_1^{1/Q^2}(x,Q^2)\ +\ 
\left( \rho^2-1 \right)^2
\left[
    \frac{3}{16}  F_1^{(0)}(x,Q^2)\
 +\ \frac{1}{16x} F_2^{(0)}(x,Q^2)
\right.							\nonumber\\
&+&
\left.
    \frac{3x}{16} F_1^{(0)\, \prime}(x,Q^2)\
 +\ \frac{x^2}{32}\, F_1^{(0)\, \prime\prime}(x,Q^2)\
 -\ \frac{1}{4}\,  h_2(x,Q^2)\
 +\ \frac{1}{8x}\, g_2(x,Q^2)
\right]							\nonumber\\
\label{eq:KP2_F2}
F_2^{1/Q^4}(x,Q^2)
&=& F_2^{1/Q^2}(x,Q^2)\ +\ 
\left( \rho^2-1 \right)^2
\left[
    \frac{23}{16} F_2^{(0)}(x,Q^2)\
 +\ \frac{3x}{8}  F_2^{(0)\, \prime}(x,Q^2)
\right.							\nonumber\\
&+&
\left.
    \frac{x^2}{32}\, F_2^{(0)\, \prime\prime}(x,Q^2)\
 -\ 3x\, h_2(x,Q^2)\
 +\ \frac{3}{4}\, g_2(x,Q^2)
\right],						\nonumber\\
F_L^{1/Q^4}(x,Q^2)
&=& F_L^{1/Q^2}(x,Q^2)\ +\ 
\left( \rho^2-1 \right)^2
\left[
    \frac{3}{16} F_L^{(0)}(x,Q^2)\		
 +\ \frac{1}{4}  F_2^{(0)}(x,Q^2)
\right.							\nonumber\\
&+&
\left.
    \frac{x}{8}    F_L^{(0)\, \prime}(x,Q^2)\
 +\ \frac{x^2}{32} F_L^{(0)\, \prime\prime}(x,Q^2)\
 -\ x h_2(x,Q^2)\
 +\ \frac{1}{2} g_2(x,Q^2)
\right],						\nonumber\\
F_3^{1/Q^4}(x,Q^2)
&=& F_3^{1/Q^2}(x,Q^2)\ +\ 
\left( \rho^2-1 \right)^2
\left[
    \frac{13}{16} F_3^{(0)}(x,Q^2)	
 +\ \frac{5x}{16} F_3^{(0)\, \prime}(x,Q^2)
\right.							\nonumber\\
&+&
\left.
    \frac{x^2}{32}\, F_3^{(0)\, \prime\prime}(x,Q^2)\
 -\ \frac{3}{4}\, h_3(x,Q^2)
\right],
\label{eq:KP2}
\end{eqnarray}
where the first $(F_i^{(0)\, \prime})$ and second
$(F_2^{(0)\, \prime\prime})$ derivatives of the structure functions
are with respect to $x$.  In fact, one can show that for a structure
function that behaves at large $x$ as $(1-x)^n$, the target mass
corrected result will vanish in the $x \to 1$ limit up to order
$1/Q^{2n-2}$ in the expansion.  For $n \approx 3$, as is typical for
nucleon structure functions, the threshold problem will therefore
appear only at order $1/Q^6$.

\begin{figure}[h]
\rotatebox{-90}{\includegraphics[width=8.5cm]{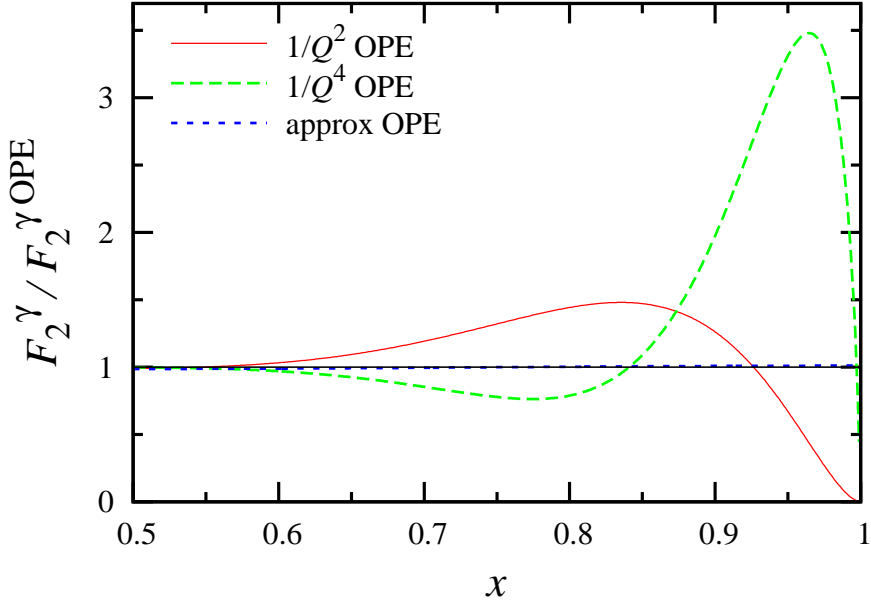}}
\caption{Ratio of the target mass corrected $F_2$ structure functions
	using the $1/Q^2$ (solid, red), $1/Q^4$ (long-dashed, green)
	and phenomenological (short-dashed, blue) OPE approximations
	compared with the exact OPE result, Eq.~(\ref{eq:OPE_F2}),
        at $Q^2 = 2$ GeV$^2$.
	Note that the phenomenological OPE approximation is almost
	indistinguishable from the exact OPE result, while the $1/Q^2$
	and $1/Q^4$ expansions deviate from this for $x \gtrsim 0.6$.}
\label{fig:OPE_comp}
\end{figure}

The accuracy of the $1/Q^2$ expansions is illustrated in
Fig.~\ref{fig:OPE_comp}, where in order to isolate the target mass
effect from the specific form of the structure function parametrization
we have taken for simplicity the form $F_2 \sim (1-x)^3$ and computed at
a fixed $Q^2 = 2$ GeV$^2$.
Both the $1/Q^2$ and $1/Q^4$ approximations are found to reproduce
the full OPE result very well up to $x \approx 0.6$, but significant
deviations are visible at larger $x$.  Furthermore, while there is
a modest improvement in the agreement with the exact result for
$0.6 \lesssim x \lesssim 0.8$ after inclusion of the $1/Q^4$ terms,
both expansions appear to break down for $x \gtrsim 0.8$.
The reliability of a low order $1/Q^2$ expansion is therefore
questionable at these $x$ values, and hence their efficacy in
removing the $x \to 1$ threshold problem.

Since the integrals in the functions $h_{2,3}$ and $g_2$ can be
time consuming to compute numerically, Schienbein {\it et al.}
\cite{TMC} isolated phenomenological analytic forms which
approximate the target mass corrected $F_2$ and $F_3$ structure
functions in Eqs.~(\ref{eq:OPE_F2}) and (\ref{eq:OPE_F3}) by
\begin{subequations}
\label{eq:approxOPE}
\begin{eqnarray}
\label{eq:approxOPE_F2}
F_2^{\rm approx}(x,Q^2)&=&\frac{(1+\rho)^2}{2\rho^3}
\left(1+\frac{3(\rho^2-1)}{\rho(1+\rho)}(1-\xi)^2\right)
F_2^{(0)}(\xi,Q^2),\						\\
F_3^{\rm approx}(x,Q^2)&=&\frac{(1+\rho)}{2\rho^2}
\left(1-\frac{(\rho^2-1)}{2\rho(1+\rho)}(1-\xi)\ln{\xi}\right)
F_3^{(0)}(\xi,Q^2).
\end{eqnarray}
\end{subequations}
These turn out to be rather good approximations to the exact results,
as Fig.~\ref{fig:OPE_comp} illustrates for the $F_2$ case.  For all
values of $x$, the phenomenological approximation (\ref{eq:approxOPE_F2})
stays within 5\% of the full OPE result.

\subsection{Collinear factorization}
\label{ssec:CF}

An alternative approach to TMCs relies on the collinear factorization
(CF) formalism \cite{EFP83,AOT94,KR02,AQ08}, which makes use of the
factorization theorem to relate the hadronic tensor for lepton--hadron
scattering to that for scattering from a parton.  Here parton
distributions are formulated directly in momentum space, avoiding the
need to perform an inverse Mellin transform to obtain the PDF from its
moments.  An advantage of the CF formalism for TMCs is that it can be
extended to other hard scattering processes, such as semi-inclusive DIS
\cite{AHM09}, where an OPE is not available; we shall demonstrate precisely
this in Sec.~\ref{sec:semi-intro}.

\begin{figure}[h]
\includegraphics[width=16.5cm]{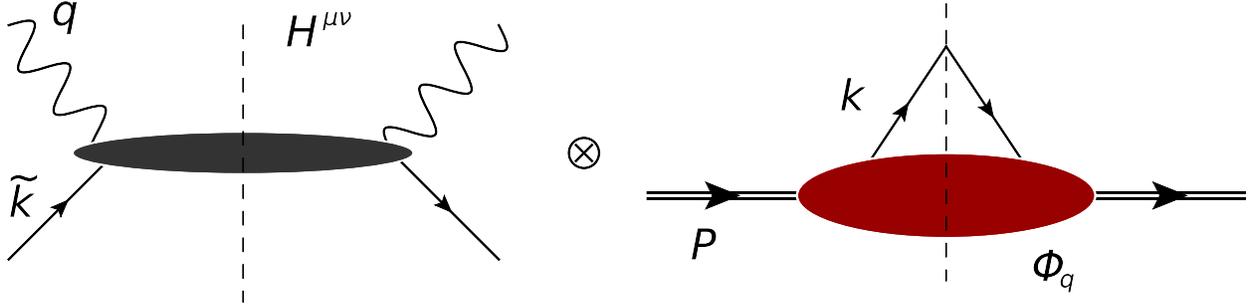}
\caption{
The factorization theorem for the $W_{\mu\nu}$ tensor is the
basis of CF formalism applied to inclusive DIS. Here, the hard
partonic scattering is represented by the left-hand diagram for
$H_{\mu\nu}$, whereas the soft correlators $\Phi_q$ encode the
nonperturbative dynamics.
}
\label{fig:CFDIS}
\end{figure}
Much as we shall illustrate in more formal detail in
Sec.~\ref{sec:semi-intro}, the CF paradigm supposes the hadronic tensor
of Eq.~(\ref{eq:WmunuC}) can be separated in momentum space as
\begin{equation}
W_{\mu\nu}(P,q)\ =\ \sum_q \int {dx \over x}\ H^q_{\mu\nu}(\tilde{k},q) 
\cdot \Phi_q(x, Q^2, M^2)\ +\ \mathcal{O}(1/Q^2)\ ,
\end{equation}
where the sum is over quark flavors and the hard amplitude $H^q_{\mu\nu}$
and quark correlator $\Phi_q$ of Eq.~(\ref{eq:quark_cors}) appear explicitly
in Fig.~\ref{fig:CFDIS}. This factorized form is obtained by expanding the quark
4-momenta in the collinear ``DIS frame''
\begin{subequations}
\begin{eqnarray}
P^\mu   &=& P^+\, \nbar^\mu 
         + \frac{M^2}{2 P^+}\, n^\mu\ ,         \\
q^\mu   &=& - \xi P^+\, \nbar^\mu 
         + \frac{Q^2}{2\xi P^+}\, n^\mu\ ,      \\
k^\mu &=& x P^+\, \nbar^\mu
         + \frac{k^2 + k^2_\perp}{2 x P^+}\, n^\mu 
         + {\bf k}_{\perp}^{\,\mu} \ ,               
\end{eqnarray}
\label{eq:DIS-CFkin}%
\end{subequations}%
about $\tilde{k}^\mu = x P^+ n^\mu$. Roughly speaking, the mass-corrected
structure functions can then be projected \cite{AQ08} from the tensor $H^q_{\mu\nu}$
using the fact that $F_i(x) = h^q_i \otimes \Phi_q(\xi)$, and noting that
the $h_i$ are parton-level analogues of the helicity structure functions $F_i(x)$

\subsubsection{Ellis, Furmanski and Petronzio}
\label{sssec:EFP}

The first study of TMCs within CF was made by Ellis, Furmanski, and
Petronzio (EFP) \cite{EFP83}, who analyzed the virtual photon-hadron
scattering amplitude using a Feynman diagram technique to expand the
hard scattering term about the collinear direction, incorporating
both parton off-shellness (or interactions) and parton transverse
momentum in twist-4 contributions \cite{Qiu90}; this is in contrast
to the logic we just sketched, which was kept strictly to twist-2 for simplicity.
Using the same notation as for the OPE TMCs above, the EFP results
for the target mass corrected structure functions are given by
\begin{subequations}
\label{eq:EFP}
\begin{eqnarray}
F_1^{\rm EFP}(x,Q^2)&=&
\frac{2}{1+\rho} F_1^{(0)}(\xi,Q^2)
+ \frac{(\rho^2-1)}{(1+\rho)^2} h_2(\xi,Q^2)\, ,	\\
F_2^{\rm EFP}(x,Q^2)&=&
\frac{1}{\rho^2} F_2^{(0)}(\xi,Q^2) + \frac{3\xi(\rho^2-1)}
{\rho^2 (1+\rho)} h_2(\xi,Q^2)\, ,			\\
F_L^{\rm EFP}(x,Q^2)&=&
F_L^{(0)}(\xi,Q^2)
+ \frac{2\xi(\rho^2-1)}{(1+\rho)} h_2(\xi,Q^2)\, ,	\\
F_3^{\rm EFP}(x,Q^2)&=&
\frac{1}{\rho} F_3^{(0)}(\xi,Q^2)+\frac{2(\rho^2-1)}
{\rho (1+\rho)^2} h_3(\xi,Q^2)\, ,
\end{eqnarray}
\end{subequations}
where again the $F_i^{(0)}$ refer to the uncorrected structure
functions, and $h_{2,3}$ are given in Eqs.~(\ref{eq:hg}).
(Note that the definition of the longitudinal structure function in
EFP differs from the usual definition (\ref{eq:FLdef}) by a factor
$x$, and the $F_2$ structure function is proportional to what EFP
call the ``transverse'' structure function, which in standard usage
is proportional to $F_1$.)
Because the massless functions $F_i^{(0)}$ are evaluated at $\xi$,
the target mass corrected structure functions will suffer from the
same threshold problem as in the OPE analysis in Eqs.~(\ref{eq:OPE}).
While the expressions in Eqs.~(\ref{eq:EFP}) were derived in
Ref.~\cite{EFP83} at leading order in $\alpha_s$, in this work
we will assume their validity also at NLO.

The prefactors for the leading terms proportional to $F_i^{(0)}$
in Eqs.~(\ref{eq:EFP}) are remarkably close to those for the leading
terms in the OPE expressions in Eqs.~(\ref{eq:OPE}). To first order
in $1/Q^2$, the leading term prefactors for $F_1$ in both OPE and EFP
reduce to $(1-x^2 M^2/Q^2)$.  Similarly, the $F_2$ prefactors both
reduce to $(1-4 x^2 M^2/Q^2)$, while those for $F_L$ reduce to 1.
For the $F_3$ structure function, however, the ${\cal O}(1/Q^2)$
prefactor is $(1-3 x^2 M^2/Q^2)$ for OPE, whereas for the EFP CF
result it is $(1-2 x^2 M^2/Q^2)$.

At leading order in the massless limit the longitudinal structure
function vanishes identically.  At NLO, however, it receives
contributions from both quark and gluon PDFs convoluted with the
respective hard coefficient functions.  For electromagnetic
scattering, for example, one has \cite{Altarelli78,Bardeen78}
\begin{equation}
F_L^{\gamma (0)}(x,Q^2)
= {\alpha_s(Q^2) \over \pi}
    \int_x^1 {dy \over y} \left( {x \over y} \right)^2
    \left\{ {4 \over 3} F_2^{\gamma (0), \rm LO}(y,Q^2)
	  + c^\gamma (y-x) g(y,Q^2)
    \right\},
\end{equation}
where $c^\gamma = 2\sum_q e_q^2$, and $F_2^{\gamma (0), \rm LO}$
is given by the leading order expression for $F_2^{\gamma (0)}$.
Similar expressions hold also for the longitudinal structure functions
associated with other electroweak currents.  In our numerical
calculations discussed below we will always compute $F_L$ at NLO, which
serves as an input to determinations of the phenomenological $R$ ratios
and PVDIS asymmetries.

It is important also to note that Eqs.~(\ref{eq:EFP}) have been
derived considering Feynman diagrams with 2 or 4 legs attached to
the hadronic correlator (see Figs.~2 and 3 of Ref.~\cite{EFP83}),
which for $M=0$ give rise to twist-2 and twist-4 contributions to
the structure functions, respectively.  For $M \neq 0$, however,
the quark and gluon equations of motion allow one to extract a
twist-2 contribution from the 4-leg diagrams, which when added
to the twist-2 target mass correction yields the full result in
Eqs.~(\ref{eq:EFP}).  It is an interesting question whether by
resumming the twist-2 parts of $n$-leg diagrams one would be able
to recover the TMC expressions (\ref{eq:EFP}).

\subsubsection{Accardi and Qiu}
\label{sssec:AQ}

In both the EFP and OPE treatments of TMCs, the resulting structure
functions are nonzero for $x>1$.  The analysis of Accardi and Qiu (AQ)
\cite{AQ08} traced this problem to baryon number nonconservation in
the handbag diagram for $M \neq 0$.  Working with 2-leg diagrams only,
in contrast to EFP who also consider 4-leg diagrams up to twist-4, the
AQ target mass corrected structure functions are given by \cite{AQ08}
\begin{subequations}
\label{eq:AQ}
\begin{eqnarray}
F_1^{\rm AQ}(x,Q^2)
&=& \widetilde{F}_1^{(0)}(\xi,Q^2),			\\
F_2^{\rm AQ}(x,Q^2)
&=& \frac{1+\rho}{2\rho^2} \widetilde{F}_2^{(0)}(\xi,Q^2),	\\
F_L^{\rm AQ}(x,Q^2)
&=& \frac{1+\rho}{2} \widetilde{F}_L^{(0)}(\xi,Q^2),	\\
F_3^{\rm AQ}(x,Q^2)
&=& \frac{1}{\rho} \widetilde{F}_3^{(0)}(\xi,Q^2).
\end{eqnarray}
\end{subequations}
Here the functions $\widetilde{F}_i^{(0)}$ are defined as
\begin{align}
\label{eq:AQ2}
\widetilde{F}_i^{(0)}(\xi,Q^2)
= \sum_f \int_\xi^{\xi/x} \!\frac{dz}{z}\,
  C_i^f\left({\xi\over z},Q^2\right) \varphi_f(z,Q^2),
\end{align}
where $C_i^f$ are the perturbatively calculable hard coefficient
functions for a given parton flavor $f$, including parton charge
factors, $\varphi_f$ are the parton densities of the nucleon,
and the sum is taken over all active flavors.
The upper limit in Eq.~(\ref{eq:AQ2}) ensures that the target mass
corrected structure functions vanish for $x > 1$, as required by
kinematics, although jet mass corrections need to be introduced in
order to render the target mass corrected functions zero at $x=1$
\cite{AQ08}.
It remains an interesting exercise to apply the same prescription
to twist-4 diagrams as in Ref.~\cite{EFP83} in order to establish a
more direct correspondence between the AQ and EFP approaches.
Of course, for $M^2/Q^2 \to 0$ the upper limit of integration in
Eq.~(\ref{eq:AQ2}) is 1, and both approaches recover the standard
factorization theorem for structure functions~\cite{CSS88}.

\subsubsection{$\xi$-scaling}
\label{sssec:xi}

When the upper limit of integration in Eq.~(\ref{eq:AQ2}) is taken to
be 1, the AQ structure functions reduce to the simple $\xi$-scaling
($\xi$-S) form introduced by Aivazis {\it et al.} \cite{AOT94} and
used by Kretzer and Reno \cite{KR02}.  The target mass corrected
structure functions in this case are simply given by
\begin{subequations}
\label{eq:xiS}
\begin{eqnarray}
F_1^{\rm \xi\mbox{-}S}(x,Q^2)
&=& F_1^{(0)}(\xi,Q^2)\, ,				\\
F_2^{\rm \xi\mbox{-} S}(x,Q^2)
&=& \frac{1+\rho}{2 \rho^2}\ F_2^{(0)}(\xi,Q^2)\, ,	\\
F_L^{\rm \xi\mbox{-} S}(x,Q^2)
&=& \frac{1+\rho}{2}F_L^{(0)}(\xi,Q^2)\, ,			\\
F_3^{\rm \xi\mbox{-} S}(x,Q^2)
&=& \frac{1}{\rho} F_3^{(0)}(\xi,Q^2)\, .
\end{eqnarray}
\end{subequations}%
Note that the form of the target mass corrected functions in
Eqs.~(\ref{eq:xiS}) closely resembles that in Eqs.~(\ref{eq:AQ}),
with the two forms equivalent at leading order.
At this order the structure functions satisfy a modified Callan-Gross
relation \cite{AQ08},
\begin{align}
\rho^2\, F_2^{\rm \xi\mbox{-} S}(x,Q^2)
= 2x F_1^{\rm \xi\mbox{-} S}(x,Q^2)\, .
\end{align}
The leading order $\xi$-scaling structure functions are also related
to the leading, ${\cal O}(1)$ terms of the OPE expressions in
Eqs.~(\ref{eq:OPE}),
\begin{equation}
F_i^{\rm OPE\, (leading)}(\xi,Q^2)
= \frac{1+\rho}{2\rho}\ F_i^{\rm \xi\mbox{-} S}(\xi,Q^2)\, ,
\end{equation}
where the prefactor, to order $1/Q^2$, is given by $(1-x^2 M^2/Q^2)$.
In fact, the $\xi$-scaling formulas (\ref{eq:xiS}) would coincide
with the EFP results in Sec.~\ref{sssec:EFP} in the absence of 4-leg 
Feynman diagrams \cite{EFP83}.

\subsection{TMC comparisons}
\label{sssec:comp}

\begin{figure}[h]
\hspace*{-0.4cm}
\rotatebox{-90}{\includegraphics[width=6.05cm]{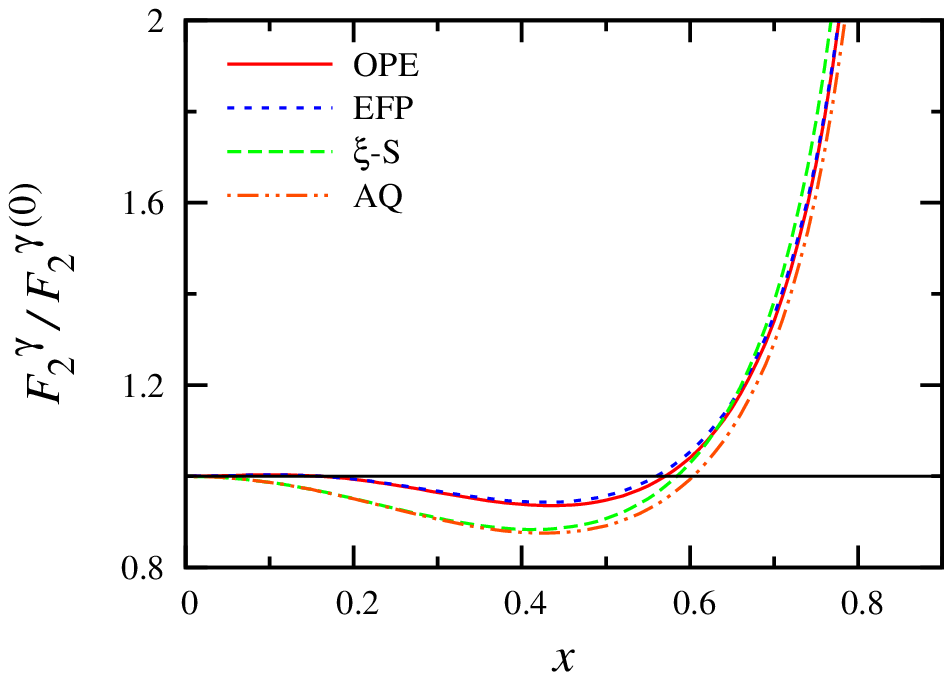}}
\rotatebox{-90}{\includegraphics[width=6.05cm]{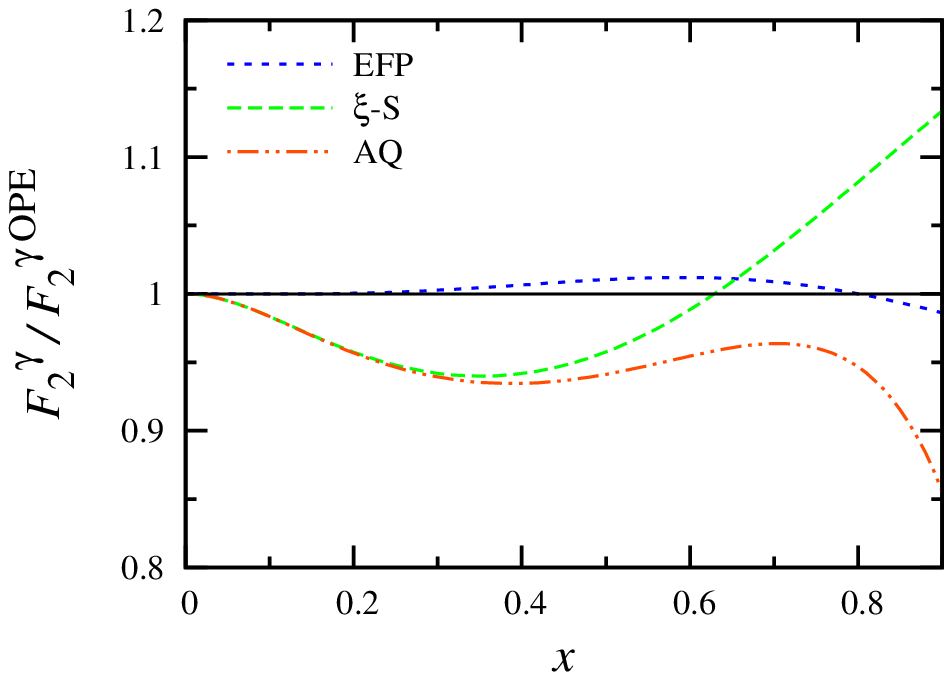}}
\caption{
Ratio of target mass corrected to uncorrected (left)
	or to OPE (right) $F_2^\gamma$ proton structure
	functions at $Q^2 = 2$~GeV$^2$ for the
	OPE (solid, red),
	EFP (short-dashed, blue),
	$\xi\mbox{-S}$ (long-dashed, green), and
	AQ (dot-dashed, orange) TMC prescriptions.
}
\label{fig:F2}
\end{figure}

The effects of the different TMC prescriptions on the $F_2^\gamma$
structure function are illustrated in Fig.~\ref{fig:F2} as a
stand-in for the analogous calculation for other quark flavor
combinations and exchange bosons as collected exhaustively in
\cite{Brady:2011uy}.

The uncorrected proton structure function $F_2^{(0)}$ (and others
required to compute parity-violating asymmetries, etc.) is
constructed from the CTEQ-Jefferson Lab (CJ) global PDF fits
\cite{CJ11}, evaluated at $Q^2 = 2$~GeV$^2$.  For each of the
structure functions the effects of TMCs become more prominent
with increasing~$x$, and naturally their magnitude decreases
at larger $Q^2$.

For $F_2^\gamma$ as given in Fig.~\ref{fig:F2}, a dip in the ratio
of corrected to uncorrected functions at $x \sim 0.4$, however,
delays the sharp rise above unity to $x \gtrsim 0.6$. More pointedly,
the EFP result agrees with the OPE to a few percent
over the entire $x$ range, and the AQ and $\xi$-scaling ratios are
almost identical for $x < 0.4$. The two sets of ratios differ by
$\lesssim 7\%$ for $x \lesssim 0.7$, before diverging somewhat as $x \to 1$.

\section{Implications for Observables}
\label{sec:obs}

Having examined the differences between the various TMC prescriptions
in $F_2^\gamma$ as an example, in the current section we consider the
effects of TMCs, and in particular their model dependence, on several
observables that will be measured in upcoming experiments.
We train special focus upon the longitudinal to transverse (LT) cross section
ratios, as well as the parity-violating deep-inelastic scattering asymmetries for the
proton and deuteron, given their prominence in Chap.~\ref{chap:ch-Q2}.

\subsection{$R = \sigma_L / \sigma_T$ ratios}
\label{sec:LTRat}


Given the importance of its phenomenology to the general control over
the asymmetry $A^{PV}$ of Chap.~\ref{chap:ch-Q2}, we first consider
mass effects in the cross section ratio $R^{\gamma N}$.
The effects of TMCs on the LT ratio are also important in
connection with establishing the low-$Q^2$ behavior of $R^{\gamma N}$
at finite $x$, so as to determine the onset of gauge invariance constraints
on the longitudinal structure function \cite{MEK05}.

\begin{figure}[h]
\hspace*{-0.4cm}
\rotatebox{-90}{\includegraphics[width=6.05cm]{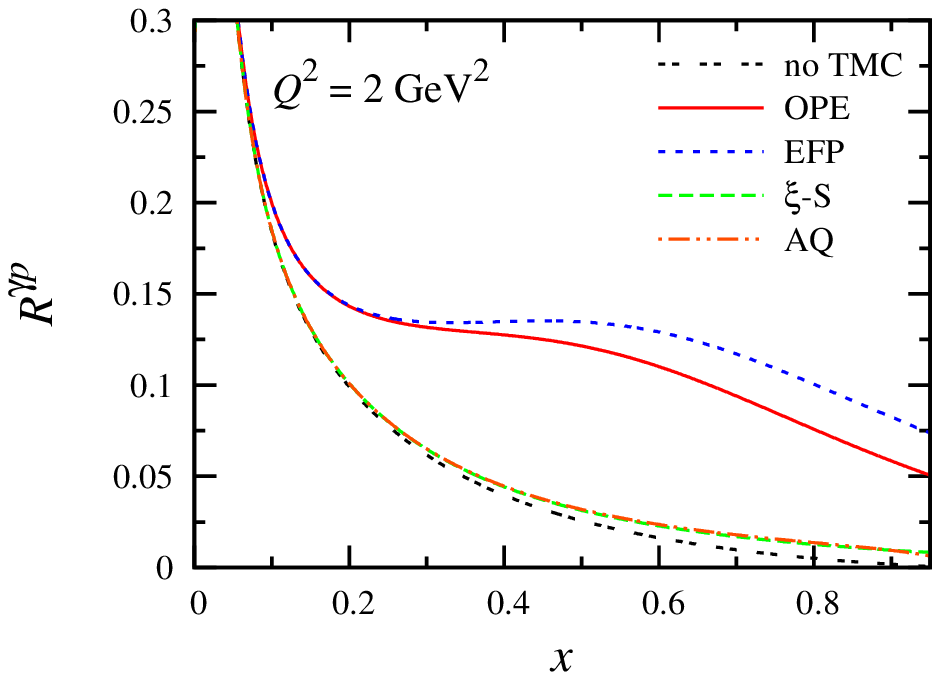}}
\rotatebox{-90}{\includegraphics[width=6.05cm]{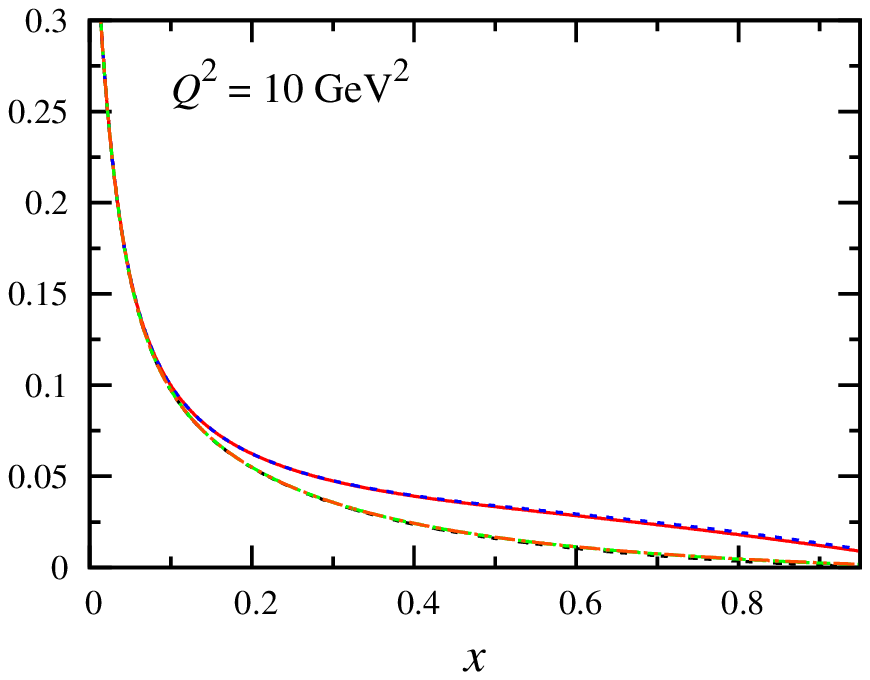}}
\caption{Longitudinal to transverse cross section ratio $R^{\gamma p}$
	for the proton at $Q^2 = 2$~GeV$^2$ (left) and 
	$Q^2 = 10$~GeV$^2$ (right), for
	no TMCs (double-dashed, black),
	the OPE (solid, red),
	EFP (short-dashed, blue),
	$\xi$-S (long-dashed, green), and
	AQ (dot-dashed, orange) TMC prescriptions.}
\label{fig:R_LT}
\end{figure}

In Fig.~\ref{fig:R_LT} we illustrate the TMC effects on $R^{\gamma p}$
for $Q^2 = 2$ and 10~GeV$^2$ for each of the TMC prescriptions considered
in Sec.~\ref{sec:TMC}. All of the TMCs increase the magnitude of the $R^{\gamma p}$ ratio, with
the AQ and $\xi$-S prescriptions having a relatively modest effect
(approximately a factor 2 for $x \approx 0.6-0.8$ at $Q^2 = 2$~GeV$^2$,
but only a few percent at $Q^2 = 10$~GeV$^2$), whereas the EFP and OPE both
modify the ratio significantly for $x \gtrsim 0.1$.  The enhancement
of $R^{\gamma p}$ for the latter is predicted to be about an order of
magnitude for $x \approx 0.6-0.8$ at $Q^2 = 2$~GeV$^2$, and still a
factor of $3-4$ at $Q^2 = 10$~GeV$^2$.

Some differences are also expected between the longitudinal to
transverse cross section ratios at NLO for processes involving
electromagnetic and weak currents.  In particular, we have already
quantified in Chap.~\ref{chap:ch-Q2} the fact that asymmetries
measured in parity-violating electron scattering are sensitive
to interference effects between $\gamma$ and $Z$ boson exchange,
and differences between the $R^\gamma$ and $R^{\gamma Z}$ LT ratios
can affect the measured asymmetries~\cite{Hobbs08, Hobbs11}.

\begin{figure}[h]
\hspace*{-0.4cm}
\rotatebox{-90}{\includegraphics[width=6.05cm]{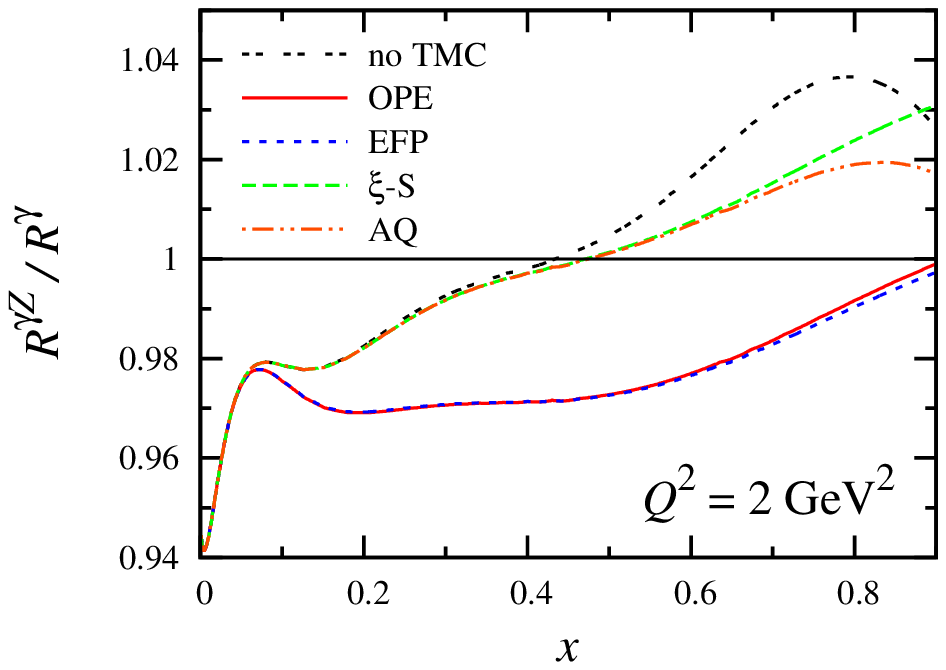}}
\rotatebox{-90}{\includegraphics[width=6.05cm]{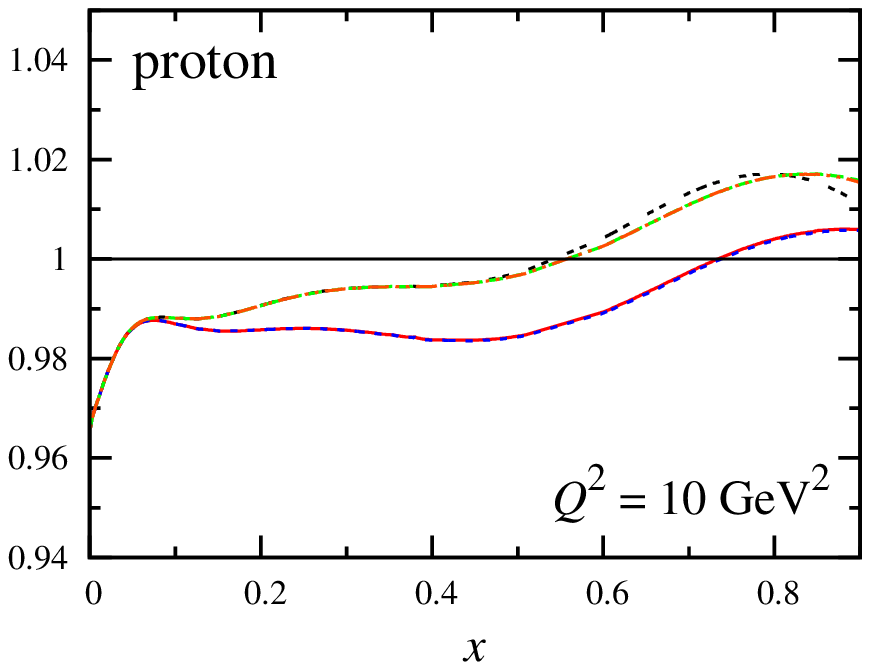}}
\caption{Ratio of $R^{\gamma Z}$ to $R^{\gamma}$ LT ratios for the
	{\it proton} computed at NLO for $Q^2 = 2$~GeV$^2$ (left)
	and $Q^2 = 10$~GeV$^2$ (right), for
	no TMCs (double-dashed, black),
	the OPE (solid, red),
	EFP (short-dashed, blue),
	$\xi$-S (long-dashed, green), and
	AQ (dot-dashed, orange) TMC prescriptions.}
\label{fig:Rp}
\end{figure}

In Fig.~\ref{fig:Rp} the ratio of the proton $R^{\gamma Z}$
to $R^\gamma$ LT ratios is shown at $Q^2 = 2$ and 10~GeV$^2$.
While at leading order both of these ratios are zero, at NLO
the different relative contributions from quark PDFs to the
electromagnetic and $\gamma Z$ interference structure functions
leads to deviations of the ratios from unity of up to $\approx 4\%$
at $Q^2 = 2$~GeV$^2$, and up to $\approx 2\%$ at $Q^2 = 10$~GeV$^2$.
The effects of the TMCs are again very small for the $\xi$-scaling
and AQ prescriptions, but more significant for the OPE and EFP
results.  Overall, the spread in the TMC predictions for the
$R^{\gamma Z}/R^\gamma$ ratio amounts to $\lesssim 4 - 5\%$ for
$x$ between 0.6 and 0.8 at $Q^2 = 2$~GeV$^2$, and $\lesssim 2\%$
at $Q^2 = 10$~GeV$^2$.
Note that the dip in the ratios at $x < 0.1$, which is insensitive
to TMCs, reflects the greater role played by gluons at low $x$,
but is mostly irrelevant for the kinematics of typical proposed
experiments \cite{BONUS12, MARATHON, SOLID}, and also more peripheral
with regard to signals of nonperturbative physics in the high $x$ valence
region.

\begin{figure}[h]
\hspace*{-0.4cm}
\rotatebox{-90}{\includegraphics[width=6.05cm]{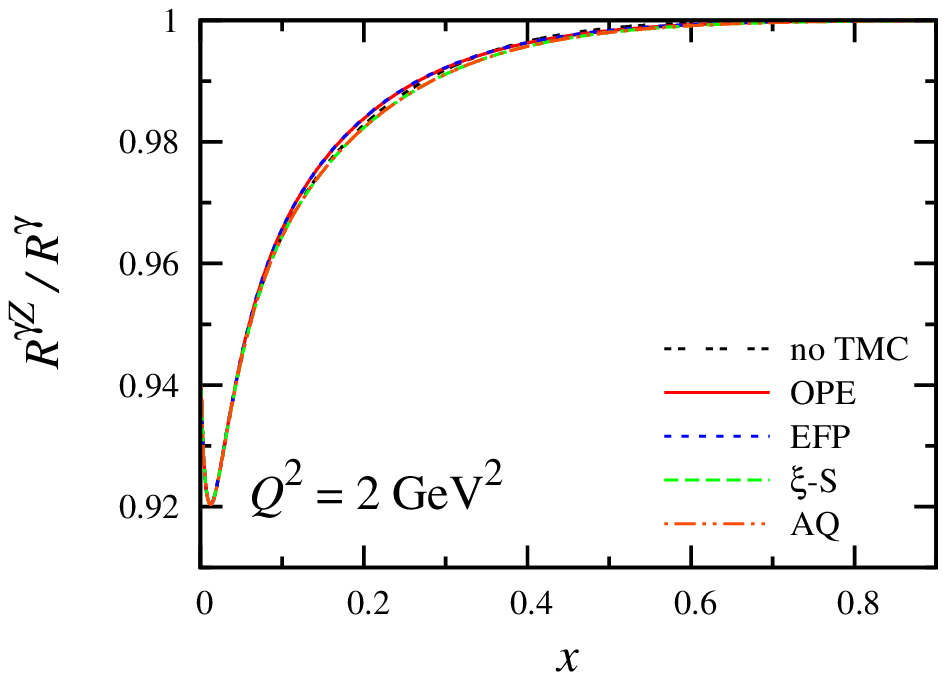}}
\rotatebox{-90}{\includegraphics[width=6.05cm]{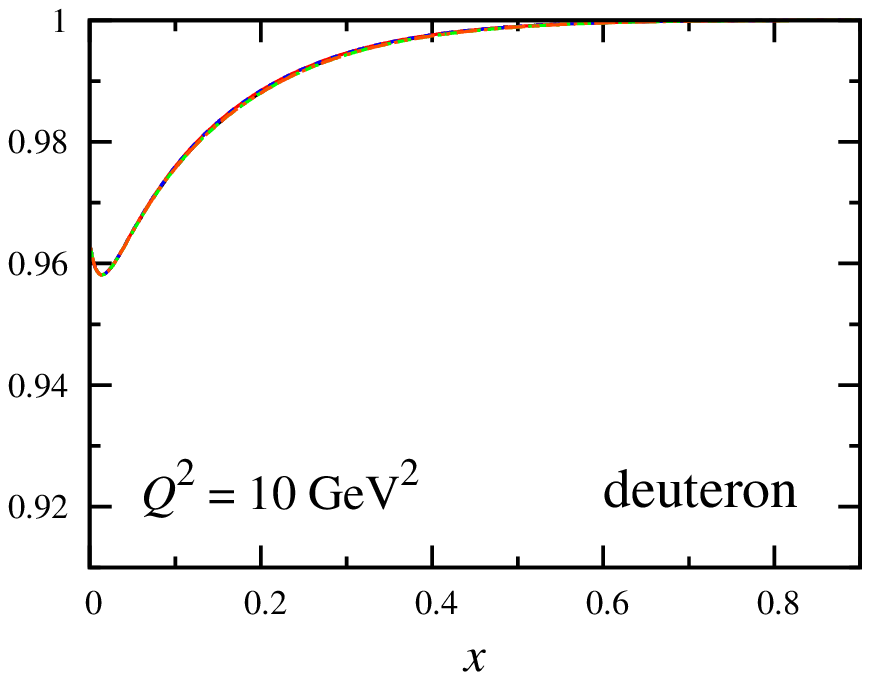}}
\caption{As in Fig.~\ref{fig:Rp}, but for the {\it deuteron}
	$R^{\gamma Z}$ to $R^{\gamma}$ LT ratio.}
\label{fig:Rd}
\end{figure}

For the case of the isoscalar deuteron target, stronger cancellations
between the quark content of $R^{\gamma Z}$ and $R^{\gamma}$ are expected
to lead to smaller deviations of their ratio from unity at large $x$; this
is perhaps unsurprising given a similar dampening of the finite-$Q^2$
effects relative to the proton as we observed in
Figs.~\ref{fig:dRgZ_Ad} \& \ref{fig:dRg_Ad}.
This behavior is indeed observed again in Fig.~\ref{fig:Rd}, where, much like like Fig.~\ref{fig:Rp},
the dip in the ratio at very low $x$ is associated with NLO gluon dominance of the LT
ratios as $x \to 0$.  At $x=0.2$, for example, the gluonic content of
$F_L$ suppresses the deuteron $R^{\gamma Z}/R^{\gamma}$ ratio by
$\approx 2\%$ for $Q^2 = 2$~GeV$^2$, and $\approx 1\%$ for
$Q^2 = 10$~GeV$^2$.
At higher $x$ the deviations decrease until the ratio approaches unity
asymptotically as $x \to 1$.  In the region of $x$ where the LT ratios
are dominated by quarks, the fact that the same isoscalar combination of
quark PDFs enters both the electromagnetic and $\gamma Z$ interference
structure functions leads to almost negligible TMC effects.
The absence of significant TMC effects in the deuteron ratio is,
as expected, even more clearly visible at the higher $Q^2$ value.
\subsection{PVDIS asymmetries}
\label{sec:tPVDIS}

In this section we examine the effects of TMCs on the PVDIS asymmetries
of the proton and deuteron, and discuss the phenomenological
implications of their uncertainties on future planned experiments.

\subsubsection{Proton asymmetry}
\label{ssec:PVDIS-p}

\begin{figure}[h]
\hspace*{-0.4cm}
\rotatebox{-90}{\includegraphics[width=6.05cm]{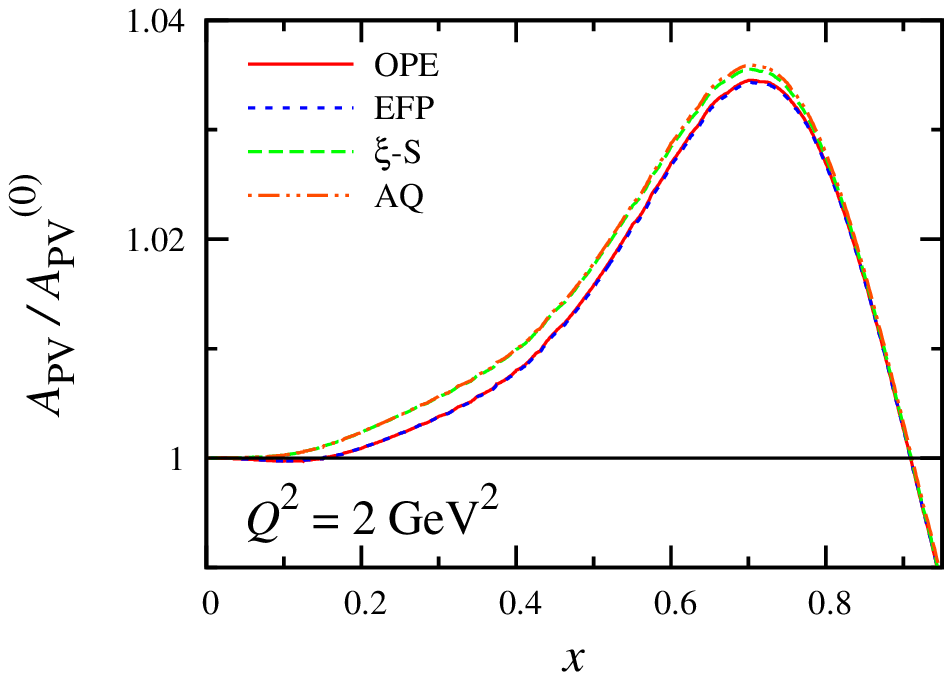}}
\rotatebox{-90}{\includegraphics[width=6.05cm]{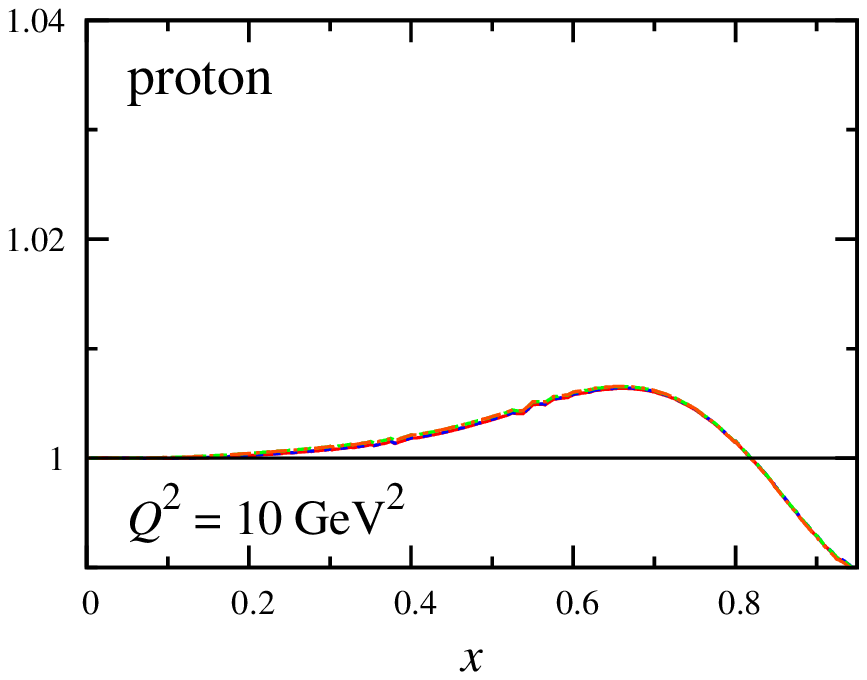}}
\caption{Ratio of target mass corrected ($A_{\rm PV}$) to uncorrected
	($A_{\rm PV}^{(0)}$) PVDIS asymmetries for the {\it proton}
	at $Q^2 = 2$~GeV$^2$ (left) and $Q^2 = 10$~GeV$^2$
	(right), for the OPE (solid, red),
	EFP (short-dashed, blue),
	$\xi$-S (long-dashed, green), and
	AQ (dot-dashed, orange) TMC prescriptions.
	Note that the AQ and $\xi$-S results are almost
	indistinguishable, as are the EFP and OPE prescriptions.}
\label{fig:PAS}
\end{figure}

The proton PVDIS asymmetry is shown in Fig.~\ref{fig:PAS} for
$Q^2=2$ and 10~GeV$^2$ in the form of the ratio of the target mass
corrected to uncorrected asymmetries.  For all prescriptions the TMC
effects are maximal at $x \approx 0.7$, where they are of the order
of $3-4\%$ at $Q^2=2$~GeV$^2$ and $\lesssim 1\%$ at $Q^2=10$~GeV$^2$.
The results are slightly higher for the $\xi$-S and AQ corrections
(which are virtually indistinguishable) than for the OPE and EFP
(which are also almost identical).  The small size of the effects is
principally due to the strong cancellation of the TMCs in the $F_1$
structure functions [effectively, the hadronic-vector part of $A^{PV}$ in Eq.~(\ref{eq:a1a3ex})],
namely,
$( F_1^{\gamma Z} / F_1^\gamma )^{\rm TMC}
 \approx 
 ( F_1^{\gamma Z} / F_1^\gamma )^{(0)}$,
even though $|F_1^{\rm TMC} / F_1^{(0)}| \gg 1$ at high $x$. 
Overall, the results indicate that the asymmetries themselves are
less sensitive to TMCs than are the LT ratios $R^{\gamma, \gamma Z}$
on which the asymmetries depend.

\begin{figure}[h]
\hspace*{-0.4cm}
\rotatebox{-90}{\includegraphics[width=6.05cm]{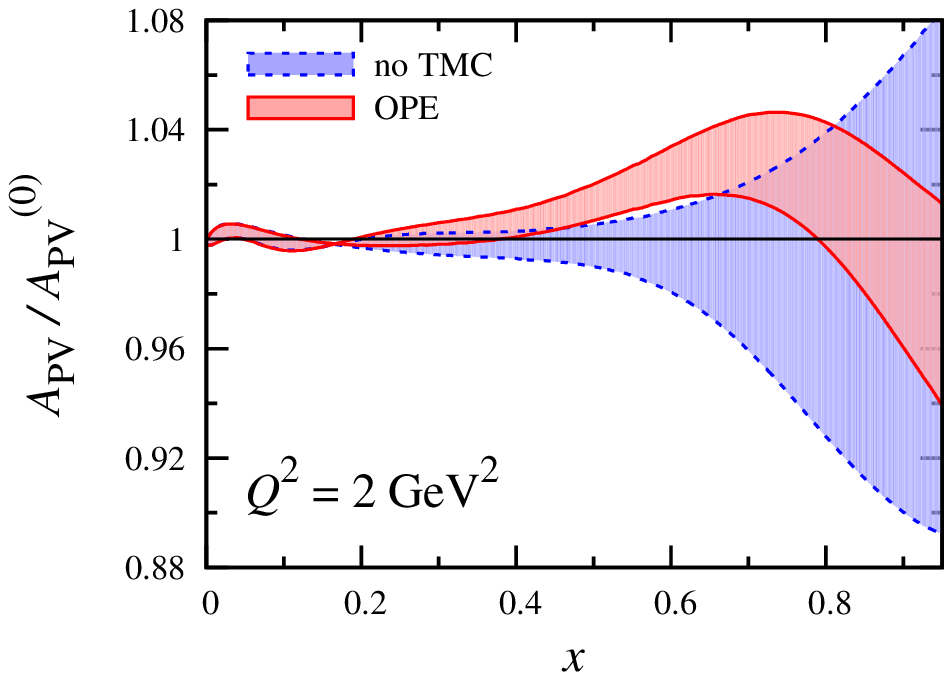}}
\rotatebox{-90}{\includegraphics[width=6.05cm]{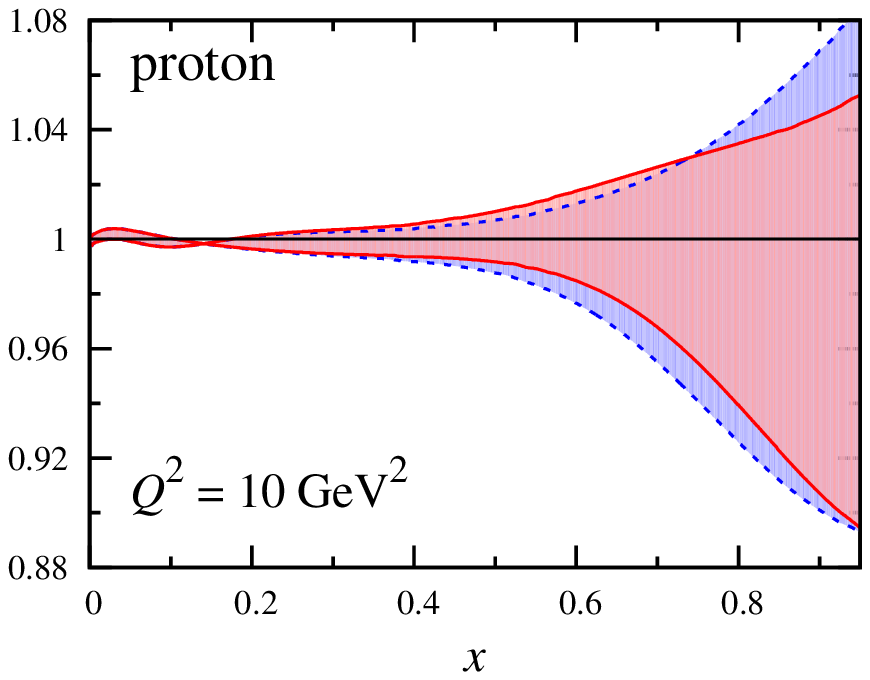}}
\caption{Proton PVDIS asymmetry $A_{\rm PV}$ at $Q^2 = 2$~GeV$^2$
	(left) and $Q^2 = 10$~GeV$^2$ (right) for
	CJ PDFs with minimal and maximal nuclear effects \cite{CJ11}
	(shaded bands), relative to the asymmetry $A_{\rm PV}^{(0)}$
	for the central CJ PDF fits, with no TMCs (dashed, blue) and
	using the OPE TMC prescription (solid, red).}
\label{fig:PASOPE}
\end{figure}

Since one of the main goals of the proton PVDIS measurements will be
to reduce the uncertainties on PDFs at large $x$, particularly on the
$d/u$ ratio, it is instructive to compare the magnitude of the TMC
effects with the expected sensitivity of the asymmetry to different
possible PDF behaviors at large $x$.  In Fig.~\ref{fig:PASOPE} we show
the proton asymmetry $A_{\rm PV}$ computed from the full range of CJ
PDFs \cite{CJ11} including minimal and maximal nuclear corrections
(shaded bands) relative to the central PDF fits.  The uncertainty band
increases with increasing $x$, reflecting the larger uncertainty on
the $d$ quark PDF at large $x$, and in the absence of TMCs ranges from
$\approx 3\%$ at $x=0.6$ to $\approx 11\%$ at $x=0.8$ for both $Q^2=2$
and 10~GeV$^2$.  This is significantly larger than the TMC uncertainty
band in Fig.~\ref{fig:PAS}, where the spread of the TMC model
predictions is $\ll 1\%$, even though the absolute target mass
effect is somewhat larger.

The effect of TMCs on the PDF uncertainty, illustrated in
Fig.~\ref{fig:PASOPE} for the OPE prescription, is to reduce the
uncertainty band at large $x$ for the lower $Q^2$ value,
with strength resultingly moving from lower $x$ to higher $x$ by the $x \to \xi$
rescaling of the structure functions.  The slightly different effects
of TMCs on the various structure functions present in the asymmetry
render the uncertainty band on $A_{\rm PV}$ more asymmetric at
$Q^2=2$~GeV$^2$.  At the higher $Q^2=10$~GeV$^2$ value, the impact
of TMCs on the uncertainty band is reduced considerably, with the
two bands (corresponding to no TMCs and the OPE TMC prescription)
approximately coinciding for all $x$.

The conclusion from the combined results of Figs.~\ref{fig:PAS} and
\ref{fig:PASOPE} is that the effect of TMCs and particularly their
uncertainties can be minimized in the $A_{\rm PV}$ ratio by measuring
the asymmetry at values of $Q^2 \sim 10$~GeV$^2$ or higher; at lower
$Q^2$, although the TMC uncertainties are not large, some residual
corrections will need to be applied in the range
$0.4 \lesssim x \lesssim 0.9$, where the TMCs are $\approx 1\%$ or
higher.

\subsubsection{Deuteron asymmetry}
\label{ssec:PVDIS-d}

As we witnessed in Chap.~\ref{chap:ch-Q2}, PVDIS on an isoscalar deuterium nucleus
features a near total cancellation of the dependence on PDFs if charge symmetry
is strictly assumed \cite{Bj78}.
In fact, it is convenient for our purposes here to recast the valence
quark region ($x \gtrsim 0.5$) deuteron asymmetry of Eq.~(\ref{eq:deut_asym})
at leading order as
\begin{equation}
A_{\rm PV}^d\
\approx\ -\left( {G_F Q^2 \over 2 \sqrt{2} \pi \alpha} \right)
{6 \over 5}
  \left( g_A^e (2 g_V^u - g_V^d)\ +\ Y_3\, g_V^e (2 g_A^u - g_A^d)
  \right)\ ,\ \ \ \ \ \ [x \gg 0]\ ,
\label{eq:APV_Bj}
\end{equation}
where the couplings $g_V^{u,d}$, etc., are again specified by Eq.~(\ref{eq:gAgV})
of Chap.~\ref{chap:ch-Q2}. We have seen that accurate measurement of deuteron PVDIS
is thus a potentially sensitive test of either the weak mixing angle $\sin^2\theta_W$
(deviations of which from its Standard Model value may signal the presence of new
physics), or more conventionally, of charge symmetry violation (CSV) in PDFs.

We have also argued that nonzero values of $\delta u$ and $\delta d$ are predicted in
various nonperturbative models of the nucleon to arise from quark mass
differences and electromagnetic effects, and can also be generated from radiative QED
corrections in $Q^2$ evolution \cite{MRSTCSV,MRSTQED,GJRQED} as we simulated in 
Chap.~\ref{chap:ch-Q2}.\ref{sec:deut-pCSV}.
If we again define charge symmetry violating PDFs by
\begin{equation}
\label{eq:dudd}
\delta u = u^p - d^n\ ,\ \ \ \ \ \ \delta d = d^p - u^n\ ,
\end{equation}
the PVDIS asymmetry (\ref{eq:APV_Bj}) in the presence of CSV
is modified according to
\begin{eqnarray}
\label{eq:CSV_g}
(2 g_{V,A}^u - g_{V,A}^d)
&\to& (2 g_{V,A}^u - g_{V,A}^d) (1 + \Delta_{V,A})\ ,
\end{eqnarray}
where the fractional CSV corrections are given by
\begin{eqnarray}
\label{eq:CSV_Delta}
\Delta_{V,A}
&=&
\left( -\frac{3}{10} + \frac{2 g_{V,A}^u + g_{V,A}^d}{2 (2 g_{V,A}^u - 
g_{V,A}^d)}
\right)
\left( \frac{\delta u - \delta d}{u + d}
\right)\ .
\end{eqnarray}
These approximate expressions serve to illustrate explicitly the role
of CSV in the PVDIS asymmetry in a fashion that explicitly manifests
the vector/axial-vector structure of Eq.~(\ref{eq:AmpII_Z}), for example;
in practice, however, the full deuteron asymmetry can be computed
including the effects of CSV at NLO, as well as sea quarks and gluons.

\begin{figure}[h]
\hspace*{-0.4cm}
\rotatebox{-90}{\includegraphics[width=6.05cm]{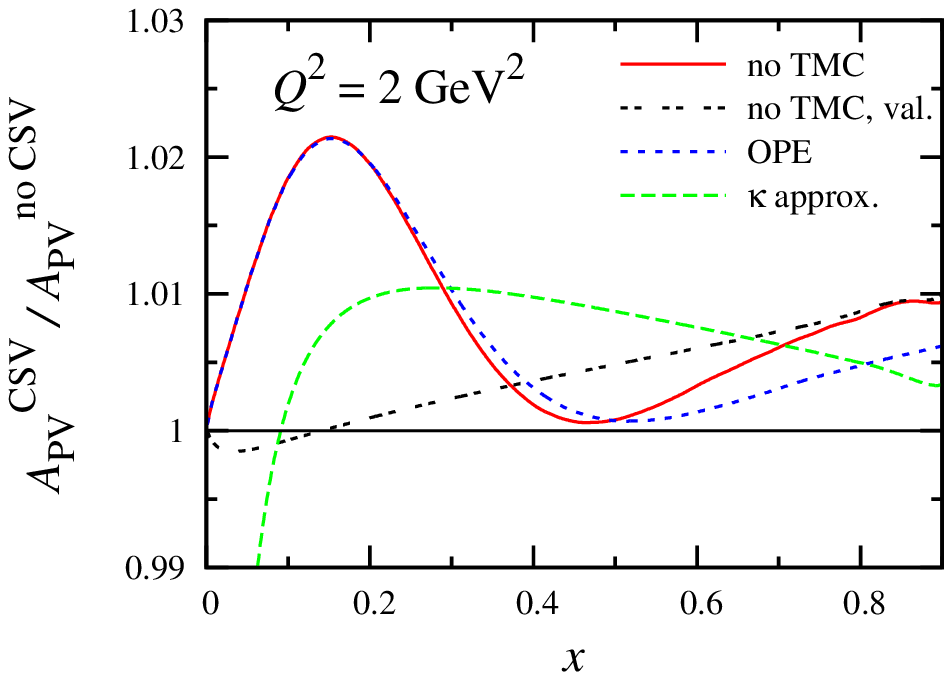}}
\rotatebox{-90}{\includegraphics[width=6.05cm]{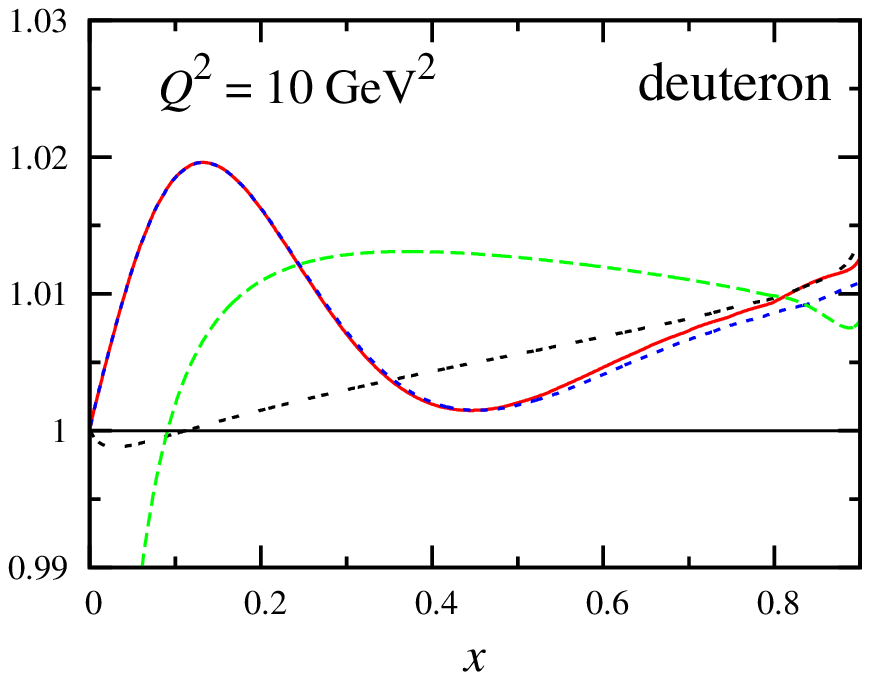}}
\caption{Deuteron PVDIS asymmetry including CSV effects, relative to the
	asymmetry with charge-symmetric PDFs, at $Q^2 = 2$~GeV$^2$
	(left) and $Q^2 = 10$~GeV$^2$ (right).
	The CSV PDFs are computed from the MRSTQED parametrization
	\cite{MRSTQED} for the full asymmetry (solid, red) and for
	the valence approximation (double-dashed, black), and from
	the $\kappa$-dependent fit (see text) in Ref.~\cite{MRSTCSV}
	(long-dashed, green).  The effects of TMCs on the full asymmetry
	with the MRSTQED PDFs are illustrated for the OPE prescription
	(short-dashed, blue).}
\label{fig:CSV}
\end{figure}

Using the MRSTQED parametrization of PDFs \cite{MRSTQED} similar to 
our description in Sec.~\ref{sec:deut-CSV},
we illustrate the effect of CSV on the deuteron asymmetry $A_{\rm PV}^d$
in Fig.~\ref{fig:CSV}.
%
%
Deviations of the full NLO result from the valence approximation of Eq.~(\ref{eq:CSVfit}) appear
already at $x \lesssim 0.7$, and the various CSV treatments differ quite markedly
at small $x$, as Fig.~\ref{fig:CSV} indicates.  Interestingly, the full
asymmetry becomes larger at smaller $x$ because of CSV effects in the
light sea quarks, which produce an asymmetry of about 2\% at
$x \approx 0.2$.  On the other hand, cleanly separating the CSV effects
from sea quark and gluon contributions, which introduce additional $x$
dependence beyond that in Eqs.~(\ref{eq:APV_Bj}), (\ref{eq:CSV_g}) and
(\ref{eq:CSV_Delta}), as well as possible differences between CSV in
valence and sea quark PDFs, becomes more challenging at small $x$.

\begin{figure}[h]
\hspace*{-0.4cm}
\rotatebox{-90}{\includegraphics[width=6.05cm]{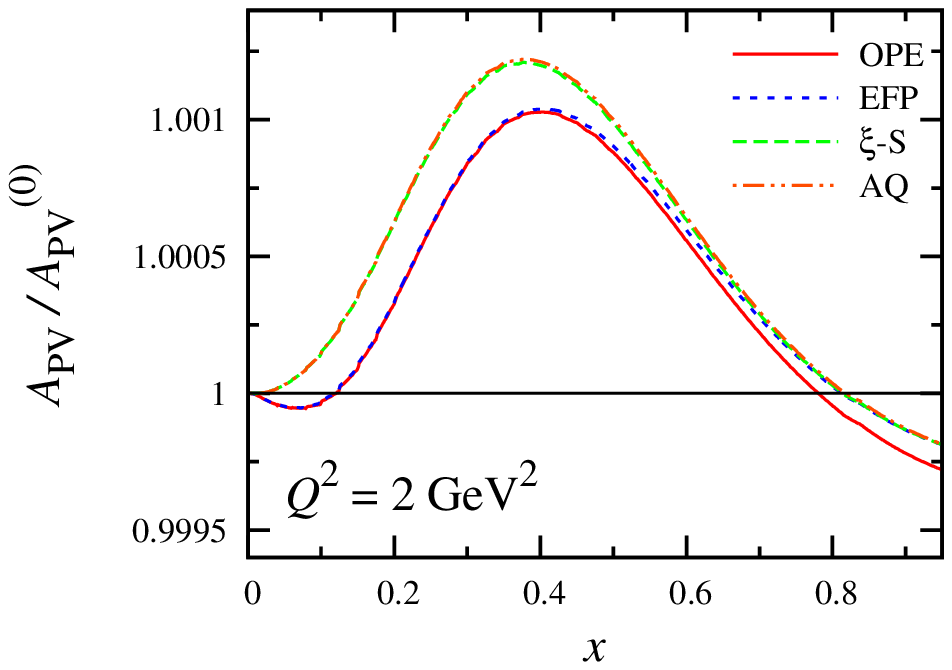}}
\rotatebox{-90}{\includegraphics[width=6.05cm]{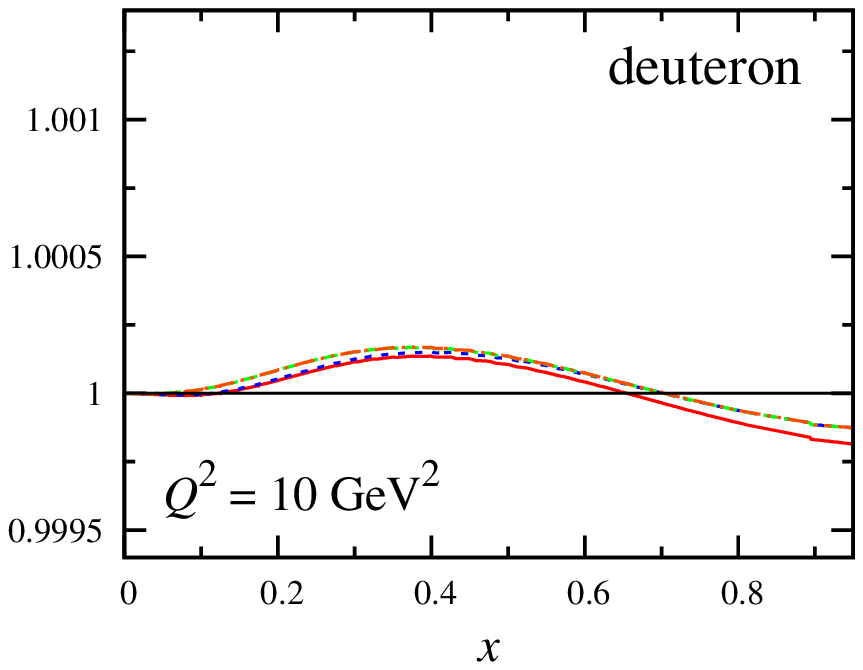}}
\caption{Ratio of target mass corrected to uncorrected PVDIS
	deuteron asymmetries $A_{\rm PV}^d$
	at $Q^2 = 2$~GeV$^2$ (left) and
	$Q^2 = 10$~GeV$^2$ (right), for the
	OPE (solid, red),
	EFP (short-dashed, blue),
	$\xi$-S (long-dashed, green), and
	AQ (dot-dashed, orange) TMC prescriptions.}
\label{fig:DAS}
\end{figure}

With sought-after CSV effects that could be $\lesssim 1-2\%$,
it is vital to quantify the impact of TMCs on the deuteron PVDIS
asymmetries and in particular the TMC prescription dependence.
The effect of TMCs on the full asymmetry relative to the
charge-symmetric asymmetry is negligible at $x \lesssim 0.5$,
but decreases the CSV signal by up to 50\% at $x \approx 0.8$,
as Fig.~\ref{fig:CSV} demonstrates for the OPE prescription.
The model dependence of the mass corrections is illustrated for the various
prescriptions in Fig.~\ref{fig:DAS}, wherein we plot the ratio
PV asymmetries with and without TMCs.
The net effect is very small, peaking at $\sim 0.1\%$ at
$x \approx 0.4$, even at the $Q^2=2$~GeV$^2$ value.
The TMC prescription dependence of this ratio is even smaller, making
it essentially negligible on the scale of a CSV signal of $\sim 1\%$.
If the target mass corrected asymmetries were calculated with the
charge symmetry violating MRSTQED PDFs, the effect would be somewhat
larger, peaking at $\sim 0.3\%$ around $x \approx 0.4$.  However,
the TMC model dependence is still negligible at around 0.05\%.
As expected, the impact of TMCs on the deuteron asymmetries at the
larger $Q^2=10$~GeV$^2$ value is considerably smaller.  It is therefore
likely that TMCs would only play a role in deuteron PVDIS measurements
if the CSV effects were on the scale of a fraction of a percent,
at which point they would not be discernible within the expected
precision of proposed experiments, \EG~\cite{SOLID}.

%% file: the-sidis.tex
\section{Semi-inclusive Hadron Mass Corrections}
\label{sec:semi-intro}

Having catalogued the physics of mass effects in fully inclusive
DIS, we turn our attention to the corresponding treatment of
semi-inclusive deep inelastic scattering (SIDIS), which has in recent
years received great attention as a tool to investigate various aspects
of hadron structure; counted highly among these is the flavor dependence
of the nucleon's parton distribution functions, polarized and unpolarized,
which may be assessed through flavor tagging of final state hadrons.
Observation of the momentum distribution of produced hadrons also
allows access to the largely unexplored transverse momentum dependent
parton distributions, which reveal a much richer landscape of the spin
and momentum distribution of quarks in the nucleon. This will be made
all the clearer by the phenomenological computations of
Chap.~\ref{chap:ch-TDIS}.

At high energies the scattering and hadronization components of the
SIDIS process factorize and the cross section can be represented
as a product of parton distribution and fragmentation functions.
In practice, however, experiments are often carried out at few-GeV
energies with $Q^2$ as low as 1~GeV$^2$, suggesting that $1/Q^2$ power
corrections must be controlled in order to determine the applicability
of partonic analyses of the data. Despite the fairly robust use of the
OPE in inclusive scattering we just observed, this method cannot be
rigorously extended to the production of hadrons in the final state.
This is because hadronic matrix elements in SIDIS are technically off-diagonal,
and hence lack the symmetric basis required for expansions of the type given
in Eq.~(\ref{eq:OPE_mat}). Due to their direct formulation in momentum space,
collinear factorization techniques, however, {\it can} be extended to SIDIS,
a fact we demonstrate in the present section.

In contrast to fully inclusive DIS, where the only mass scale entering the
problem is that of the target hadron, in SIDIS finite-$Q^2$ corrections
arise from both the target mass and the mass of the produced hadron.
For generality we shall refer to their combined effects as
``hadron mass corrections'' (HMCs).
Accordingly, we use the CF framework to derive the mass corrections to
the SIDIS cross section at finite $Q^2$, and systematically investigate
their implications at kinematics relevant to contemporary experiments.
In Sec.~\ref{ssec:CFkin} -- \ref{ssec:CFhad} we review the formal aspects
of the collinear approach and discuss their application to semi-inclusive
hadron production. To expose the origin of the corrections we work at leading
order in $\alpha_s$.

In Sec.~\ref{ssec:HMC-results} we explore the relative importance of the
HMCs numerically, and evaluate the size of the corrections in the
cross sections and fragmentation functions at various kinematics.
To assess their possible impact on data analyses, we also compare
the magnitude of the HMCs at kinematics typical of modern facilities
like Jefferson Lab with experimental errors from recent experiments.

%
\subsection{Collinear kinematics} 
\label{ssec:CFkin}

We begin the discussion of SIDIS at finite values of the photon
virtuality $Q^2$ by defining the relevant kinematics and momentum
variables in a collinear frame, and introduce the hadronic tensor
computed in a covariant parton model.
Collinear factorization is then performed in the leading order
approximation in which the produced hadron is effectively collinear
with the scattered parton, which more directly reveals the effects
of hadron masses on the cross sections and fragmentation functions.

\begin{figure}[h]
\includegraphics[height=5.5cm]{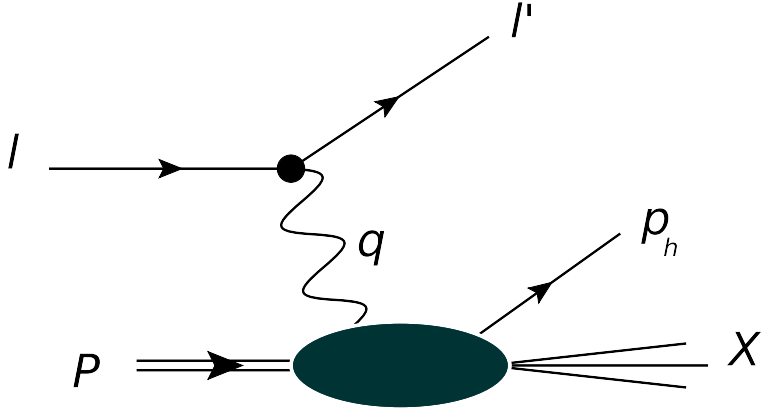}
\caption{
Unlike fully inclusive DIS, some of the hadronic final state is directly
observed in SIDIS --- in our case, a single produced hadron (\EG~a pion or kaon)
of momentum $p_h$.
}
\label{fig:SIDIS-kin}
\end{figure}
By definition, a collinear frame in Minkowski space is defined by any two
4-vectors. The intersection of the plane where they lie with the light-cone
defines the light-cone $4$-vectors $\nbar^\mu$ and $n^\mu$, which satisfy
$n^2=\nbar^2=0$ and $\nbar \cdot n = 1$. In SIDIS the hadronic tensor depends
on the $4$-vectors $P^\mu$, $q^\mu$ and $p_h^\mu$ that in turn define three possible
collinear frames denoted $(P,q)$, $(p_h,q)$ and $(p_h,P)$. More pointedly, these
correspond to the $4$-momenta of the struck nucleon $(P)$, virtual exchange photon
$(q)$, and final state, or `produced,' hadron $(p_h)$; the basic relationship among
these parameters is illustrated schematically in Fig.~\ref{fig:SIDIS-kin}.

The $(P,q)$ frame is the only frame that can be defined in DIS
and is the one used in this work;
the $(p_h,q)$ frame on the other hand is unique to
semi-inclusive hadron production in $e^+ e^-$ collisions;
and finally the $(p_h,P)$ frame is typically preferred for analysis
of transverse momentum dependent parton distributions in SIDIS.

In terms of the vectors $P$, $q$ and $p_h$ one can define two
fragmentation invariants,
\begin{align}
    z_h & = \frac{P \cdot p_h}{P \cdot q} \ , \qquad
  \eta_h = \frac{2 p_h \cdot q}{q^2}\, ,
\label{eq:zhzx}
\end{align}
which together with $x_B$, $Q^2$, $M^2$ and $m_h^2$ form a complete
set of scalar Lorentz invariants in SIDIS. Note that for the purposes
of the present section, we refer to the Bjorken limit parameter $x$
of previous chapters as $x_B$, so as to prevent confusion with an analogous
parton-level variable to be introduced shortly.
Because the variable $\eta_h$ is defined independently of the target
momentum, the effects of the final state hadron mass will decouple from
those of the target mass in all reference frames.
In contrast, $z_h$ is defined in terms of both the target and produced
hadron momenta, so that the target and hadron mass effects here will be
entangled.

The light-cone fractional momentum $\xi$ (Nachtmann variable) and
the fragmentation variable $\zeta_h$ are defined in terms of the
plus and minus components of the momenta as
\begin{align}
     \xi = - \frac{q^+}{P^+}\ ,   \qquad
 \zeta_h = \frac{p_h^-}{q^-} \ .
 \label{eq:xizetah}   
\end{align}
We use these definitions in all three frames; however, in each frame
the light-cone vectors (and therefore the plus and minus components of
the 4-momenta) will be different.
In the following we discuss each of the three collinear frames and the
consequences within each frame of the choice of fragmentation invariant.

In the DIS frame $(P, q)$ used to compute in subsequent sections, the 4-momenta
of the target nucleon ($P$), virtual photon ($q$) and produced hadron $h$ ($p_h$)
can be decomposed in terms of light-cone unit vectors $n$ and $\nbar$ as \cite{EFP}
\begin{subequations}
\begin{eqnarray}
P^\mu   &=& P^+\, \nbar^\mu 
         + \frac{M^2}{2 P^+}\, n^\mu\ ,		\\
q^\mu   &=& - \xi P^+\, \nbar^\mu 
         + \frac{Q^2}{2\xi P^+}\, n^\mu\ ,	\label{eq:kin1b} \\
p_h^\mu &=& \frac{\xi m_{h\perp}^2}{\zeta_h Q^2} P^+\, \nbar^\mu
	 + \frac{\zeta_h Q^2}{2 \xi P^+}\, n^\mu 
	 + p_{h\perp}^{\,\mu} \ ,		\label{eq:kin1c}
\end{eqnarray}
\label{eq:kinematics}%
\end{subequations}%
where as usual $M$ is the target nucleon mass, $Q^2=-q^2$, and the light-cone
vectors satisfy $n^2 = \nbar^2 = 0$ and $n \cdot \nbar=1$.
Here we define light-cone components of any 4-vector $v$ by
$v^+ = v \cdot \nbar = (v_0 + v_z)/\sqrt{2}$ and
$v^- = v \cdot n     = (v_0 - v_z)/\sqrt{2}$. 
The momenta $P$ and $q$ are chosen to lie in the same plane as $n$
and $\nbar$; particularly advantageous for the analysis of inclusive DIS, we
designate this the $(P,q)$ collinear frame.

The nucleon plus-momentum $P^+$ can be interpreted as a parameter for
boosts along the $z$-axis, connecting the target rest frame to the
infinite-momentum frame; the target rest frame ($P^+=M/\sqrt{2}$) and
the Breit frame ($P^+=Q/(\sqrt{2}\xi$)) are part of this family of frames. 
The transverse momentum 4-vector of the produced hadron
$p_{h\perp}^{\,\mu}$ satisfies
$p_{h\perp} \cdot n = p_{h\perp} \cdot \nbar = 0$,
and we define the transverse mass squared as
$m_{h\perp}^2 = m_h^2 - p_{h\perp}^2$,
where $m_h$ is the produced hadron mass, and the transverse
4-vector squared is
$p_{h\perp}^2 = -\bm{p}_{h\perp}^{\,2}$.

In the chosen collinear frame the variable $\xi = - q^+ / P^+$ 
defined in Eq.~\eqref{eq:kin1b} coincides with the finite-$Q^2$
Nachtmann scaling variable \cite{Nachtmann,Greenberg},
\begin{align}
\xi = \frac{2 x_B}{1 + \sqrt{1 + 4 x_B^2 M^2/Q^2}}\ ,
\label{eq:xi}
\end{align}
which in the Bjorken limit ($Q^2 \to \infty$ at fixed $x_B$)
reduces to the Bjorken scaling variable $x_B=Q^2/2P\cdot q$.
The scaling fragmentation variable $\zeta_h = p_h^- / q^-$ defined
in Eq.~\eqref{eq:kin1c} is related to the fragmentation invariant
$z_h = P \cdot p_h / P \cdot q$ by
\begin{subequations}
\label{eq:zeta_h}
\begin{align}
\zeta_h
  & = \frac{z_h}{2} \frac{\xi}{x_B}
  \left( 1 + \sqrt{1 - \frac{4 x_B^2 M^2 m_{h\perp}^2}{z_h^2\ Q^4}} 
  \right)\, ,
\label{eq:zeta_h1}
\end{align}
and the positivity of the argument in the radical in
Eq.~(\ref{eq:zeta_h1}) is ensured by the condition $E_h \geq m_{h\perp}$,
which imposes
\begin{align}
z_h \geq z_h^{\rm min} = 2 x_B \frac{M m_h}{Q^2} \ .
\end{align}
One can also define $\zeta_h$ in terms of the invariant
$\eta_h = 2p_h \cdot q / q^2$ by
\begin{align}
\zeta_h
  & = \frac{\eta_h}{2} 
  \left( 1 + \sqrt{1 + \frac{4 m_{h\perp}^2}{\eta_h^2\ Q^2}} 
  \right)\ ,
\label{eq:eta_h2}
\end{align}
\end{subequations}
which is convenient for discriminating between the target and current
fragmentation hemispheres in hadron production.
Note that in the target rest frame $z_h = E_h/\nu$ is the usual
ratio of the produced hadron to virtual photon energies. 
In the Breit frame $\eta_h=p_{hz}/q_z$ is the ratio of the longitudinal 
components of the hadron and photon energies, which can be used to
define the current ($\eta_h>0$) and target ($\eta_h<0$) hemispheres
for hadron production.
In the Bjorken limit one has $\zeta_h \to z_h \to \eta_h$.

Conservation of 4-momentum and baryon number impose an upper limit
on the $x_B$ variable,
\begin{gather}
  x_B \leq \left( 1 + {m_h^2 + 2 M m_h \over Q^2} \right)^{-1}
      \equiv x_B^{\rm max}\, ,
\label{eq:xBlim}
\end{gather}
which corresponds to the exclusive production of a nucleon and
a hadron $h$ in the final state.
Similarly the limits on the fragmentation variable $\zeta_h$ are
given by
\begin{gather}
  \frac{\xi}{1-\xi} \frac{M^2}{Q^2} \leq \zeta_h 
    \leq 1 + \xi \frac{M^2}{Q^2}\, ,
\label{eq:zetahlim}
\end{gather}
where the lower limit corresponds to diffractive production of the
hadron $h$, and the upper limit reflects the fragmentation threshold,
which approaches unity in the Bjorken limit.

As mentioned, the $(P,q)$ frame of Eq.~(\ref{eq:kinematics}) is hardly
a unique selection, and we might equally well compute in an alternative
collinear frame; these choices lead of course to slightly modified scaling
parameters, as we summarize here.

For the sake of comparison and completeness, the $(p_h,q)$ frame used in Ref.~\cite{Albino}
defines the external SIDIS vectors as
\begin{subequations}
\begin{eqnarray}
  P^\mu   &=& P^+ \nbar^\mu
         + \frac{M_\perp^2}{2 P^+} n^\mu
         + P_{\perp}^{\,\mu} \ ,                  \\
  q^\mu   &=& - \xi P^+ \nbar^\mu
         + \frac{Q^2}{2\xi P^+} n^\mu\ ,        \\
  p_h^\mu &=& \frac{m_{h}^2}{\zeta_h Q^2/\xi} P^+ \nbar^\mu
         + \zeta_h \frac{Q^2}{2 \xi P^+} n^\mu \ ,
\label{eq:kinematics_phq}
\end{eqnarray}
\end{subequations}
where $M_\perp^2 = M^2 - P_\perp^2 = M^2 + \bm{P}_\perp^2$ is the
transverse mass of the target nucleon.
The Nachtmann variable in this case is given by
\begin{align}
  \xi = \frac{2 x_B}{1 + \sqrt{1 + 4 x_B^2 M_\perp^2/Q^2}}\ ,
\end{align}   
which, in contrast to its definition in the $(P,q)$ frame,
depends explicitly on the transverse mass of the target nucleon.
Furthermore, in terms of the fragmentation invariant $z_h$,
the finite-$Q^2$ fragmentation variable $\zeta_h$ is given by
\begin{align}
  \zeta_h  = \frac{z_h}{2} \frac{\xi}{x_B}
    \left(
      1 + \sqrt{1 - 4 \frac{x_B^2}{z_h^2} \frac{M_\perp^2 m_h^2}{Q^4}}
    \right)\ ,  
  \label{eq:zeta_h-zh_phq} 
\end{align}
or in terms of $\eta_h$ by
\begin{align}
  \zeta_h  = \frac{\eta_h}{2}
    \left(
      1 + \sqrt{1 + 4 \frac{1}{\eta_h^2} \frac{m_h^2}{Q^2}}
    \right)\ .
  \label{eq:zeta_h-zx_phq}
\end{align}

Lastly, we point out that the external vectors in the $(p_h,P)$ frame of Ref.~\cite{Mulders}
are specified by
\begin{subequations}
\begin{eqnarray}
  P^\mu   &=& P^+ \nbar^\mu
         + \frac{M^2}{2 P^+} n^\mu \ ,         \\
  q^\mu   &=& - \xi P^+ \nbar^\mu
         + \frac{Q_\perp^2}{2\xi P^+} n^\mu
         + q_{\perp}^{\,\mu} \ ,                  \\
  p_h^\mu &=& \frac{m_{h}^2}{\zeta_h Q^2/\xi} P^+ \nbar^\mu  
         + \zeta_h \frac{Q^2}{2 \xi P^+} n^\mu \ ,
\label{eq:kinematics_php}
\end{eqnarray}
\end{subequations}
where $Q_\perp^2 = Q^2 - q_\perp^2 = Q^2 + \bm{q}_\perp^2$
is the transverse mass of the virtual photon.
The Nachtmann variable in this frame depends explicitly on $Q_\perp^2$,
\begin{align}
  \xi = \frac{Q_\perp^2}{Q^2}
  \frac{2 x_B}{1 + \sqrt{1 + 4 x_B^2 M^2 Q_\perp^2/Q^4}}\ ,
\end{align}
and the finite-$Q^2$ fragmentation variable is given by
\begin{align}
  \zeta_h  = \frac{z_h}{2} \frac{\xi}{x_B} \frac{Q^2}{Q_\perp^2}
    \left(
      1 + \sqrt{1 - 4 \frac{x_B^2}{z_h^2} \frac{M^2 m_h^2}{Q^4}}
    \right)\ ,
  \label{eq:zeta_h-zh_php}
\end{align}   
or
\begin{align}
  \zeta_h  = \frac{\eta_h}{2} \frac{Q^2}{Q_\perp^2}
    \left(
      1 + \sqrt{1 + \frac{4}{\eta_h^2} \frac{m_h^2 Q_\perp^2}{Q^4}}
    \right)\ .
  \label{eq:zeta_h-zx_php}
\end{align}

%

While the three frames discussed here are generally distinct,
up to leading order in $1/Q^2$ the vectors $P$, $q$ and $p_h$
lie in the same plane and the frames actually coincide.
Comparing the $(P,q)$ and $(p_h,q)$ frames, for example, the
differences between the transverse momenta and scaling variables
can be expressed as
\begin{subequations}
\begin{eqnarray}
\bm{p}_{h\perp} &= \bm{P}_\perp^* + {\cal O}(\bm{P}_\perp^{*2}/Q^2)\, ,\\
\xi & = \xi^* + {\cal O}(\bm{P}_\perp^{*2}/Q^2)\, ,		       \\
\zeta_h & = \zeta_h^* + {\cal O}(\bm{P}_\perp^{*2}/Q^2)\, ,
\end{eqnarray}
\end{subequations}
where the asterisks ($^*$) label quantities in the $(p_h,q)$ frame.
Similar relations are applicable also for the parton fractional 
momentum $x$ and the hadron fractional momentum $z$.
At leading order in collinear factorization one has
$\bm{p}_{h\perp} = 0$ and the frames are manifestly equivalent.
Moreover, since $\langle \bm{p}^2_{h\perp} \rangle \ll Q^2$ for
$\bm{p}_{h\perp}$-integrated cross sections at next-to-leading order
the differences between the collinear frames should remain small.  
For the sake of future work, it will be important to check more thoroughly whether,
and in what kinematic range, this approximation is valid;
this is especially true for unintegrated cross sections, where differences
between frames become relevant and must be better quantified.


\begin{figure}[h]
\includegraphics[height=8cm]{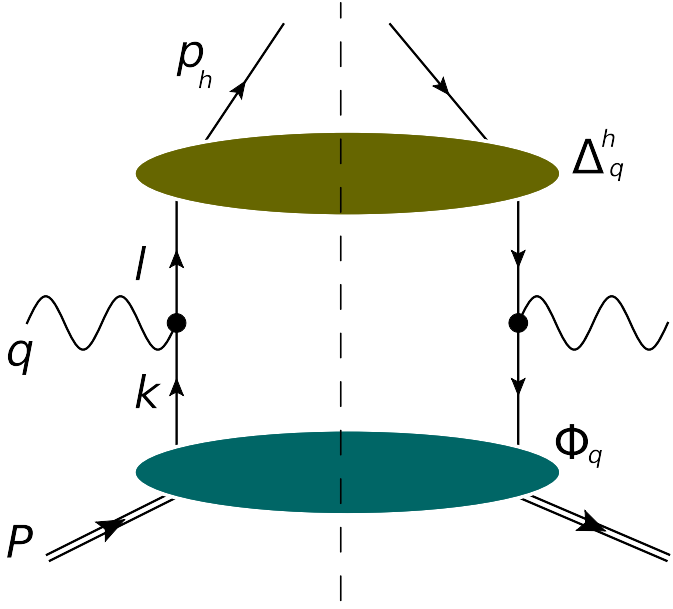}
\caption{Parton-level kinematics of semi-inclusive deep inelastic lepton--nucleon
	scattering at leading order, producing a final state hadron $h$.
	The momenta of the target nucleon ($P$), virtual photon ($q$),
	incident ($k$) and scattered quarks ($l$), and the produced
	hadron ($p_h$) are labeled explicitly, with $\Phi_q$ and $\Delta^h_q$
	denoting the correlators relevant to the quark distribution
	and fragmentation functions.  The vertical dashed line
	represents the cut of the forward amplitude.}
\label{fig:1}
\end{figure}

\vspace*{0.15cm}
{\bf \it Parton kinematics in collinear factorization.}
\vspace*{0.15cm}

Having charted formal issues related to frame choices, we proceed with
the analysis using the decomposition in the $(P,q)$ frame of Eq.~(\ref{eq:kinematics}).
At the partonic level the SIDIS process at leading order in $\alpha_s$ is illustrated in Fig.~\ref{fig:1}.
It occurs by means of scattering from a quark carrying a light-cone
momentum fraction $x = k^+ / P^+$, which then fragments (nonperturbatively) to a hadron $h$
carrying a light-cone momentum fraction $z = p_h^-/l^-$, where $k$ and
$l$ are the 4-momenta of the initial and scattered quarks.
At higher orders the hard scattering can also take place from a gluon,
and additional partons might be created in the collision.

The parton momenta $k$ and $l$ can be parametrized in terms of the
light-cone vectors $n$ and $\nbar$ as
\begin{subequations}
\begin{eqnarray}
k^\mu &=& xP^+\, \bar{n}^\mu
       + \frac{k^2 + k_\perp^2}{2 x P^+}\, n^\mu
       + k_\perp^\mu\ ,				\\
l^\mu &=&  \frac{l^2 + l_\perp^2}{2 p_h^-/z}\, \bar{n}^\mu
       + \frac{p_h^-}{z}\, n^\mu
       + l_\perp^\mu\ ,
\end{eqnarray}
\end{subequations}
with the parton transverse momentum 4-vectors $k_\perp$ and
$l_\perp$ orthogonal to $n$ and $\nbar$. 
In collinear factorization the hard scattering amplitude is expanded
around on-shell and collinear momenta $\tilde k$ and $\tilde l$,
\begin{subequations}
\begin{eqnarray}
  \tilde k^\mu &=& xP^+\, \bar{n}^\mu
       + \frac{\tilde k^2}{2 x P^+}\, n^\mu \, \\
  \tilde l^\mu &=&  \frac{\tilde l^2 + p_{h\perp}^2/z^2}{2 p_h^-/z}\,
		    \bar{n}^\mu
       + {p_h^- \over z}\, n^\mu
       + {p_{h\perp}^\mu \over z}\ ,
\end{eqnarray}
\end{subequations}
where the initial and final collinear parton ``masses'' $\tilde k^2$
and $\tilde l^2$ are kept for generality.

Defining the invariant $\hat x = -q^2/2\tilde k\cdot q$ as the partonic
analog of the Bjorken variable $x_B$, at finite $Q^2$ one has
\begin{align}
\hat x = {\xi \over x}
         \left( 1 + {x \over \xi} {\tilde k^2 \over Q^2} \right)\, .
\label{eq:xf}
\end{align}
Using the formal methods of Ref.~\cite{AQ} one can show that for
SIDIS cross sections integrated over $p_{h \perp}$, $\hat x$ is
constrained to be in the range
\begin{align}
  1 + \frac{m_h^2}{\zeta_h Q^2}
    - \frac{\tilde k^2}{Q^2}
      \left( 1 - \frac{\xi m_h^2}{x \zeta_h Q^2} \right)
  \leq \, \frac{1}{\,\hat x\,} \, \leq 
  \frac{1}{x_B}
    \left( 1 - x_B \frac{2 M m_h + \tilde k^2}{Q^2} \right)\ ,
\label{eq:xflimits}
\end{align}
where the lower limit arises from the minimum of the current jet mass,
and the upper limit corresponds to collinear spectators with minimal mass.
These limits agree with the limit on $x_B$ in Eq.~\eqref{eq:xBlim} for
any $\tilde k^2 \geq x (\zeta_h-1)Q^2/\xi$.
For the fragmentation process one finds analogous limits on $\zeta_h$,
\begin{align}
  \zeta_h\ \leq\ {1 \over z}\, \zeta_h\
	   \leq\ 1 + {\xi \over x} {\tilde k^2 \over Q^2} \ ,
\label{eq:wlimits}
\end{align}
which agrees with the limit in Eq.~\eqref{eq:zetahlim},
provided that $\tilde k^2 \leq x M^2$. 
The requirement that the collinear parton masses be independent
of the parton momentum ({\it viz.}, independent of $x$) implies
$\tilde k^2 \leq 0$.
Combined with the above lower limit on $\tilde k^2$, this naturally
leads to a collinear expansion around a massless initial state parton,
$\tilde k^2 = 0$.

The choice of $\tilde l^2$ is made by considering the cross section
at leading order in $\alpha_s$.
Four-momentum conservation for the hard scattering, together with
the choice $\tilde k^2 = 0$, leads to the relations
$x = \xi (1 + \tilde l^2/Q^2) \equiv \xi_h$ and $z = \zeta_h$.
Clearly $z$ falls within the kinematic limits \eqref{eq:wlimits}.
However, in order for $x$ to respect the limits \eqref{eq:xflimits}
we choose $\tilde l^2 = m_h^2/\zeta_h$, in which case
\begin{align} 
\xi_h\ =\ \xi\, \left( 1 + {m_h^2 \over \zeta_h\, Q^2} \right)\, .
\label{eq:xi_h}
\end{align}
While larger values of $\tilde l^2$ would also allow $x$ to fall
within the limits \eqref{eq:xflimits}, this choice is the closest to
the physical quark mass.

We stress that our prescription for the collinear parton masses
$\tilde k^2$ and $\tilde l^2$ is dictated by the external kinematic
limits in Eqs.~\eqref{eq:xBlim} and \eqref{eq:zetahlim}, which are
independent of the parton model and collinear factorization
approximations.
As discussed in~\cite{AQ}, this is crucial when considering
cross sections close to the kinematic limits, such as at large $x_B$
or large $z_h$.
However, as we shall see in the next section, the SIDIS cross section
can also receive non-negligible corrections at small $x_B$ since
$\xi_h > \xi \approx x_B$.
This is qualitatively different from the behavior of the target mass
corrections in inclusive DIS, which are always suppressed at small $x_B$
\cite{AQ}.

\subsection{Leading order cross sections}
\label{ssec:CFhad}

In collinear factorization the hadron tensor at leading order, to
which we restrict the rest of our analysis, can be written as
\begin{align}
2 M W^{\mu\nu}(P,q,p_h)
= \sum_q e_q^2 \int d^4k\ d^4l\ \delta^{(4)}(\tilde k + q - \tilde l)\
  {\rm Tr}[ \Phi_q(P,k)\, \gamma^{\mu}\,
	    \Delta_q^h(l,p_h)\, \gamma^{\nu} ]\ ,
\label{eq:Wmunu}
\end{align}
where the sum is taken over quark flavors $q$, and the correlators
$\Phi_q$ and $\Delta_q^h$ encode the relevant quark distribution and
fragmentation functions, respectively \cite{Collins,Collins2,Mulders}.
According to our prescription for the collinear momenta, the
$\delta$-function depends on the collinear momenta $\tilde k$
and $\tilde l$, so that integrations over $dk^-\, d^2k_{\perp}$
and $dl^+\, d^2l_{\perp}$ act directly on the correlators $\Phi$
and $\Delta$.
The leading twist part of the cross section can then be extracted 
by retaining the $\nbslash$ and $\nslash$ components in the Dirac
structure expansion of the integrated correlators,
\begin{subequations}
\begin{align}
  \int dk^- d^2k_{\perp}\, \Phi_q(P,k)
	& = \frac12 f_q(x) \nbslash + \ldots\ ,		\\
  \int dl^+ d^2l_{\perp}\, \Delta_q^h(l,p_h)
	& = \frac{1}{2z} D_q^h(z) \nslash + \ldots\ ,
\end{align}
\end{subequations}
where the dots indicate contributions of higher twist \cite{TMD}.
The nonperturbative quark distribution function $f_q(x)$ and
quark-to-hadron fragmentation function $D_q^h(z)$ are explicitly
defined as
\begin{subequations}
\begin{align}
f_q(x)
    &= \frac12 \int dk^- d^2k_\perp\, \Tr\left[ \gamma^+ 
    \Phi_q(P,k)\right]_{k^+=xP^+}\ 		\nonumber \\
    & \stackrel{\text{LC}}{=}
    \frac{1}{2} \int \frac{dw^-}{2\pi} e^{i xP^+ w^-}
    \langle N | \overline{\psi}_q(0)\, \gamma^+\, \psi_q(w^- n)
    | N \rangle\ , \\
D_q^h(z)
    &= \frac{z}{2} \int dl^+ d^2l_\perp\,
	   \Tr\left[\gamma^-\Delta_q^h(l,p_h)\right]_{l^-=p_h^-/z}
						\nonumber \\
    &\stackrel{\text{LC}}{=}
    \frac{z}{2} \sum_X \int \frac{dw^+}{2\pi} e^{i(p_h^-/z)w^+}
    \langle 0 | \psi_q(w^+ n) |h, X \rangle
    \langle h,X | \overline{\psi}_q(0) \gamma^-| 0 \rangle\ , 
\end{align}
\end{subequations}
where ``LC'' denotes use of the light-cone gauge, and the
fragmentation function is normalized such that
$\sum_h \int_0^1 dz\, z\, D_q^h(z) = 1$ \cite{Mulders}. 
By construction,\footnote{Actually, by behavior under charge conjugation
and crossing symmetry, the fragmantation functions $D_q^h(z)$
can be related to $f_q(x)$ by the {\it DLY relation} \cite{Drell:1969jm}:
\begin{equation}
D_q^h(z)\ =\ (-1)^{2(s_q+s_h)+1}\ (2s_h+1)\ {z \over \gamma_q}\ f_q(x=1/z)\ ,
\nonumber
\end{equation}
where $s_q, s_h$ are the quark and hadron spins, and $\gamma_q$ the spin-color degeneracy
of quark $q$.} the probability functions $D_q^h(z)$ are equivalent
to the liklihood for a quark of species $q$ to ``fragment''
(a nonperturbative process) into a hadron $h$ as a function of
the invariant momentum fraction $z$.

From Eq.~(\ref{eq:Wmunu}) the energy-momentum conserving
$\delta$-function can be decomposed along the plus, minus,
and transverse components of the light-cone momentum.
The plus and minus components yield a product of $\delta$-functions,
which when integrated impose
\begin{subequations}
\begin{align}
\int dk^+\ \delta^{(+)} \left(\tilde{k} + q - \tilde{l}\right)\ \rightarrow\ 
xP^+ + q^+ - {z \tilde{l}^2 \over 2 p^-_h}\ =\ 0\ , \\
\int dl^-\ \delta^{(-)} \left(\tilde{k} + q - \tilde{l}\right)\ \rightarrow\ 
{\tilde{k}^2 \over 2xP^+} + q^- - {p^-_h \over z}\ =\ 0\ ;
\end{align}
\label{eq:CF_delta}
\end{subequations}
factorizing the hard partonic process about $\tilde{k}^2 =0$ and
$\tilde{l}^2 = m^2_h/\zeta_h$ as just argued, and evaluating Eq.~(\ref{eq:CF_delta})
using the $(P,q)$ definitions of Eq.~(\ref{eq:kinematics})
fixes $x = \xi_h$ and $z = \zeta_h$. Concordantly, the transverse
$\delta$-function components constrain the transverse momentum of the
scattered quark to vanish, restricting the produced hadrons to be purely
longitudinal, $p_{h\perp} = z\, l_\perp=0$.
Hadrons with nonzero transverse momentum can be generated from
higher order perturbative QCD processes, or from intrinsic
transverse momentum in the parton distribution functions as with
transverse momentum dependent distributions \cite{TMD},
but are not considered in this thesis.
The resulting hadron tensor in the presence of hadron mass effects,
\begin{align}
2M W^{\mu\nu}(P,q,p_h)
= \frac{\zeta_h}{4} \sum_q e_q^2\ 
  \delta^{(2)}(\bm{p}_\perp)
  \Tr \left[ \nbslash\gamma^{\mu}\nslash\gamma^{\nu} \right]
  f_q(\xi_h) D_q^h(\zeta_h)\, ,
\label{eq:TMC_hadtens}
\end{align}
is then factorized into a product of parton distribution and
fragmentation functions evaluated at the finite-$Q^2$ scaling
variables $\xi_h$ and $\zeta_h$, instead of $x_B$ and $z_h$ as
would be obtained in the massless case, and recovered from
Eq.~\eqref{eq:TMC_hadtens} in the Bjorken limit.

Finally, the SIDIS cross section is computed by contracting the
hadron tensor with the analogous lepton tensor of Eq.~(\ref{eq:lep}),
leading to
\begin{align}
\sigma\ \equiv\ \frac{d\sigma}{dx_B\, dQ^2\, dz_h}
&=\ \frac{2\pi\alpha_s^2}{Q^4} \frac{y^2}{1-\varepsilon}
    \frac{d\zeta_h}{dz_h}
    \sum_q e_q^2\, f_q(\xi_h,Q^2)\, D_q^h(\zeta_h,Q^2)\ ,
\label{eq:dsigTMC}
\end{align}
where the dependence of the functions on the scale $Q^2$ is made
explicit, and the Jacobian
$d\zeta_h/dz_h = (1 - M^2\xi^2/Q^2)
	       / (1-\xi^2 M^2 m_h^2/\zeta_h^2 Q^4)$.
In Eq.~\eqref{eq:dsigTMC} the variable $y$ defined as
$y = P\cdot q / P\cdot p_\ell$, where $p_\ell$ is the lepton momentum,
represents the fractional energy transfer from the lepton to the hadron in the
target rest frame ($y=\nu/E$, with $E$ the lepton energy), and
$\varepsilon = (1 - y - y^2 \gamma^2/4)
	     / (1 - y + y^2 [1/2+\gamma^2/4])$
is the ratio of longitudinal to transverse photon flux, with
$\gamma^2 = 4 x_B^2 M^2/Q^2$.
The cross section differential in $\eta_h$ can be obtained using
$d\zeta_h/d\eta_h = 1/(1+m_h^2/\zeta_h^2 Q^2)$ instead of
$d\zeta_h/dz_h$.
It is interesting to observe that since $\xi_h$ depends explicitly
on $m_h$ and $\zeta_h$ depends on $z_h$ and $x_B$, at finite $Q^2$
the scattering and fragmentation parts of the cross section
(\ref{eq:dsigTMC}) are not independent.

As a final remark we note that at the maximum allowed $x_B$ for
SIDIS, Eq.~(\ref{eq:xBlim}), the value of $\xi_h$ is smaller than  
$\xi_h (x_B=x_B^{\rm max}) < 1$.
As in the case of inclusive DIS \cite{AQ}, the SIDIS cross section
therefore does not vanish as $x_B \to x_B^{\rm max}$, which is yet another
manifestation of the well-known threshold problem \cite{Schienbein}.
On the other hand, from Eq.~\eqref{eq:wlimits} the fragmentation
variable $\zeta_h \leq 1$, and no threshold problem appears in the 
fragmentation function since $D(\zeta_h) \to 0$ as $\zeta_h \to 1$.
In the next section we shall examine the phenomenological consequences
of the finite-$Q^2$ rescaling of the SIDIS cross section numerically.

\subsection{HMC Phenomenology}
\label{ssec:HMC-results}

Using the hadron mass corrected expressions for the SIDIS cross
section derived above, 
we next explore the dependence of the cross sections and fragmentation
functions on the fragmentation variable $z_h$, for various $x_B$ and
$Q^2$ values and for different final state hadron masses.
We then compare the relative size of the HMCs with the experimental
uncertainties from a recent SIDIS experiment at Jefferson Lab,
as well as with higher energy data from the European Muon Collaboration
(EMC) and HERA.

\begin{figure}[h]
\vspace*{1.5cm}
\includegraphics[height=9cm]{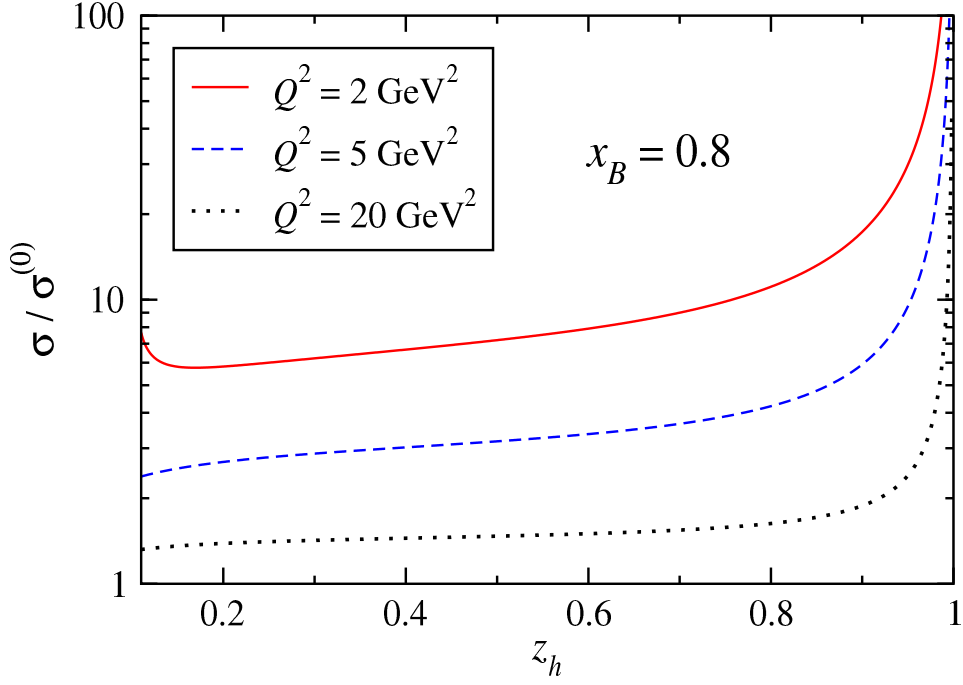}
\caption{Ratio of cross sections $\sigma/\sigma^{(0)}$ for
        semi-inclusive charged-pion production (($\pi^+ + \pi^-)/2$)
        as a function of $z_h$ at several $Q^2$ values for
        $x_B = 0.8$.}
\label{fig:2}
\end{figure}


To illustrate most directly the effects of the HMCs, in
Fig.~\ref{fig:2} we consider charged pion production (average of
$\pi^+$ and $\pi^-$) and plot as a function of $z_h$, for different
$Q^2$ at $x_B = 0.8$, the ratio of the full cross section $\sigma$ in
Eq.~(\ref{eq:dsigTMC}) to the cross section $\sigma^{(0)}$,
defined by taking the massless limit for the scaling variables
$\sigma^{(0)} \equiv \sigma(\xi_h \to x_B, \zeta_h \to z_h)$ and
setting $d\zeta_h/dz_h=1$.
For the numerical computations we use the leading order CTEQ6L
parton distributions \cite{PDF} and the KKP leading order
fragmentation functions \cite{KKP}, unless otherwise specified.
The ratio at $x_B = 0.8$ is strongly enhanced by nearly an
order-of-magnitude at $Q^2=2$~GeV$^2$ for $z_h \lesssim 0.7$, but
then becomes practically divergent in the limit $z_h \to 1$. This
is of course an artifact of computing correction ratios with respect
to ``massless'' cross sections that vanish at $z_h = 1$ while the
full cross section $\sigma$ remains finite due to rescaling by
$(\xi_h,\ \zeta_h)$.
The effect is naturally smaller at higher $Q^2$ values, but the
rise at high $z_h$ is a common feature for all kinematics.

\begin{figure}[h]
\vspace*{1.5cm}
\includegraphics[height=9cm]{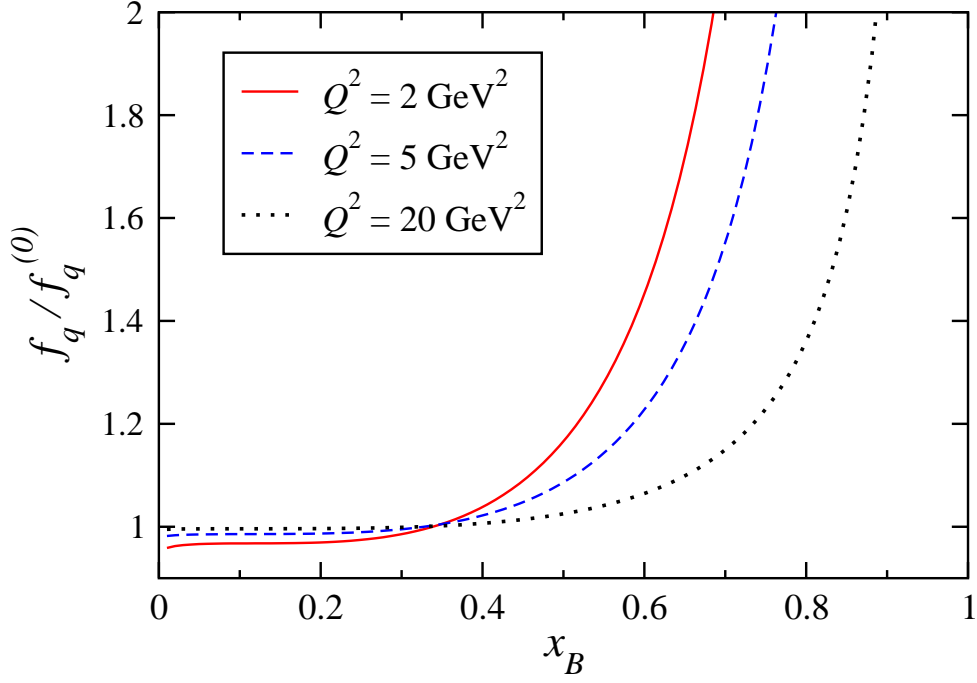}
\caption{Ratio of the hadron mass corrected isoscalar parton
	distribution function $f_q(\xi_h)$ for $q=u+d$ to the
	massless limit distribution $f_q^{(0)}$ as a function
	of $x_B$, for $m_h = m_\pi$ and $\zeta_h = 0.2$. }
\label{fig:3}
\end{figure}

On the other extreme of the spectrum, the small upturn in the ratios at low $z_h$ for the lowest $Q^2$ in
Fig.~\ref{fig:2} can be understood from the interplay between the
finite-$Q^2$ kinematics and the shape of the fragmentation function.
Assuming the fragmentation function is smooth, it is trivial to expand the
ratio of corrected to uncorrected functions in a Taylor series as
\begin{equation}
\frac{D(\zeta_h)}{D(z_h)} \approx
1 + \frac{D^\prime(z_h)}{D(z_h)} (\zeta_h - z_h)\, .
\label{eq:TMCdif}
\end{equation}
The $z_h$ dependence of the HMCs arising in the fragmentation function 
is mostly determined by the negative shift in the fragmentation 
variable ($\zeta_h-z_h$) and by the local rate of change over $z_h$
of the fragmentation function.
The pion fragmentation function generally behaves as a negative power
of $z_h$ at small $z_h$, and the negative slope drives the ratio of
corrected to uncorrected fragmentation functions upward as
$z_h \to z_h^{\rm min}$, where $|\zeta_h-z_h|$ is maximum.
For kaons and protons the slope of the form factor can be positive,
which would suppress the mass corrected cross section in the vicinity
of $z_h^{\rm min}$. 
In the limit $z_h \to 1$, on the other hand, the ratio
$\sigma/\sigma^{(0)}$ becomes divergent for any kinematics and any
hadron species because the cross section $\sigma^{(0)} \propto D(z_h)$
vanishes, while the rescaled cross section remains finite.

At very small values of $z_h$ the factor $(1+m_h^2/\zeta_h Q^2)$ in
the definition of $\xi_h$ in Eq.~(\ref{eq:xi_h}) can render $\xi_h$
larger than $x_B$, suppressing the $\xi_h$-rescaled parton distributions
relative to their asymptotic limit and driving $\sigma/\sigma^{(0)}$
slightly below unity.
As discussed below, for heavier hadrons this effect will be more
pronounced. 
The effect of the $\xi_h$ rescaling on the SIDIS cross section is 
illustrated explicitly in Fig.~\ref{fig:3}, where we show the ratio 
of the isoscalar parton distribution functions $f_q$, $q=u+d$,
with [$f_q=f_q(\xi_h)$] and without [$f_q^{(0)}=f_q(x_B)$] hadron mass
corrections, as a function of $x_B$ for $\zeta_h=0.2$ and $m_h=m_\pi$.
At $Q^2 = 2$~GeV$^2$ the mass corrected parton distribution is
several times larger than the uncorrected one at $x_B=0.8$, and
even at $Q^2 = 20$~GeV$^2$ the HMC is some 50\%, with the effect
increasing dramatically as $x_B \to 1$.
This sharp rise is analogous to that in inclusive DIS as seen in
Fig.~\ref{fig:F2}, and arises from $\xi_h$ being smaller than $x_B$
when the latter is large.
In contrast, the $\xi_h$ rescaling effect becomes quite small at
$x_B \lesssim 0.3$ for all the $Q^2$ considered, and in fact drives
the ratio below unity, as discussed above.

\begin{figure}[h]
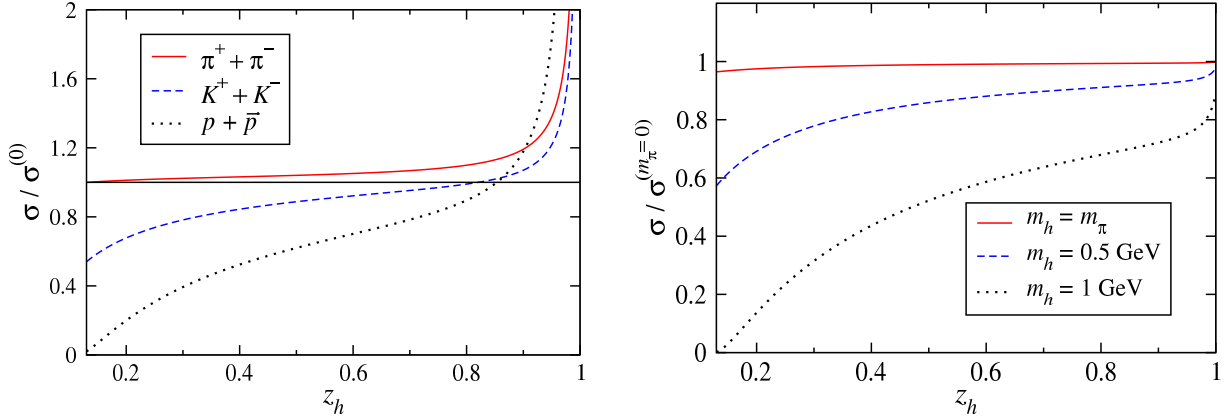

\vspace*{1cm}
\includegraphics[height=5.5cm]{sidis-Fig/Fig4aZ.eps}\ \ \ \ \
\includegraphics[height=5.5cm]{sidis-Fig/Fig4bZ.eps}
\caption{(Left) Dependence of the ratio of SIDIS cross sections
	$\sigma/\sigma^{(0)}$ with and without HMCs for different
	produced hadrons, $h=\pi^+ + \pi^-$, $K^+ + K^-$ or
	$p + \bar p$.
	(Right) Ratio of cross sections for $h=\pi^+ + \pi^-$ for
	different values of the pion mass, relative to the massless
	cross section.
	In both cases the kinematics chosen are $x_B = 0.3$ and
	$Q^2 = 5~$GeV$^2$.}
\label{fig:4}
\end{figure}

The relative importance of HMCs for different produced hadron
species is illustrated in the left panel of Fig.~\ref{fig:4}, where the
ratio $\sigma/\sigma^{(0)}$ is shown as a function of $z_h$
for $x_B = 0.3$ and $Q^2 = 5~$GeV$^2$.
Over the range $0.3 \lesssim z_h \lesssim 0.8$ the HMCs yield
an upward correction of $\lesssim 10\%$ for the pions,
but a downward correction of $\lesssim 20\%$ and $\lesssim 40\%$
for kaons and protons/antiprotons, respectively.
At lower $z_h$ the cross section ratio for the heavier hadrons
decreases dramatically because of the large suppression of
the parton distribution from the $(1+m_h^2/\zeta_h Q^2)$ factor
in $\xi_h$, which overwhelms any other small-$z_h$ effect.

Note that in the left-hand side of Fig.~\ref{fig:4} the appropriate fragmentation function
for each produced hadron species has been used, which introduces a
flavor dependence in the HMC because of the different fragmentation
function shapes for each hadron.
To isolate the effects of the hadron mass alone, on the RHS of Fig.~\ref{fig:4}
the ratios of cross sections computed with charged pion fragmentation
functions and masses $m_h = m_\pi$ (= 0.139~GeV), 0.5~GeV and 1~GeV
are shown relative to the cross section with $m_\pi=0$, for which
$\zeta_h = z_h \xi / x_B$.
One can see that in general increasing the hadron mass suppresses the
cross section because of the $\xi_h$ scaling, and the inversion of the
HMC hierarchy in the LHS of Fig.~\ref{fig:4} going from low to high $z_h$ is
due to the increasingly negative large-$z$ slope of the fragmentation
functions for kaons and protons.
While the differences at the physical pion mass are very small,
for larger hadron masses $\sim 1$~GeV the effects can be quite
significant at $z_h \lesssim 0.4$ even for $Q^2$ values of
several~GeV$^2$.

\vspace*{0.15cm}
{\it Experimental implications.}
\vspace*{0.15cm}

\begin{figure}[h]
\vspace*{1.5cm}
\includegraphics[height=9cm]{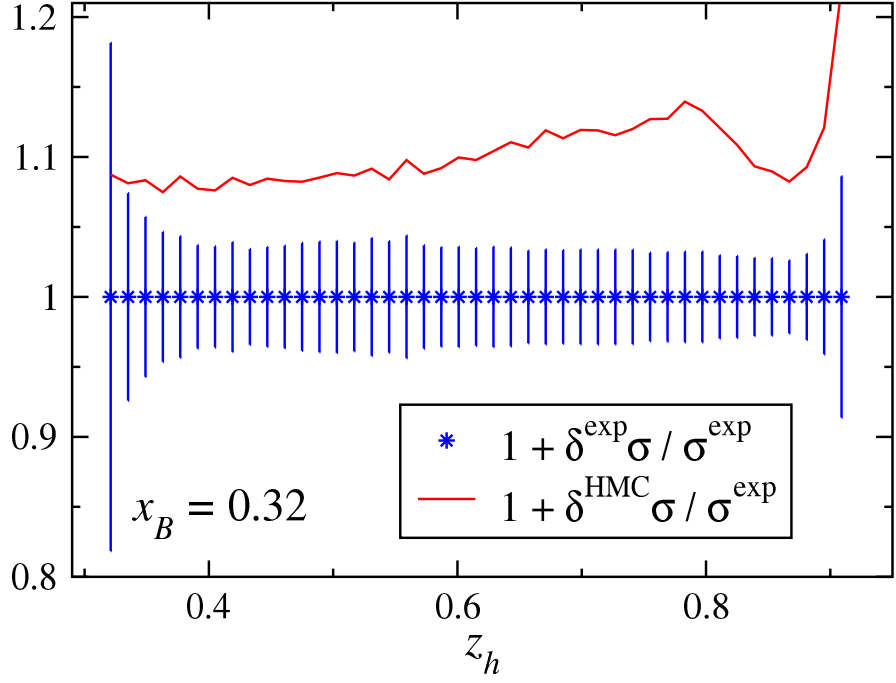}
\caption{Comparison of the hadron mass correction to the SIDIS cross
	section for charged hadron production, $\delta^{\rm HMC}\sigma$,
	relative to the experimental cross section,
	$\delta^{\rm exp} \sigma$, with the relative experimental
	uncertainty as a function of $z_h$ for the
	Jefferson Lab experiment E00-108 \cite{HallC} at
	$Q^2 \sim 2.3$ GeV$^2$ and $x_B=0.32$.}
\label{fig:6}
\end{figure}

The importance of the hadron mass corrections for experimental
cross sections is examined in Fig.~\ref{fig:6}, where we compare
the calculated difference
$\delta^{\rm HMC}\sigma \equiv \sigma - \sigma^{(0)}$ with the 
experimental uncertainties $\delta^{\rm exp}\sigma$, normalized to
the central values of the cross section for charged hadron production 
from Jefferson Lab \cite{HallC}. As the JLab measurements are generally
dominated by the semi-inclusive production of pions, $\xi_h\approx\xi$,
and HMCs generally produce upward shifts relative to data.
For the specific Jefferson Lab experiment E00-108 \cite{HallC} 
of Fig.~\ref{fig:6}, $Q^2 \sim 2.5$ GeV$^2$, with $x_B = 0.32$, and the 
mass effects are approximately $2$ times larger than the experimental
statistical  errors.
This illustrates the potentially significant impact that HMCs can have
on leading-twist analyses of SIDIS data at moderate and large $x_B$ and
low $Q^2$.
To avoid these effects one would either need to go to smaller $x_B$ or 
larger $Q^2$ as might, for example, be afforded by the 12~GeV energy upgrade 
at Jefferson Lab.
Alternatively, since the HMCs are calculated and model independent, 
lower $Q^2$ and higher $x_B$ data will still yield useful leading twist 
information provided the mass corrections are accounted for.

At higher energies, we find HMCs to fixed-angle measurements
by the EMC \cite{EMC2} at large $x_B$ values to be
negligible due to suppression by $Q^2$, which increases with $x_B$.
Were these experiments conducted at smaller angles, however, it is 
likely that HMCs would become important.

Similarly, measurements at small $x_B \sim 0.001$ and $Q^2\gtrsim 12$~GeV$^2$ have
been performed by the H1 collaboration \cite{Adloff} at HERA, and the
data presented in terms of the fragmentation invariant $\eta_h$.
The phenomenology of HMCs is markedly different in terms of $\eta_h$
from that discussed thus far in terms of $z_h$ because of the 
different functional forms for $\zeta_h$ in Eqs.~(\ref{eq:zeta_h}), 
which constrains $\zeta_h > \eta_h$, and because of the Jacobian
$d\zeta_h / d\eta_h$.
In their analysis of the H1 data, Albino {\it et al.} \cite{Albino}
included the $\zeta_h$ rescaling of the fragmentation process, but
neglected the effects of the target mass, which would be problematic
for heavier hadrons such as kaons and protons.
The H1 Collaboration measured charged hadron multiplicities,
dominated by pions ($\sim$60\%), with smaller contributions from
kaons ($\sim$30\%) and protons ($\sim$10\%).
In the measured $Q^2$ range the $m_h^2/Q^2$ term in $\xi_h$ is
therefore strongly suppressed and at the typically low $x_B$ values
one has $\xi \approx x_B$, so that overall we find the HMCs to be
similar to those in Ref.~\cite{Albino}.
However, for identified kaons, and especially protons, the SIDIS cross
section would be more strongly suppressed compared to the results of 
Ref.~\cite{Albino} because $\xi_h \approx x_B (1+m_h^2/Q^2)$ is 
significantly larger than $x_B$.
This suppression may be non-negligible for the extraction of kaon
and proton fragmentation functions from small-$x_B$ data.
It therefore seems patent, particularly given the dramatic results of
Fig.~\ref{fig:6}, that the nonperturbative mass of final state
hadrons are a necessary analytical consideration in various processes;
as always, this is especially true in the limit of small $Q^2$ and
$x \sim 1$.  
%
%

%% file: the-charm.tex


An improved grasp of the role played by heavy quarks is of critical importance
for fully understanding the transition from the pQCD dynamics
described in Chap.~\ref{chap:ch-intro} to the physics of color-neutral bound
states. In particular, the constituent mass of the charm quark,
$m_c \sim 1.3$ GeV, places the threshold for its production at the upper
periphery of the nucleon excitation/resonance region discussed in some
detail for elastic scattering in Chap.~\ref{chap:ch-DIS}.\ref{sec:Compt}. Moreover, threshold
effects in charm production are a necessary consideration for determinations
of scaling violations and $Q^2$ dependence in QCD global analyses, as well
as for computations of background processes in precision searches for novel
physics outside the Standard Model.

\begin{figure}[h]
\includegraphics[width=6.5cm]{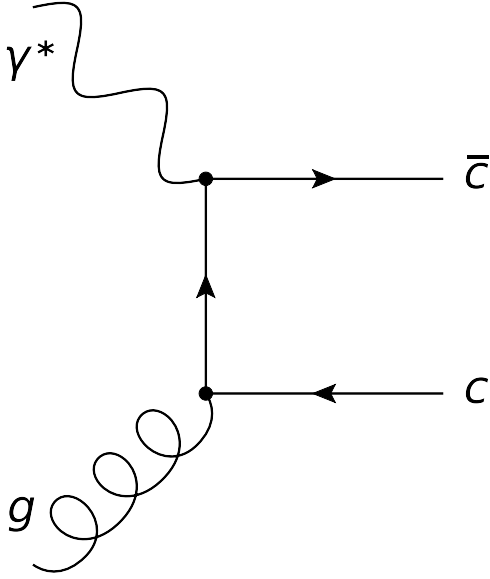}
\caption{
At leading order, photon-gluon fusion predominates at $Q^2 \sim m^2_c$ leading
to the contribution to $F^c_2(x,Q^2)$ given by Eq.~(\ref{eq:PGF}).
}
\label{fig:PGF}
\end{figure}
In conventional analyses, charm is usually incorporated into the nucleon's DIS
structure functions by means of standard pQCD under the assumption that
$c(x, Q^2 \le m^2_c) = \bar{c}(x, Q^2 \le m^2_c) \equiv 0$ below the physical
charm production threshold $Q^2 = m^2_c$. Any charm produced then for $Q^2 > m^2_c$
enters {\it extrinsically} via perturbatively calculable diagrams, with the dominant leading order (LO) mechanisms
coming from the gluon splitting and bremsstrahlung graphs depicted in Fig.~\ref{fig:QCDsplt},
as well as from the boson-gluon fusion, or photon-gluon fusion process displayed in
Fig.~\ref{fig:PGF}. For the former, we already saw in Chap.\ref{chap:ch-Q2}.\ref{sec:deut-pCSV} that
generic DGLAP evolution stipulates convolutions of the charm distributions with pQCD splitting functions [Eq.~(\ref{eq:QCD_DGLAP})]
according to Eq.~(\ref{eq:DGLAP}). This procedure is numerically adequate at LO for
large virtualities --- \IE~$Q^2 \gg m^2_c$, but at moderate kinematics
characterized by $Q^2 \sim m^2_c$, photon-gluon fusion is a more reliable approximation.\footnote{We
should also point out the existence of a variety of `interpolation' schemes \cite{SMT99,Stavreva:2012bs} that track
the scale-dependent contributions of each mechanism, such that ``massless'' DGLAP dominates
at $Q^2 \gg m^2_c$, photon-gluon fusion at $Q^2 \sim m^2_c$, and in-between, some admixture of the two.}
Computing Fig.~\ref{fig:PGF} by standard Feynman rules determined by Eq.~(\ref{eq:QCD}), it can
be shown that \cite{SMT99}
\begin{subequations}
\begin{align}
\label{eq:PGFI}
F^c_{2,\ {\rm PGF}}(x,Q^2)\ &=\ {\alpha_s(\mu^2) \over 9\pi}\ \int_x^{z'}\ {dz \over z}\ C^{\rm PGF}(z,Q^2,m^2_c)
\cdot xg\left({x \over z}, \mu^2 \right)\ , \\
C^{\rm PGF}(z,Q^2,m^2_c)\ &=\ 4 \left\{1-2z+2z^2+4z(1-3z){m^2_c \over Q^2} - 8z^2 \left({m^2_c \over Q^2}\right)^2 \right\}\
\ln \left({1+\beta \over 1-\beta}\right) \nonumber\\
&+\ \beta \left\{ 32z(1-z) -4 -16z(1-z) {m^2_c \over Q^2} \right\}\ ; \hspace*{0.5cm} \beta \defeq \sqrt{1-{4z \over (1-z)} {m^2_c \over Q^2}}\ ;
\end{align}
\label{eq:PGF}
\end{subequations}
in Eq.~(\ref{eq:PGFI}), to account for effects at the charm threshold, the upper bound on the convolution must be
$z' = 1 \Big/ \Big(1 + 4m^2_c / Q^2 \Big)$.

While this methodology and its extension to arbitrary order in perturbation theory are capable of predicting experimental
results (especially at low $x$ and high $Q^2$) with admirable accuracy, potential nonperturbative effects may conspire to
generate charm at more moderate $Q^2$ and higher $x$, signaling an {intrinsic} charm component of the nucleon.
This would represent a fundamental contribution to the bound state structure of the nucleon apart from its known make-up
consisting of a conventional mix of valence content with the light quark sea.

To estimate the plausibility and significance of this hypothesis, we first briefly review the existing literature concerning
nonperturbative or intrinsic charm (IC) in the nucleon in Sec.~\ref{sec:5q}. With this in mind, we then present in
Secs.~\ref{sec:amp}--\ref{sec:results} the recently-published results of a comprehensive effective field theory calculation
\cite{Hobbs13}, which for the first time immediately connects intrinsic charm predictions to the $SU(4)$ hadronic spectrum.
Finally, in Sec.~\ref{sec:GA} we constrain the model parameters of our formalism using the technology of a recently completed
QCD global analysis \cite{Globe}.

\section{Five-quark models of nucleon structure}
\label{sec:5q}

\begin{figure}[h]
\includegraphics[width=16.2cm]{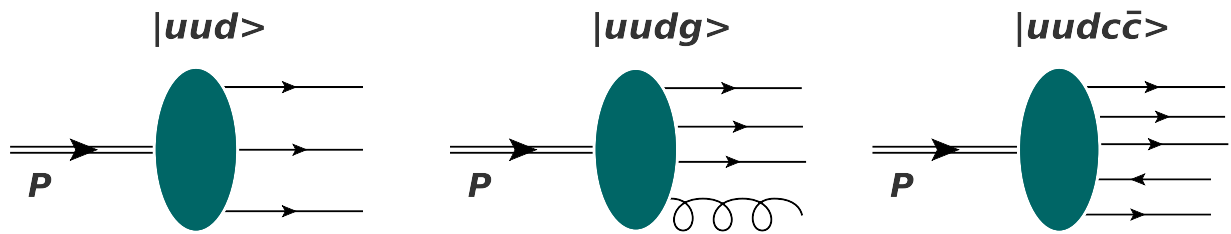}
\caption{
In the most basic conceptions, an intrinsic charm component appears in the higher states
of a Fock space expansion of the proton wavefunction.
}
\label{fig:ch-Fock}
\end{figure}

In this section we review models of intrinsic charm based on particular
five-quark Fock state components of the nucleon wave function.
We will focus on models that describe the process by which a nucleon
initially containing three light valence quarks transitions to a four
quark plus one antiquark state containing a charm-anticharm quark pair
as depicted in the final panel of Fig.~\ref{fig:ch-Fock}.
%

\subsection{Scalar five-quark models}
\label{ssec:scalar5q}

Before our calculation in \cite{Hobbs13}, the various five-quark pictures
essentially all involved the assumption of point-like interactions taken from
na\"ive scalar field theories. Here we discuss some of the more archetypal options.

\vspace*{0.15cm}
{\it The BHPS model.}
\vspace*{0.15cm}

The earliest (and simplest) model for producing intrinsic charm was proposed over
30 years ago by Brodsky, Hoyer, Peterson, and Sakai (BHPS) \cite{BHPS}.
In the infinite momentum frame (IMF) the probability for a
proton with mass $M$ to make the transition $p \to uudc\bar{c}$ (or to
a five-quark state containing any heavy quark pair) shown in the rightmost process
of Fig.~\ref{fig:ch-Fock} involves an `old-fashioned'
perturbation theory energy
denominator that can be rendered in terms of the masses $m_i$ and
momentum fractions $x_i$ of its constituents,
\be
P(p \to uudc\bar{c})
\sim \left[ M^2 - \sum_{i=1}^5 \frac{k_{\perp i}^2 + m_i^2}{x_i}
     \right]^{-2}\ .
\label{eq:BHPSeq}
\ee
Here, $k_{\perp i}$ is the transverse momentum of quark $i$, and the
heavy quarks in the $c \bar{c}$ pair are assigned indices $4$ and $5$.

For simplicity, the BHPS calculation assumed a point coupling for the
$c \bar{c}$ production vertex, and neglected the effect of transverse
momentum in the five-quark transition amplitudes.  With the additional
assumption that the charm mass is much greater than the nucleon and
light quark masses, the probability for producing a single charm quark
can be derived analytically,
\be
P(x) = \frac{N x^2}{2}
	 \left[ \frac{(1-x)}{3}\left( 1 + 10x + x^2 \right)
		+ 2 x (1+x) \ln(x)
	 \right]\ ,
\label{eq:charmprob}
\ee
with the normalization $N$ fixed by the overall charm quark
probability in the proton. In Eq.~(\ref{eq:charmprob}), we have
made the replacement $x_5 \to x$ to go from Eq.~(\ref{eq:BHPSeq}).

Using the analytic expression (\ref{eq:charmprob}) one obtains a
``valence-like'' (\IE~non-singular in the $x \rightarrow 0$ limit)
charm quark distribution that is significant for $0.1 \leq x \leq 0.5$.
The valence-like shape arises from the structure of the energy
denominator in Eq.~(\ref{eq:BHPSeq}), which for large quark masses
$m_i$ favors configurations with large momentum fractions $x_i$. This
feature is common to all similar five-quark models and invariably results
in a valence-like heavy quark distribution.

It is also possible to compute the charm quark probability numerically
without assuming that the charm quark mass is much greater than all
other masses, though with realistic masses the charm quark distribution turns
out to be similar to the analytic form of Eq.~(\ref{eq:charmprob}).
Note also that since the charm and anticharm probabilities enter
Eq.~(\ref{eq:BHPSeq}) symmetrically, the charm production mechanism
in this model will produce equal probabilities for $c$ and $\bar{c}$,
not unlike the pQCD contributions computed in, \EG~Eq.~(\ref{eq:PGF}).
Although the BHPS model is rather simplistic, it nevertheless
provides a useful benchmark to test intrinsic
charm and anticharm distributions obtained from other prescriptions.

\vspace*{0.15cm}
{\it The Pumplin model.}
\vspace*{0.15cm}

In a detailed study of intrinsic heavy quark probabilities, Pumplin
\cite{Pum05} considered a series of models for the Fock space
wave function on the light-front for a proton to make a transition to
a four quark plus one antiquark system, with the heavy $q\bar{q}$ pair
composed of either charm or bottom quarks.
A simplified case was studied where a point scalar particle of mass
$m_0$ couples with strength $g$ to $N$ scalar particles with masses
$m_1, m_2, \ldots, m_N$.  The light-front Fock space probability
density $dP$ for such a process then takes the form
\cite{Pum05}
\bea
dP
&=& \frac{g^2}{(16\pi^2)^{N-1}(N-2)!}\,
    \prod_{j=1}^N dx_j\, \delta\left( 1-\sum_{j=1}^N x_j\right)
    \int_{s_0}^\infty ds\, \frac{(s-s_0)^{N-2}}{(s-m_0^2)^2}\,
    |F(s)|^2\ ,
\label{eq:dPFock}
\eea
where $s_0 = \sum_{j=1}^N m_j^2/x_j$, and the form factor $F(s)$
serves to suppress contributions from high-mass states.
If one neglects the effects of transverse momentum and the factors
of $1/x_j$ in Eq.~(\ref{eq:dPFock}), and assumes a point form factor
$F(s) = 1$, then in the limit that the charm mass is much larger than
all other masses one recovers the distribution in the BHPS model
\cite{BHPS}.

To incorporate the effects of the finite size of the nucleon,
Pumplin considered both an exponential form factor,
\be
|F(s)|^2 = \exp \left[-(s-m_0^2)/\Lambda^2 \right]\ ,
\label{eq:expFF}
\ee
and a power-law suppression factor,
\be
|F(s)|^2 = {1 \over (s + \Lambda^2)^n}\ , 
\label{eq:powerFF}
\ee
with $\Lambda$ a cutoff mass regulator.
Fixing the overall normalization to be a constant, the resulting
shape of the charm quark momentum distribution with the power-law
suppression, for a reasonable choice of $n=4$, turns out to be softer
than the BHPS prediction for a range of cutoffs, $\Lambda = 2-10$~GeV
\cite{Pum05}.
For the exponential suppression, the shape depends somewhat more
strongly on the cutoff parameter, with the distribution being harder
than the BHPS result for smaller $\Lambda$ values and softer for
larger $\Lambda$.
All of the resulting charm distributions are valence-like, however,
with significant tails even beyond $x \approx 0.4$.

While these simple five-quark models give some qualitative insights
into the possible generation of intrinsic charm at large $x$,
they retain a high degree of dependence on the model parameters,
whose connection with the underlying QCD dynamics is unclear.
Furthermore, it is also not obvious how one could constrain
the parameters phenomenologically by comparing deep-inelastic
scattering with other observables, for instance.
In the next section we discuss an alternative approach which may
offer greater promise for relating intrinsic charm distributions
to inputs determined from independent reactions.

\subsection{Meson--baryon models}
\label{sec:MBM}

A hybrid class of intrinsic charm models that involves both quark
and hadron degrees of freedom, and makes some unique and testable
predictions for the $c$ and $\bar c$ distributions in the nucleon,
are meson--baryon models (MBMs). Such models attempt to quantify the
fluctuations of the nucleon to states with a virtual meson $M$
plus baryon $B$,
\be
|N\rangle\
=\ \sqrt{Z_2}\, \left| N \right.\rangle_0\
+\ \sum_{M,B} \int\! dy\, d^2\bm{k}_\perp\,
   \phi_{MB}(y,k^2_\perp)\,
   | M(y,\bm{k}_\perp); B(1-y,-\bm{k}_\perp) \rangle\ ,
\label{eq:Fock}
\ee 
where $\left| N \right.\rangle_0$ is the ``bare'', three-quark
nucleon state, and $Z_2$ is the wave function renormalization.
The function $\phi_{MB}(y,k^2_\perp)$ gives the probability
amplitude for the physical nucleon to be in a state consisting
of a virtual meson $M$ with longitudinal momentum fraction $y$
and transverse momentum $\bm{k}_\perp$, and a baryon $B$ with
longitudinal momentum fraction $1-y$ and transverse momentum 
$-\bm{k}_\perp$.
The total invariant mass squared of the meson--baryon system
$s_{MB}$ can be written in the IMF as
\be
s_{MB}(y,k^2_\perp)
= \frac{k^2_\perp + m_M^2}{y} + \frac{k^2_\perp + M_B^2}{1-y}\ ,
\label{eq:CoM_En}
\ee
where $m_M$ and $M_B$ are the meson and baryon masses, respectively.
If the meson--baryon terms include states containing charm quarks,
the resulting probability distributions for anticharm and charm
quarks in the nucleon can be written in the form of convolutions,
\begin{subequations}
\label{eq:mesoncloud}
\bea
\bar{c}(x)
&=& \sum_{M,B}\,
    \Big[ \int_x^1 \frac{dy}{y}\,
	  f_{MB}(y)\, \bar{c}_M\Big(\frac{x}{y}\Big)
	+ \int_x^1 \frac{d\bar y}{\bar y}\,
	  f_{BM}(\bar{y})\, \bar{c}_B\Big(\frac{x}{\bar{y}}\Big)
   \Big]\ ,
\label{eq:mesoncloud_cb}			\\
c(x)
&=& \sum_{B,M}\,
    \Big[ \int_x^1 \frac{d\bar y}{\bar y}\,
	  f_{BM}(\bar y)\, c_B\Big(\frac{x}{\bar y}\Big)
	+ \int_x^1 \frac{dy}{y}\,
	  f_{MB}(y)\, c_M\Big(\frac{x}{y}\Big)
    \Big]\ ,
\label{eq:mesoncloud_c}
\eea
\end{subequations}%
where $\bar y \equiv 1-y$, and for ease of notation we have omitted
the dependence of the distributions on the scale $Q^2$.

In analogy with the quark-gluon splitting functions of pQCD defined
in Eq.~(\ref{eq:DGLAP}), Eqs.~(\ref{eq:mesoncloud}) involve the splitting
functions $f_{MB}(y)$ for a nucleon to fluctuate to meson $M$ with
fraction $y$ of the proton's momentum, and a spectator baryon $B$.
The charm and anticharm distributions in the baryon $B$ are denoted
by $c_B(z)$ and $\bar{c}_B(z)$, respectively, and carry a fraction
$z=x/\bar y$ of the baryon's momentum.
On the same grounds, $f_{BM}(\bar y)$ represents the splitting function for a
nucleon fluctuating into a baryon $B$ with fraction $\bar y$ of the
proton's momentum, with a spectator meson $M$.  The quark distributions
inside the meson $M$ are denoted by $c_M(z)$ and $\bar{c}_M(z)$,
respectively.
If the charm quark resides exclusively in the baryon, with the
anticharm in the meson (as is consistently the case in the present chapter),
Eqs.~(\ref{eq:mesoncloud}) simplify further since
\bea
c_M(x)\ \to\ 0,\ \ \ 
\bar{c}_B(x)\ \to\ 0\ .
\label{eq:ccbar-simp}
\eea

The splitting functions in Eqs.~(\ref{eq:mesoncloud}) are related
to the probability amplitudes $\phi_{MB}$ by
\bea
f_{MB}(y)
&=& \int_0^\infty d^2\bm{k}_\perp \,|\phi_{MB}(y,k^2_\perp)|^2\
 =\ f_{BM}(\bar y)\ ,
\label{eq:fphi}
\eea
where the reciprocity relation in the second equality arises
from the conservation of $3$-momentum at the $MBN$ vertex 
\cite{Zoller92, MT93, Holtmann96}.
This can be shown to be satisfied explicitly in the infinite momentum
frame (or on the light-front), but is violated in covariant calculations
\cite{MT93, Holtmann96, MST94} in the presence of $MBN$
form factors (or other ultraviolet regulators) which do not exhibit
the $y \leftrightarrow \bar y$ symmetry of the amplitudes $\phi_{MB}$
\cite{Bur13}.

The convolution equations (\ref{eq:mesoncloud}) allow the symmetries
of the splitting functions to be represented in terms of moments of
the parton distributions $C^{(n)}$ and $\overline C^{(n)}$, defined as
\begin{subequations}
\label{eq:c_mom}
\bea
\overline C^{(n)}
&=& \int_0^1 dx\, x^n\, \bar c(x)\
 =\ \sum_{M,B}\, {\cal F}_{MB}^{(n)}\, \overline C_M^{(n)}\ ,	\\
C^{(n)}
&=& \int_0^1 dx\, x^n \,c(x)\
 =\ \sum_{B,M}\, {\cal F}_{BM}^{(n)}\, C_B^{(n)}\ ,
\eea
\end{subequations}%
where
\begin{subequations}
\label{eq:f_mom}
\bea
{\cal F}_{MB}^{(n)}
&=& \int_0^1 dy\, y^n\, f_{MB}(y)\ ,		\\
{\cal F}_{BM}^{(n)}
&=& \int_0^1 d\bar y\, \bar y^n\, f_{BM}(\bar y)\ ,
\eea
\end{subequations}%
are the $n$-th moments of the splitting functions.
The corresponding moments of the $\bar c$ and $c$ distributions
in the meson $M$ and baryon $B$ are denoted $\overline C_M^{(n)}$
and $C_B^{(n)}$, respectively.
In particular, the lowest moment of the splitting functions
gives the average multiplicity of mesons $M$ and baryons $B$,
\bea
\langle n \rangle_{MB}
&\equiv& {\cal F}_{MB}^{(0)}\ =\ {\cal F}_{BM}^{(0)}\ ,
\label{eq:recip1}
\eea
which reflects global charge conservation, while conservation
of momentum implies that the momentum fractions
$\langle y \rangle_{MB} \equiv {\cal F}_{MB}^{(1)}$ and
$\langle y \rangle_{BM} \equiv {\cal F}_{BM}^{(1)}$ are
related by
\bea
\langle y \rangle_{BM}\ +\ \langle y \rangle_{MB}
&=& \langle n \rangle_{MB}\ .
\label{eq:conserv}
\eea

In contrast to the five-quark models discussed in Sec.~\ref{sec:5q},
in which the $x$ dependence of the $c$ and $\bar c$ distributions
was identical, in the MBM the distributions of heavy quarks and
antiquarks in the nucleon are generally expected to be different.
Indeed, since the $c$ in the baryon and $\bar c$ in the meson
reside in rather different local environments, an asymmetry
$c(x) \ne \bar{c}(x)$ is almost unavoidable. In fact, the experimental
observation of $c(x) \ne \bar{c}(x)$, as might be extracted from
charge asymmetries in charmed hadron production processes, is a
trademark signal for nonperturbative charm \cite{Catani04}.
Of course, since the proton has no net charm, the lowest moments
of $c$ and $\bar c$ must cancel; however, all higher moments will
be nonzero,
\be
C^{(0)} - \overline C^{(0)} = 0\ ,\ \ \ \ \ \ \
C^{(n)} - \overline C^{(n)} \neq 0\ \ (n \geq 1)\ ,
\label{eq:c_moments}
\ee
with the result of Eq.~(\ref{eq:RSach}) ensuring the first relation for
all $Q^2$ --- up to NLO corrections. In general, however, the realization of
Eq.~(\ref{eq:recip1}) guarantees the proton to have no net charm at the
nonperturbative scale $Q^2 = m^2_c$.

Because quarks and antiquarks possess opposite intrinsic parities,
parity conservation will require that the quark wave functions respect
overall parity conservation.  For example, if the initial proton state
is treated as three constituent $(uud)$ quarks in $S$-wave orbitals
and a $c \bar{c}$ pair is added, then the placement of a charm quark in
an $S$ state necessitates that anticharm be in an odd-parity configuration.
In the MBM, this behavior is accommodated by proper use of physical
vertices that correctly account for the spin degrees of freedom of
the relevant fields. As a rule, models that treat quarks as scalar point-like
particles, for example, will therefore not satisfy these constraints
\cite{Pum05, Pum07}.

\begin{table}[bt]
\caption{Lowest mass meson--baryon Fock states of the proton containing
	charm and anticharm quarks.  For each state the isospin $I$,
	spin $J$ and parity $P$ are listed for the meson and baryon,
	together with the masses.}
\centering
\begin{tabular}{l c l c}
\hline\hline
Baryon  & \ $I(J^{P})$\ \ \ \ \ & \ \ Meson & $I(J^{P})$	\\ 
						[0.5ex]\hline
$p\, (938)$            & $\frac{1}{2}\, (\frac{1}{2}^+)$
                                               & $J/\psi(3097)$\ \ \
	& $0\, (1^-)$		\\
$\Lambda_c^+(2286)$    & $0\, (\frac{1}{2}^+)$ & $\bar{D}^0(1865)$
	& $\frac{1}{2}\, (0^-)$	\\
                       &                       & $\bar{D}^{*0}(2007)$
	& $\frac{1}{2}\, (1^-)$	\\
$\Sigma_c^+(2455)$     & $1\, (\frac{1}{2}^+)$ & $\bar{D}^0(1865)$
	& $\frac{1}{2}\, (0^-)$	\\
                       &                       & $\bar{D}^{*0}(2007)$
	& $\frac{1}{2}\, (1^-)$	\\
$\Sigma_c^{++}(2455)$  & $1\, (\frac{1}{2}^+)$ & $D^-(1870)$
	& $\frac{1}{2}\, (0^-)$	\\
                       &                       & $D^{*-}(2010)$
	& $\frac{1}{2}\, (1^-)$	\\
$\Sigma_c^{*+}(2520)$  & $1\, (\frac{3}{2}^+)$ & $\bar{D}^0(1865)$
	& $\frac{1}{2}\, (0^-)$	\\
                       &                       & $\bar{D}^{*0}(2007)$
	& $\frac{1}{2}\, (1^-)$	\\
$\Sigma_c^{*++}(2520)$ & $1\, (\frac{3}{2}^+)$ & $D^-(1870)$
	& $\frac{1}{2}\, (0^-)$	\\
                       &                       & $D^{*-}(2010)$
	& $\frac{1}{2}\, (1^-)$	\\ [1ex]
\hline
\end{tabular}
\label{table:mass_spect}
\end{table}

In the present analysis, we consider various meson--baryon states
containing charm quarks that could contribute to the intrinsic
charm in the proton, as summarized in Table~\ref{table:mass_spect}.
These include the SU(4) octet isoscalar $\Lambda_c$ and isovector
$\Sigma_c$ baryons, and the decuplet $\Sigma_c^*$, while for the
mesons, the pseudoscalar $D$ and vector $D^*$ mesons are included.
In addition, the state involving a proton fluctuation to
$p + J/\psi$, where both the $c$ and $\bar{c}$ reside in
the $J/\psi$, was considered in Ref.~\cite{Pum05}.
Although this has a combined mass which is actually lower
than all the other charmed meson--baryon configurations,
its contribution should be strongly suppressed by the OZI rule.\footnote{
Using the model we present in the following sections, one can estimate
the numerical strength of the $pp J/\psi$ coupling using, \EG~PHENIX
data at $s=\sqrt{200}$ GeV \cite{Adler:2003qs}. Taking from this a
typical $J/\psi$ production cross section of $d\sigma^{J/\psi}/dy \sim 1 \mu\mathrm{b}$
at $y \sim 0.4$, we estimate the $J/\psi$ contribution to be suppressed by a
factor of 10$^{-4}$ relative to the dominant mode.}

To constrain the model parameters in the calculations ---
namely, the hadronic couplings and form factor cutoffs ---
we use phenomenological input from $DN$ and $\bar{D} N$
scattering analyses \cite{Sib01, Hai07, Hai08, Hai11},
together with inclusive charmed baryon production data in
$pp$ collisions. We discuss the formal aspects of these considerations
in Sec.~\ref{sec:mb} after first deriving in the next section
the splitting functions $f_{MB}(y)$ for the various configurations
in Table~\ref{table:mass_spect}, as well as the associated
distributions $c_B(z), \bar{c}_M(z)$ within them.
%
%
\section{Amplitudes for IC}
\label{sec:amp}

{\it Derivation of meson--baryon splitting functions.}
\vspace*{0.15cm}

The essential ingredients of the two-step meson-baryon models
are amplitudes formulated separately at hadron-
and quark-level, which then undergo convolution per
Eq.~(\ref{eq:mesoncloud}). Here we outline the technical details of
the derivations of the splitting functions for the dissociation of a
nucleon with $4$-momentum $P$ into a meson $M$ with momentum $k$
and baryon $B$ with momentum $p$.  We consider dissociations into
the SU(4) octet isoscalar $\Lambda_c$ and isovector $\Sigma_c$
baryons, and the decuplet $\Sigma_c^*$ baryon, accompanied by the
charmed pseudoscalar $D$ and vector $D^*$ mesons.  The transitions
to specific isospin states are obtained using appropriate isospin
transition factors, as discussed in Sec.~\ref{sec:mb}.

The contribution of a specific meson--baryon component to the
nucleon hadronic tensor $W_{\mu\nu}^N$ is
defined\footnote{It is convenient
to define the tensors
  $\widetilde{g}_{\mu\nu} = -g_{\mu \nu} + q_{\mu} q_{\nu}/q^2$, and
  $\widetilde{P}_{\mu} = P_\mu - P \cdot q\, q_\mu/q^2$,
which appear explicitly in the decomposition of $T_{\mu\nu}$ as in
Eq.~(\ref{eq:OPE_genII}).
}
in terms of the contributions $\delta^{[MB]} F_{1,2}^N$
to the structure functions as \cite{Beringer:1900zz}
\bea
\delta^{[MB]} W_{\mu\nu}^N(P,q)
&=& \widetilde{g}_{\mu \nu}\, \delta^{[MB]} F_1^N\
 +\ \frac{\widetilde{P}_\mu \widetilde{P}_\nu}{P \cdot q}\,
    \delta^{[MB]} F_2^N\ ,
\label{eq:had_tens}
\eea
where $q$ is the $4$-momentum of the external electromagnetic
field.

For reasons that will become clear in the subsequent discussion, we
find it advantageous to compute strictly forward-propagating diagrams
in the IMF. Within the framework of time-ordered perturbation theory evaluated
in IMF kinematics ($P_L \to \infty$), with intermediate state
particles on their mass-shells but off their ``energy-shells'', the
standard decomposition for the momentum variables is \cite{DLY70}
\begin{subequations}
\label{eq:kin-TOPT}
\bea
P_0 &=& P_L + \frac{M^2}{2P_L} + {\cal O}\left(\frac{1}{P_L^2}\right)\ ,
\\
k_0 &=& |y| P_L + \frac{k^2_\perp + m_M^2}{2 |y| P_L} 
	+ {\cal O}\left(\frac{1}{P_L^2}\right)\ ,
\\ 
p_0 &=& |1 - y| P_L + \frac{k^2_\perp + M_B^2}{2 |1 - y| P_L}
	+ {\cal O}\left(\frac{1}{P_L^2}\right)\ ,
\eea  
\end{subequations}%
for the energies, and
\begin{subequations}
\label{eq:kin-TOPT_mom}
\bea
\bm{k} &=& |y| \bm{P} + \bm{k_\perp}\ ,
\\
\bm{p} &=& |1-y| \bm{P} - \bm{k_\perp}\ ,
\eea
\end{subequations}%
for the $3$-momenta, with $\bm{k}_\perp \cdot \bm{P} = 0$.
In the non-vanishing forward limit one has
$y \in [0,1]$, such that $|1-y| = (1-y)$.

Obeying the standard rules \cite{Weinberg:1966jm} for computing forward-moving TOPT diagrams such
as Fig.~\ref{fig:f_MN}, we find on purely general grounds
\begin{equation}
\delta^{[MB]} W_{\mu \nu}^N\ =\ \int {d^3{\bf k} \over (2\pi)^3 (2P_0) (2k_0)^2}\,
\frac{g^2_{MBN}({\bf k})}{(P_0 - p_0 - k_0)^2} N^{MB}_{\mu \nu}\ ;
\end{equation}
it is straightforward to show that the perturbation theory energy denominator can be rewritten as
$(P_0 - p_0 - k_0) = (M^2 - s_{MB}) \Large/ 2P_L$, where the center-of-mass energy $s_{MB}$ is as
defined in Eq.~(\ref{eq:CoM_En}). It is in fact simpler to translate to the space spanned by
the parameters $(y, k^2_\perp)$; keeping track of the phase space factor
$d^3{\bf k} \rightarrow dy dk^2_\perp$, and using the kinematical definitions of Eq.~(\ref{eq:kin-TOPT}),
we find that the generic one-loop diagram for the scattering from the meson $M$ evaluates to
\bea
\delta^{[MB]} W_{\mu\nu}^N
&=& \frac{g^2_{DBN}}{16\pi^2} \int_0^1 dy \int_0^\infty 
  \frac{dk^2_\perp}{y(1-y)}
  \frac{|F(s)|^2} {(M^2 - s)^2}\
  N^{MB}_{\mu\nu}\ ,
\label{eq:del_MB}
\eea
where $y$ is the longitudinal momentum fraction carried by the meson,
$F(s)$ is the $MBN$ hadronic factor, with the invariant mass squared
of the $MB$ system $s$ defined in Eq.~(\ref{eq:CoM_En}).
The tensor $N^{MB}_{\mu\nu}$ is computed from the spin trace of the
appropriate meson and baryon propagators and vertices, with explicit
forms given below.

In addition, we shall find that the expression in Eq.~(\ref{eq:del_MB}) is generally
evaluated in terms of inner products $P \cdot k$, $P \cdot p$ and $P \cdot k$ computed
from Eqs.~({\ref{eq:kin-TOPT}) and (\ref{eq:kin-TOPT_mom}) as
\begin{subequations}
\begin{align}
P \cdot k &=\ {1 \over 2 y}\ \Big( k^2_\perp + m^2_M + y^2 M^2 \Big)\ , \\
P \cdot p &=\ {1 \over 2 (1-y)}\ \Big( k^1_\perp + M^2_B + (1-y)^2 M^2 \Big)\ , \\
p \cdot k &=\ k^2_\perp\ +\ {y (k^2_\perp + M^2_B) \over 2 (1-y)}\ +\  {(1-y) \cdot (k^2_\perp + m^2_M) \over 2 y}\ ;
\end{align}
\label{eq:inn-prod}
\end{subequations}
the corrections to the structure functions $\delta^{[MB]} F_{1,2}^N$
follow after equating the coefficients of the tensors in
Eqs.~(\ref{eq:had_tens}) and (\ref{eq:del_MB}).
The corrections to the $c$ and $\bar c$ distributions
in Eqs.~(\ref{eq:mesoncloud}) may then be extracted from
$\delta^{[MB]} F_{1,2}^N$ using parton model relations
analogous to Eq.~(\ref{eq:EW-SFs}) as we now demonstrate.

\begin{figure}[t]
\includegraphics[width=8cm]{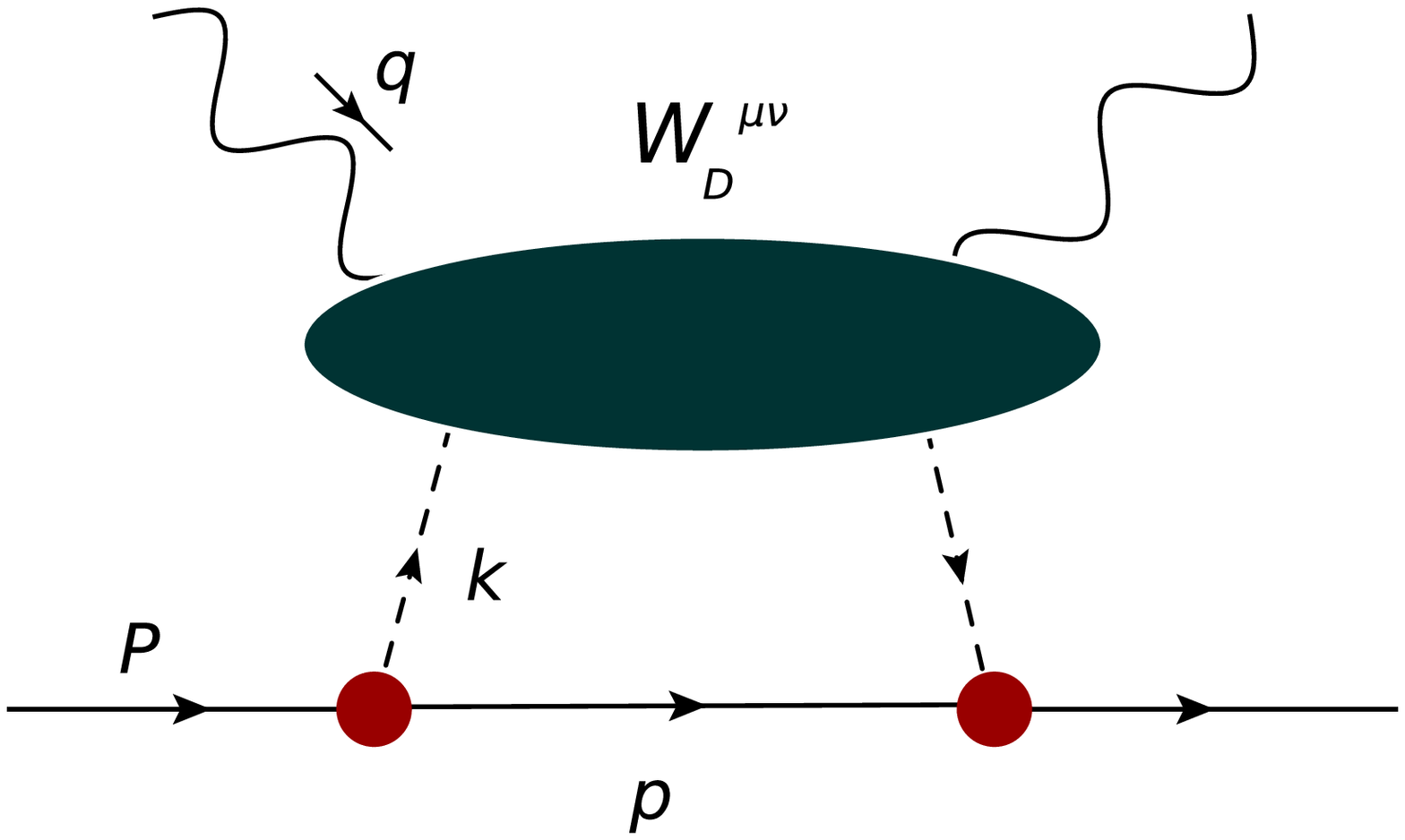} \ \ \ \
\includegraphics[width=8cm]{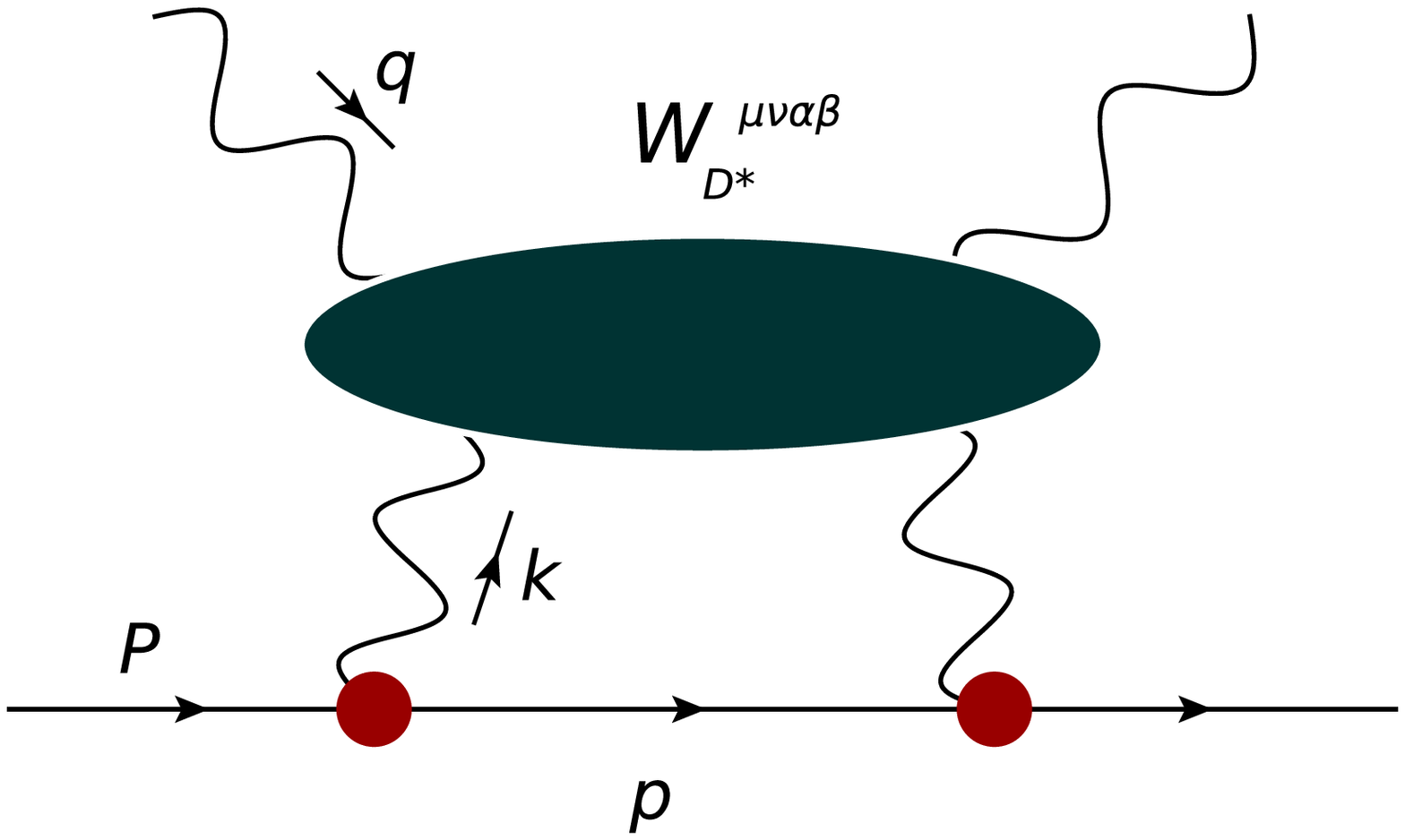}
\caption{
Diagrams for the $p \rightarrow DB$ (left) and $p \rightarrow D^*B$ (right)
processes.
}
\label{fig:f_MN}
\end{figure}

\underline{\em $N \to DB$ splitting} 

The dissociation of a nucleon to a spin-1/2 charmed baryon
$B = \Lambda_c$ or $\Sigma_c$ and a pseudoscalar $D$ meson is
derived from the effective hadronic lagrangian \cite{Hai11}
\bea
{\cal L}_{DBN}
&=& i g\, \bar{\psi}_N\, \gamma_5\, \psi_B\, \phi_D\ +\ \mathrm{h.c.}\ ,
\label{eq:Lagrangian}
\eea
where $\psi_N$ and $\psi_B$ are the nucleon and charmed baryon
fields, respectively, $\phi_D$ is the spin-0 $D$ meson field,
and the coupling constant $g \to g_{DBN}$.
Treating the propagation of the meson and baryon fields as
appropriate for point particles, the trace factor\footnote{The trace
calculations of this section can be performed using standard identities
for the Dirac matrices, which satisfy a Clifford algebra defined by
$\{\gamma^\mu, \gamma^\nu\} \defeq 2g^{\mu\nu}$. Also, in the conventional
way, we take the hermitian operator $\gamma^5 \defeq i \gamma^0 \dots \gamma^3$,
such that $(\gamma^5)^2 \equiv \mathbb{I}$, and $\{\gamma^\mu,\gamma^5\} \equiv 0$.
With these we can write the useful trace identities
\begin{align}
\mathrm{Tr}[\gamma^\mu]\ &=\ 0\ , \hspace*{1cm} \mathrm{Tr}[\gamma^\mu \gamma^\nu] \equiv 4g^{\mu\nu}\ , \nonumber\\
\mathrm{Tr}[\gamma^\mu \gamma^\nu \gamma^\alpha \gamma^\beta] &\equiv
4\ \Big(g^{\mu\nu} g^{\alpha\beta}\ +\ g^{\mu\beta} g^{\nu\alpha}\ -\ g^{\mu\alpha} g^{\nu\beta} \Big)\ . \nonumber
\end{align}
} $N^{DB}_{\mu\nu}$ can be written as
\begin{eqnarray}
N_{\mu\nu}^{DB}
&=& \frac{1}{2}\ \sum_{s,s'}\ \mathrm{Tr}
    \left[ \bar{u}^N_s(P) (i\gamma_5) u^B_{s'}(p)\, W^D_{\mu\nu}(k,q)\,
	   \bar{u}^B_{s'}(p) (i\gamma_5) u^N_s(P)
    \right]					\nonumber\\
&=& \frac{1}{2} \mathrm{Tr}
    \left[ i\gamma_5 (\Psl + M) i\gamma_5\, W^D_{\mu\nu}(k,q)\,
	   (\psl + M_B)
    \right]					\nonumber\\
&=& -\frac{1}{2} \mathrm{Tr}
    \left[ (\gamma_5)^2 (-\Psl + M)\, (\psl + M_B) \right]\,
    \widetilde{g}_{\mu\nu} F_1^D\, +\, \dots   \nonumber\\
&=& \left( 2 P \cdot p - 2 M M_B \right)\,
    \widetilde{g}_{\mu\nu} F_1^D\, +\, \dots\ ,
\label{eq:NLD-trace}
\end{eqnarray}
where $W_D^{\mu\nu}$ is the hadronic tensor for the $D$ meson,
with a form similar to that in Eq.~(\ref{eq:had_tens}), and
$F_1^D$ is the corresponding structure function which depends
on the $\bar c$ distribution in $D$. Note also that to reduce
the first line of Eq.~(\ref{eq:NLD-trace}) we used the property
of the nucleon/baryon spinors that
\begin{equation}
\sum_s u_s (P) \cdot \bar{u}_s(P)\ =\ (\Psl + M)\ {u \bar{u} \over 2M}\ ,
\hspace*{0.5cm} u \bar{u} \equiv 2M\
\end{equation}
--- a step we shall perform implicitly in the subsequent computations of this
section. Using Eq.~(\ref{eq:inn-prod}), and equating the coefficients of $\widetilde{g}_{\mu\nu}$
in Eqs.~(\ref{eq:had_tens}), (\ref{eq:del_MB}) and
(\ref{eq:NLD-trace}) then yields the convolution expression in
Eq.~(\ref{eq:mesoncloud_cb}) in which we identify the hadronic splitting
function as
\begin{eqnarray}
f_{DB}(y)
&=& T_B
    \frac{g^2}{16\pi^2} \int{ dk_\perp^2 \over y (1-y) }
    { |F(s)|^2 \over (s - M^2)^2 }
    \left[ \frac{k_\perp^2 + (M_B - (1 -y) M)^2}{1-y} \right]\ ,
\label{eq:DLsplit}
\end{eqnarray}
where for ease of notation we have used for the coupling constant
  $g \to g_{DBN}$ and for the $DB$ invariant mass
  $s \to s_{DB}$.
The isospin transition factor $T_B$ is given by
\begin{equation}
T_B = 1 + \delta_{t_B,+1}\ ,
\label{eq:TB}
\end{equation}
in which the third component of the isospin of the charmed baryon is
$t_B =  0$ for $B = \Lambda_c^+$ and $\Sigma_c^+$, and
$t_B = +1$ for $B = \Sigma_c^{++}$.
The states described by the splitting function $f_{DB}(y)$
include the lowest-mass configuration
	$\bar{D}^0 \Lambda_c^+$,
as well as the isovector charmed baryon states
	$\bar{D}^0 \Sigma_c^+$ and
	$D^- \Sigma_c^{++}$.

Performing an analogous calculation for the recoil process involving
scattering from the baryon $B$ confirms the symmetry relation
(\ref{eq:fphi}), which follows from the global charge and momentum
conservation relations in Eqs.~(\ref{eq:recip1}) and (\ref{eq:conserv}).
Specifically, we evaluate the recoil analogue of the left diagram in
Fig.~\ref{fig:f_MN}, with scattering from an interacting baryon of $4$-momentum
$p$. Much as before, the trace algebra yields
\begin{eqnarray}
N_{\mu\nu}^{BD}
&=& \frac{1}{2} \mathrm{Tr}
    \left[ i\gamma_5 (\Psl + M) i\gamma_5\, W^B_{\mu\nu}(k,q)\,
	   (\psl + M_B)
    \right]					\nonumber\\
%
%
&=& \left( 2 P \cdot p - 2 M M_B \right)\,
    \widetilde{g}_{\mu\nu} F_1^B\, +\, \dots\ ,
\label{eq:NDL-trace}
\end{eqnarray}
though we must evaluate this using the proper recoil kinematics. These
are now
\begin{subequations}
\label{eq:kin-TOPT_recoil}
\bea
P_0 &=& P_L + \frac{M^2}{2P_L} + {\cal O}\left(\frac{1}{P_L^2}\right)\ ,
\\
p_0 &=& |\bar{y}| P_L + \frac{k^2_\perp + M_B^2}{2 |\bar{y}| P_L} 
	+ {\cal O}\left(\frac{1}{P_L^2}\right)\ ,
\\ 
k_0 &=& |1 - \bar{y}| P_L + \frac{k^2_\perp + m_M^2}{2 |1 - \bar{y}| P_L}
	+ {\cal O}\left(\frac{1}{P_L^2}\right)\ ,
\eea  
\end{subequations}%
for the $0$-components, while for the $3$-momenta we require
\begin{subequations}
\label{eq:kin-TOPT_mom_recoil}
\bea
\bm{p} &=& |\bar{y}| \bm{P} + \bm{k_\perp}\ ,
\\
\bm{k} &=& |1-\bar{y}| \bm{P} - \bm{k_\perp}\ ,
\eea
\end{subequations}%
with $\bm{k}_\perp \cdot \bm{P} = 0$ as before, and an appropriate
redefinition of the baryon-meson invariant mass must be made vis-\'a-vis Eq.~(\ref{eq:CoM_En}):
$s_{MB}(y, k^2_\perp) \rightarrow \bar{s}_{BM}(\bar{y}, k^2_\perp)$. Performing a series
of deductions with these definitions similar to those used to obtain
Eq.~(\ref{eq:DLsplit}) leads to
\begin{eqnarray}
f_{BD}(\bar{y})
&=& T_B
    \frac{g^2}{16\pi^2} \int{ dk_\perp^2 \over \bar{y} (1-\bar{y}) }
    { |F(\bar{s})|^2 \over (\bar{s} - M^2)^2 }
    \left[ \frac{k_\perp^2 + (M_B - \bar{y} M)^2}{\bar{y}} \right]\ ,
\label{eq:LDsplit}
\end{eqnarray}
in which the reciprocity of Eq.~(\ref{eq:fphi}) is clearly manifest. Similar
operations with the following amplitudes also confirm such symmetry
relations in each case.

\underline{\em $N \to D^* B$ splitting} 

For the interaction between the nucleon and spin-1/2 baryon with a
vector $D^*$ meson, the effective lagrangian is given by \cite{Hai11}
\bea
{\cal L}_{D^* B N}
&=& g\, \bar{\psi}_N \gamma_\mu\, \psi_B\, \theta_{D^*}^\mu\
 +\ \frac{f}{4 M}
    \bar{\psi}_N \sigma_{\mu\nu} \psi_B\, F_{D^*}^{\mu \nu}\
 +\ \mathrm{h.c.}\ ,
\label{eq:L-S1}
\eea
where $\theta_{D^*}^\mu$ is the vector meson field, with field
strength tensor
$F_{D^*}^{\mu\nu} = \partial^\mu \theta_{D^*}^\nu
		  - \partial^\nu \theta_{D^*}^\mu$,
the tensor operator $\sigma_{\mu\nu} = (i/2) [\gamma_\mu,\gamma_\nu]$,
and the vector and tensor couplings are $g \to g_{D^* B N}$
and $f \to f_{D^* B N}$.
In this case the trace factor $N_{\mu\nu}^{D^* B}$ is given by
\bea 
N_{\mu\nu}^{D^* B}
&=& \frac{1}{2} \mathrm{Tr}
    \bigg[
    (\Psl + M)
    \Big( g \gamma^\alpha
	+ \frac{f}{2M} (\Delta^\alpha - \gamma^\alpha \Dsl)
    \Big)
    (\psl + M_B)					\nonumber\\
& & \hspace*{2.2cm} \times
    \Big( g \gamma^\beta
	- \frac{f}{2M} (\Delta^\beta - \gamma^\beta \Dsl)
    \Big)
    \bigg]\, W^{D^*}_{\mu\nu\alpha\beta}(k,q)		\nonumber\\
&=& \bigg( g^2 G_v + \frac{gf}{M} G_{vt} + \frac{f^2}{M^2} G_t
    \bigg)\, \widetilde{g}_{\mu\nu} F_1^{D^*}\ +\ \dots\ ,
\label{eq:S1-tr2}
\eea 
where $\Delta = P - p$, and the kinematical factors $G_v$, $G_{vt}$
and $G_t$ are given below in Eqs.~(\ref{eq:vectABC-app}).
The rank-4 tensor for the interacting $D^*$ meson can be expressed
in the form \cite{MT93}
\bea
W^{D^*}_{\mu\nu\alpha\beta}(k,q)
&=& \Big(
    \widetilde{g}_{\mu\nu} F_1^{D^*}
    + \frac{\widetilde{k}_\mu \widetilde{k}_\nu}{m^2_{D^*}} F_2^{D^*}
    \Big)\,
    \widetilde{g}_{\alpha\beta}\ ,
\label{eq:rank-4}
\eea
with $F_{1,2}^{D^*}$ the corresponding vector meson structure functions.

Thus for the dissociation of the proton to a charmed vector meson
$D^* = \bar{D}^{*0}$ or $D^{*-}$ and spin-1/2 charmed baryon, the
corresponding splitting function is given by a sum of vector ($G_v$),
tensor ($G_t$) and vector-tensor interference ($G_{vt}$) terms,
\bea 
f_{D^* B}(y)
&=& T_B \frac{1}{16\pi^2} \int{ dk_\perp^2 \over y (1-y) }
    { |F(s)|^2 \over (s - M^2)^2 }			\nonumber\\
& & \times
    \left[
	g^2\, G_v(y,k_\perp^2)\
     +\ {g f \over M}\, G_{vt}(y,k^2_\perp)\
     +\ \frac{f^2}{M^2}\, G_t(y,k^2_\perp)
    \right]\ ,
\label{eq:spin1-fMB}
\eea 
where 
\begin{subequations}
\label{eq:vectABC-app}%
\bea
\hspace*{-3.5cm}G_v(y,k_\perp)
&=& {1 \over 2} \mathrm{Tr} \Big[ (\Psl + M) \gamma^\alpha (\psl + M_B) \gamma^\beta \Big]\,
    \tilde{g}_{\alpha\beta}                            \nonumber\\
&=& - 6 M M_B\
 +\ \frac{4(P \cdot k) (p \cdot k)}{m_D^2}\
 +\ 2 P \cdot p\ ,
\eea
\bea
G_{vt}(y,k_\perp)
&=& {1 \over 4} \mathrm{Tr} \Big[ (\Psl + M) (\Delta^\alpha - \gamma^\alpha \Dsl) (\psl + M_B) \gamma^\beta \nonumber\\
& & \hspace*{1.5cm} -\ (\Psl + M) \gamma^\alpha  (\psl + M_B) (\Delta^\beta - \gamma^\beta \Dsl) \Big]\, \tilde{g}_{\alpha\beta} \nonumber\\
&=& 4(M + M_B)(P \cdot p - M M_B)			\nonumber\\
& &
 -\ \frac{2}{m_D^2}
    \left[ M_B (P \cdot k)^2
	 - (M + M_B)(P \cdot k)(p \cdot k)
	 + M (p \cdot k)^2
    \right]\ ,
\eea
\bea
G_t(y,k_\perp)
&=& {1 \over 8} \mathrm{Tr} \Big[ (\Psl + M) (\Delta^\alpha - \gamma^\alpha \Dsl) (\psl + M_B) (\gamma^\beta \Dsl - \Delta^\beta) \Big]\,
    \tilde{g}_{\alpha\beta}                            \nonumber\\
&=& -(P \cdot p)^2\
 +\ (M + M_B)^2\, P \cdot p\
 -\ M M_B (M^2 + M_B^2 + M M_B)				\nonumber\\
& &
 +\ \frac{1}{2m_D^2}
    \Big[ (P \cdot p - M M_B) [(P-p) \cdot k]^2
	- 2 (M_B^2 P\cdot k - M^2 p\cdot k) [(P-p) \cdot k]	
\nonumber\\
& & \hspace*{1cm}
	+ 2 (P \cdot k) (p \cdot k) (2P \cdot k - M_B^2 - M^2)
    \Big]\ ,
\eea
\end{subequations}%
and where $p$ is the $4$-momentum of the baryon,
and the inner products $P \cdot p$, $P \cdot k$ and $p \cdot k$
have again been computed explicitly in Eq.~(\ref{eq:inn-prod}).
The splitting function $f_{D^* B}(y)$ describes transitions
to the states
        $\bar{D}^{*0} \Lambda_c^+$,
        $\bar{D}^{*0} \Sigma_c^+$ and
        $D^{*-} \Sigma_c^{++}$,
and the isospin transition factor $T_B$ is as in Eq.~(\ref{eq:TB}).
As before, for compactness we have again used the shorthand notation
for the couplings
  $g \to g_{D^* B N}$ and
  $f \to f_{D^* B N}$, with
  $s \to s_{D^* B}$.

\begin{figure}
\includegraphics[width=8cm]{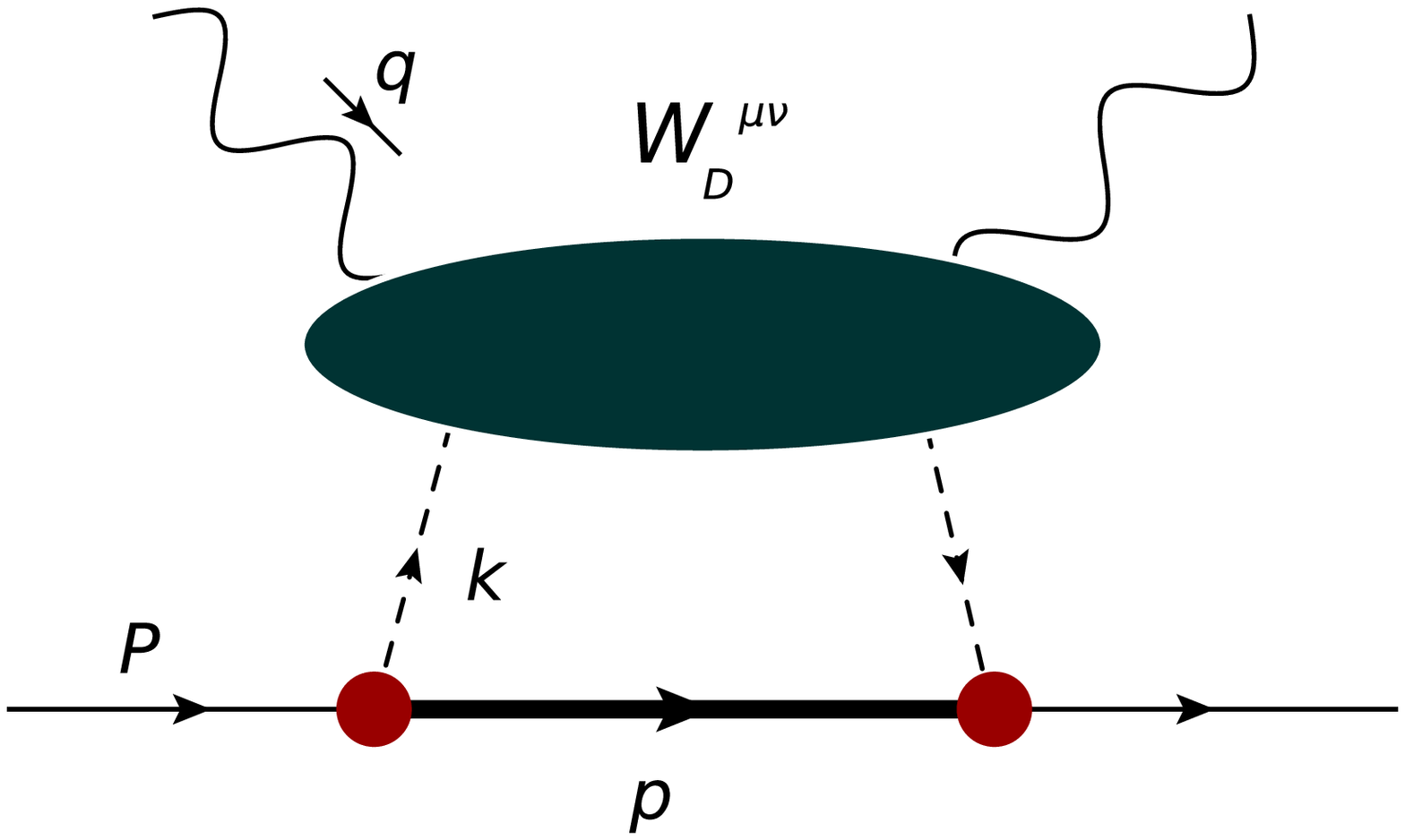} \ \ \ \
\includegraphics[width=8cm]{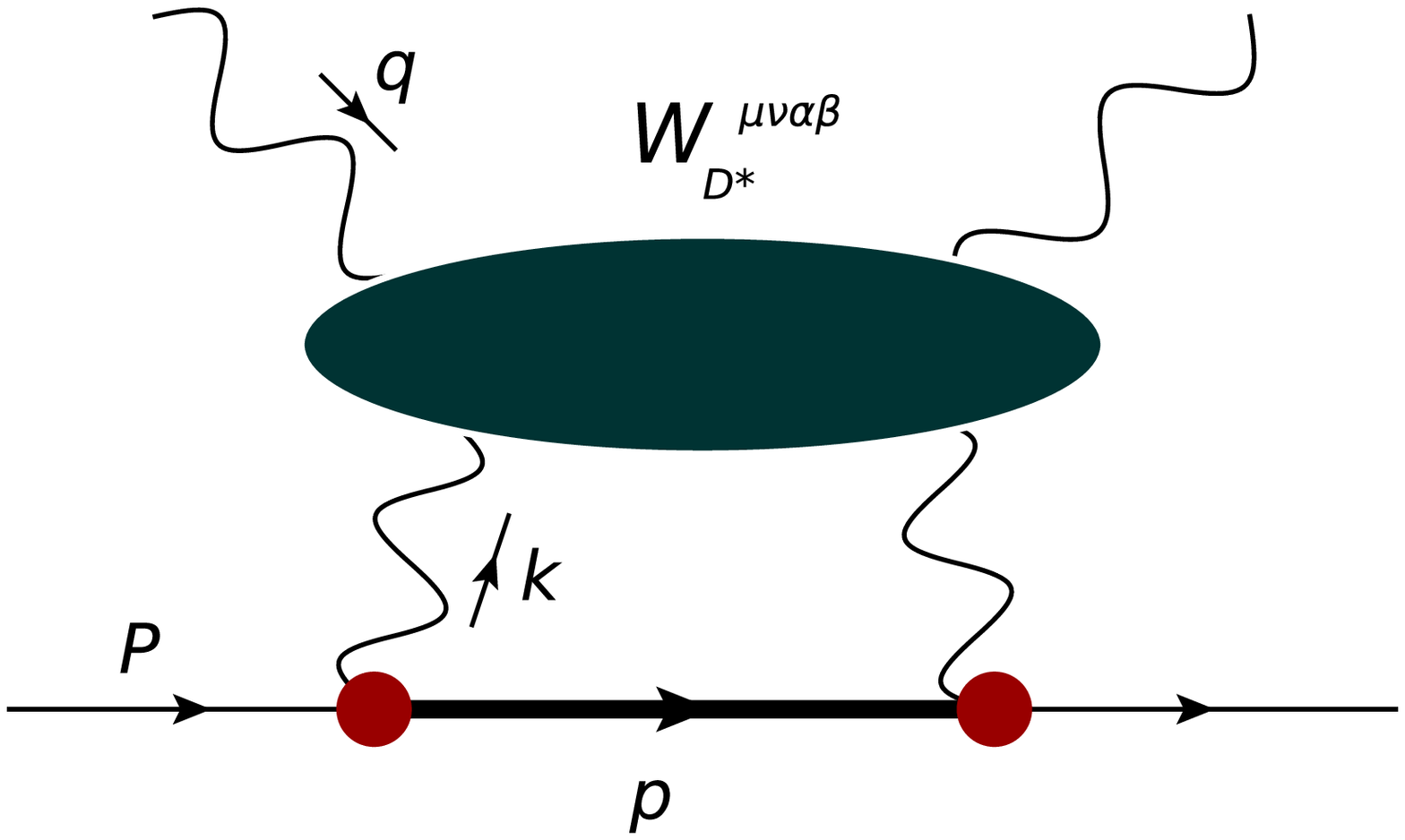}
\caption{
Diagrams for the $p \rightarrow DB^*$ (left) and $p \rightarrow D^*B^*$ (right)
processes.
}
\label{fig:f_MDel}
\end{figure}

\pagebreak

\underline{\em $N \to D B^*$ splitting} 

The interaction of a nucleon with a spin-3/2 charmed baryon
$B^* = \Sigma_c^*$ and a pseudoscalar meson can be calculated using the
lagrangian \cite{Hai11}
\bea
{\cal L}_{D B^* N}
&=& \frac{f}{m_D}
    \Big( \bar{\Psi}_{B^*}^\mu\, \psi_N\, \partial_\mu \phi_D\
       +\ \bar{\psi}_N \Psi_{B^*}^\mu \partial_\mu \phi_D
    \Big)\ ,
\label{eq:piD-Lagr}
\eea
where $\Psi_{B^*}^\mu$ is the Rarita-Schwinger spinor-vector field,
and the coupling $f \to f_{D B^* N}$.
From Eq.~(\ref{eq:piD-Lagr}) and the LHS diagram of Fig.~\ref{fig:f_MDel}, the
trace factor tensor is
\bea 
N_{\mu\nu}^{D B^*}
&=& \frac{1}{2} \mathrm{Tr}
    \bigg[ (\Psl + M)\,
	   \Lambda^{\alpha\beta}_{B^*}(p)\,
	   \Delta_\alpha \Delta_\beta\, W_{\mu\nu}^D(k,q)
    \bigg]						\nonumber \\
&=& \frac{4}{3}
    \left( P \cdot p + M M_{B^*} \right)
    \left( \frac{(p \cdot \Delta)^2}{M_{B^*}^2} - \Delta^2 \right)
    \widetilde{g}^{\mu\nu}\, F_1^D\ +\ \dots\ ,
\label{eq:piD-trace}
\eea 
where the energy projector for the spinor-vector is
\bea
\Lambda^{\alpha\beta}_{B^*}(p)
&=& (\psl + M_{B^*})
    \left(- g^{\alpha\beta}
          + {\gamma^\alpha \gamma^\beta \over 3}
          + {\gamma^\alpha p^\beta - \gamma^\beta p^\alpha \over 3M_{B^*}}
          + {2\, p^\alpha p^\beta \over 3 M_{B^*}^2}
    \right)\ .
\eea
The inner products in Eq.~(\ref{eq:piD-trace}) can once again be
worked out in terms of $y, k^2_\perp$ using the definitions in Eq.~(\ref{eq:inn-prod})
with the replacement $B \to B^*$, which leads directly to the desired
hadronic amplitude for fluctuations to spin-3/2 charmed
baryons $B^* = \Sigma_c^{* +}$ and $\Sigma_c^{* ++}$.  Namely, for the
dissociations of a proton to states with a spin-0 $D$ meson, 
	$\bar{D}^0 \Sigma_c^{* +}$ and
	$D^- \Sigma_c^{* ++}$,
the splitting function is given by
\bea
f_{D B^*}(y)
&=& T_{B^*}
    \frac{g^2}{16\pi^2} \int{ dk_\perp^2 \over y (1-y) }
    { |F(s)|^2 \over (s - M^2)^2 }			\nonumber\\
& & \times
    { \left[ k_\perp^2 + (M_{B^*} - (1-y) M)^2 \right]
      \left[ k_\perp^2 + (M_{B^*} + (1-y) M)^2 \right]^2
    \over 6 M_{B^*}^2 (1-y)^3}\ ,
\label{eq:DS*-split}
\eea
with 
  $g \to g_{D B^* N}$ and
  $s \to s_{D B^*}$.
The isospin transition factor $T_{B^*}$ here is similar to that
in Eq.~(\ref{eq:TB}), but with the third component of the charmed
baryon isospin
$t_{B^*} =  0$ for $B^* = \Sigma_c^{* +}$ and
$t_{B^*} = +1$ for $B^* = \Sigma_c^{* ++}$.

\underline{\em $N \to D^* B^*$ splitting} 

Finally, for the nucleon splitting to a spin-3/2 charmed baryon $B^*$
coupled to a vector meson $D^*$ the effective hadronic lagrangian is
given by \cite{Hai11}
\bea
{\cal L}_{D^* B^* N}
&=& \frac{f}{m_{D^*}}
    i \Big( \bar{\Psi}_{B^* \nu} \gamma^5 \gamma_{\mu} \psi_N
	  - \bar{\psi}_N \gamma^5 \gamma_{\mu} \Psi_{B^* \nu} 
      \Big) F_{D^*}^{\mu\nu}\ ,
\label{eq:DsSigS-Lagr}
\eea
where $f \to f_{D^* B^* N}$.
This yields the resulting trace tensor
\bea 
N_{\mu\nu}^{D^* B^*}
&=& \frac{1}{2}
    \mathrm{Tr}
    \bigg[
	(\Psl + M)
	\gamma^5 \gamma^\alpha\,
	\Lambda_{B^*}^{\alpha'\beta'}(p)\,
	\gamma^5 \gamma^\beta
    \bigg]
    \mathcal{G}_{\alpha\beta\alpha'\beta'}\,
    \widetilde{g}_{\mu\nu} F_1^{D^*}\ +\ \dots\ ,
\label{eq:D*S*-trace1}
\eea 
for which we define
\bea
\mathcal{G}_{\alpha\beta\alpha'\beta'}
&=& \Delta_{\alpha\beta}\,   \widetilde{g}_{\alpha' \beta'}
  - \Delta_{\alpha\beta'}\,  \widetilde{g}_{\alpha' \beta}
  - \Delta_{\alpha'\beta}\,  \widetilde{g}_{\alpha  \beta'}
  + \Delta_{\alpha'\beta'}\, \widetilde{g}_{\alpha  \beta}\ ,
\eea
with $\Delta_{\alpha\beta} \equiv \Delta_\alpha \Delta_\beta$,
the other expressions having been defined above.
Evaluating the trace in Eq.~(\ref{eq:D*S*-trace1}) and equating
coefficients of the $\widetilde{g}_{\mu\nu}$ terms then leads to
the convolution relation with the splitting function for the
fluctuations to states with $D^*$ mesons and spin-3/2 baryons $B^*$
given by
\bea 
\label{eq:RS-fMB}
f_{D^* B^*}(y)
&=& T_{B^*}
    \frac{g^2}{m_{D^*}^2 16\pi^2}
    \int{ dk_\perp^2 \over y(1-y) }
    { |F(s)|^2 \over (s - M^2)^2 }			\\
& &
 -\ \left[
    \frac{4 M M_{B^*}}{3}
    \Big( 2 M_{B^*}^2 + M M_{B^*} + 2 M^2 \Big)
     - \frac{4 M M_{B^*}}{3 m_{D^*}^2} ((P-p) \cdot k)^2
    \right.						\nonumber\\
& &
 -\ \frac{4}{3m_{D^*}^2}
    \Big( M_{B^*}^2 (P \cdot k)^2 + M^2 (p \cdot k)^2 \Big)
     + \frac{4 P \cdot p}{3}
	\Big( 2 M_{B^*}^2 + 4 M M_{B^*} + M^2 \Big)	\nonumber\\
& & \left.
 +\ \frac{4 P \cdot p}{3m_{D^*}^2}(p \cdot k)^2
	 \Big( 1-\frac{M^2}{M_{B^*}^2} \Big)
    - 4 (P \cdot p)^2
        \left(
	  1 - \frac{2 (P \cdot k)(p \cdot k)}{3m_{D^*}^2 M_{B^*}^2}
	    - \frac{P \cdot p}{3 M_{B^*}^2}
	\right)
    \right]\ ,						\nonumber
\eea
where
  $g \to g_{D^* B^* N}$,
  $s \to s_{D^* B^*}$,
and the inner products in Eq.~(\ref{eq:RS-fMB}) are given
by Eq.~(\ref{eq:inn-prod}) after the replacements
$D \to D^*$ and $B_c \to \Sigma_c^*$.
%

\vspace*{0.15cm}
{\it Charm content of charmed baryons and mesons.}
\vspace*{0.15cm}

In the literature estimates of the distributions of heavy quarks in
heavy hadrons have been made using the heavy quark limit \cite{MT97},
and within a scalar constituent quark model \cite{Pum05}.
Collecting and building upon several features of these approaches in constructing
a relativistic quark model with the correct spin degrees of freedom,
we compute results analogous to the hadronic splittings derived above.
Again we apply the time-ordered perturbation theory framework in the IMF,
but now at parton-level, defining the IMF momentum fraction
$\hat{y} = \hat{k}_L/P_L$ to be the ratio of the longitudinal
momentum of the constituent quark or antiquark ($\hat{k}_L$) to
that of the parent charmed meson or baryon ($P_L$).
Convolution with the leading twist point-like structure of constituent
quarks gives distributions as functions of the quark-level Bjorken
limit variable, denoted here by $z$ to prevent confusion with quark
distributions in the proton.
In the following we summarize the vital features of the derivations of
$\bar c$ distributions in $D$ and $D^*$ mesons, and the $c$ distributions
in the $\Lambda_c$ and $\Sigma_c^*$ baryons.
The numerical results of this relativistic quark--spectator model vis-a-vis 
our eventual meson-baryon model shall be presented systematically in Sec.~\ref{sec:cinc}.
\begin{figure}[h]
\includegraphics[width=15cm]{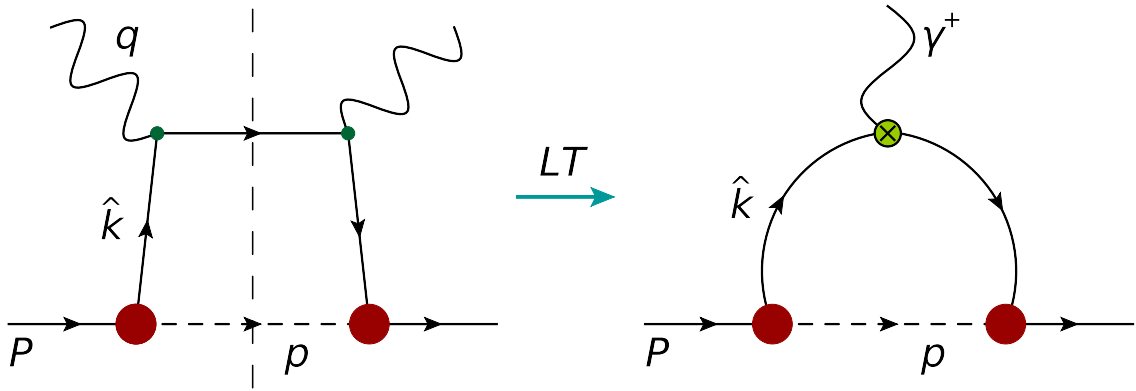}
\caption{
The leading twist (LT) reduction of the quark-level handbag diagram
(left), to a form in which the quark-photon interaction enters explicitly
via the $\gamma^+$ structure (right).
}
\label{fig:qua_LT}
\end{figure}

\underline{\em $\bar c$ in $D$} 

To model the distribution of a $\bar c$ quark in the pseudoscalar
$D$ meson we consider the effective lagrangian describing the
coupling of the $D$ to the $\bar c$ and a light quark $q$,
\bea
{\cal L}_{\bar{c} q D}
&=& i g\, \bar{\psi}_{\bar c}\, \gamma_5\, \psi_q\, \phi_D\
 +\ \mathrm{h.c.}\ ,
\label{eq:Lagrangian-q}
\eea
where $\psi_q$ and $\psi_{\bar c}$ are the quark $q$ and $\bar c$
fields, and the effective coupling constant is now $g \to g_{\bar{c} q D}$.
The contribution to the hadronic tensor of the $D$ meson from scattering
off the $\bar c$ quark with a spectator light quark $q$ can be written
in analogy with Eq.~(\ref{eq:del_MB}) for the hadronic calculation,
\bea
\bar{c}_D (z)
&=& \frac{N_D}{16\pi^2}
    \int_0^\infty \frac{d\hat{k}^2_\perp}{z(1-z)}
    \frac{|G(\hat{s})|^2}{(m_D^2 - \hat{s})^2}\,
    \widehat{T}^{\bar{c} q}\ ,
\label{eq:del_ctau}
\eea
where $\hat{k}_\perp$ denotes the internal quark transverse momentum
in the $D$ meson and $z$ is the Bjorken scaling variable for the quark
inside the $D$ meson. 
Note that the total invariant mass squared of the $\bar{c} q$ pair is defined
[see also Eq.~(\ref{eq:CoM_En})] for the corresponding invariant mass
of the meson--baryon system) as
\be
\hat{s}(z,\hat{k}^2_\perp)
= \frac{m_{\bar c}^2 + \hat{k}^2_\perp}{z}
+ \frac{m_q^2 + \hat{k}^2_\perp}{1-z}\ ,
\label{eq:s-hat}
\ee
where $m_{\bar{c}}$ is the constituent anticharm quark mass,
$m_q$ is the mass of the (light) spectator quark.

The trace factor $\widehat{T}^{\bar{c} q}$ can then be computed from the
quark-level ``handbag'' diagram, yielding
\bea
\widehat{T}^{\bar{c} q}
&=& \frac{1}{4 \hat{k}^+}
    \mathrm{Tr}
    \left[
	i \gamma_5
	(\hat{\ksl} + m_{\bar c})\,
	\gamma^+\,
	(\hat{\ksl} + m_{\bar c})\,
	(-i \gamma_5)
	(-\hat{\psl} + m_q)
    \right]\ ,						\nonumber\\
&=& 2\, \big( \hat{p} \cdot \hat{k} + m_{\bar c} m_q \big)\ ,
\label{eq:N_c-Lc1}
\eea
which follows from the on-mass-shell condition in time-ordered
perturbation theory, \mbox{$\hat{k}^2 = m^2_{\bar{c}}$}, with
$\hat{p}$ the $4$-momentum of the spectator quark. Following
the replacement shown in Fig.~\ref{fig:qua_LT}, the $\gamma^+$
structure arises from reducing the hard scattering amplitude
to its leading twist approximation \cite{Mulders:1992za},
\begin{equation}
  \gamma^{\mu} (\hat{\ksl} + \qsl + m_{\bar{c}}) \gamma^{\nu}\,
   \, \delta \left( [\hat{k} + q]^2 - m_{\bar c}^2 \right)\, \rightarrow \,
  {\gamma^+ \over 2 \hat{k}^+}\,\, \delta \left( 1 - {z \over \hat{y}} \right)\ ,
\label{eq:Mulders}
\end{equation}
where $\hat{y}$ is the parton fraction of the hadron momentum,
after equating the coefficients of $g^{\mu\nu}$ and selecting
the $+$ component of the external photon current.

The result for $\bar{c}_D(z)$ is then
obtained by inserting the expression for $\widehat{T}^{\bar{c} q}$
in Eq.~(\ref{eq:N_c-Lc1}) into Eq.~(\ref{eq:del_ctau}), and using the
IMF momenta as in Eq.~(\ref{eq:inn-prod}) but with the replacements
$y \to z$, $M \to M_D$, $m_D \to m_{\bar{c}}$, and $M_B \to m_q$.
This yields the desired distribution of a relativistic $\bar c$ quark in
a pseudoscalar $D$ meson, with a spectator $u$ or $d$ quark, in
analogy with the $p \to D \Lambda_c$ splitting function in
Eq.~(\ref{eq:DLsplit}). Specifically, the manipulations of Eq.~(\ref{eq:N_c-Lc1})
demonstrate that using a pseudoscalar meson--quark--antiquark
vertex parametrized by the structure $\gamma_5 G(\hat{s})$, where
$G(\hat{s})$ is the $D$-$\bar c$-$q$ vertex function ($q=u, d$),
gives the distribution
\begin{eqnarray}
\bar{c}_D(z)
&=& \frac{N_D}{16\pi^2} \int_0^\infty 
    \frac{d\hat{k}_\perp^2}{[z (1-z)]^2}
    \frac{|G(\hat{s})|^2}{(\hat{s} - m_D^2)^2}
    \Big[ \hat{k}^2_\perp + (z\, m_q + (1-z)\, m_{\bar{c}})^2 \Big]\ ,
\label{eq:cbar_Dbar}
\end{eqnarray}
in which the integration is over the transverse momentum $\hat{k}_\perp^2$
of the interacting heavy quark, and $z$ is the Bjorken scaling variable
of the heavy quark inside the charmed hadron. Here and in the following
computations, the overall normalization $N_D$ is determined by the valence
condition,
\be
\int_0^1 dz\, \bar{c}_D(z) = 1\ .
\label{eq:cbarDNorm}
\ee
\begin{figure}[h]
\vspace*{-0.9cm}
\includegraphics[width=8cm]{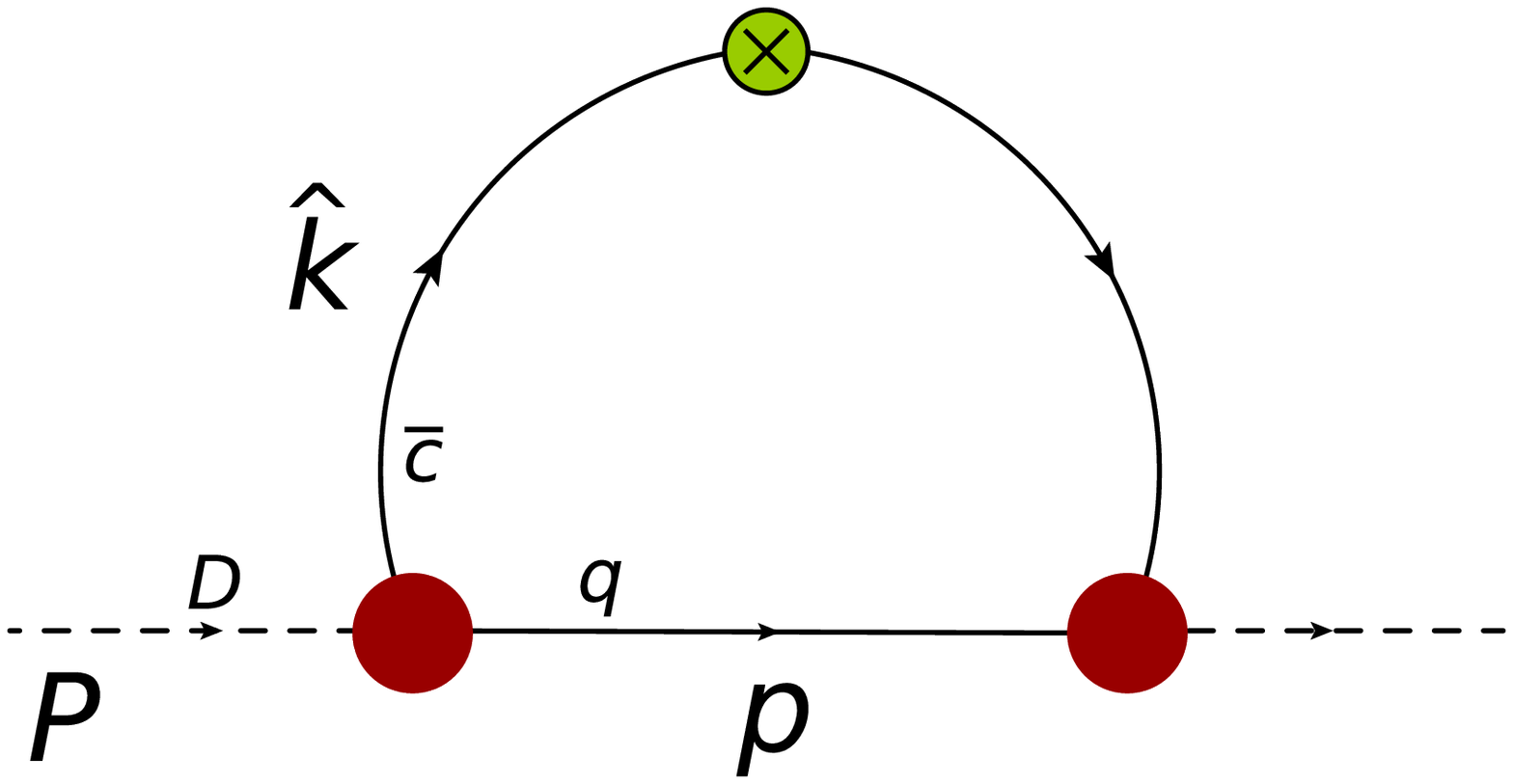} \ \ \ \
\includegraphics[width=8cm]{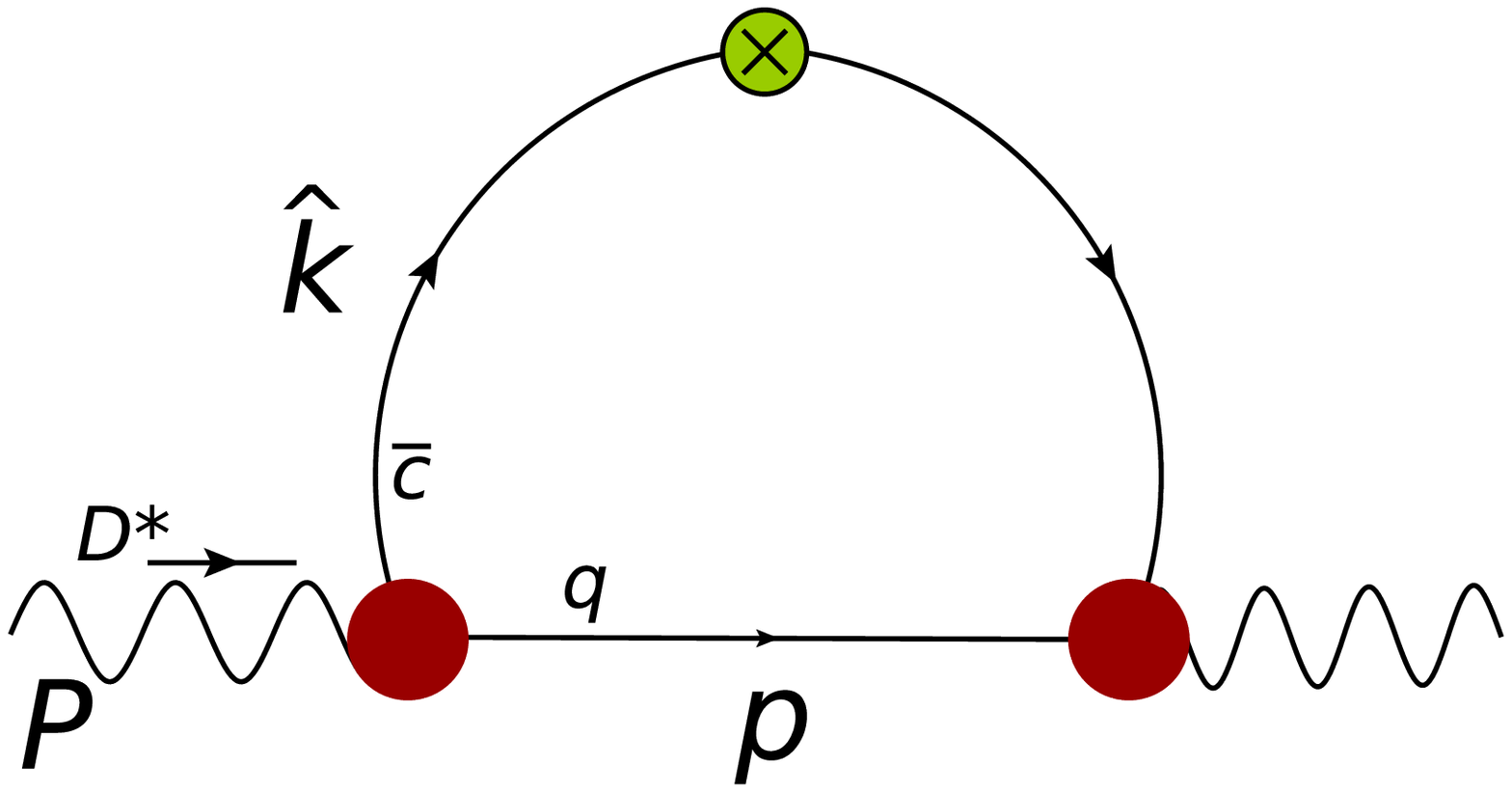}
\caption{
Diagrams for anticharm quark distributions inside charmed mesons.
}
\label{fig:cb_Mes}
\end{figure}

\underline{\em $\bar c$ in $D^*$} 

For the $\bar c$ distribution in a vector $D^*$ meson, there
exist in principle both the Dirac and Pauli couplings of the
$D^*$ to a quark and antiquark, as for the $D^* B$ splitting
function in Eq.~(\ref{eq:spin1-fMB}).  To reduce the number
of free parameters in the calculation, we make the simplifying
assumption that the $D^*$--quark--antiquark coupling is governed
by a purely vector interaction, $\gamma_\alpha\, G(\hat{s})$.
The result is that the following simple vector form is chosen for
the lagrangian describing the $\bar{c} q D^*$ interaction,
\bea
{\cal L}_{\bar{c} q D^*}
&=& g\, \bar{\psi}_{\bar c}\, \gamma_\mu\, \psi_q\, \theta_{D^*}^\mu\
 +\ \mathrm{h.c.}\ ,
\label{eq:cD*-Lagr}
\eea
where $g \to g_{D^* \bar{c} q}$.  This yields the trace factor
\bea 
\widehat{T}^{(\bar{c} q)^*}
&=& \frac{1}{4 \hat{k}^+}
    \mathrm{Tr}
    \left[
	(-\hat{\psl} + m_q)
	\gamma^\alpha
	(\hat{\ksl} + m_{\bar c})\,
	\gamma^+\,
	(\hat{\ksl} + m_{\bar c})\,
	\gamma^\beta \right] \widetilde{g}_{\alpha\beta} \nonumber\\
&=& 4
    \left(
	  \frac{(P \cdot \hat{p})(P \cdot \hat{k})}{m_{D^*}^2}
	+ \frac{3}{2} m_{\bar c}\, m_q
	+ \frac{\hat{p} \cdot \hat{k}}{2}
    \right)\ .
\label{eq:N-cbD*}
\eea 
Applying the same procedure as for the $\bar c$ distribution inside
the pseudoscalar $D$ meson, one immediately arrives at the distribution
of a $\bar c$ quark in the $D^*$ meson, which is given by
\bea 
\bar{c}_{D^*}(z)
&=& \frac{N_{D^*}}{16\pi^2} \int_0^\infty 
    \frac{d\hat{k}_\perp^2}{[z (1-z)]^2}
    \frac{|G(\hat{s})|^2}{(\hat{s} - m_{D^*}^2)^2}
    \Biggl[
       \left( \frac{\hat{k}^2_\perp + m_q^2}{m_{D^*}^2} + (1-z)^2 \right)
       \left( \hat{k}^2_\perp + m_{\bar c}^2 + z^2\, m_{D^*}^2    \right)
\nonumber\\
& & \hspace*{3cm}
     +\ \hat{k}^2_\perp\
     +\ \left( z\, m_q + (1-z)\, m_{\bar c} \right)^2\
     +\ 4z(1-z)\, m_q m_{\bar c}
    \Biggr]\ , 
\label{eq:cbar_D-st}
\eea 
with the normalization factor $N_{D^*}$ again determined
by the valence quark number conservation condition in
Eq.~(\ref{eq:cbarDNorm}).
As with the $\bar c$ distribution in the $D$ meson, for point
interactions the integral in Eq.~(\ref{eq:cbar_D-st}) would be
divergent, in this case linearly in $\hat{k}_\perp^2$.
This behavior can be controlled with vertex form factors $G(\hat{s})$
as we shall discuss in greater detail in Sec.~\ref{sec:cinc}.

\underline{\em $c$ in $\Lambda_c$, $\Sigma_c$} 

The charm quark distributions inside charmed baryons are obtained
from an expression analogous to that in Eq.~(\ref{eq:del_ctau}),
\bea
c_B (z)
&=& \frac{N_B}{16\pi^2}
    \int_0^\infty \frac{d\hat{k}^2_\perp}{z(1-z)}
    \frac{|G(\hat{s})|^2}{(m_B^2 - \hat{s})^2}\,
    \widehat{T}^{c [qq]}\ ,
\label{eq:c_baryon}
\eea
where $\widehat{T}^{c [qq]}$ is the corresponding trace factor for
the scattering from the $c$ quark with a spectator diquark $[qq]$
in the intermediate state.
For spin-$1/2$ baryons, we consider the scalar interaction between
the $c$ and $[qq]$ quarks and the charmed baryon, given by the
lagrangian
\bea
{\cal L}_{c [qq] B}
&=& g\, \bar{\psi}_B\, \psi_c\, \phi_{[qq]}\ +\ \mathrm{h.c.}\ ,
\label{eq:c-Lamb-Lagr}
\eea
with $g \to g_{c [qq] B}$, and $\phi_{[qq]}$ is the field of the scalar 
diquark.  The trace factor $\widehat{T}^{c [qq]}$ can be explicitly
derived from the left diagram of Fig~\ref{fig:c_Bary} as
\bea 
\widehat{T}^{c [qq]}
&=& \frac{1}{4 \hat{k}^+}
    \mathrm{Tr}
    \left[
	(\Psl + M_B) (\hat{\ksl} + m_c) \gamma^+ (\hat{\ksl} + m_c)
    \right]  \nonumber\\
&=& 2 \left( P \cdot \hat{k} + m_c\, M_B \right)\ ,
\label{eq:N_c-Lc2}
\eea 
giving the net result for the charm quark distribution inside
a spin-$1/2$ baryon $B$ ($B = \Lambda_c$ or $\Sigma_c$) with a scalar
$qq$ spectator:
\bea
c_B(z) &=& \frac{N_B}{16\pi^2} \int_0^{\infty} 
  \frac{d\hat{k}_\perp^2}{z^2 (1-z)}
  \frac{|G(\hat{s})|^2}{(\hat{s} - M_B^2)^2}
  \Big[ \hat{k}^2_\perp + (m_c + z M_B)^2 \Big]\ ,
\label{eq:c_in_Lambda}
\eea
where for the charm quark mass we take $m_c = m_{\bar{c}}$,
and the invariant mass squared of the quark--diquark system here
is defined as
\be
\hat{s}(z,\hat{k}^2_\perp)
= \frac{m_c^2 + \hat{k}^2_\perp}{z}
+ \frac{m_{qq}^2 + \hat{k}^2_\perp}{1-z}\ .
\label{eq:s-hatB}
\ee
The functional form of the $B$-$c$-$qq$ vertex function $G(\hat{s})$
will be specified as for the $D$-$\bar c$-$q$ function in the
phenomenological schemes to be outlined in Sec.~\ref{sec:cinc}, and the
normalization constant $N_B$ is now determined from an analogous valence
charm quark number condition to that in Eq.~(\ref{eq:cbarDNorm}),
\be
\int_0^1 dz\, c_B(z) = 1\ .
\label{eq:cLambdacNorm}
\ee
Note that for the $\Sigma_c^{++}$ baryon, the $uu$ spectator diquark
has spin 1, so that the calculation of its $c$ quark distribution
here is approximated by neglecting the diquark's spin structure.
In principle, it is straightforward to include both spin-0 and spin-1 diquark
contributions in analogy with the spin structures discussed
for the anticharm distributions; however, because the overall contribution
from the dissociation of the proton to $D^- \Sigma_c^{++}$ is at least an
order of magnitude smaller than for $\bar{D}^{*0} \Lambda_c^+$,
this will have negligible effect on the numerical results.

\underline{\em $c$ in $\Sigma_c^*$} 

Lastly, for the charm density of the spin-$3/2$ $B^*$ baryons we adopt the following
lagrangian,
\bea 
{\cal L}_{c [qq]^* B^*}
&=& \frac{g}{m_{[qq]^*}} \
    i \left(
	\bar{\Psi}_{B^* \nu} \gamma_{\mu} \psi_c
      - \bar{\psi}_c \gamma_{\mu} \Psi_{B^* \nu}
    \right)
    F_{[qq]^*}^{\mu\nu}\ ,
\label{eq:cSig*-Lagr}
\eea 
with $g \to g_{c [qq]^* B^*}$, which correctly gives the parities
of the physical $B^*$ and quark fields.  Again, the field strength
tensor here has the form
  $F_{[qq]^*}^{\mu \nu}
  = \partial^{\mu} \theta_{[qq]^*}^{\nu}
  - \partial^{\nu} \theta_{[qq]^*}^{\mu}$,
where $\theta_{[qq]^*}$ denotes the (spin-1) axial-vector diquark.
The trace factor in this case is found to be
\bea 
\widehat{T}^{c [qq]^*}
&=& \frac{1}{4 m^2_{[qq]^*} \hat{k}^+}
    \mathrm{Tr}
    \left[
	\Lambda_{B^*}^{\beta'\alpha'}(P)
	\gamma^\alpha
	(\hat{\ksl} + m_c)\, \gamma^+\, (\hat{\ksl} + m_c)
	\gamma^\beta
    \right]
    \mathcal{G}_{\alpha\beta\alpha'\beta'}\ .
\label{eq:cSig*-trace1}
\eea 
After re-indexing, $\mathcal{G}_{\alpha\beta\alpha'\beta'}$ is
given by Eq.~(\ref{eq:D*S*-trace1}), the exchange boson carries
the $4$-momentum $\hat{\Delta} = P - \hat{k}$, and the metric
tensor of the massive, spin-1 diquark is
  $\widetilde{g}^{\mu\nu} = 
    -g^{\mu\nu} + P^{\mu} P^{\nu}/m^2_{[qq]^*}$.
Computing the trace and contractions of Eq.~(\ref{eq:cSig*-trace1})
and evaluating the result with the appropriate kinematic definitions
analogous to Eq.~(\ref{eq:inn-prod}), the charm distribution in
spin-3/2 $\Sigma_c^*$ baryons emerges. Here, the charm quark is always
accompanied by a spin-1 diquark, such that after incorporating the fully
relativistic Rarita-Schwinger structure for the spin-3/2 state,
the charm quark distribution in the $\Sigma_c^*$ is given by
\bea
c_{B^*}(z)
&=& \frac{N_{B^*}}{12\pi^2 m_{qq}^2} \int_0^\infty
    \frac{d\hat{k}_\perp^2}{z (1-z)}
    \frac{|G(\hat{s})|^2}{(\hat{s} - M_{B^*}^2)^2}
    \Biggl( (\hat{k} \cdot \hat{\Delta})(P \cdot \hat{\Delta})
	  + 2 m_c M_{B^*} \hat{\Delta}^2
\nonumber\\
&+& \frac{1}{m_{qq}^2}
    \Bigl[
      m_c M_{B^*} (\hat{p} \cdot \hat{\Delta})^2
    - (\hat{p} \cdot \hat{\Delta})
      \Bigl( (\hat{p} \cdot \hat{\Delta})(\hat{k} \cdot \hat{p})
	   - (P \cdot \hat{\Delta})(\hat{k} \cdot \hat{p})
	   - (\hat{k} \cdot \hat{\Delta})(P \cdot \hat{p})
      \Bigr)
\nonumber\\ 
&-&   (P \cdot \hat{p})(\hat{k} \cdot \hat{p}) \hat{\Delta}^2
     - \frac{P \cdot \hat{k}}{M_{B^*}^2} 
       \Bigr(
	 (P \cdot \hat{p})^2 \hat{\Delta}^2
	- m_{qq}^2 (P \cdot \hat{\Delta})^2
	- 2 (P \cdot \hat{\Delta})
	    (\hat{p} \cdot \hat{\Delta})
	    (P \cdot \hat{p})
       \Bigr)
    \Bigr] 
    \Biggr)\ ,
\nonumber\\
\label{eq:c_Sigma-st}
\eea 
in which $\hat{p}$ is the momentum of the spectator diquark $qq$,
and $\hat{\Delta} \equiv P - \hat{k}$ [see Eq.~(\ref{eq:inn-prod})].

\begin{figure}[t]
\includegraphics[width=8cm]{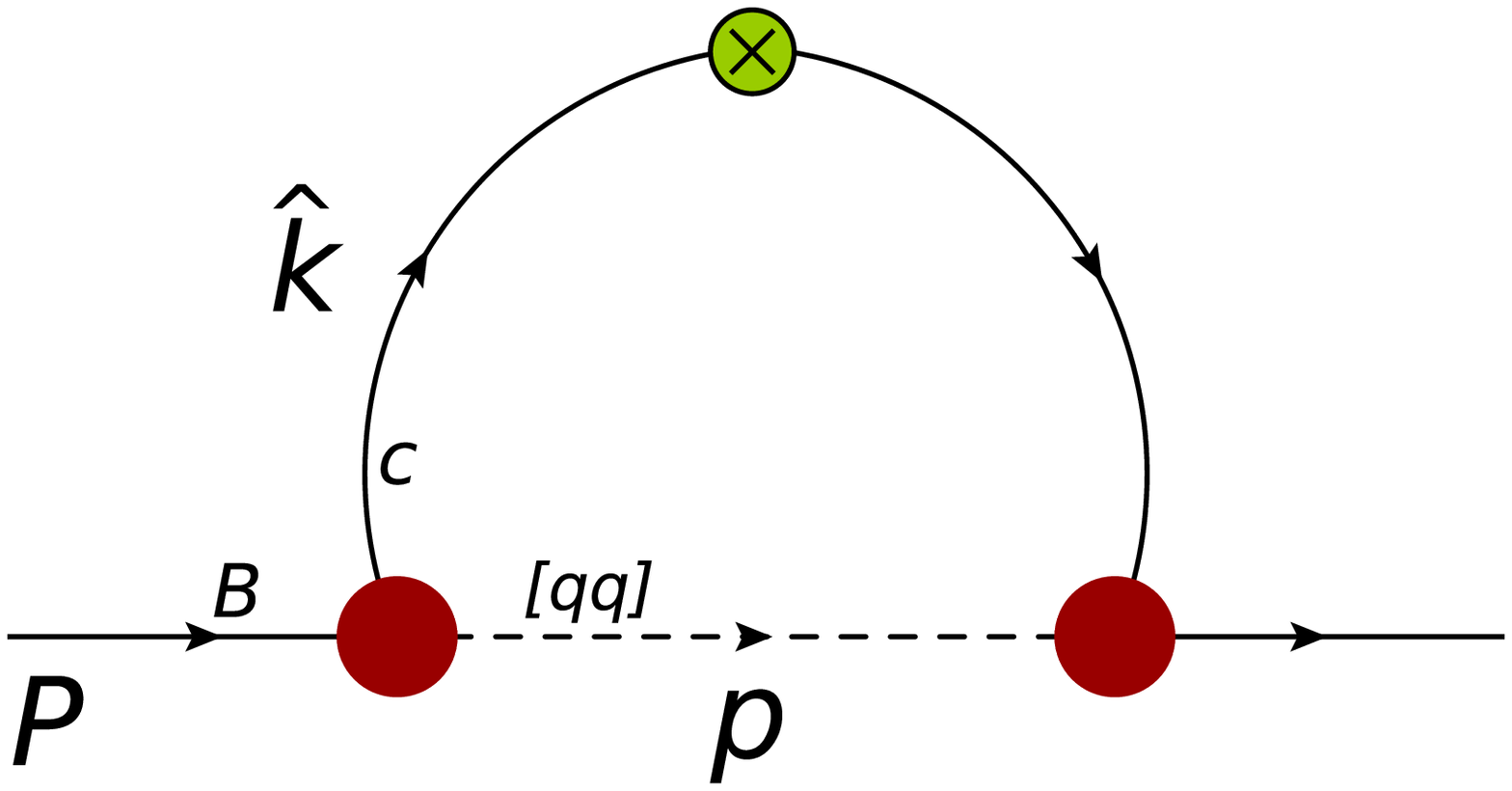} \ \ \ \
\includegraphics[width=8cm]{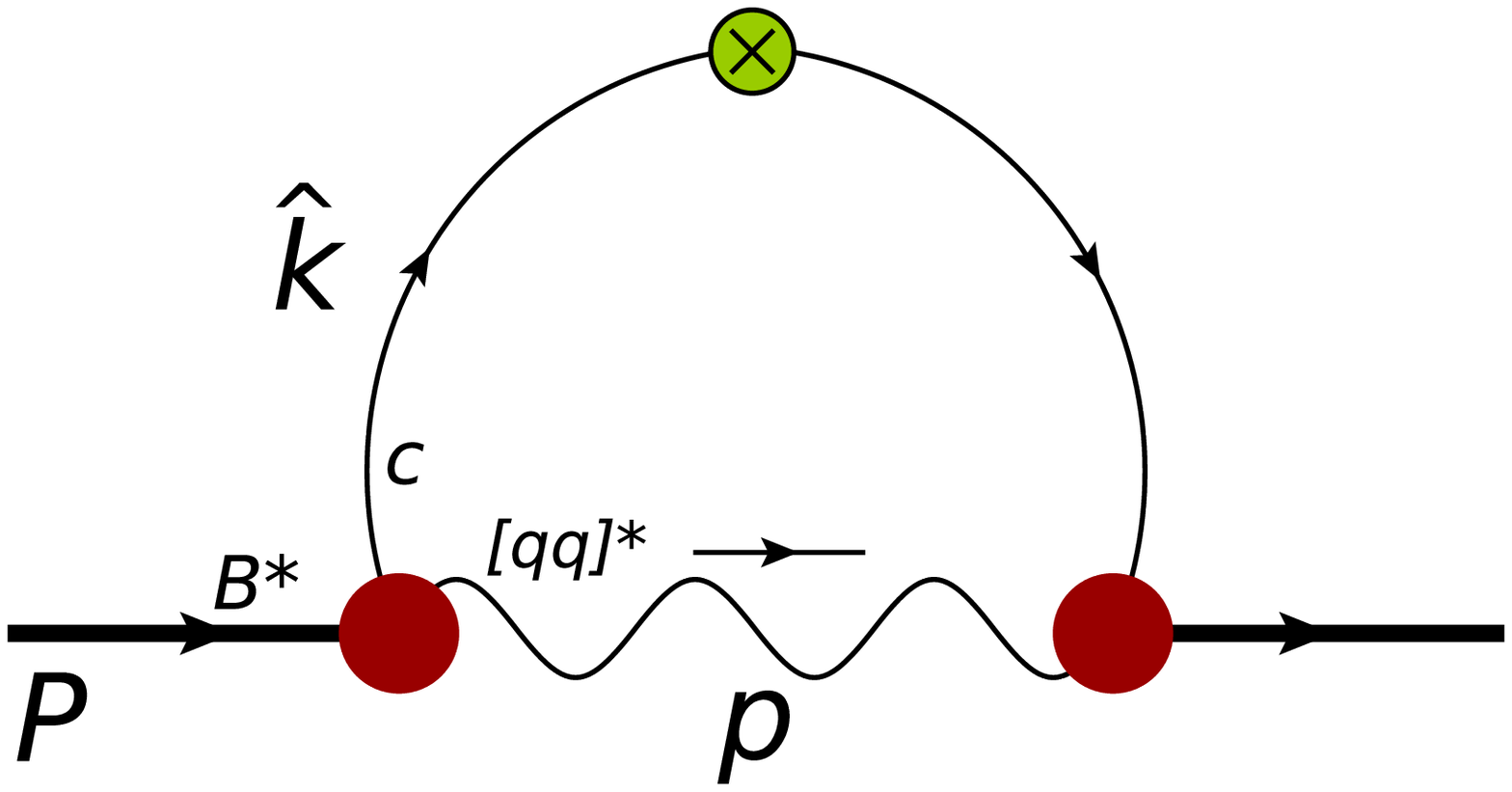}
\caption{
Diagrams for anticharm quark distributions inside baryons.
}
\label{fig:c_Bary}
\end{figure}
%

\subsection{Hadronic Probability Distributions}
\label{sec:mb}

Thus, for a given meson--baryon state $MB$, the splitting function can
be evaluated as in Eq.~(\ref{eq:fphi}) in terms of an integral over
the transverse momentum of the square of the probability amplitude
$\phi_{BM}(y,k^2_\perp)$ which was defined in Eq.~(\ref{eq:Fock}).
The desired probability amplitudes can then be taken from time-ordered
perturbation theory as we have just done at the start of Sec.~\ref{sec:amp}.
\begin{figure}[t]
\vspace*{1.4cm}
\includegraphics[height=9cm]{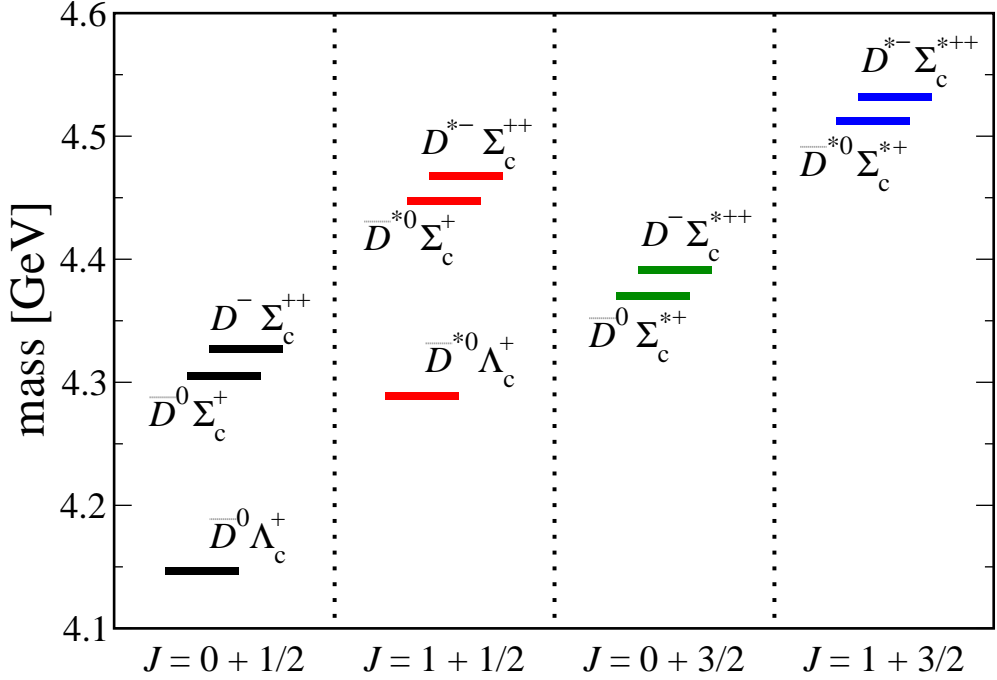}
\caption{
	Spin configurations ($J=$ meson + baryon spin) and masses
	for the spectrum of charmed hadron configurations included
	in the present MBM calculation.}
\label{fig:Spectrum}
\end{figure}
The full model is generated from an incoherent sum over the various
states detailed in Table~\ref{table:mass_spect} and illustrated in
Fig.~\ref{fig:Spectrum} (with the exception of the $p\, J/\psi$ state).
In particular, this requires all the hadronic splitting functions derived above
for $p \to MB$ fluctuations in the spin transitions
\begin{equation}
\bm{\frac{1}{2}} \ \ \ \rightarrow \ \ \
	 \bm{0} \bigoplus \bm{\frac{1}{2}}, \ \ \
	\bm{1} \bigoplus \bm{\frac{1}{2}}, \ \ \
	\bm{0} \bigoplus \bm{\frac{3}{2}}, \ \ \
        \bm{1} \bigoplus \bm{\frac{3}{2}}\ .
\end{equation}
We take numerical values for the couplings to each of these states from the lowest-order
effective hadronic lagrangian for each transition as we describe momentarily, after
a brief discussion of the phenomenology of the vertex calculation.

\vspace*{0.15cm}
{\it Vertex regularization.}
\vspace*{0.15cm}

At large transverse momenta, the invariant mass goes like $s \sim k_\perp^2$,
making integrals of the type given in Eq.~(\ref{eq:DLsplit}) logarithmically divergent.
A simple way to regulate these divergences is through wave function suppression factors $F(s)$,
which act to dampen the ultraviolet contributions.
If we make the reasonable assumption that the quark model symmetries used to fix the hadronic
coupling constants $g$ (discussed below) are not strongly broken, the scheme used to regulate
these transverse divergences is is in fact the primary source of model dependence.
The ambiguity as to which choice of functional form, etc., is most appropriate has led to a
diverse collection of approaches, many of which possess unique numerical and phenomenological
advantages.

Throughout the present analysis, we use an especially convenient parametrization in the form of
an exponential function of $s$,
\begin{equation}
F(s) = \exp[-(s-M^2)/\Lambda^2]\ ,
\label{eq:FF_exp}
\end{equation}
which has the merit of possessing simple normalization properties
on-shell, although multipole, or power-law, functional forms as
in Eq.~(\ref{eq:powerFF}) would also suffice.

Moreover, as we catalogued in Sec.~\ref{sec:amp}, both the hadronic and quark-level amplitudes
are evaluated in an IMF/light front framework; at the same time, in an explicitly covariant
formulation, $t$-channel parametrizations of the form factor may possess more general consistency.
Such a prescription specifies the splitting functions directly in terms of the Mandelstam variable
$t = -k^2$ (here, the exchanged mass of the virtual meson), leading to, \EG
\begin{equation}
f_{M B}(y) = {c_I \, g^2_{MBN} \over 16 \pi^2} \,\, y \int_{t_{min}}^\infty dt
\left( t + [M_B - M]^2 \over (t + m_M^2)^2 \right) \,\, F^2(t)\ ,
\label{eq:cov-piN}
\end{equation}
for the simplest pseudoscalar process $p \rightarrow MB$, corresponding to Eq.~(\ref{eq:DLsplit}).
Typical choices for vertex suppression then usually involve multipole forms such as
\begin{equation}
F(t) = \left( \Lambda^2 + m_M^2 \over \Lambda^2 - t \right)^n\ , \,\,\,\, n \in \{1,\, 2\}\ .
\label{eq:cov-Ft}
\end{equation}
Despite the frame-independence conferred by the covariant approach, several shortfalls of this method
prevent us from using it extensively here; perhaps the greatest of these is the explicit lack of symmetry
between scattering and recoil processes typified by the $y \rightarrow \bar{y} \equiv 1-y$ reciprocity relation
of Eq.~(\ref{eq:fphi}), which prevails in the IMF framework. As we have seen, this property in the IMF enforces
$3$-momentum conservation in interactions determined by the effective $\mathcal{L}_{MBN}$ of
Sec.~\ref{sec:amp}. It can be shown, however, that $f_{M B}(y) \neq f_{B M}(\bar{y})$ in amplitudes like those of
Eqs.~(\ref{eq:cov-piN} \& \ref{eq:cov-Ft}) --- a fact which implies that covariant schemes generally fail to uphold
explicit momentum conservation at hadronic vertices.

As mentioned, the splitting functions also depend upon the magnitude of the coupling
$g$ to each meson--baryon state, which we take from baryon--baryon
scattering models extended to the charm sector.  For simplicity,
we assume a universal exponential form factor for all couplings,
where the resulting scale factor $\Lambda$ is varied
to fit charmed baryon production in hadronic interactions,
as discussed below.
(Note that the exponential expression used to define $F(s)$ in
Eq.~(\ref{eq:FF_exp}) differs by the square of the form factor
appearing in Eq.~(\ref{eq:expFF}); the two can of course be related
by a simple rescaling of the cutoff mass $\Lambda^2 \to 2 \Lambda^2$.)


\vspace*{0.15cm}
{\it Constraints from inclusive charmed hadron production.}
\vspace*{0.15cm}

To calculate the contributions of the various charmed mesons and baryons
listed in Table~\ref{table:mass_spect} requires the couplings of these
states to the proton.  In this analysis we take the coupling constants
from boson-exchange models that were originally applied to pion-nucleon
interactions \cite{Machleidt87}, and later generalized to $KN$
scattering \cite{HDHPS95}.  In the extension to the strange sector,
the relevant couplings are taken from non-strange analyses with SU(3)
arguments used to incorporate the corresponding strange particles.
The off-shell behavior of the amplitudes is typically regulated by
a multipole form factor of the type
\begin{equation}
F(t) = \left( \frac{\Lambda^2 + m_M^2}{\Lambda^2 - t} \right)^n\ ,
\label{eq:t-chan_FF}
\end{equation}
where $t$ is the usual Mandelstam variable for the squared momentum
transfer of the exchanged meson with mass $m_M$.  A monopole form factor
($n=1$) is most often used for low-spin states, while for higher-spin
states a dipole form factor ($n=2$) is typically employed to damp the
higher powers of momentum that enter into the transition amplitudes.

The extension of meson--baryon couplings from the non-strange to the
strange sector has generally been quite successful phenomenologically.
Continuing this program further to the charm sector, Haidenbauer
{\it et al.} \cite{Hai07, Hai08, Hai11} used SU(4) symmetry arguments
to describe exclusive charmed hadron production in $\bar{D}N$ and $DN$
scattering within a one-boson-exchange framework.
We fix the couplings for the spin-1/2 charmed baryons $\Lambda_c$ and
$\Sigma_c$ to those found in Ref.~\cite{Hai11}, as summarized in
Table~\ref{table:couplings}.  For the couplings to spin-3/2 states
$\Sigma_c^*$, we take the couplings from those obtained for the
analogous strange states by Holzenkamp \EA~\cite{Holzenkamp89}.
The signs of the couplings are related to the value of the $\pi NN$
coupling, for which we use $g_{\pi NN}/\sqrt{4\pi} = -3.795$.

\begin{table}[t]
\caption{Charm-sector coupling constants, deduced from $DN$
	and $\bar D N$ scattering analyses \cite{Hai11} for
	the spin-1/2 charmed baryons $\Lambda_c$ and $\Sigma_c$,
	and by extending the SU(3) sector analysis of
	Ref.~\cite{Holzenkamp89} for the spin-3/2 $\Sigma_c^*$
	states.}
\centering
\begin{tabular}{l c c}				\hline\hline
Vertex 				& $g_{MBN}/\sqrt{4\pi}$\ \ \ \ \
				& $f_{MBN}/\sqrt{4\pi}$	\\
[0.5ex] \hline
$p \to \bar{D}^0 \Lambda_c^+$ 	& 3.943    & ---      \\
$p \to \bar{D}^{*0}\Lambda_c^+$	& 1.590    & 5.183    \\
$p \to \bar{D}^0 \Sigma_c^+,\
	D^- \Sigma_c^{++}$ 	& 0.759    & ---      \\
$p \to \bar{D}^{*0}\Sigma_c^+,\
	D^{*-} \Sigma_c^{++}$ 	& 0.918    & $-2.222$ \\
$p \to \bar{D}^0 \Sigma_c^{*+},\
	D^- \Sigma_c^{*++}$	& $-0.193$ & ---      \\
$p \to \bar{D}^{*0}\Sigma_c^{*+},\
	D^{*-} \Sigma_c^{*++}$	& $-1.846$ & ---      \\
[1ex] \hline
\end{tabular}
\label{table:couplings}
\end{table}

The remaining parameters of the model are the form factor cutoffs,
$\Lambda$, which could in principle be constrained for the various
meson--baryon vertices by data from exclusive or inclusive charmed
baryon production.  In practice, such data are rather limited,
and the most direct constraints come from inclusive $\Lambda_c$
production in proton--proton scattering, $p p \to \Lambda_c X$,
measured by the R680 Collaboration at the ISR \cite{Chauvat:1987kb}.
Since it is currently not possible to constrain the cutoffs for
the individual charmed meson--baryon configurations, we assume a
universal exponential form factor cutoff as in Eq.~(\ref{eq:FF_exp})
for all the fluctuations listed in Table~\ref{table:couplings},
and tune $\Lambda$ to best reproduce the shape and normalization
of the inclusive $\Lambda_c$ production cross section data.
This will place an upper bound on $\Lambda$ and the magnitude
of the charmed meson--baryon contribution in the MBM.

Within the same one-boson exchange framework as adopted in the $DN$
and $\bar{D} N$ scattering analyses \cite{Hai07, Hai08, Hai11},
the contribution from charmed meson exchange to the differential
cross section for inclusive baryon production in $pp$ scattering
can be written \cite{Holtmann96}
\begin{equation}
E \frac{d^3\sigma}{d^3\bm{p}}\
=\ \frac{\bar{y}}{\pi}
   \frac{d^2\sigma}{d\bar{y}\, dk_\perp^2}\
=\ \frac{\bar{y}}{\pi} \sum_M
   \left| \phi_{BM}(\bar{y},k_\perp^2) \right|^2\,
   \sigma_{\rm tot}^{Mp}(sy)\ ,
\label{eq:diff3_CS}
\end{equation}
where $E$ is the energy of the proton beam, and the sum over
$M$ includes incoherent contributions from processes involving
the exchange of meson $M$ leading to a final baryon $B$.
The total meson--proton cross section $\sigma_{\rm tot}^{Mp}$
here is evaluated at the meson energy $s y$, with $s$ being
the total $pp$ invariant mass squared.
For the case of $\Lambda_c^+$ production, the sum is
restricted to the $\bar{D}^0$ and $\bar{D}^{*0}$ mesons.
The $k_\perp^2$-integrated cross section for $\Lambda_c^+$
production is then given by
\begin{equation}
\frac{d\sigma}{d\bar{y}}
= \sum_{M=D,D^*}
  f_{\Lambda_c^+ M}(\bar{y})\,
  \sigma_{\rm tot}^{Mp}(sy)\ .
\label{eq:diff_CS}
\end{equation}
Note that in Ref.~\cite{Cazaroto13} this cross section is defined
with an additional factor $(\pi/\bar y)$ on the right hand side.
For the total charmed meson--proton cross section
$\sigma_{\rm tot}^{Mp}$ we take a constant value, as suggested
by the analysis of pion-nucleon scattering \cite{Holtmann96},
where $\sigma_{\rm tot}^{\pi p} \approx \sigma_{\rm tot}^{\rho p}$.
Adopting a similar approach to the strange and charmed meson cross
sections, we have
\begin{equation}
\sigma_{\rm tot}^{Dp}\
\approx\ \sigma_{\rm tot}^{D^*p}\
\approx\ \sigma_{\rm tot}^{\bar{K}p}\
\approx\ (20 \pm 10) \ \mathrm{mb}\ ,
\label{eq:ISR_Mp}
\end{equation}
where the value of the $\bar{K}p$ total cross section is taken
from Ref.~\cite{Beringer:1900zz}, and we assign a conservative
50\% uncertainty on the central value.

\begin{figure}[t]
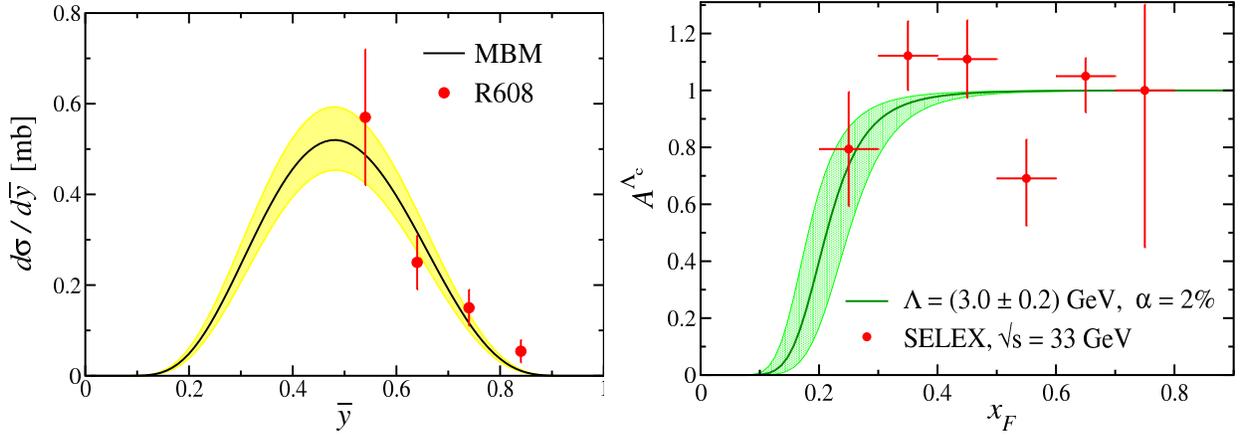

\vspace*{1cm}
\includegraphics[width=8.0cm]{IC-Fig/Fig-2a.eps}
\includegraphics[width=8.2cm]{IC-Fig/Fig-2b.eps}
\caption{
	(Left)
	Differential cross section $d\sigma/d\bar y$ for the inclusive
	charm production reaction $pp \to \Lambda_c^+ X$ as a function
	of the momentum fraction $\bar y$ carried by the $\Lambda_c^+$.
	The MBM cross section (solid) is computed using the central value
	for the $Dp$ cross section $\sigma^{Dp}_{\rm tot} = 20$~mb,
	and the resulting error band (shaded) represents the
	purely statistical uncertainty.  The data (red circles)	are
	from the R608 collaboration at the ISR \cite{Chauvat:1987kb}.
	(Right)
	Charge asymmetry $A^{\Lambda_c}$ for
	$\Lambda_c^+/\bar\Lambda_c^-$ production in the MBM (solid),
	using the $\xF$ dependence of the $\bar\Lambda_c$ cross section
	in Eq.~(\ref{eq:SELEX-iii}), compared with data from the
	SELEX Collaboration \cite{Garcia:2001xj}.}
\label{fig:R608_CS}
\end{figure}

Using Eqs.~(\ref{eq:diff_CS}) and (\ref{eq:ISR_Mp}), the calculated
cross section is shown in the left-hand panel of Fig.~\ref{fig:R608_CS}
at the kinematics of the $\Lambda_c^+$ production data from the R608 Collaboration
\cite{Chauvat:1987kb} at the ISR.  The kinematical coverage of the
ISR data was restricted to $k_\perp \le 1.1$~GeV, which we impose
in the computed cross section.  For the central value of the $Dp$
total cross section, $\sigma_{\rm tot}^{Dp} = 20$~mb, the best fit
value of the cutoff parameter is found to be
	$\Lambda = (2.89 \pm 0.04)$~GeV,
which gives a good fit to both the overall normalization and the
shape of the inclusive $\Lambda_c^+$ data.  Including the uncertainty
in the $Dp$ cross section from Eq.~(\ref{eq:ISR_Mp}), the cutoff becomes
	$\Lambda = (3.0 \pm 0.2)$~GeV.

In calculating the inclusive $\Lambda_c^+$ production
cross section we have included the possibility that higher-mass
baryons such as $\Lambda_c^*$ and $\Sigma_c^*$ are produced
and subsequently decay to a $\Lambda_c^+$, using the relevant
branching ratios for the decays.  With the coupling constants
from Table~\ref{table:couplings}, we find that the dominant
contribution to inclusive $\Lambda_c$ production arises from
the state $\bar{D}^{*0} \Lambda_c^+$.  This is in contrast to
earlier analyses, where the largest contribution was assumed
to be from the lowest-energy $\bar{D}^0 \Lambda_c^+$ state.
As we discuss below, this will have significant ramifications
for intrinsic charm production in electromagnetic reactions.

We must point out that the original data from Ref.~\cite{Chauvat:1987kb}
were recorded in terms of the variable
    \mbox{$\xF = 2 p_{\Lambda}^0 / \sqrt{s}$},
which in general differs from the momentum fraction $\bar y$
that scales the calculations of the MBM.
It can be shown, however, that at high energies
(\IE~$s \gg M_B^2$, $k_\perp^2$), one has $\bar y \to \xF$, which
we used to create the abscissa of the left panel in Fig.~\ref{fig:R608_CS}.

Somewhat more recently the SELEX Collaboration at Fermilab \cite{Garcia:2001xj}
produced data on the charge asymmetry for inclusive $\Lambda_c^+$
and $\bar\Lambda_c^-$ production in the scattering of 540~GeV protons
from copper and carbon targets,
\be
A^{\Lambda_c}(\xF)
= { \sigma^{\Lambda_c}(x_F) - \sigma^{\bar{\Lambda}_c}(x_F)
    \over
    \sigma^{\Lambda_c}(x_F) + \sigma^{\bar{\Lambda}_c}(x_F) }\ ,
\label{eq:SELEX-i}
\ee
where $\sigma^{\Lambda_c}(x_F) \equiv d\sigma^{\Lambda_c}/d\xF$.
While the contribution to the production of $\Lambda_c^+$ can be
calculated in the MBM, the computation of the asymmetry $A^{\Lambda_c}$
requires in addition an estimate of the $\bar{\Lambda}_c$ cross
section.  Following Ref.~\cite{Garcia:2001xj} we approximate this
using a simple monomial parametrization.
Furthermore, we assume that the $\Lambda_c^+$ cross sections can be
written as the sum of valence and sea components, with the generation
of the former described by the nonperturbative MBM and dominating at
intermediate and high values of $\xF$, and the latter concentrated
at small $\xF$,
\be
\frac{d\sigma^{\Lambda_c}}{d\xF}
= \frac{d\sigma^{\Lambda_c}_{\rm (val)}}{d\xF}
+ \frac{d\sigma^{\Lambda_c}_{\rm (sea)}}{d\xF}\ ,
\label{eq:SELEX-ii}
\ee
where
\begin{subequations}
\label{eq:SELEX-iii}
\begin{eqnarray}
\frac{d\sigma^{\Lambda_c}_{\rm (val)}}{d\xF}
&\approx& \sigma_0 \sum_M f_{\Lambda_c M}(\xF)\ ,
\label{eq:SELEX-iiia}					\\
\frac{d\sigma^{\Lambda_c}_{\rm (sea)}}{d\xF}
&\equiv& \frac{d\sigma^{\bar{\Lambda}_c}}{d\xF}\
 \approx\ \bar\sigma_0 (1-\xF)^{\bar{n}}\ .
\label{eq:SELEX-iiib}
\end{eqnarray}%
\end{subequations}%
In Eq.~(\ref{eq:SELEX-iiia}) the factor $\sigma_0$ corresponds to
the total meson--proton cross section in Eq.~(\ref{eq:diff_CS}),
which we take to be independent of the flavor and spin of the
meson, as in Eq.~(\ref{eq:ISR_Mp}), while $\bar\sigma_0$ is a
normalization parameter for the corresponding $\bar{\Lambda}_c$
production cross section.
Using Eqs.~(\ref{eq:SELEX-iii}), the asymmetry in Eq.~(\ref{eq:SELEX-i})
can then be written
\be
A_{\Lambda_c}(\xF)
= \frac{ \sum_M f_{\Lambda_c M}(\xF)}
       { \sum_M f_{\Lambda_c M}(\xF)
	 + 2 \alpha (1-\xF)^{\bar{n}} }\ ,
\label{eq:SELEX-iv}
\ee
where
    $\alpha = \bar\sigma_0 / \sigma_0$
is the ratio of the sea to valence contributions to the
$\Lambda_c$ cross sections.
For $\bar{\Lambda}_c$ production induced by $\Sigma^-$
beams, the SELEX Collaboration found for the exponent
    $\bar{n} \approx 6.8$,
which we assume also for the $x_F$ dependence of the proton
induced cross section in Eq.~(\ref{eq:SELEX-iiib}).
Using the MBM cutoff parameter $\Lambda = (3.0 \pm 0.2)$~GeV,
a good fit to the SELEX charge asymmetry data can then be
obtained with $\alpha \approx 2.0\%$, as displayed in the
the right panel of Fig.~\ref{fig:R608_CS}.
We should note, however, that $\Lambda_c$ charge asymmetry
data are rather sensitive to the form of the $\bar{\Lambda}_c$
cross section, so that agreement with the SELEX data should
not be considered as a stringent test of the MBM; rather,
with an appropriate choice of parameter $\alpha$ the model
is able to accommodate the empirical results. We therefore
regard this reasonable description of SELEX data as a rough
consistency argument on behalf of the MBM we have constructed,
given the plausible behavior assumed for the sea contribution
to $\Lambda_c$ production in Eq.~(\ref{eq:SELEX-iiib}).

Having thus constrained the scale parameter for the meson--baryon
form factor by the inclusive $\Lambda_c$ production data,
and with the coupling constants for the various meson--baryon
states given in Table~\ref{table:couplings}, we are now able to
compute the meson--baryon splitting functions in Eq.~(\ref{eq:fphi}),
which we consider in the following section.

\vspace*{0.15cm}
{\it Phenomenology of charmed meson--baryon splitting functions.}
\vspace*{0.15cm}

The complete set of the four basic splitting functions representing
the dissociation of a proton to charmed meson--baryon states
$p \to D B$ (pseudoscalar meson + octet baryon),
    $D^* B$ (vector meson + octet baryon),
    $D B^*$ (pseudoscalar meson + decuplet baryon) and
  $D^* B^*$ (vector meson + decuplet baryon)
is illustrated in Fig.~\ref{fig:f_MB}.  The functions are shown
for the neutral $\bar{D}^0$ and $\bar{D}^{*0}$ mesons, and all
the other charge states in Table~\ref{table:couplings} can be
obtained using appropriate Clebsch-Gordan coefficients.
For the best fit value of the universal cutoff parameter
$\Lambda = 3$~GeV from the inclusive $\Lambda_c^+$ production
data, the $\bar{D}^{*0} \Lambda_c^+$ contribution is found to be
dominant --- an order of magnitude larger than the corresponding
$\bar{D}^0 \Lambda_c^+$ and $\bar{D}^{*0} \Sigma_c^{*+}$ contributions.
The $\bar{D}^0 \Sigma_c^{*+}$ contribution is two orders of magnitude
smaller still, and effectively plays no role in the phenomenology.
To a good approximation, therefore, one can represent the total charm
distribution in the proton by the single $\bar{D}^{*0} \Lambda_c^+$ state.

\begin{figure}[t]
\vspace*{1.4cm}
\includegraphics[height=9cm]{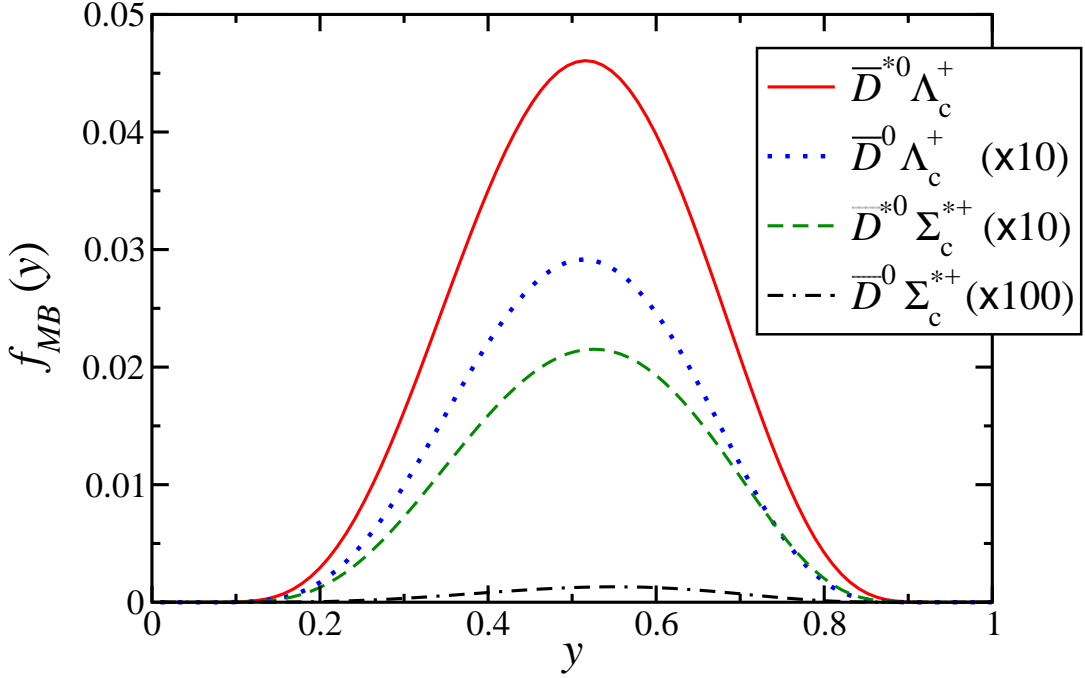}
\caption{
	Splitting functions for the four basic dissociations
	of a proton into charmed meson--baryon states, for the
	spin-1 meson + spin-1/2 baryon state
	$\bar{D}^{*0} \Lambda_c^+$
	  (red solid),
	spin-0 meson + spin-1/2 baryon state
	$\bar{D}^0 \Lambda_c^+$
	  (scaled $\times 10$, blue dotted),
	spin-1 meson + spin-3/2 baryon state
	$\bar{D}^{*0} \Sigma_c^{*+}$
	  (scaled $\times 10$, green dashed), and
	spin-0 meson + spin-3/2 baryon state
	$\bar{D}^0 \Sigma_c^{*+}$
	  (scaled $\times 100$, black dot-dashed).
	A universal exponential cutoff mass $\Lambda = 3$~GeV is
	used with the couplings from Table~\ref{table:couplings}.}
\label{fig:f_MB}
\end{figure}

Again as indicated in Fig.~\ref{fig:f_MB}, the shapes of the various charmed meson--baryon
distributions $f_{MB}(y)$ are interestingly rather similar, peaking just above
$y = 1/2$.  This is in contrast to the distributions in the light
flavor sector, where the corresponding $\pi N$ splitting function
is considerably more skewed in~$y$~\cite{MT93, Holtmann96}.
The skewedness arises from the large difference in mass between
the pion and nucleon in the dissociation, whereas the masses of
both the charmed meson and baryon are of the order $\sim 2$~GeV.
This is also one reason why the lowest mass $\pi N$ configuration
is the dominant one in the pion sector (a fact to which we will return in
Chap.~\ref{chap:ch-TDIS}), unlike the lowest mass
charmed state $\bar{D}^0 \Lambda_c^+$, which as Fig.~\ref{fig:f_MB}
indicates gives a significantly smaller contribution than the
$\bar{D}^{*0} \Lambda_c^+$.
The dominance of the SU(2) flavor sector by the $\pi N$ state is
ensured by the relatively large energy gap between higher mass
configurations involving $\rho$ mesons or $\Delta$ baryons,
whereas no significant energy gap exists between the various states
in the charm sector.

To explore further the origin of the dominance of the
$\bar{D}^{*0} \Lambda_c^+$ contribution, we note the relatively
strong coupling to the vector meson state, particularly for the
tensor coupling term, as seen in Table~\ref{table:couplings},
where the tensor to vector coupling ratio is
    $f_{D^* \Lambda_c N} / g_{D^* \Lambda_c N} = 3.26$
\cite{Hai07, Hai08, Hai11}.
This is analogous to the large tensor coupling for the $\rho$
meson in one-boson exchange models of the $NN$ interaction
\cite{Machleidt87}, where in the Bonn-J{\"u}lich model,
for instance, one has an even larger tensor/vector ratio,
    $f_{\rho NN} / g_{\rho NN} = 6.1$
\cite{Hohler75}.
%
Using the charm couplings from Table~\ref{table:couplings},
the tensor contribution clearly dominates over the vector term.
This feature is preserved even if one uses the SU(2) couplings from
the $\rho$ exchange in the $NN$ analysis instead of the SU(4) couplings
\cite{Hai07, Hai08, Hai11} (but with the same charm hadron masses).
In particular, since the SU(4) vector coupling
    $g_{D^* \Lambda_c N}^2/4\pi = 2.53$
is around 5 times larger than that found for the $\rho$ from
$NN$ analyses,
    $g_{\rho NN}^2/4\pi = 0.55$
\cite{Hohler75}, the vector contribution to the charm splitting
function is significantly larger than for the SU(2) coupling case.
This is compensated somewhat by the $\sim 2$ times smaller SU(4)
tensor/vector ratio, making the total contribution to the charmed
vector meson splitting function $f_{\bar{D}^{*0} \Lambda_c^+}(y)$
similar.


At the effective lagrangian level, the large tensor contribution
is associated with the additional momentum dependence induced
by the derivative coupling in the tensor interaction, which is
a general feature of couplings to states with higher spin
[see Eq.~(\ref{eq:L-S1})].  This additional momentum dependence
can have a significant impact on the relative importance of various
charmed meson--baryon transitions, as is evident from the form of
the splitting function in Eq.~(\ref{eq:spin1-fMB}).
The effect of the momentum dependence of the meson--baryon vertices
on the splitting functions can be illustrated even more dramatically
by considering the normalizations
    $\langle n \rangle_{MB} = \int dy\, f_{MB}(y)$
as a function of the cutoff $\Lambda$.
These are displayed in the left panel of Fig.~\ref{fig:f_MB_int} for the four
charmed states shown in Fig.~\ref{fig:f_MB}, together with the
sum over all contributions.
At the best fit value of $\Lambda \sim 3$~GeV, the lowest mass
vector state $\bar{D}^{*0} \Lambda_c^+$ makes up around 70\%
of the total charm normalization of
    $\langle n \rangle_{MB}^{\rm (charm)} = 2.40\%$.
Including the uncertainty on the cutoff (indicated by the shaded
band), the total charm normalization ranges from $\approx 1.04\%$
to $\approx 4.87\%$.

\begin{figure}[t]
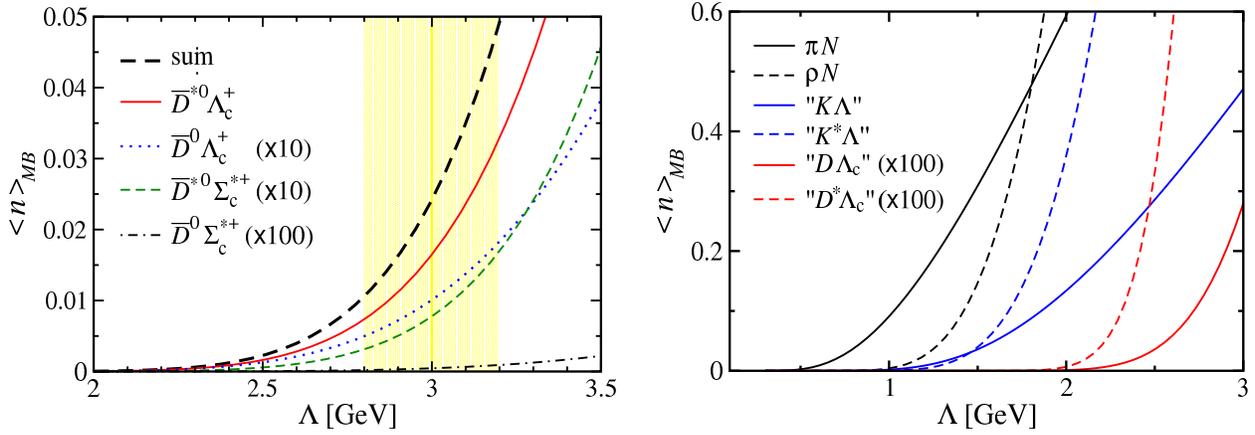

\vspace*{1cm}
\includegraphics[width=8.1cm]{IC-Fig/Fig-5a.eps} \ \ \
\includegraphics[width=7.9cm]{IC-Fig/Fig-5b.eps}
\caption{
	(Left)
	Normalizations $\langle n \rangle_{MB}$ of the charmed
	meson--baryon splitting functions as a function	of the
	form factor cutoff $\Lambda$, for the states
	$\bar{D}^{*0} \Lambda_c^+$
	  (red solid),
        $\bar{D}^0 \Lambda_c^+$
	  (scaled $\times 10$, blue dotted),
        $\bar{D}^{*0} \Sigma_c^{*+}$
	  (scaled $\times 10$, green dashed), and
	$\bar{D}^0 \Sigma_c^{*+}$
	  (scaled $\times 100$, black dot-dashed),
	as well as the sum of all contributions (black dashed).
	The (yellow) shaded band represents the uncertainty on the cutoff
	obtained from fits to inclusive $\Lambda_c^+$ production data.
	(Right)
	Normalizations of the splitting functions to pseudoscalar
	(solid) and vector (dashed) mesons computed with
	SU(2) sector ($\pi N$ and $\rho N$) masses (black),
	SU(3) masses, denoted by ``$K \Lambda$'' and
	``$K^* \Lambda$'' (blue), and
	SU(4) masses, denoted by ``$D \Lambda_c$'' and
	``$D^* \Lambda_c$'' (scaled $\times 100$, red),
	all for the same SU(2) couplings.}
\label{fig:f_MB_int}
\end{figure}

While the variation of the charm splitting functions with the choice
of SU(4) or SU(2) couplings is in reality relatively mild, a more significant
effect is seen for the dependence of the splitting functions on the hadron masses.
In Fig.~\ref{fig:f_MB_int}(b) the normalizations
$\langle n \rangle_{MB}$ of splitting functions to pseudoscalar
and vector mesons are illustrated for the light-quark, strange
and charmed sectors, using the SU(2) coupling constants for
$\pi NN$ and $\rho NN$ listed above.
The curves labeled ``$K \Lambda$'' and ``$K^* \Lambda$'' are
obtained from the $\pi N$ and $\rho N$ splitting functions by
replacing the pion and recoil baryon masses with the appropriate
kaon and hyperon masses, and those labeled ``$D \Lambda_c$''
and those labeled ``$D^* \Lambda_c$'' are obtained by using the
corresponding charmed meson and baryon masses.
For small values of the cutoff, the normalizations of the
pseudoscalar meson--baryon states is larger than for the vector
meson states, but with increasing $\Lambda$ the contributions
from the vector meson states eventually dominate.
With increasing hadron masses the cross-over point between
the pseudoscalar and vector meson states occurs at progressively
smaller $\Lambda$ values.  Neglecting differences between the
coupling constants (which are small if quark model symmetries
are assumed), the size of hadronic masses relative to the cutoff
scale $\Lambda$ is the main determinant of the balance between
the pseudoscalar and vector states for a given flavor sector.

The best fit value $\Lambda = 3$~GeV for the charmed splitting
functions corresponds to a region where the vector meson term
clearly dominates over the pseudoscalar meson contribution.
Had we found a significantly softer cutoff $\Lambda \sim 1$~GeV,
the pseudoscalar contribution would have dominated, although for
such values the total charm contribution would be negligible.
Since only a single charmed baryon production cross section was
available to constrain the charm splitting functions, only a
single parameter $\Lambda$ could be determined.
The existence of data for various charmed channels, on the other
hand, would allow the cutoffs to be determined for individual
meson--baryon states.  This could in principle lead to hard form
factor cutoffs for some states and soft cutoffs for others,
which would affect the degree to which the charmed vector meson
states dominate the splitting functions.
The results of our MBM calculations imply that the production of
charmed mesons in $pp$ reactions would occur almost entirely through
$D^*$ mesons, with subsequent decays of $D^*$ to $D$ mesons.

\subsection{Models for constituent quarks}
\label{sec:cinc}


\vspace*{0.15cm}
{\it Anticharm in charmed mesons.}
\vspace*{0.15cm}

For point particles, the ultraviolet behavior of the $\hat{k}_\perp^2$
integration would be logarithmically divergent for the $\bar{c}_D(z)$
distribution in Eq.~(\ref{eq:cbar_Dbar}), just as we discussed for the
hadronic probability distributions.
As before, this divergence may be regulated by defining the vertex function
$G(\hat{s})$ to suppress contributions from large parton momenta.
Following Sec.~\ref{sec:mb}, we might use, for example, an exponential
functional dependence on $\hat{s}$,
\bea
G(\hat{s}) &=& \exp\left[-(\hat{s}-m_D^2)/\hat{\Lambda}^2\right]\ ,
\label{eq:G_exp}
\eea
with $\hat{\Lambda}$ serving the role of a corresponding momentum
cutoff on the partonic quark-antiquark system.

At low momenta, on the other hand, a mass singularity can arise
in the energy denominator $(\hat{s} - m_D^2)^{-2}$ in the
infrared limit ($\hat{k}_\perp^2 \to 0$) for physical quark masses
$m_q$ and $m_{\bar c}$.  A simple solution adopted by Pumplin
\cite{Pum05} was to assume an artificially large effective
mass for the anticharm quark, $m_{\bar c}^{\rm eff}$, and a large
constituent quark mass for the spectator $u$ or $d$ quark,
$m_q^{\rm eff}$, such that
\bea
m_{\bar c}^{\rm eff} + m_q^{\rm eff} &>& m_D\ .
\label{eq:mceff}
\eea
In our numerical analysis we fix the effective charm mass to be
$m_{\bar c}^{\rm eff} = 1.75$~GeV and the light constituent quark
mass $m_q^{\rm eff} = M/3 = 0.31$~GeV, similar to that used in
Ref.~\cite{Pum05}, which is sufficient to remove the
propagator singularity.

An alternative method to avoid the pole is to utilize a form factor
that simulates confinement by directly canceling the singular
denominator, similar to that advocated in Ref.~\cite{MST94}.
A form that satisfies this is
\bea
G(\hat{s}) &=& (\hat{s} - m_D^2)\,
	       \exp\left[-(\hat{s}-m_D^2)/\hat{\Lambda}^2\right]\ .
\label{eq:FF_con}
\eea
An attractive feature of this form of the vertex function is that it
permits any values of the quark masses to be used, allowing the partons
to be confined without the need for {\it ad hoc} constraints to avoid
singularities through judicious choice of effective quark masses.

\begin{figure}[t]
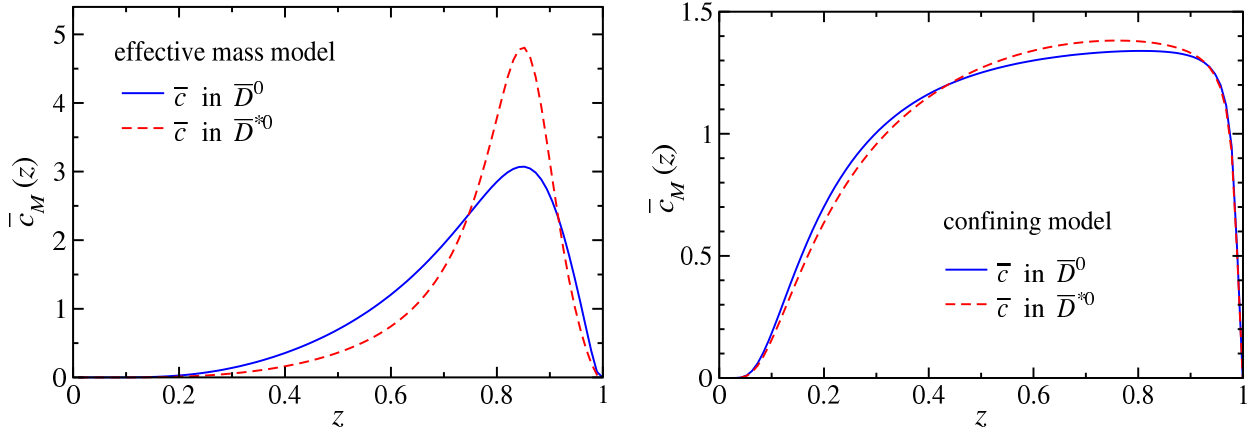

\vspace*{1cm}
\includegraphics[width=8cm]{IC-Fig/Fig-6a.eps} \ \ \
\includegraphics[width=8cm]{IC-Fig/Fig-6b.eps}
\caption{
	Anticharm quark distributions in charmed $D$ (solid)
	and $D^*$ (dashed) mesons, within the
	effective mass model (left), with the vertex form
	factor in Eqs.~(\ref{eq:G_exp}) and (\ref{eq:mceff}),
	and
	confining model (right), with the form factor in
	Eq.~(\ref{eq:FF_con}).}
\label{fig:QDF_mc}
\end{figure}

The results for the $\bar{c}$ distributions in the $D$ and $D^*$ mesons
as computed formally in Eqs.~(\ref{eq:cbar_Dbar} \& \ref{eq:cbar_D-st})
are illustrated in Fig.~\ref{fig:QDF_mc} for both types of vertex
functions $G(\hat{s})$. In the absence of empirical constraints on PDFs
in charmed mesons, the partonic cutoff $\hat\Lambda$ is a free parameter.
However, since for heavy quarks the typical masses of the intermediate
states ($D B$ or $\bar c q$) are comparable, to a first approximation we
can fix $\hat\Lambda$ to the meson--baryon cutoff, $\hat\Lambda=\Lambda$.
In the effective mass model, Eqs.~(\ref{eq:G_exp}) and (\ref{eq:mceff}),
the peak of the anticharm distribution in $z$ reflects the fraction of
the meson mass carried by the $\bar{c}$ quark. For both the $D$ and
$D^*$ mesons, the $\bar{c}$ distribution peaks at $z \sim 0.85$, with
the latter being slightly narrower. The distributions in the confining
model, Eq.~(\ref{eq:FF_con}), also peak at similarly large momentum
fractions, but are significantly broader.

A numerical feature of the effective mass model is the presence of the
energy denominator $\propto (\hat{s} - m_D^2)^{-2}$, which largely
determines the qualitative shapes of the $\bar c$ distributions.
For specific mass choices, $\hat{s} - m_D^2$ is minimized at a unique
value of $z$, resulting in the strongly-peaked shapes observed in
the effective mass model.  In the confining model, on the other hand,
the energy denominator responsible for this $z$ dependence is suppressed
directly such that the resulting distribution shapes no longer possess
pronounced maxima. In the effective charm model, however, the closer
the energy denominator approaches its pole value, the more ``singular''
the behavior at the distribution maximum; as such, if we fix the charm
and spectator masses according to Eq.~(\ref{eq:mceff}), the energy
denominator approaches the zero pole for heavier hadron masses,
producing the more sharply peaked distributions seen in 
Fig.~\ref{fig:QDF_mc}(a).

\vspace*{0.15cm}
{\it Charm in charmed baryons.}
\vspace*{0.15cm}

The calculation of the charm quark distributions in charmed baryons
proceeds in similar fashion to that for the $D$ and $D^*$ mesons,
but is more involved since the spectator system consists of two
(or more) particles.  In practice, however, one can simplify the
calculation by treating the spectator $qq$ system as an effective
``diquark'' with a fixed mass $m_{qq}$.
For spin-1/2 charmed baryons, in general the spectator diquark state
can have either spin~0 or spin~1, with corresponding scalar and
pseudovector vertex functions describing the momentum dependence.
The spin of the spectator diquark can affect the spin and flavor
dependence of the associated parton distribution; for example,
the suppression of the $d/u$ ratio in the proton at large $x$ is
usually attributed to a higher energy of the spin-1 diquark in
the proton compared with the spin-0 diquark \cite{MT97, Close88}.
Since here we are concerned with the total effect on the charm quark
distribution, rather than the flavor dependence, it will be sufficient
to consider only the leading contribution arising from the scalar
spectators, for which we take an effective mass of $m_{qq} = 1$~GeV.

\begin{figure}[t]
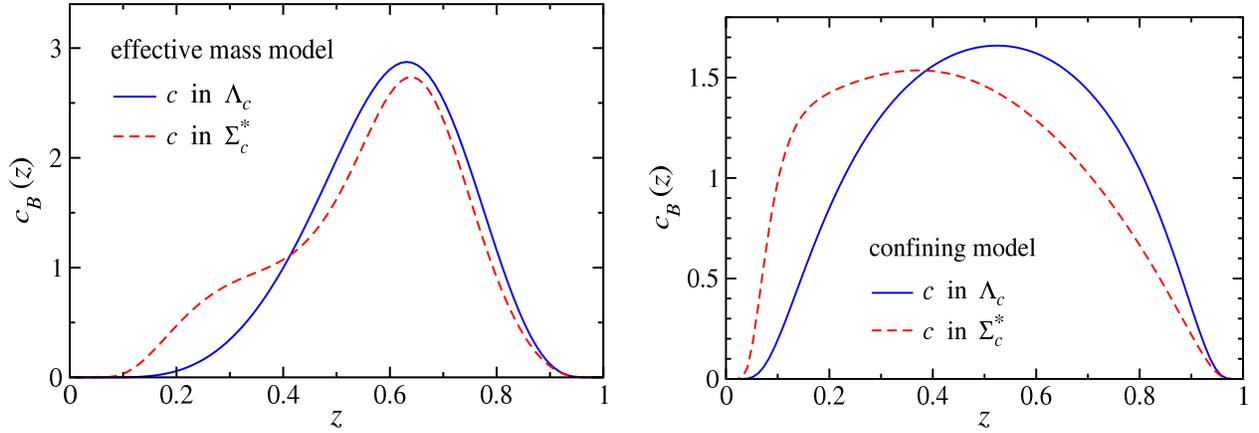

\vspace*{1cm}
\includegraphics[width=8cm]{IC-Fig/Fig-7a.eps} \ \ \
\includegraphics[width=8cm]{IC-Fig/Fig-7b.eps}
\caption{
	Charm distributions in the charmed $\Lambda_c$ (solid)
	and $\Sigma_c^*$ (dashed) baryons, within
	the effective mass model (left), and
	the confining model (right).}
\label{fig:QDF}
\end{figure}

The resulting $c$ quark distributions in the spin-1/2 and
spin-3/2 charmed baryons as generated by our formulas in
Eqs.~(\ref{eq:c_in_Lambda} \& \ref{eq:c_Sigma-st}) are
illustrated by Fig.~\ref{fig:QDF}; where relevant, we use the same numerical
values for the masses and cutoffs as in the $\bar c$ calculation
in the charmed mesons above. Compared with the $\bar d$
distributions in $D$ and $D^*$, the $c$ quark PDFs are peaked
at somewhat smaller values of $z$. In the effective mass model
for the $B$-$c$-$qq$ vertex function, both the $c$ distributions
in $\Lambda_c$ and in $\Sigma_c^*$ are maximal at $z \approx 0.6-0.65$,
with a relatively narrow distribution in $z$. The bulge in the
$c$ distribution in the $\Sigma_c^*$ baryon is associated with
the more complicated spin algebra compared with the $\Lambda_c$.
The $c$ distributions with the confining model vertex function
are once again somewhat broader, peaking at smaller $z$ values,
$z \approx 0.55$ for the $\Lambda_c$ baryon and $z \approx 0.4$ for
the $\Sigma_c^*$, with the latter having a sharp drop off at $z \to 0$.
The broader distributions here are generated by the suppression of the
energy denominator $(\hat{s} - M_B^2)^{-2}$, which is mostly responsible
for the strongly-peaked distributions in the effective mass model.
In all cases the distributions have been normalized to respect the
valence quark number sum rule, as in Eq.~(\ref{eq:cLambdacNorm}).

Having assembled the various ingredients for the calculation
of the convolution expressions in Eqs.~(\ref{eq:mesoncloud}),
in the next section we gather these inputs at last to compute the
$c$ and $\bar c$ distributions in the nucleon.

\section{Numerical results}
\label{sec:results}

Combining the distributions of $c$ and $\bar c$ quarks in the
charmed mesons and baryons discussed in the previous section
with the splitting functions summarized in Sec.~\ref{sec:mb},
we present the resulting $c$ and $\bar c$ distributions in
the physical nucleon. We consider in Sec.~\ref{ssec:c_in_MBM}
contributions to the intrinsic charm PDFs $c(x)$ and $\bar{c}(x)$
from the various meson--baryon configurations in
the MBM, as well as the dependence of the results on the models for the
charm distributions inside the charm hadrons. To better control the
systematic uncertainties within the calculation, we also compare our
results with other prescriptions for intrinsic charm distributions
in Sec.~\ref{ssec:approx}, and conclude the section with comparisons
relative to measurements of the charm structure function $F^c_2$.

\subsection{Intrinsic charm in the MBM}
\label{ssec:c_in_MBM}

The contributions to the charm and anticharm quark distributions
in the nucleon from various meson--baryon states are presented
in Fig.~\ref{fig:PS_mc}, using as benchmark the confining model for the PDFs
in the charmed hadrons (\ref{eq:FF_con}) with a mass parameter
$\Lambda = 3$~GeV.  The contributions correspond to the same
configurations as in Fig.~\ref{fig:f_MB}, namely, the dominant
  $\bar{D}^{*0} \Lambda_c^+$ state, the
  $\bar{D}^0 \Lambda_c^+$ and
  $\bar{D}^{*0} \Sigma_c^{*+}$ states, as well as the (negligible)
  $\bar{D}^0 \Sigma_c^{*+}$ contribution.
As expected from the magnitudes of the splitting functions in
Fig.~\ref{fig:f_MB}, the $\bar{D}^{*} \Lambda_c^+$ state produces
the dominant meson--baryon contribution.

\begin{figure}[t]
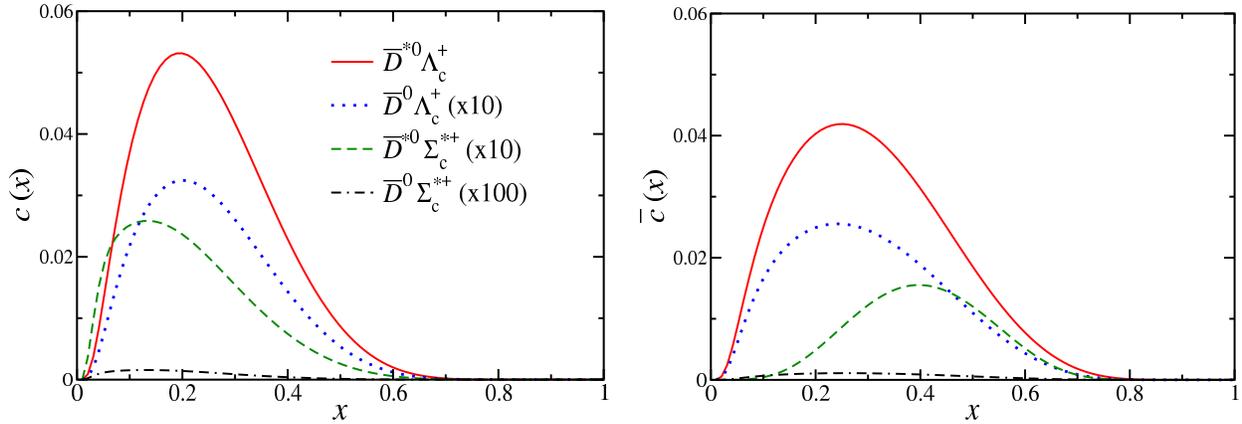

\vspace*{1cm}
\includegraphics[width=8cm]{IC-Fig/Fig-8a.eps}\ \ \
\includegraphics[width=8cm]{IC-Fig/Fig-8b.eps}
\caption{
	Charm (left) and anticharm (right) quark distributions
	in the nucleon in the MBM, with contributions from the
	meson--baryon configurations as in Fig.~\ref{fig:f_MB}:
        $\bar{D}^{*0} \Lambda_c^+$ (red solid),
        $\bar{D}^0 \Lambda_c^+$ (scaled $\times 10$, blue dotted),
        $\bar{D}^{*0} \Sigma_c^{*+}$ (scaled $\times 10$, green dashed),
	and
	$\bar{D}^0\Sigma_c^{*+}$ (scaled $\times 100$, black dot-dashed).}
\label{fig:PS_mc}
\end{figure}

Summing over all the contributions listed in Fig.~\ref{fig:Spectrum},
the total $xc$ and $x\bar{c}$ distributions are shown in the left panel of
Fig.~\ref{fig:c-c+}, at the input scale $Q^2 = m_c^2$ and evolved
to $Q^2=50$~GeV$^2$ (which is typical for charm structure function
measurements). In both cases, the dominant $\bar{D}^{*0} \Lambda_c^+$
contribution consistently accounts for approximately 70\% of the total.
An additional characteristic feature of our MBM evident in the left-hand
side of Fig.~\ref{fig:c-c+} is the relative ``hardness'' of the $x\bar{c}(x)$
distribution over $xc(x)$ --- a property that holds for every meson--baryon
configuration in the MBM, reflecting the fact that the charm quark represents
a larger fraction of the total mass of the meson than of the baryon.
Since the peak in the charm distribution in a hadron is related to the
fraction of the hadron mass carried by the charm quark, the resulting
distribution of $\bar{c}$ in the $\bar{D}$ meson will typically be
harder than that for the $c$ in the $\Lambda_c$.
While it is possible to make the intrinsic $c$ distribution as hard
as the $\bar{c}$ distribution in convolution models, doing so generally
requires rather unnatural parton distributions inside the baryon and meson
states.

\begin{figure}[h]
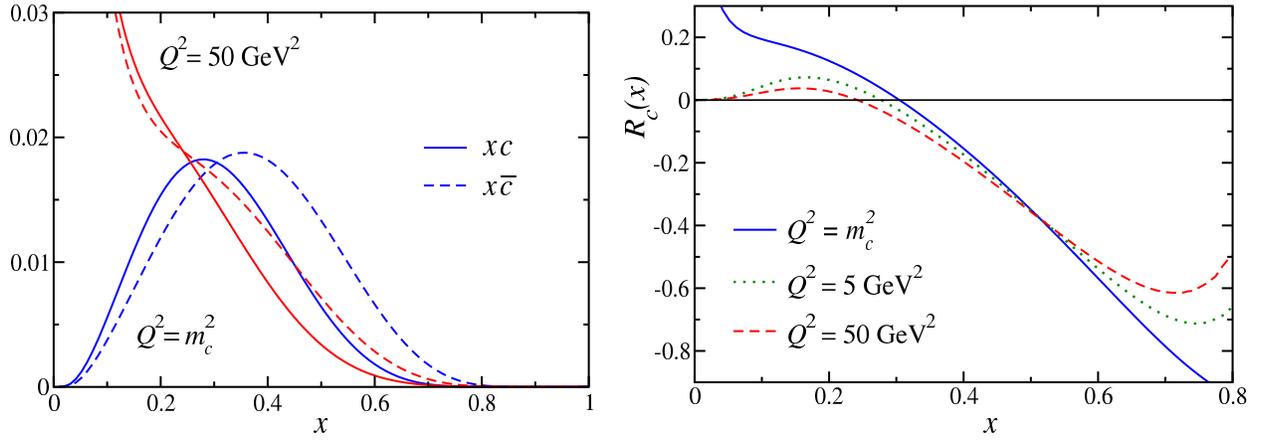

\vspace*{1cm}
\includegraphics[width=7.8cm]{IC-Fig/Fig-9a.eps}\ \ \
\includegraphics[width=8.3cm]{IC-Fig/Fig-9b.eps}
\caption{
	(Left)
	Total $xc$ (solid lines) and $x\bar{c}$ (dashed lines)
	distributions in the MBM with the confining model for
	the PDFs in the charmed hadrons, Eq.~(\ref{eq:FF_con}), at
	$Q^2 = m_c^2$ (blue) and evolved to
	$Q^2 = 50$~GeV$^2$ (red).
	(Right)
	Corresponding charm--anticharm asymmetry
	$R_c(x) = (c(x)-\bar{c}(x))/(c(x)+\bar{c}(x))$ at
	$Q^2 = m_c^2$ (solid),
	$Q^2 = 5$~GeV$^2$ (dotted), and
	$Q^2 = 50$~GeV$^2$ (dashed).}
\label{fig:c-c+}
\end{figure}

To quantify the magnitude of the nonperturbative charm in the nucleon,
we can compute the total proton momentum carried by charm and anticharm
quarks,
\bea
P_c &=& C^{(1)} + \overline{C}^{(1)}\ ,
\label{eq:cmom_tot}
\eea
where the moments $C^{(1)}$ and $\overline{C}^{(1)}$ are defined
in Eqs.~(\ref{eq:c_mom}). For the confining model distributions
in the charmed mesons and baryons, the momentum fraction at the
input model scale $Q^2 = m^2_c$ is found to be
  $P_c = 1.34^{\, +1.35}_{\, -0.75}\, \%$
for the cutoff mass parameter $\Lambda = (3.0 \pm 0.2)$~GeV
obtained from the inclusive $\Lambda_c$ production data,
Sec.~\ref{sec:mb}.  Again we note that these first moments
differ numerically from the charm multiplicities also mentioned in
Sec.~\ref{sec:mb}, where we found
$\langle n \rangle_{MB}^{\rm (charm)}
= 2.40^{\, +2.47}_{\, -1.36}\, \%$.
The strong dependence of the total momentum on $\Lambda$
stems from the controlling influence of the dominant meson--baryon
splitting function on the hadronic form factor, as seen in the left panel
of Fig.~\ref{fig:f_MB_int}.
Contrastingly, the BHPS model normalized to a 1\% charm probability in the nucleon
yields a corresponding momentum fraction of $P_c = 0.57$\%; our MBM
at $\Lambda = 3$~GeV therefore predicts about twice the intrinsic charm momentum
as the BHPS model.

Valence quark normalization requires that the first moment of
$c-\bar{c}$ vanishes, as in Eq.~(\ref{eq:c_moments}), which follows
for any splitting function that obeys the reciprocity relation,
Eq.~(\ref{eq:fphi}).  Higher moments, on the other hand, are not
required to vanish.  In fact, the magnitude of the $c-\bar{c}$
asymmetry can be quantified in terms of the difference of the
second moments (momentum carried by charm and anticharm quarks),
\bea
\Delta P_c &=& C^{(1)} - \overline{C}^{(1)}\ .
\label{eq:cmom_dif}
\eea
At the model scale $Q^2 = m_c^2$, we find
  $\Delta P_c = -(0.13^{\, +0.14}_{\, -0.08})\, \%$
for $\Lambda = (3.0 \pm 0.2)$~GeV.
The overall negative values of $\Delta P_c$ follow
from the behavior in the MBM that the $\bar c$ distribution
is harder than the $c$ as we just argued.

The momentum imbalance of anticharm quarks compared to charm
can be estimated from the ratio of the difference $\Delta P_c$
to the sum $P_c$, for which we find $\Delta P_c/P_c \approx -10\%$.
As a function of $x$, however, the imbalance is not uniformly
distributed.  Defining the ratio
\bea
R_c(x) &=& \Big( c(x) - \bar{c}(x) \Big) \Big/  \Big( c(x) + \bar{c}(x) \Big)\ ,
\label{eq:Rx}
\eea
we see from the RHS of Fig.~\ref{fig:c-c+} that the relative asymmetry
can exceed 50\% at large values of $x$ ($x \gtrsim 0.5$).
The $Q^2$ dependence of the ratio indicates relatively mild
effects over the large range considered (up to $Q^2=50$~GeV$^2$),
with the slope of the asymmetry becoming slightly more shallow
with increasing $Q^2$.
Note that the ratio $R_c$ is nonzero at $x=0$ at the model scale,
but perturbative evolution forces $R_c(x=0)$ to vanish
at large $Q^2$ due to the growth of the denominator $c+\bar c$.

\subsection{Comparison with other models}
\label{ssec:approx}

While some of the features of the nonperturbative $c$ and $\bar c$
distributions in the MBM are relatively robust, such as the generally
harder $x$ dependence compared with the perturbatively generated
distributions and the presence of a $c-\bar{c}$ asymmetry, the
detailed $x$ dependence does depend on the specifics of the model.

To estimate the model dependence of the calculated $c$ and $\bar c$
PDFs, we compare the results obtained in the previous section, using
the splitting functions and quark distributions from Sec.~\ref{sec:amp}
and the confining prescription for the PDFs in the charmed hadrons of
Sec.~\ref{sec:cinc}, with distributions computed under various assumptions
and approximations.

\begin{figure}[h]
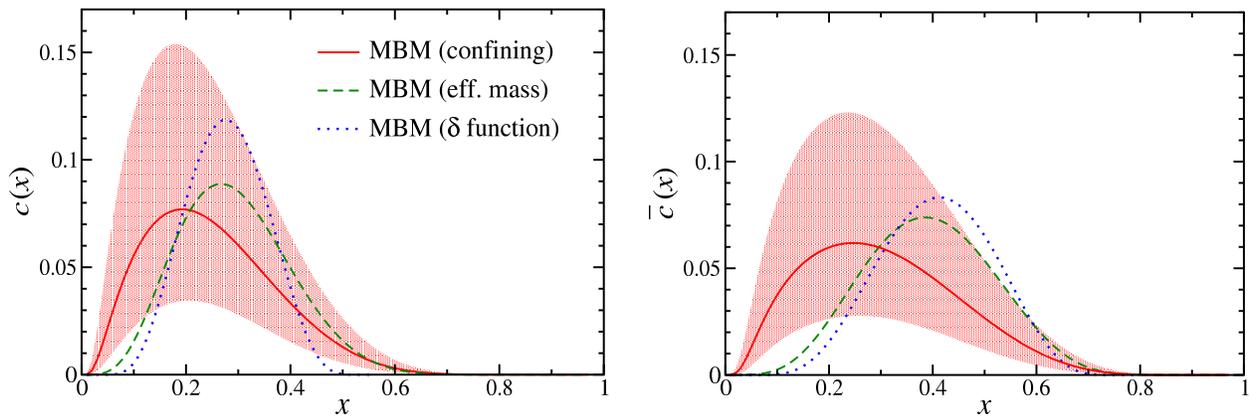

\vspace*{1cm}
\includegraphics[width=8cm]{IC-Fig/Fig-10a.eps} \ \ \
\includegraphics[width=8cm]{IC-Fig/Fig-10b.eps}
\caption{
	Model dependence of the charm distributions in the nucleon
	for $c(x)$ (left) and $\bar{c}(x)$ (right), for the
	MBM with the confining model for the PDFs in charmed hadrons
	(red solid), the effective mass model (green dashed),
	and the $\delta$ function model (blue dotted).
	All distributions use a common value for the cutoff mass of
	$\Lambda = (3.0 \pm 0.2)$~GeV, with the uncertainty band
	shown for the confining model.}
\label{fig:cnaive2}
\end{figure}

Within the same MBM framework, if one instead uses the effective mass model
as defined in Eq.~(\ref{eq:mceff}) for the (anti-)charm distributions in
intermediate states, the resulting $c$ and $\bar c$ distributions in the nucleon
are slightly harder, especially for the $\bar c$, as Fig.~\ref{fig:cnaive2}
illustrates. This generally follows from the shape of the $\bar{c}_M$
distribution in the confining and effective mass models in
Fig.~\ref{fig:QDF_mc}, where the latter is more strongly peaked at large
values of the parton momentum fraction.

The corresponding value of the total nucleon momentum carried
by charm and anticharm quarks in the effective mass model is
  $P_c = 1.67^{\, +1.70}_{\, -0.94}\, \%$
for cutoff masses $\Lambda = (3.0 \pm 0.2)$~GeV,
and
  $\Delta P_c = -(0.24^{\, +0.28}_{\, -0.14})\, \%$
for the momentum asymmetry, with the resulting momentum imbalance
$\Delta P_c/P_c \approx -15\%$.  The $c-\bar{c}$ asymmetry in this
model is therefore more pronounced than in the confining model.

In a more simplified approach, the $c$ and $\bar c$ distributions
inside the charmed hadrons were approximated in Ref.~\cite{MT97}
by $\delta$ functions centered at the $x$ values corresponding to
the fraction of the hadron mass carried by the constituent charm
or anticharm quark,
\begin{equation}
c_B(x) = \delta(x - x_B) \hskip 0.3cm {\rm and} \hskip 0.3 cm 
\overline{c}_M(x) = \delta(x - x_M).
\label{eq:cdelta}
\end{equation}
From Eqs.~(\ref{eq:mesoncloud}), the charm and anticharm distributions
in the nucleon are then given directly as sums over the various
meson--baryon splitting functions,
\begin{subequations}
\label{eq:cnaive}
\bea
c(x)
&=& \sum_{B,M}\, \frac{1}{x_B}\, f_{BM}\left(\frac{x}{x_B}\right)\ , \\
\bar{c}(x)
&=& \sum_{M,B}\, \frac{1}{x_M}\, f_{MB}\left(\frac{x}{x_M}\right)\ .
\eea
\end{subequations}
This is especially easy to see if we use an equivalent form for the relevant
convolution in Eqs.~(\ref{eq:mesoncloud}); for instance for anticharm\footnote{
We use an identity for the transformation properties of the Dirac $\delta$-function:
\begin{equation}
\delta(x - x_1 x_2)\ \equiv\ {1 \over |-x_2|} \cdot \delta(x_1 - x/x_2)\ . \nonumber
\end{equation}
}:
\begin{align}
\bar{c}(x)\ &=\ \sum_{M,B}\ \int dx_1 \int dx_2\ f_{MB}(x_1)\ \bar{c}_M(x_2)\ \delta(x - x_1 x_2) \nonumber\\
&=\ \sum_{M,B}\ \int_0^1 {dx_2 \over x_2}\ f_{MB}\left( {x \over x_1} \right) \delta(x_2 - x_M)\ ,
\end{align}
whence comes the expression for $\bar{c}(x)$ in Eq.~(\ref{eq:cnaive}).

Since the masses of the charm quark [$m_c = {\cal O}(1.5~{\rm GeV})$]
and the $D$ mesons [$m_D = {\cal O}(\mbox{1.8--2}~{\rm GeV})$] are
similar, as a first approximation one can take $x_M \approx 1$.
Similarly, for the fractional mass of the $c$ quark in the charmed
baryon, the approximation $x_B \approx 2/3$ was utilized \cite{MT97}.
In a somewhat more sophisticated approach, one can choose $x_M$
and $x_B$ to minimize the $\hat{s}$-dependent energy denominator,
which depends on the combination
  $m_c^2/x + m_{\rm spec}^2/(1-x)$,
where the spectator mass $m_{\rm spec}$ corresponds to the light quark
mass $m_{u,d}$ for a meson, and to an effective diquark mass $m_{qq}$
for a baryon.  Choosing $m_c = 1.3$~GeV, $m_{u,d} = 0.313$~GeV and
$m_{qq} = 1$~GeV, one has
\be
x_B\, =\, \frac{m_c}{m_c + m_{qq}} \approx 0.57\ ,\ \ \ \ \ \
x_M\, =\, \frac{m_{\bar c}}{m_{\bar c} + m_{u,d}} \approx 0.81\ .
\label{eq:mB}
\ee
For the best fit form factor cutoff mass $\Lambda = 3$~GeV, the
momentum carried by charm in this $\delta$ function approximation
model is $P_c = 1.66$\%, which is slightly greater than in the
MBM confining or effective mass models.

In Fig.~\ref{fig:cnaive2} we compare the $c$ and $\bar{c}$
distributions in the MBM obtained using the confining model
PDFs in the charmed hadrons with those computed from the
effective mass model and $\delta$ function approximations,
with a common cutoff mass $\Lambda = (3.0 \pm 0.2)$~GeV.
The MBM confining model distributions are generally softer
than those in the effective mass and $\delta$ function schemes,
with the confining model giving a slightly broader shape, and
the $\delta$ function model having the narrowest distribution.
Within the uncertainty bands of the parameters (for clarity
we have only shown the uncertainty band for the confining model
in Fig.~\ref{fig:cnaive2}), the distributions are indeed compatible with
each other. In all three models the anticharm distributions are clearly
harder than the charm, so that the qualitative features of
the ratio $R_c$ in Fig.~\ref{fig:c-c+} are largely retained.
Interestingly, the $\delta$ function model gives an $x$ dependence
for the charm PDFs that closely resembled the shape of the
effective mass model distributions for $\Lambda = 3$~GeV.
This feature may be exploited in simplified calculations that aim
simply to approximate general features of nonperturbative charm
distributions.

\vspace*{0.15cm}
{\it The charm structure function $F^c_2$.}
\vspace*{0.15cm}

Having explored the model dependence of the total intrinsic
$c$ and $\bar{c}$ distributions in the nucleon, we can now
directly confront the results with measurements of the charm
structure function, $\Fcc$.  This will provide additional
constraints on the model parameters, complementing those of
the inclusive $\Lambda_c$ production in $pp$ scattering
discussed in Sec.~\ref{sec:mb}.

Thought the calculations of intrinsic charm in this analysis have been normalized
to inclusive charm production data in $pp$ collisions as discussed
in Sec.~\ref{sec:mb}, the results may also be confronted with data
on the charm structure function $\Fcc$ obtained from measurements of
charm production cross sections in deep-inelastic lepton scattering.
In comparing with experimental measurements of $\Fcc$, in addition
to intrinsic charm arising from nonperturbative fluctuations of the
nucleon into states with 5 or more quarks, one must also consider the
``extrinsic'' charm arising from gluon radiation to $c\bar c$ pairs,
which is described by perturbative QCD evolution. As we have already
seen in the discussion of high-$Q^2$ CSV, these effects at LO in $\alpha_s$
are embodied by the diagrams in Fig.~\ref{fig:QCDsplt} and
Eqs.~(\ref{eq:QCD_DGLAP}).

To lowest order in the strong coupling constant $\alpha_s$,
the charm structure function is straightforwardly related to
the $c$ and $\bar c$ parton distributions in the nucleon,
\be
\Fcc (x,Q^2) = \frac{4x}{9}\left[ c(x,Q^2) + \bar{c}(x,Q^2) \right]\ .
\label{eq:F2cQM}
\ee
When combining the two charm contributions, it is necessary to assign
a scale $Q_0^2$ at which the nonperturbative charm is generated,
and then evolve this to the relevant $Q^2$ for comparison with experiment.
While the absolute scale of the intrinsic contribution is a
characteristic ingredient of the model (in our case, the MBM),
it is customary to set this to the effective charm quark mass,
$Q_0^2 = m_c^2 = 1.69$~GeV$^2$.

\begin{figure}[ht]
\vspace*{1.4cm}
\includegraphics[height=9cm]{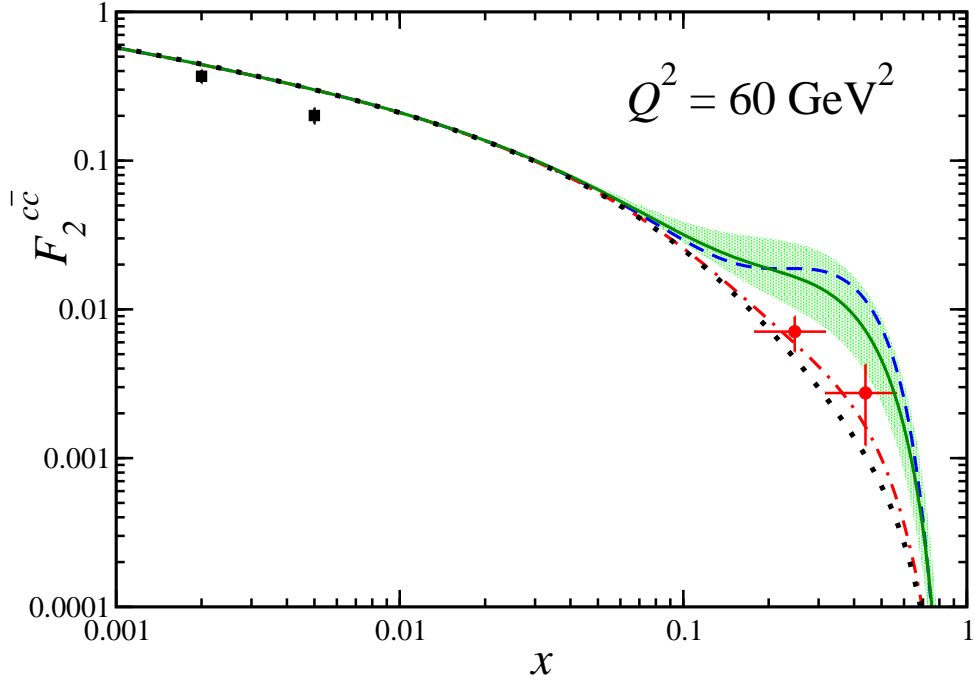}
\caption{
	Charm structure function $\Fcc$ at
	60~GeV$^2$.
	The perturbative QCD calculation (black dotted line)
	is compared with nonperturbative charm contributions
	in the MBM using the
	confining model with cutoff $\Lambda=(3.0 \pm 0.2)$~GeV
	  (green solid line and shaded band),
	confining model with $\Lambda=2.5$~GeV
	  (red dot-dashed line), and
	effective mass model with $\Lambda=3.0$~GeV
	  (blue dashed line).
	The data are from H1 and ZEUS (black squares)
	and EMC (red circles).}
\label{fig:F2c-MBM_DC-EMC}
\end{figure}

The calculated charm structure function $\Fcc (x,Q^2)$ is shown
in Fig.~\ref{fig:F2c-MBM_DC-EMC} at $Q^2 = 60$~GeV$^2$, and compared with
data from the H1 \cite{H1} and ZEUS \cite{ZEUS} Collaborations at HERA and
with higher-$x$ data from EMC \cite{EMC3}.
While the extrinsic charm distribution is generated completely
perturbatively by evolving an initial ``zero distribution''
$c(x,Q^2_0) = \bar{c}(x,Q^2_0) \equiv 0$ with the next-to-leading
order evolution code from Ref.~\cite{Miyama96},
nonperturbative models for $\Fcc$ of course start with the MBM
predictions at the scale $Q_0^2$ as just described.
Since the experimental $Q^2$ values are large compared to $m_c^2$,
standard massless QCD evolution in the form of the {\it Variable Flavor
Number Scheme} is appropriate.

As has been suggested previously \cite{Brodsky:1995vr}, at $Q^2 = 60$~GeV$^2$
the perturbative results usually underestimate the EMC data at high $x$,
and the addition of a nonperturbative contribution raises the total
$\Fcc$ in rough proportion to the amount of intrinsic charm assumed in the
model. However, for our MBM with the confining model vertex functions,
$\Lambda = (3.0 \pm 0.2)$~GeV generally leads to overestimation of the EMC $\Fcc$ data
at large $x$ and high $Q^2$, being marginally consistent with the data at the
lower edge of the error band.
Using instead the MBM with the effective mass model vertex
functions and the same cutoff $\Lambda = 3.0$~GeV, the peak
at large $x$ is still more pronounced, and hence overestimates the
EMC data to a slightly greater extent.
Lowering the cutoff to $\Lambda = 2.5$~GeV for the confining model,
the resulting $\Fcc$ is in better agreement with the data,
producing a smaller overestimate of the lower-$Q^2$ data points
and resulting in a better fit to the $Q^2 = 60$~GeV$^2$ data. 
Note that for such a small cutoff the average charm normalization
$\langle n \rangle_{MB}^{\rm (charm)} \lesssim 0.5\%$, which would
significantly underestimate the inclusive $\Lambda_c^+$ production
data (see Fig.~\ref{fig:R608_CS}).
%

%% file: the-GA.tex
\section{A Global Analysis of IC}
\label{sec:GA}

Having constructed a systematic framework for the analysis of intrinsic charm
in Secs.~\ref{sec:5q}--\ref{sec:results}, we turn now to a more comprehensive
effort \cite{Globe} to constrain its overall normalization via QCD global analysis.
There have in fact already been several attempts of this sort.

In an earlier study \cite{Pum07}, Pumplin, Lai and Tung carried
out a global fit to high-energy data with phenomenological PDFs
including an intrinsic charm component at the starting scale $Q_0 = m_c$.
Using several different phenomenological forms for the intrinsic
charm (including the BHPS model, Eq.~(\ref{eq:charmprob}),
a scalar MBM employing Eq.~(\ref{eq:dPFock}), and a
``sea-like'' charm model), the magnitude of the charm contribution
was varied until a substantial increase in the $\chi^2$ was found
with the set of global high-energy data.
The analysis found that the global fits could accommodate charm
momentum fractions of $P_c \approx 2\%$ in the BHPS and scalar MBM
models and $P_c \approx 2.5\%$ in the sea-like model at the 90\%
confidence level, which are significantly larger than the constraints
from the EMC $\Fcc$ data, and at the upper boundary of the range
allowed by the ISR R608 $\Lambda_c^+$ production data.
The more recent update \cite{Dulat13} that includes NNLO corrections
finds $P_c \leq 1.5\%$ for the sea-like model and $P_c \leq 2.5\%$
for the BHPS model at the scale $Q_0$.

We should note, however, that while the analysis in Ref.~\cite{Pum07}
fitted the precision low-$x$ charm structure function data from
H1 and ZEUS, it {\em did not} include the EMC $\Fcc$ data at large $x$
\cite{Aubert:1982tt}, which results in tighter constraints on the
magnitude of the intrinsic charm as we now show by explicit computation.
Moreover, while the very same EMC data are occasionally mentioned \cite{Brodsky:1995vr}
as a leading evidence for intrinsic charm, they have never been incorporated
systematically into an analysis such as the present one.

Like these previous studies, \EG~\cite{Pum07,Dulat13}, we find it most
natural to fit the total momentum carried by the nonperturbative $c(x),\ \bar{c}(x)$
contained in five-quark Fock states latent in the foregoing meson-baryon
model. Explicitly, this is
\begin{equation}
\langle x \rangle_{\rm IC} \equiv \int_0^1 dx \, x \, [c(x) + \bar{c}(x)]\ .
\label{eq:momfrac}
\end{equation}
We saw that the distribution shapes can be reasonably well-determined by the relativistic
wavefunctions employed to describe contributions from the $SU(4)$ spectrum shown
in Fig.~\ref{fig:Spectrum} (after fitting the momentum cutoff $\Lambda$ to the
$pp \rightarrow \Lambda_c X$ data of Fig.~\ref{fig:R608_CS}); this procedure imposes
$\Lambda = (3.0 \pm 0.2)$ GeV, but perhaps less clearly limits the overall normalization
of IC in the proton as evidenced by the sizable uncertainty extracted in
Sec.~\ref{ssec:c_in_MBM} for the relevant moments: 
\begin{equation}
\langle n \rangle_{MB}^{\rm (charm)}
= 2.40^{\, +2.47}_{\, -1.36}\, \%\ , \hspace{1.5cm}
P_c = 1.34^{\, +1.35}_{\, -0.75}\, \%\ .
\end{equation}

That being the case, it is useful to extend the analysis of preceding sections
by considering the world's $F_2^c$ and other high-energy
scattering data, and performing a new global QCD analysis
along the lines of the recent JR14 fit \cite{JR14}.
Unlike the previous global analyses \cite{Pum07, Dulat13} which
placed more stringent cuts on the data ($Q^2 \gtrsim 4$~GeV$^2$
and $W^2 \gtrsim 12$~GeV$^2$), excluding, for instance, all fixed
target cross section measurements from SLAC \cite{SLAC},
we include all available data sets with $Q^2 \geq 1$~GeV$^2$
and $W^2 \geq 3.5$~GeV$^2$.

In order to have simple analytic forms to serve the QCD fit of the
IC normalization, we follow Ref.~\cite{Pum05} and provide compact
three-parameter fits to the $c$ and $\bar c$ PDFs in the MBM
computed using several different models for the charm quark
distributions in the charmed mesons and baryons, including the
confining model, effective mass model, and the $\delta$ function
model. The parameters matching the best fits to each of these models
generated at $\Lambda = 3$ GeV are then given in Table \ref{table:phen-fits}.
The parametric form for the charm distributions in the nucleon
for the confining and effective mass models is taken to be
\begin{subequations}
\label{eq:conf-fit}
\bea
c(x)
&=& C^{(0)}\, A\, x^\alpha (1-x)^\beta\ ,	\\
\bar{c}(x)
&=& C^{(0)}\, \bar{A}\, x^{\bar\alpha} (1-x)^{\bar\beta}\ ,
\eea
\end{subequations}
where the normalization constants
  $A = 1/B(\alpha+1,\beta+1)$
and
  $\bar{A} = 1/B(\bar\alpha+1,\bar\beta+1)$,
with $B$ the Euler beta function, ensure that the distributions
are normalized to $C^{(0)}$.

\begin{table}[h]
\caption{Best fit parameter values for the $c$ and $\bar c$
	distributions in the nucleon in Eqs.~(\ref{eq:conf-fit})
	and (\ref{eq:deltafn-fit}) in the MBM for a central
	cutoff mass $\Lambda=3.0$~GeV.}
\centering
\begin{tabular}{c|c c c}			\hline\hline
$c$, $\bar c$ fit
	& confining
	& effective
	& $\delta$ function			\\
parameters
	& model
	& mass model
	& model					\\ \hline
$A$	
	&\ \ \ $1.720 \times 10^2$\ \ \
	&\ \ \ $1.052 \times 10^2$\ \ \
	&\ \ \ $2.638 \times 10^5$\ \ \		\\
$\alpha$
	& 1.590
	& 3.673
	& 4.266					\\
$\beta$
	& 6.586
	& 10.16
	& 4.485					\\ \hline
$\bar A$	
	&\ \ \ $7.404 \times 10^1$\ \ \
	&\ \ \ $4.160 \times 10^0$\ \ \
	&\ \ \ $2.463 \times 10^4$\ \ \		\\
$\bar\alpha$
	& 1.479
	& 4.153
	& 5.003					\\
$\bar\beta$
	& 4.624
	& 6.800
	& 4.857					\\ \hline	
\end{tabular}
\label{table:phen-fits}
\end{table}

For the $\delta$ function model, where the $c$ and $\bar c$ PDFs
in the charmed hadrons are given by $\delta$ functions in $x$,
it is more convenient to parametrize the distributions in
Eqs.~(\ref{eq:cnaive}) as
\begin{subequations}
\label{eq:deltafn-fit}
\bea
c(x)
&=& C^{(0)}\, A\, x^\alpha (x_B - x)^\beta\,
    \theta(x_B - x)\ ,					\\
\bar{c}(x)
&=& C^{(0)}\, \bar{A}\, x^{\bar\alpha} (x_M - x)^{\bar\beta}\,
    \theta(x_M - x)\ ,
\eea
\end{subequations}
with $x_B$ and $x_M$ given by Eq.~(\ref{eq:mB}). Note that in each
of these prescriptions, it is simple to use Eq.~(\ref{eq:momfrac}) to
go between the specific $C^{(0)}$ preferred by fits and the values
of $\langle x \rangle_{\rm IC}$ plotted in subsequent figures.

The guiding philosophy of QCD global analysis \cite{Brock:1993sz} is the validity
of the {\it factorization theorem}, which posits the separability of QCD-mediated
processes into hard and soft components. For instance, in the case of the
electromagnetic structure functions of a hadron $h$, 
\begin{equation}
F^\gamma_{i h}(x, Q^2)\ =\ \sum_f\ \int_0^1\ {d\xi \over \xi}\ 
C^{\gamma f}_i \left( {x \over \xi}, {Q^2 \over \mu^2}, {\mu^2_F \over \mu}, \alpha_S(\mu^2) \right)\
\cdot\ \phi_{f/h}(\xi, \mu^2_F, \mu^2)\ ,
\end{equation}
suggests that the $F^\gamma_{i h}(x, Q^2)$ contain information on the universal parton densities\linebreak
$\phi_{f/h}(\xi, \mu^2_F, \mu^2)$ of flavor $f$; these in turn may be found with high
precision given an accurate computation of the hard process encoded in the coefficient
function $C^{\gamma f}_i$. The presence of the renormalization and factorization scales
$\mu,\ \mu_F$ demands that we must elect to use a particular scheme, upon which our results
will depend. In the present treatment we consistently compute in the $\overline{\rm MS}$ scheme.

For the QCD analysis we use the framework of the JR14 global analysis
\cite{JR14}, in which the total $F_2$ structure function is given by
\begin{equation}
F_2 = F_2^{\rm light} + F_2^{\rm heavy}\ ,
\label{eq:F2def}
\end{equation}
where $F_2^{\rm light}$ denotes the light-quark ($u$, $d$, $s$)
contributions, and $F_2^{\rm heavy}$ includes contributions
from the heavy $c$ and $b$ quarks.
The charm structure function is further decomposed into a
perturbative part, $F_2^{c\bar c}$, and a nonperturbative (IC)
component, $F_2^{\rm IC}$,
\begin{equation}
F_2^c = F_2^{c\bar c} + F_2^{^{\rm IC}}\ .
\label{eq:F2c} 
\end{equation}
The perturbative contribution is computed in the fixed-flavor number
scheme (FFNS) from the photon-gluon fusion process,
\begin{equation}
F_2^{c\bar c}(x,Q^2,m_c^2)
= \frac{Q^2 \alpha_s}{4\pi^2 m_c^2}
  \sum_i \int\frac{dz}{z}\
  \hat\sigma_i(\eta,\xi)\ \cdot\ f_i\Big(\frac{x}{z}, \mu\Big)\ ,
\label{eq:F2cc}
\end{equation}
where $\hat\sigma_i$ is the hard scattering cross section for
the production of a $c\bar c$ pair from a parton of flavor $i$,
with $i = u, d, s$ or $g$, and $f_i$ is the corresponding parton
distribution, both calculated to NLO [${\cal O}(\alpha_s)$] accuracy.
The partonic cross section $\hat\sigma_i$ is evaluated as a function
of the variables $\xi = Q^2/m_c^2$ and $\eta = Q^2 \cdot (1-z)\big/(4m^2_c z) - 1$; the
PDF is computed at the factorization scale $\mu_F^2 = 4 m_c^2 + Q^2$,
where the charm mass is $m_c = 1.3$~GeV at threshold.

On the other hand, $F_2^{^{\rm IC}}$ of Eq.~(\ref{eq:F2c}) is actually
specified only at the charm production threshold by the MBM of the previous
sections; whereas before we relied upon a brute-force integration
of the DGLAP equations to render Eq.~(\ref{eq:F2cQM}) at the $Q^2$ of
the EMC measurements, we now require a more thorough scheme to complement
the photon-gluon FFNS used in Eq.~(\ref{eq:F2cc}). For this purpose, we
turn to the ans\"atz of Hoffmann and Moore described in \cite{Hoffmann:1983ah}.

In this method, the scale dependence of the IC distributions parametrized in
Eqs.~(\ref{eq:conf-fit}) is incorporated through target and quark mass effects as implemented
in the OPE --- much as described in Sec.~\ref{ssec:OPE}. A version of Eq.~(\ref{eq:Nacht}), slightly
altered to depend on the heavy quark mass, therefore rescales the intrinsic distributions
as
\begin{align}
c(x)\ \,\,\,\, \rightarrow \,\,\,\, c(\xi_c, \gamma_c)\ &=\ c(\xi_c) - {\xi_c \over \gamma_c}\ c(\gamma_c)\ \iff\ \xi_c < \gamma_c \\
&= 0\ \iff\ \xi_c \ge \gamma_c\ ,
\end{align}
where the modified Nachtmann variable is now
\begin{equation}
\xi_c\ =\ {2ax \over \left( 1 + \sqrt{1 + 4 x^2 M^2 \Big/ Q^2} \right) }\ , \hspace*{1.5cm}
a = {1 \over 2}\ \Big( 1 + \sqrt{1 + 4 m^2_c \Big/ Q^2} \Big)\ .
\end{equation}
$\gamma_c$ serves as another means of circumventing the OPE threshold problem, in
this case by simply terminating input distributions at $\xi_c < \gamma_c$, where
the latter is just $\xi_c$ evaluated at $x = \hat{x} \defeq Q^2 \Big/ (Q^2 + 4m^2_c + M^2)$.

With these expressions, we can write the full OPE result: 
\begin{align}
\label{eq:F2c_HM}
F_2^{^{\rm IC}}\ &=\ {4x^2 \over 9(1 + 4 x^2M^2/Q^2)^{3/2}}\
\left\{ {1 + 4m^2_c/Q^2 \over \xi_c}\ \Big( c(\xi_c, \gamma_c) + \bar{c}(\xi_c, \gamma_c) \Big)\ +\ 3 \hat{g}(\xi_c, \gamma_c) \right\}\ , \\
&\hat{g}(\xi_c, \gamma_c)\ =\ {2x M^2/Q^2 \over (1+4x^2M^2/Q^2)}\ \int_{\xi_c}^{\gamma_c}\ {dt \over t}\
\left( c(t,\gamma_c) + \bar{c}(t,\gamma_c) \right)\ \\
& \hspace*{4.5cm} \times \left[1 + 2xtM^2/Q^2 + 2xM^2/(tQ^2) \right] \cdot \left(1 - {m^2_c \over t^2 M^2} \right)\ . \nonumber
\end{align}

The expression in Eq.~(\ref{eq:F2c_HM}), together with NLO corrections from
gluon loops and bremsstrahlung, is the primary input to the global analysis by
which we constrain $\langle x \rangle_{\rm IC}$. For a detailed description of the
choice of data and kinematic cuts we refer to the JR14 analysis; however,
whenever possible the original cross section data were used rather than
structure functions and ratios of structure functions extracted from those
cross sections. Since data at relatively low $Q^2$ were used in this
global fit, higher-twist corrections were employed for the low-$Q^2$ data. For
deuteron data nuclear corrections were supplied by the CJ group
\cite{Accardi:2011fa}, while for data on heavier nuclei nuclear corrections
were used from nDS09 \cite{deFlorian:2003qf}.
\begin{figure}[h]
\hspace*{-0.4cm}
\includegraphics[width=8.5cm]{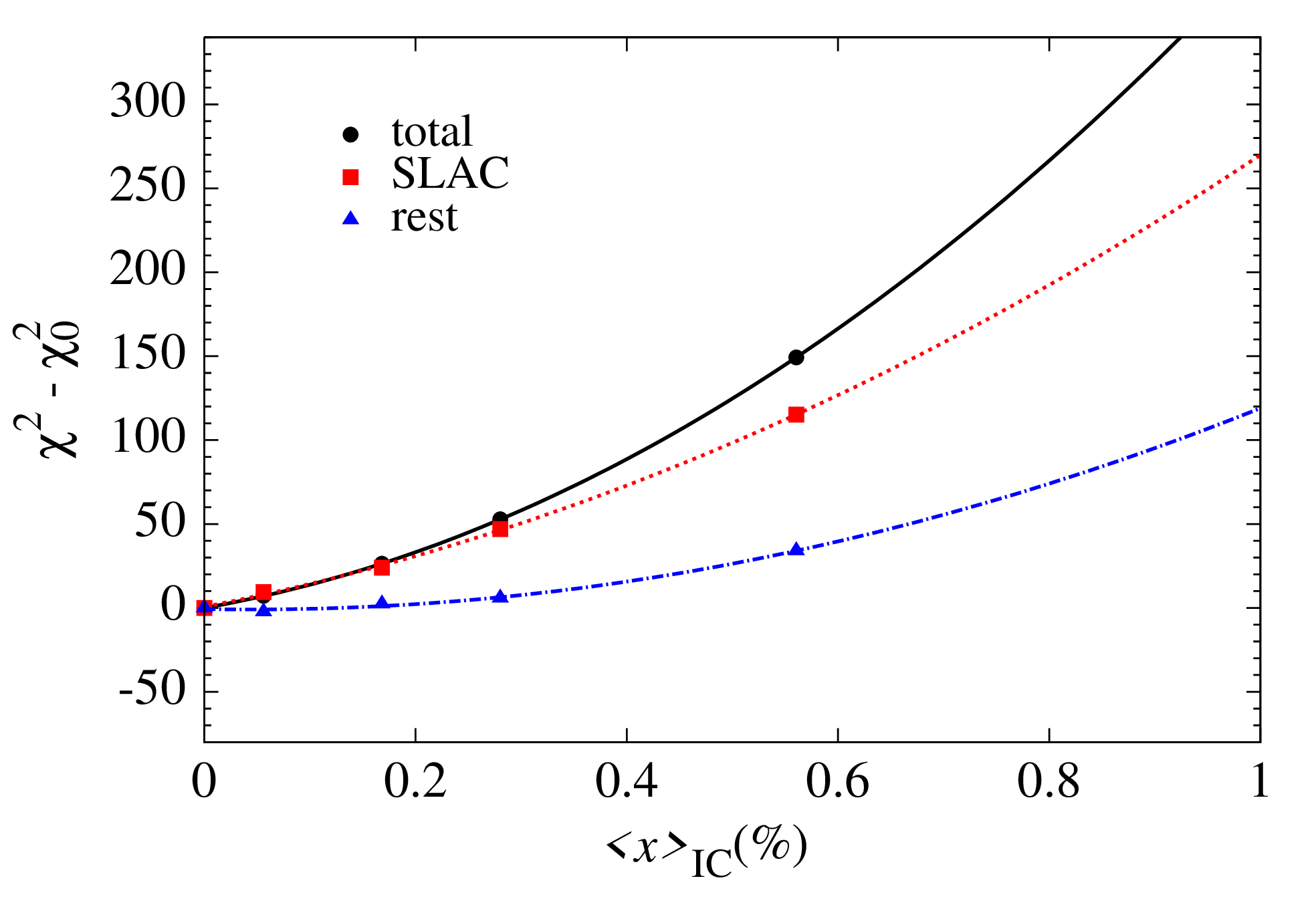}
\includegraphics[width=8.5cm]{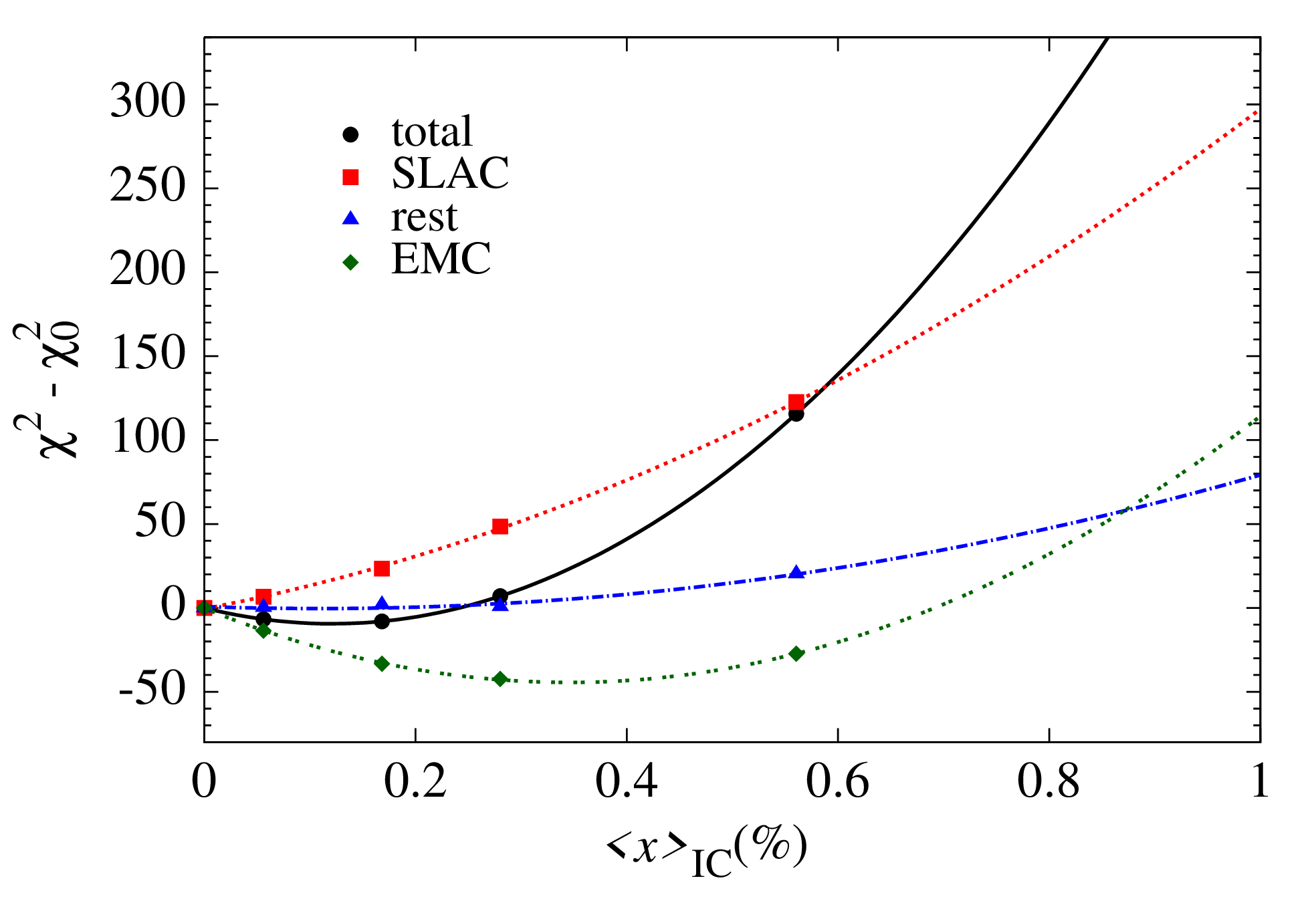}
\caption{
Contributions of various data sets to $\chi^2$ vs.~$\langle x \rangle_{\rm IC}$.
Solid circles: total $\chi^2$;
red squares: contribution from SLAC data~\cite{Whitlow:1991uw}; blue triangles:
all other data sets. EMC $F^c_2$ data is excluded in the left panel.
The right panel includes EMC charm data, and the resulting
contribution to $\chi^2$ is represented by open green squares.
}
\label{fig:scans}
\end{figure}

Our results for intrinsic charm are summarized in Fig.~\ref{fig:scans}, where
the solid circles show the total $\chi^2$ for the global fit vs.~the
fraction of the proton momentum carried by intrinsic charm. The left
panel shows the result excluding the EMC inclusive charm leptoproduction data,
while the lower panel includes EMC charm data. Results were calculated using
the confining approximation of Table~\ref{table:phen-fits} as a proxy for the other models,
which yield qualitatively similar constraints to $\langle x \rangle_{\rm IC}$.
The upper limit allowed for momentum carried by charm quarks is a
fraction of a percent; namely, the $\chi^2$ profile for the `Total' fit {\it without}
EMC $F_2^{c\bar{c}}$ data in the top panel of Fig.~\ref{fig:scans} excludes an IC of
$\langle x \rangle_{\rm IC} \ge 0.1$ at the $5\sigma$ level. This high level of exclusion
is largely due to the constraints provided by SLAC data \cite{Whitlow:1991uw}.
In the absence of the SLAC data -- that is, for the `Rest' fits shown as blue dot-dashed lines
in Fig.~\ref{fig:scans} -- results are comparable to earlier CTEQ findings \cite{Pum07}
when statistical tolerances are considered. Namely, $\langle x \rangle_{\rm IC} = 0.1 \%$ is contained
within the $1\sigma$ interval. On the other hand, the $2\sigma$ and $3\sigma$ confidence levels
permit IC up to $\langle x \rangle_{\rm IC} = 0.2 \%$ and $0.3 \%$, respectively. If the $\Delta \chi^2 = 100$
statistical tolerance of CTEQ is used, these limits must be multiplied by $10$ to give, for instance,
$\langle x \rangle_{\rm IC} = 2.0 \%$ at the $2\sigma$ C.L., in agreement with \cite{Pum07}.

Thus, if the EMC data are to be neglected, the minimum of the $\chi^2$ curve occurs at
$\langle x \rangle_{\rm IC} = 0$. On the contrary, including EMC data gives a best value
$\langle x \rangle_{\rm IC} = 0.13 \pm 0.04$\%, also at the $1\sigma$ level. Clearly this
is a much smaller upper limit on IC than obtained by the CTEQ analysis from Dulat
\EA~\cite{Dulat13}, who found that their global analysis allowed IC momentum up
to 2.5\%.

This sharp discrepancy is understandable in light of the content of Fig.~\ref{fig:scans}.
The red squares show the contribution to $\chi^2$ from SLAC $ep$ and $ed$ data
\cite{Whitlow:1991uw}, while the blue triangles show the contribution from
all the remaining data. The rapid take-off of the red curves highlights the fact that
the limits on intrinsic charm are dominated by the SLAC DIS data --- especially those at
low-$Q^2$ and high-$x$. But this is of course precisely the kinematic regime at which one
would anticipate the effects of nonperturbative heavy quarks to be most sizable. Contrastingly,
the kinematic cuts of the CTEQ and MSTW global fits \cite{Tung:2006tb,Martin:2009iq} are
targeted at high energy collider processes typified by the LHC, and consequently emphasize data
at substantially lower $x$ and higher $Q^2$.
%
%

The global fits, either with or without the EMC charm data,
agree very well with the inclusive charm electroproduction
measurements at HERA \cite{Abramowicz:1900rp}; our results are essentially
identical with those given in Fig.~10 of the JR14 analysis
\cite{JR14}. The HERA data are at very small
$x$, a region where intrinsic charm is relatively unimportant.
On the other hand, the global fits here do poorly in reproducing the EMC
inclusive charm data, as evidenced by the fact that the fit quality degrades
sharply to $\chi^2/\mathrm{d.o.f.} = 4.3$ for those measurements.
In fact, regardless of whether the fits are or are not directly constrained by
the EMC charm data, they are not well fit --- a finding which suggests that the EMC points, being
older and considerably less precise than the HERA data, have
significant tension with the latter. This fact necessarily calls into
question the reliability of such data for the purposes of extracting
a nonperturbative component of $F^c_2$ at higher $x$.
%
%

%% file: the-TDISa.tex


In spite of its early discovery \cite{Lattes:1947mw} and numerous decades of related phenomenology,
much remains unknown regarding the structure and interactions of the pion.  
Nowhere is this quite as evident as in the still-perplexing behavior of the pion electromagnetic form
factor $F_\pi(Q^2)$, which is traditionally accessed via exclusive processes such as electroproduction
and defined as
\begin{subequations}
\begin{align}
\label{eq:pi-defI}
q^2 < 0\ &\implies\ \lan \pi^+(k')| J_\mu(0) | \pi^+(k) \ran\ =\ e (k_\mu +k'_\mu)\ F_\pi(q^2)\ , \\ 
q^2 > 0\ &\implies\ \lan \pi^+(k') \pi^-(k) | J_\mu(0) | 0 \ran\ =\ e (k'_\mu - k_\mu)\ F_\pi(q^2)\ ,
\label{eq:pi-defII}
\end{align}
\end{subequations}
where as usual, $Q^2 = -q^2$, the 4-momenta are $q=k'-k$ in Eq.~(\ref{eq:pi-defI}), and $q=k+k'$ in
Eq.~(\ref{eq:pi-defII}), and a normalization condition ensures $F_\pi(0)=1$.
For spacelike $Q^2 > 0$, $F_\pi(Q^2)$ is indeed calculable in the context of pQCD \cite{Chang:2013nia}, which suggests
a very simple form for its asymptotic behavior; namely,
\begin{equation}
Q^2 \gg \Lambda^2_{QCD}\,\,\,\,\, :\ \,\,\,\, Q^2 F_\pi(Q^2)\ \sim\ 16 \pi\ \alpha_s(Q^2)\ f^2_\pi \omega^2_\pi\ ,
\label{eq:pi-QCD}
\end{equation}
in which the pion decay constant is $f_\pi = 93$ MeV, and $\omega_\pi$ depends upon the behavior of the valence
{\it parton distribution amplitude} of the pion, and approaches unity at sufficiently large $Q^2$.
\begin{figure}[h]
\hspace*{-0.4cm}
\includegraphics[height=7.2cm,angle=90]{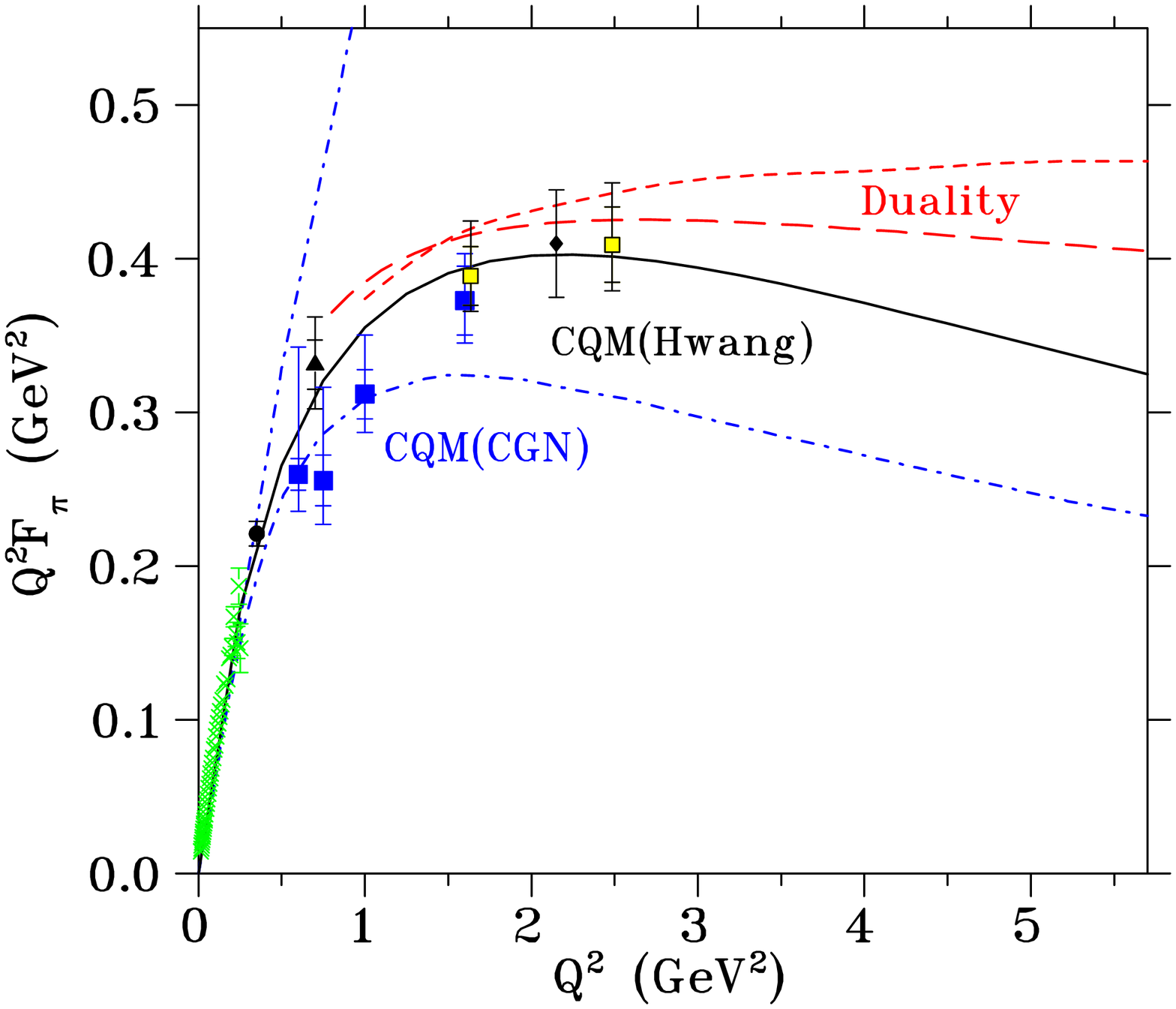}
\raisebox{-0.07\height}{\includegraphics[height=7.2cm]{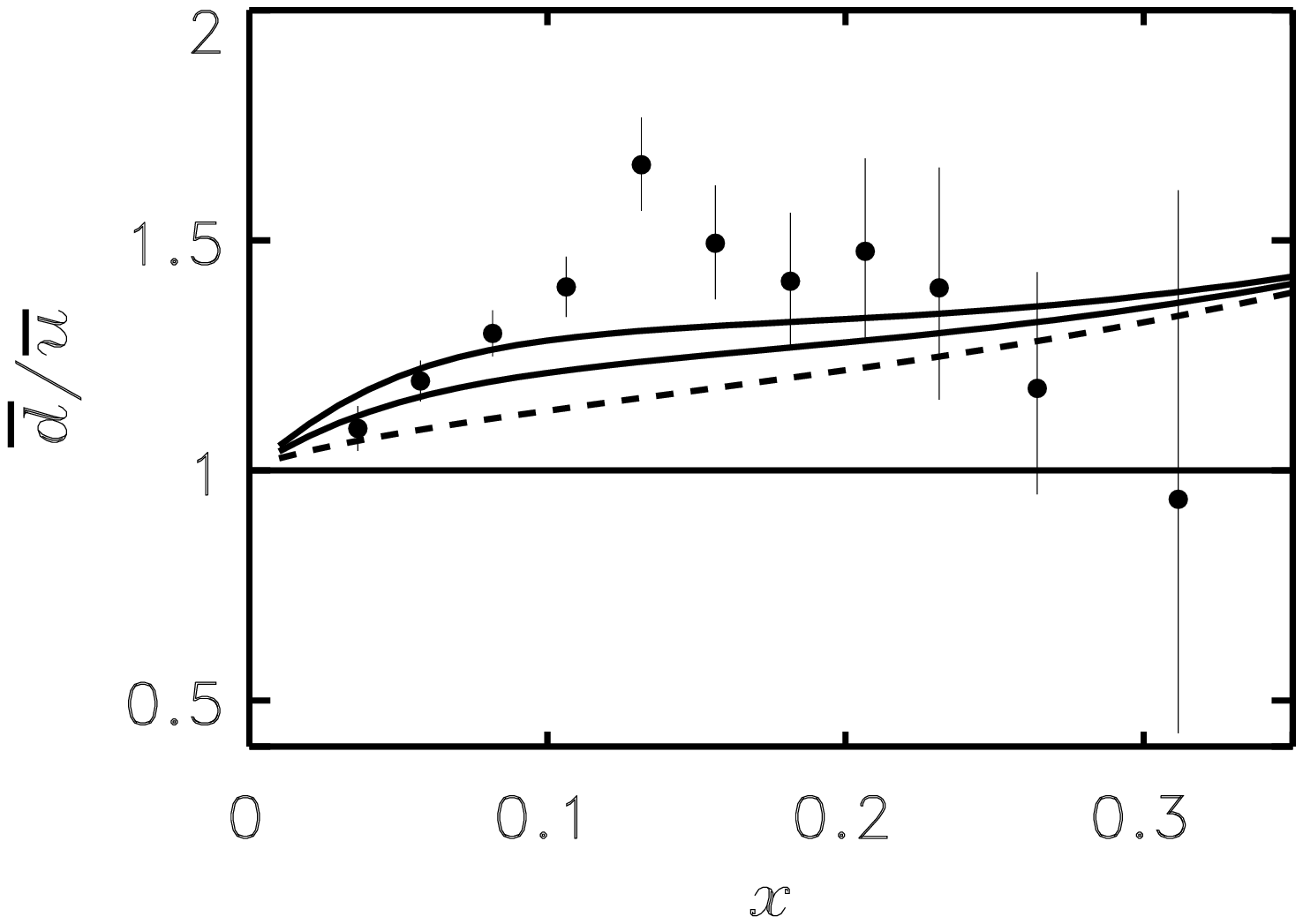}}
\caption{
(Left) As illustrated in this plot from \cite{Huber:2008id}, the behavior
of the spacelike pion form factor has been the object of many modeling
efforts, given the ambiguity of the $Q^2$ scale at which the pQCD result
of Eq.~(\ref{eq:pi-QCD}) becomes an accurate description. (Right) As we
shall demonstrate, the pion cloud of the nucleon is responsible for potential
flavor asymmetries in the light quark sea as computed in
\cite{Melnitchouk:1998rv}.
}
\label{fig:pi-mot}
\end{figure}
Of course there has been no shortage of experimental determinations of $F_\pi(Q^2)$
as shown, for instance, in the left panel of Fig.~\ref{fig:pi-mot}, but these have
typically been at rather shallow values of $Q^2$, and therefore not adequately near
the expected transition region at which the onset of the pQCD prediction of Eq.~(\ref{eq:pi-QCD})
should begin. Rather, at these intermediate kinematics ($2 \lesssim Q^2 \lesssim 5$ GeV$^2$)
there exists a menu of discrepant model predictions that necessitate various proposed and
upcoming measurements \cite{TDIS}. Typically, these measurements seek to access the
$F_\pi(Q^2)$ form factor at intermediate $Q^2$ via extrapolation to a physical $t = m^2_\pi$ pole
based upon pion electroproduction measurements off the proton obeying
\begin{equation}
{d\sigma \over dt}\ \sim\ g^2_{\pi NN}\ \left[{-t \over (t-m^2_\pi)^2}\right]\ Q^2 F^2_\pi(Q^2)\ ,
\label{eq:tchan-pi}
\end{equation}
which results from a ``pion cloud'' picture of nucleon structure similar to what we present shortly.

Moreover, through the dynamics of the Sullivan process \cite{Sullivan:1971kd} as described in the preceding
analysis of Chap.~\ref{chap:ch-charm} for intrinsic charm, the pion is expected to contribute crucially
to various nonperturbative aspects of the nucleon's flavor and spin decomposition. In particular,
the parton distributions in the nucleon are expected to receive contributions
of the form
\begin{subequations}
\begin{align}
\delta q(x)\ &=\ \Big\{ \delta^{[\pi N]} q(x)\ +\ \delta^{[N \pi]} q(x) \Big\}\ +\ \dots\ , \\
\delta \bar{q}(x)\ &=\ \Big\{ \delta^{[\pi N]} \bar{q}(x)\ +\ \delta^{[N \pi]} \bar{q}(x) \Big\}\ +\ \dots\ .
\end{align}
\label{eq:pi_cl}
\end{subequations}
Among other things, the differing flavor content of the charge states in the isovector pion triplet implies that
the combinations contained in Eq.~(\ref{eq:pi_cl}) would naturally induce flavor asymmetries such as
$\bar{d} / \bar{u} \neq 1$, plotted in the right panel of Fig.~\ref{fig:pi-mot}. In fact, it was the observation
of strong violations of the Gottfried Sum Rule defined in Eq.~(\ref{eq:Gott}) by NMC \cite{Amaudruz:1991at} that
called attention to the experimental significance of light quark asymmetries and gave added currency to modeling efforts
aimed at nonperturbative effects in the nucleon quark sea. This is simple to see by unpacking the expression in
Eq.~(\ref{eq:Gott}) using the QPM under the assumption of partonic charge symmetry ({\it unlike} the analyses of
Chap.~\ref{chap:ch-Q2}.\ref{sec:deut-pCSV}); using the electromagnetic quark-level expressions of Eq.~(\ref{eq:EW-SFs}),
we get
\begin{align}
S_G\ &=\ \int\ {dx \over x}\ \Big( F^p_2(x, Q^2) - F^n_2(x, Q^2) \Big) \nonumber\\
&=\ \sum_q\ e^2_q \int dx\ \left\{ q^p_v(x) + 2\bar{q}^p(x)\ -\ \Big(q^n_v(x) + 2\bar{q}^n(x) \Big) \right\} \nonumber\\
&= {1 \over 3}\ \left\{ 1\ +\ 2 \int_0^1 dx [\bar{u}(x) - \bar{d}(x)] \right\}\ ,
\label{eq:Gott-quark}
\end{align}
such that the original NMC finding $S_G = 0.240 \pm 0.016$ suggests an $SU(2)$ flavor symmetry breaking pattern that
favors $\bar{d}$ over $\bar{u}$, much as suggested by the right panel of Fig.~\ref{fig:pi-mot}. The direction of the
asymmetry in Eq.~(\ref{eq:Gott-quark}) is at least qualitatively in line with expectations based upon cloud
mechanisms in the proton of the form $p \rightarrow (\pi^+ n)$, given the positively charged pion's valence quark content
$\pi^+ = (u\bar{d})$.

Thus, in both examples embodied by Eqs.~(\ref{eq:tchan-pi}) and (\ref{eq:pi_cl}), the pion cloud of the nucleon figures
prominently --- \IE~through the possible contribution to exclusive electroproduction off the meson
cloud of the nucleon for the former, and through the role of the Sullivan process in electron-nucleon DIS in the
case of the latter. The realization of the importance of this mechanism makes clear that its phenomenology should be
probed in further detail, with a special premium on precise data, which even now remain fairly elusive.

\begin{figure}[ht]
\includegraphics[width=7cm]{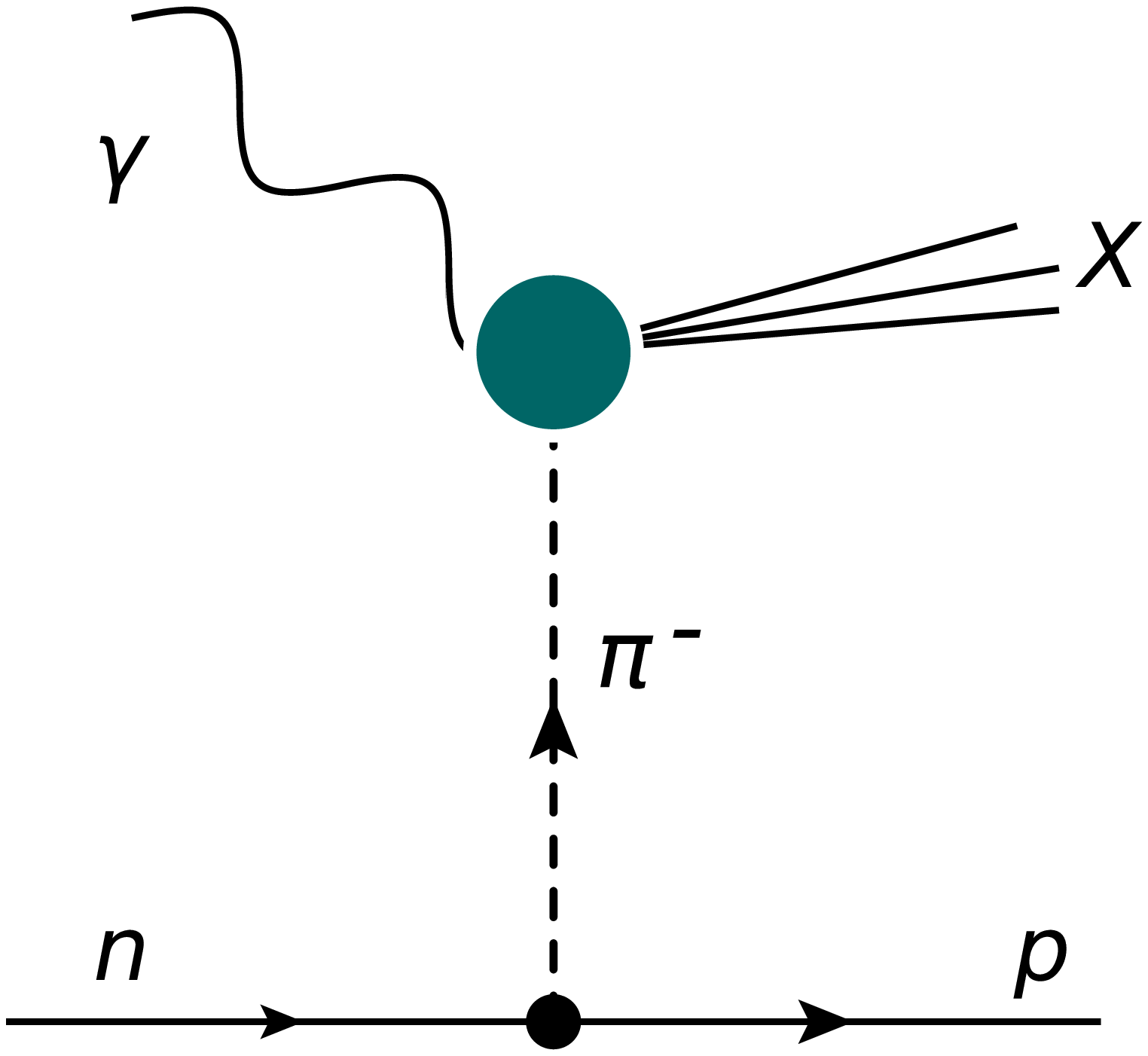} \ \ \ \ \ \ \ \ \ \ \ \ \ \ \ 
\includegraphics[width=7.5cm]{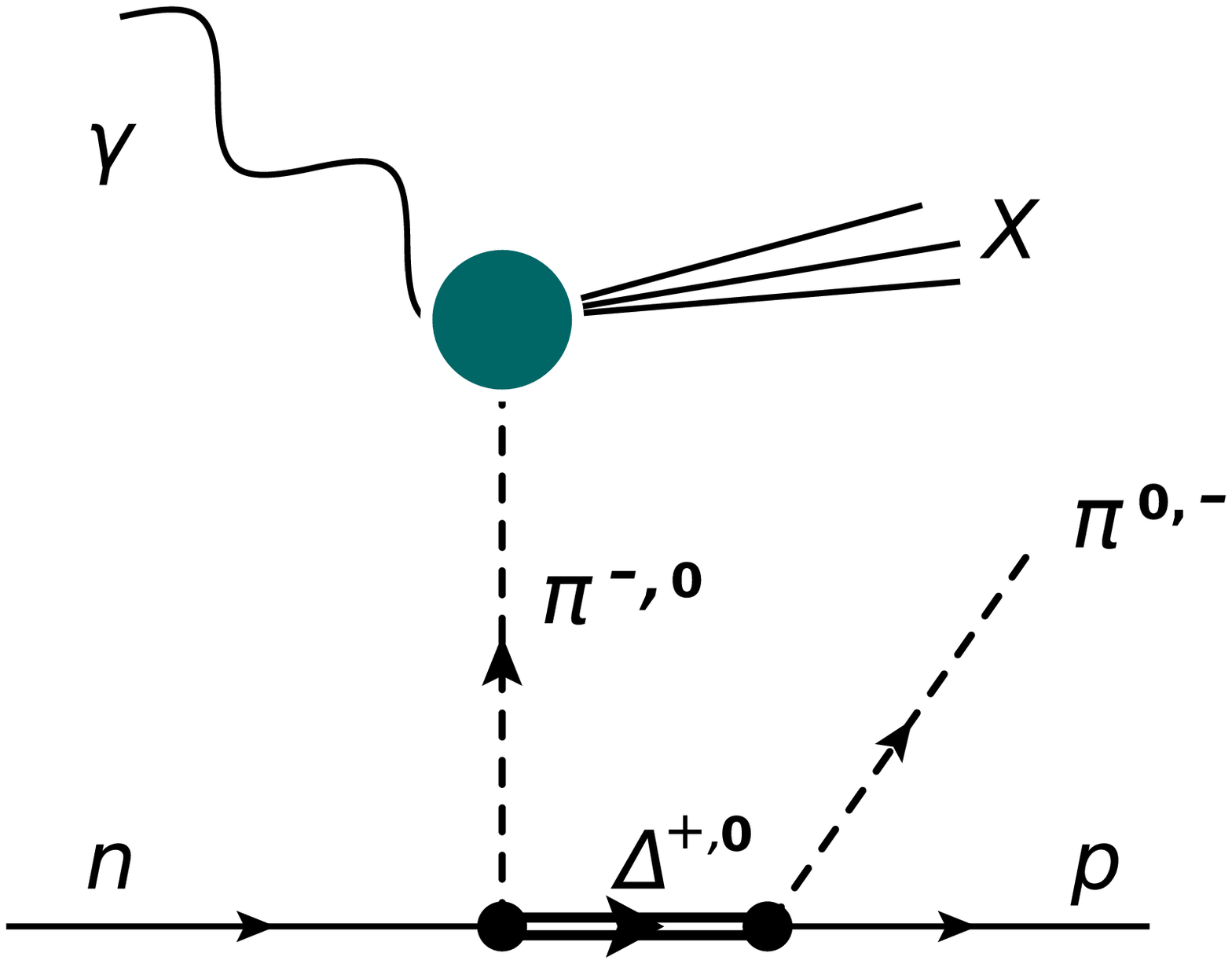}
\caption{
   Representative processes contributing to single-inclusive production
of forward protons off the neutron.}
\label{fig:feyn}
\end{figure}
As it happens, for specific regions of kinematics, the observation of low-momentum
recoil protons in the semi-inclusive reaction $e N \to e N X$ depicted (for select possibilities) in
Fig.~\ref{fig:feyn} can
reveal features associated with correlated $q\bar q$ pairs in the
nucleon, \IE~the nucleon's pion cloud as just described.
In particular, at low values of the $4$-momentum transfer squared
$t \equiv k^2 = (p-p')^2 \sim m^2_\pi$, where $p$ and $p'$ are the initial and
final nucleon $4$-momenta, the cross section displays behavior
characteristic of pion pole dominance --- akin to the behavior implied by
the relation in Eq.~(\ref{eq:tchan-pi}) for exclusive electroproduction.
In the DIS process at low $t$, contributions from exchanges of non-pseudoscalar quantum
numbers ($J^P = 0^-$), such as those of the $\rho$ meson and its isosinglet counterpart the $\omega$,
are suppressed. That this must be the case is made plain in the analysis surrounding the
right panel of Fig.~\ref{fig:f_MB_int}, whereby we argued that the prodigious energy gap between
the $\pi$ and $\rho$ masses suppresses the contribution of the latter to the $SU(2)$ Fock space
expansion. Furthermore, the direct fragmentation of the scattered quark, or spectator ``diquark''
system that remains after a quark is pulled out of the nucleon, generally produces
a considerably flatter $t$ dependence, which is qualitatively
different from behavior near the pion pole.

This general behavior should permit a precise extraction of the pion cloud's role in DIS from the
nucleon, and we perform here some illustrative phenomenological computations that make this
case; we do so by first exploring for context inclusive DIS in Sec.~\ref{sec:iDIS},
before performing in Sec.~\ref{sec:TDISii} analogous calculations of semi-inclusive cross sections, where
hadronic tagging of the final state may yield the desired information.

\section{Pion cloud for inclusive DIS}
\label{sec:iDIS}

Separate from its identity as a hadronic bound state, the pion may also
be understood according to its status as the pseudo-Goldstone boson
\cite{GellMann:1968rz} associated with spontaneous breaking of QCD's
chiral symmetry. That nature confers upon chiral perturbation theory
($\chi$PT) as formulated in terms of pionic degrees of freedom considerable
power in elucidating the dynamics of pion-nucleon interactions.

$\chi$PT specifies the appropriate pseudovector lagrangian for the pion-nucleon
interactions \cite{Burkardt:2012hk} as
\begin{eqnarray}
{\cal L}_{\pi N}
&=& {g_A \over 2 f_\pi}\,
    \bar\psi_N \gamma^\mu \gamma_5\,
        \bm{\tau} \cdot \partial_\mu \bm{\pi}\, \psi_N\
 -\ {1 \over (2 f_\pi)^2}\,
    \bar\psi_N \gamma^\mu\, \bm{\tau} \cdot
        (\bm{\pi} \times \partial_\mu \bm{\pi})\, \psi_N\ ,
\label{eq:piN-PV}
\end{eqnarray}
where the first term generates characteristic ``rainbow'' diagrams representing
the dominant piece of the one-loop contribution to the nucleon wavefunction
due to its ($N \rightarrow \pi N$) dissociations in the Sullivan process \cite{Sullivan:1971kd}.
This contribution allows us to write the $SU(2)$ counterpart to Eq.~(\ref{eq:Fock}) that we
encountered in the $SU(4)$ flavor sector:
\begin{equation}
|N\rangle\ =\ \sqrt{Z_2}\, \left| N \right.\rangle_0\
+\ \sum_{M,B} \int\! dz\, f_{MB}(z)\, | M(z); B(1-z) \rangle\ ,
\end{equation}
where the sum over states $(M,B)$ now runs over light hadrons represented by multiplets
like the $s=0$ entries of Fig.~\ref{fig:SU3_mult}; the resulting contribution to the
inclusive $F_2$ structure function of the nucleon from scattering
off a virtual pion emitted from the nucleon is therefore
\begin{equation}
F_2^{(\pi N)}(x)
= \int_x^1 dz\, f_{\pi N}(z)\ \cdot\ F_{2\pi}\Big(\frac{x}{z}\Big),
\label{eq:conv}
\end{equation}
where $z = k^+/p^+$ is the light-cone momentum fraction of the
initial nucleon carried by the interacting pion. For reasons of
more general compatibility with the standing literature related to 
semi-inclusive production of final state protons, we resort to $z$,
rather than $y$ for the light front momentum fraction $k^+/p^+$ as we
had in Chap.~\ref{chap:ch-charm}. We point out that the second term of
Eq.~(\ref{eq:piN-PV}) produces four-point pion-nucleon vertices responsible
for ``bubble'' diagrams that do not participate at $z \neq 0$; since the
present study is mainly concerned with observable effects at finite
$z$, we dispense with these contributions altogether. While this may be the
case, the bubble diagrams generated by the second term of Eq.~(\ref{eq:piN-PV}) 
{\it do} in fact contribute for $z \equiv 0$ --- an important and
largely uninvestigated consideration of consequence to sum rules like
Eq.~(\ref{eq:Gott-quark}).

In the infinite momentum frame, the light-front definition $z = k^+/p^+$
coincides with the longitudinal momentum fraction, whereas in the rest
frame of the target nucleon used in the subsequent treatment, $z$ is expressed
as
\begin{align}
z = \Big( &k_0 + |{\bm k}| \cos\theta \Big) \Big/ M\ , \nonumber\\ 
&k_0 = M - \sqrt{M^2+{\bm k}^2}\ ,
\end{align}
where yet again $M$ is the mass of the nucleon, $k_0$ the pion energy,
and $\theta$ is the angle between the vector $\bm k$ and the
$z$-axis (which is equal to the angle between the recoil proton
momentum ${\bm p}'$ and the photon direction).
For ease of notation, we also suppress the explicit dependence of
the structure functions on the scale $Q^2$.

In this setting,
the function $f_{\pi N}(z)$ gives the light-cone momentum distribution
of pions in the nucleon [similar to Eq.(\ref{eq:DLsplit})],
\begin{equation}
f_{\pi N}(z)
= c_I \frac{g_{\pi NN}^2}{16 \pi^2}
  \int_0^{\infty} \frac{dk^2_\perp}{(1-z)}
  \frac{G_{\pi N}^2}{z \ (M^2 - s_{\pi N})^2}
  \left( \frac{k^2_\perp + z^2 M^2}{1-z} \right),
\label{eq:piN-SF}
\end{equation}
where $k_\perp$ is the transverse momentum of the pion,
$g_{\pi NN}$ is the $\pi NN$ coupling constant,
and the isospin factor
$c_I=1$ for $\pi^0$ ($p \to p \pi^0$ or $n \to n \pi^0$) and
$c_I=2$ for $\pi^\pm$ ($p \to n \pi^+$ or $n \to p \pi^-$).
The function $G_{\pi N}$ parametrizes the momentum dependence
of the $\pi NN$ vertex function, which, due to the finite size
of the nucleon, suppresses contributions from large-$|\bm k|$
configurations.
Similar expressions may of course be written for various other contributions,
such as those of $\rho$ mesons or involving $\Delta$ baryons in an intermediate state
(\EG~the right panel of Fig.~\ref{fig:feyn}):
we have already developed the required technology in Chap.~\ref{chap:ch-charm} with
Eqs.~(\ref{eq:spin1-fMB} \& \ref{eq:DS*-split}). However, because of the
small mass of the pion, the $\pi N$ configuration is expected
to be the dominant one as we demonstrate below through explicit calculation.
Also, as in Chap.~\ref{chap:ch-charm} the variable
$s_{\pi N} = (k_\perp^2 + m_\pi^2)/z + (k_\perp^2 + M^2)/(1-z)$
of Eq.~(\ref{eq:piN-SF}) represents the total squared center of mass energy of the
intermediate $\pi N$ system, and is related to the pion virtuality $t$ by
$t-m_\pi^2 = z \cdot (M^2 - s_{\pi N})$.

\begin{figure}[t]
\vspace*{1.1cm}
\hspace*{-0.25cm}
\includegraphics[width=8cm]{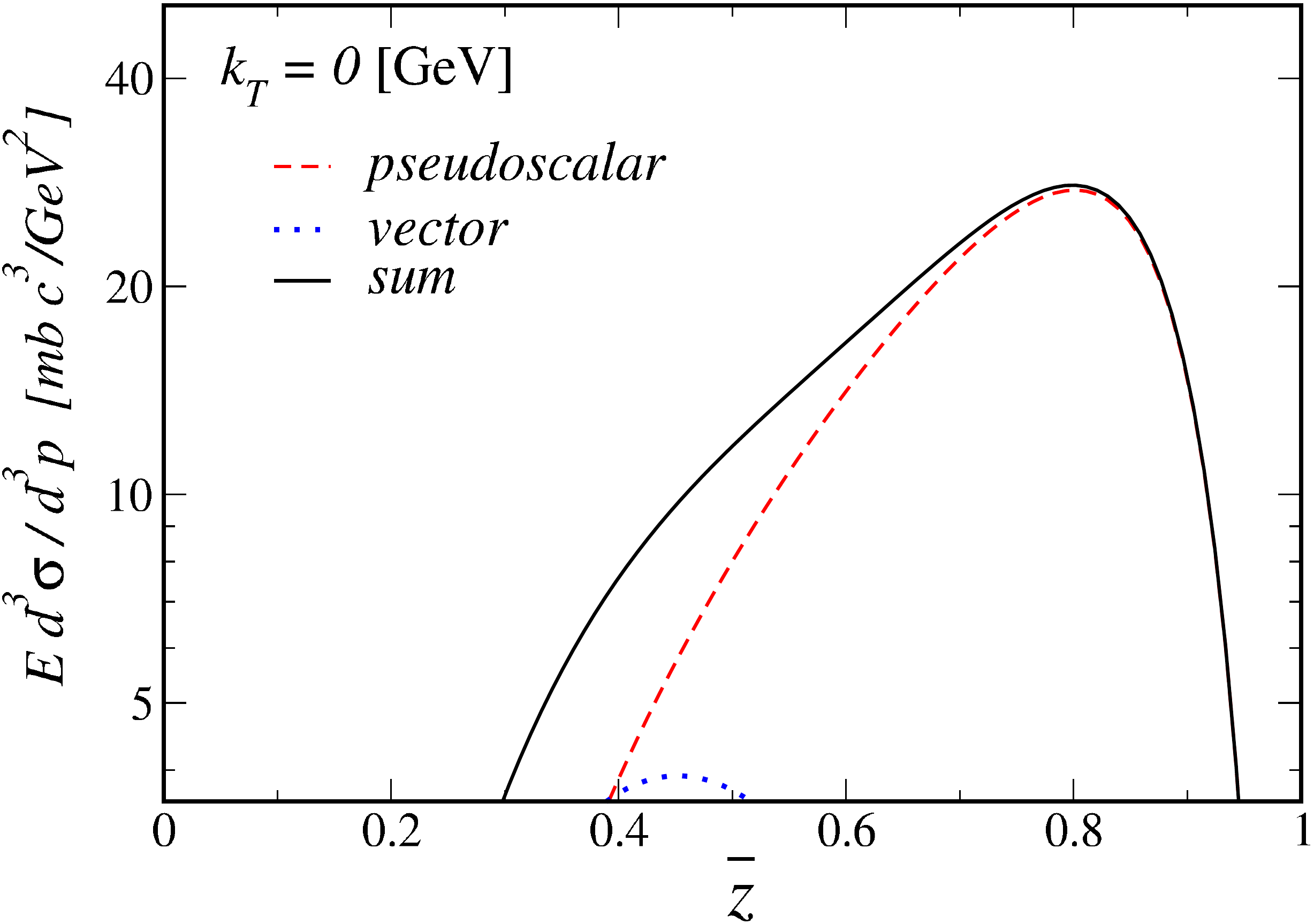} \ \ \ \ \ \
\includegraphics[width=7.8cm]{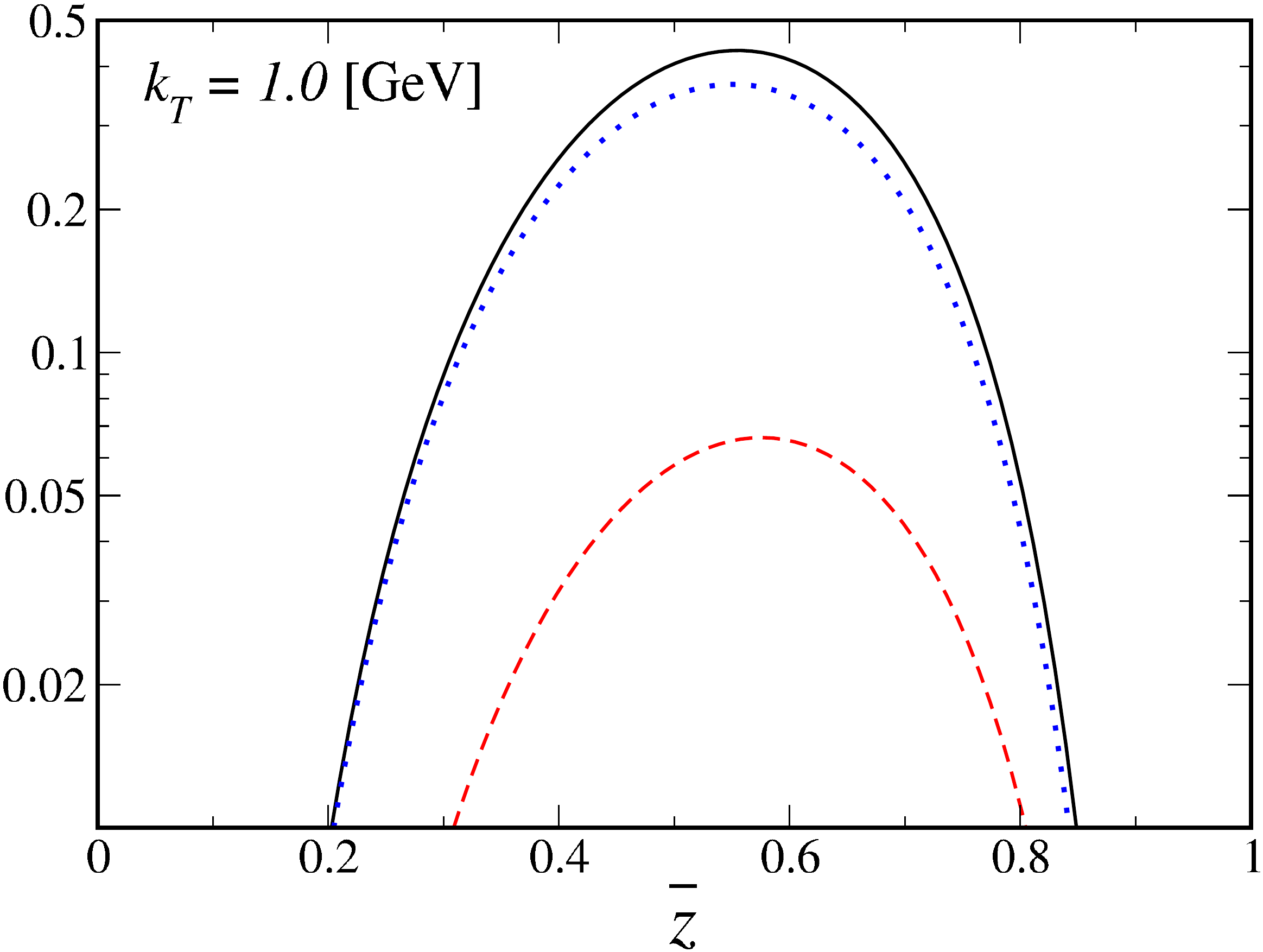}
\caption{Typical spectra for the differential cross section
	$E d^3\sigma/d^3p'$ in the $pp \to nX$ reaction
	for transverse momentum $k_\perp=0$ (left)
	and $k_\perp=1$~GeV (right), as a function of
	the light-cone momentum fraction $\bar z \equiv 1-z$.
	The pseudoscalar $\pi$ (red dashed lines) and vector
	$\rho$ (blue dotted lines) contributions, and their
	sum (black solid lines), are indicated explicitly.}
\label{fig:HSS}
\end{figure}

The form factor $G_{\pi N}$ (or more generally $G_{MN}$ for a
meson $M$) can be constrained by comparing the pion cloud
contributions with data on inclusive $pp \to nX$ scattering
(akin to the ISR $\Lambda_c$ production data used to fit
the analogous model parameters of Chap.~\ref{chap:ch-charm}.\ref{sec:mb}),
as performed by Holtmann {\it et al.} \cite{Holtmann96}.
For the purpose of this penultimate chapter, we use the parametric
form
\begin{equation}
G_{\pi N} = \exp\left[(M^2 - s_{\pi N})/\Lambda^2\right],
\label{eq:FF}
\end{equation}
where $\Lambda$ is the form factor cutoff parameter.
(Note that in Ref.~\cite{Holtmann96} a parametrization of
the form $\exp[(M^2 - s_{\pi N})/2\Lambda^2]$ is used,
so that the corresponding cutoffs there are smaller by a
factor of $\sqrt{2}$.)
An illustration of the typical spectra for the differential
cross section $E d^3\sigma/d^3p'$ in the $pp \to nX$ reaction
arising from $\pi$ and $\rho$ exchange is shown in
Fig.~\ref{fig:HSS} as a function of the light-cone momentum
fraction $\bar z \equiv 1-z$ carried by the final nucleon,
for two values of the transverse momentum $k_\perp$.
For small $k_\perp$ the $\pi$ exchange contribution clearly
dominates at all $\bar z$, while at larger momenta the
contributions from heavier mesons such as the $\rho$ become
more important. Again, the specifics of this momentum dependent
balance between pseudoscalar and vector exchange mechanisms is
largely accounted for by the profile shown in the right panel of
Fig.~\ref{fig:f_MB_int}.

\begin{figure}[t]
\vspace*{1cm}
\includegraphics[width=7.9cm]{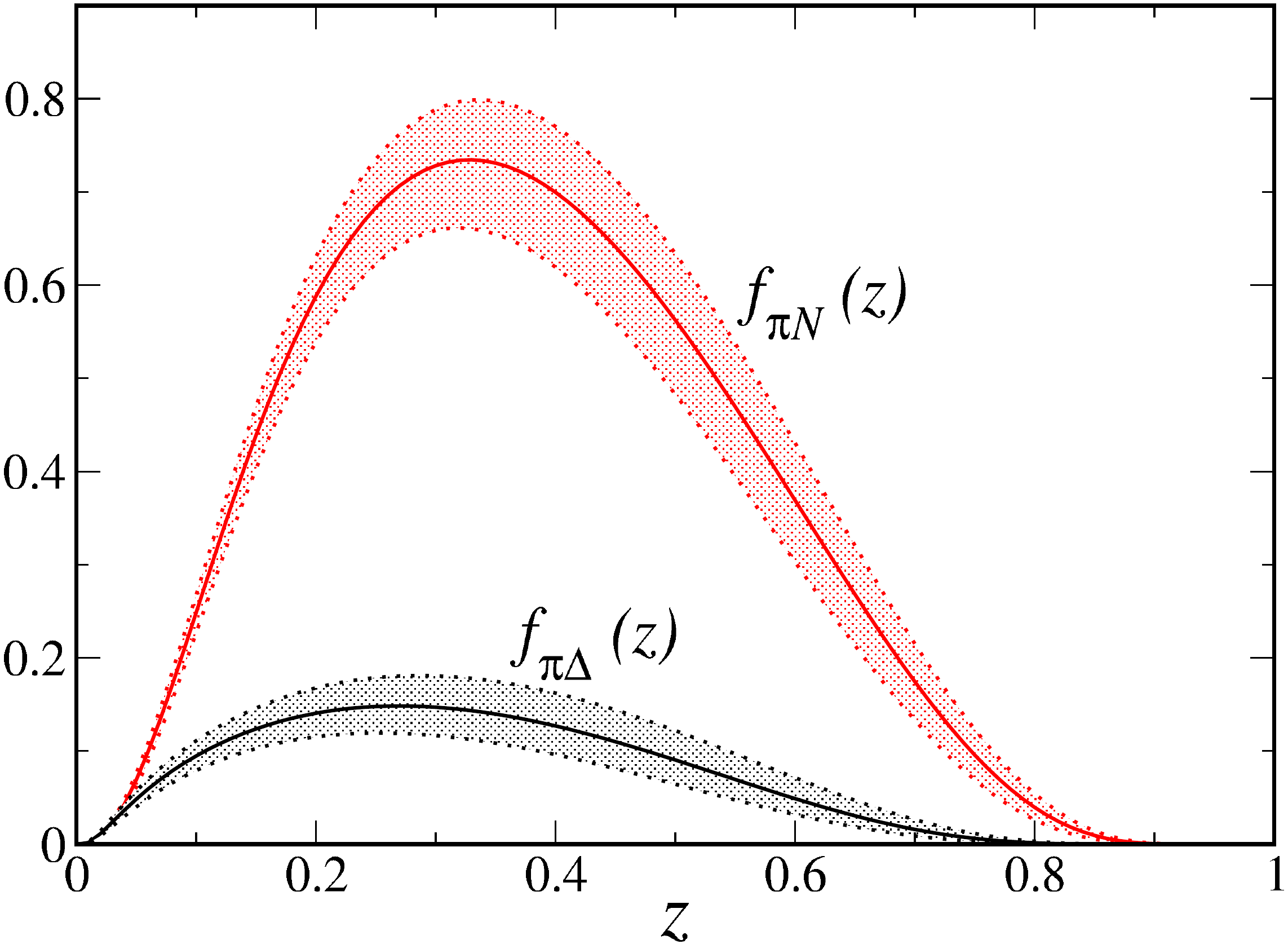} \ \ \ \ \ \ \ \ \
\includegraphics[width=7.6cm]{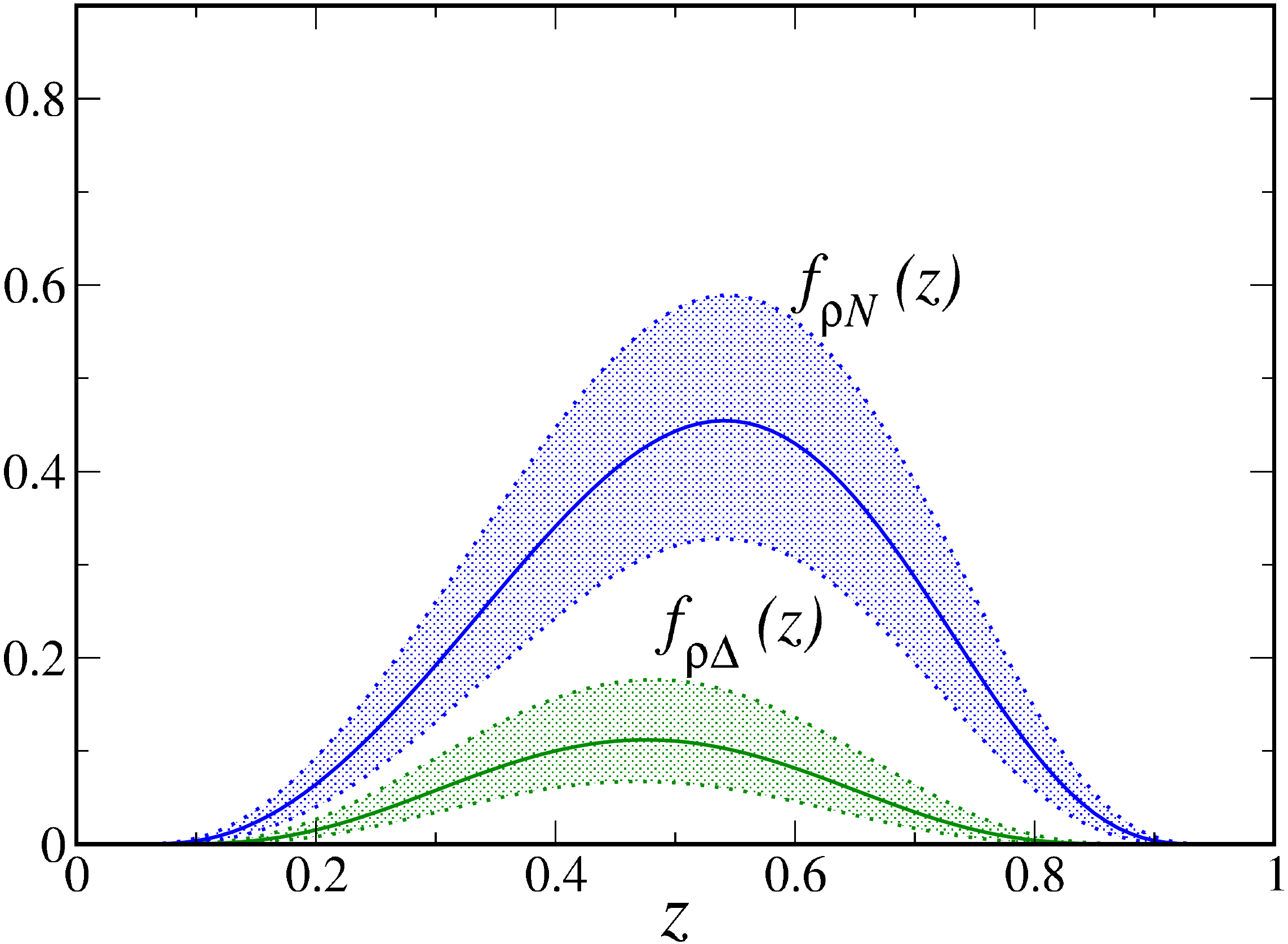}
\caption{Light-cone momentum distributions of the pion,
	$f_{\pi N}$ and $f_{\pi \Delta}$ (left)
	and the $\rho$ meson, $f_{\rho N}$ and $f_{\rho \Delta}$
	(right), as a function of the meson light-cone
	momentum fraction $z$.
	The error bands correspond to the cutoff parameter ranges
	as given in the text.}
\label{fig:fpi-rho}
\end{figure}
\begin{figure}[ht]
\vspace*{1.4cm}
\includegraphics[width=7.9cm]{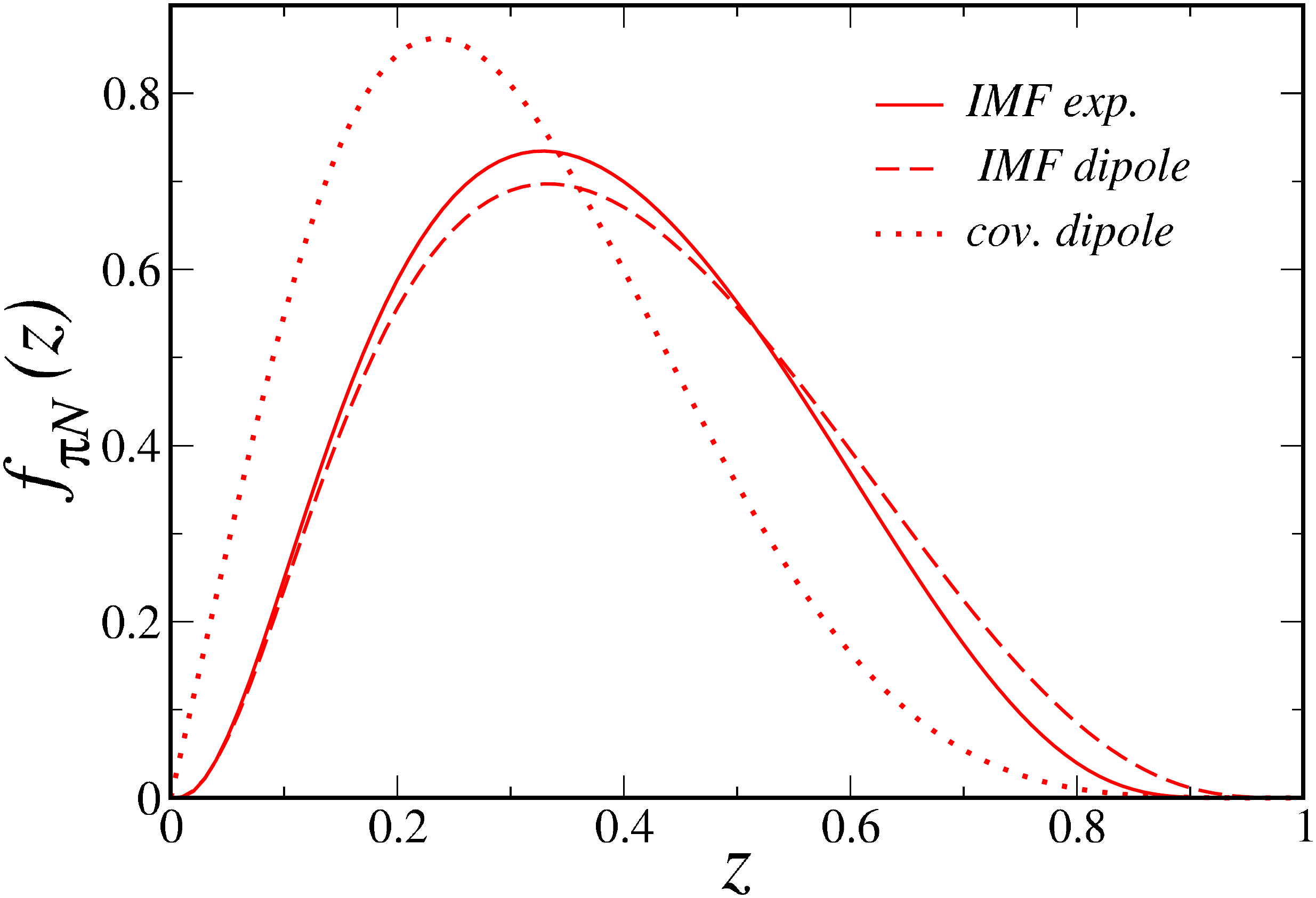} \ \ \ \ \ \
\includegraphics[width=7.6cm]{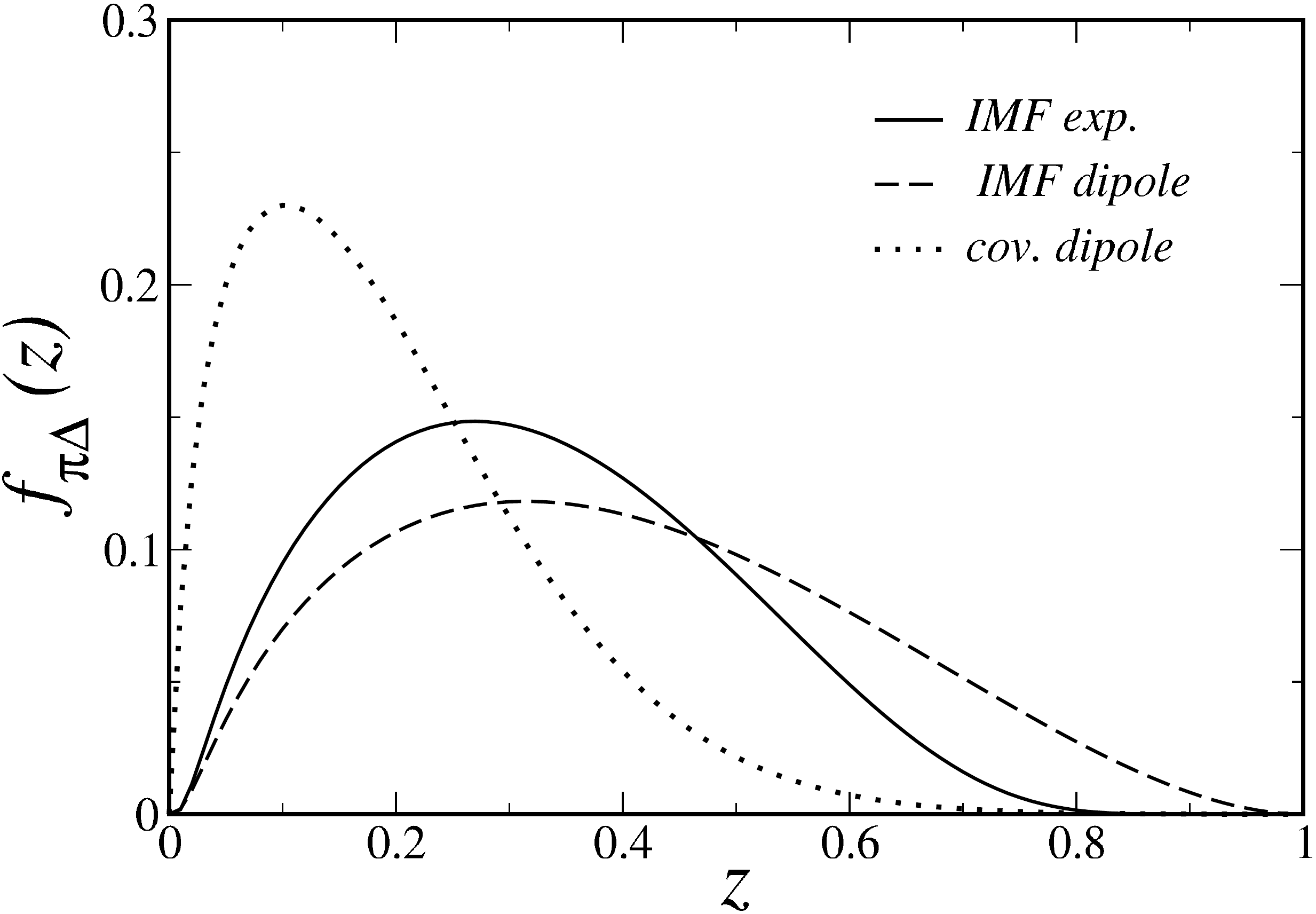}
\caption{Light-cone momentum distributions for the $\pi N$
	(left) and $\pi \Delta$ (right) intermediate
	states, for several different functional forms of the
	form factor $G$ in Eq.~(\ref{eq:piN-SF}):
	``IMF'' refers to $s$-dependent forms such as in
	Eq.~(\ref{eq:FF}), while ``cov'' denotes a form factor
	that depends only on the variable $t$.}
\label{fig:FFdep}
\end{figure}

Using cutoff parameters constrained by the inclusive hadronic
$pp \to n X$ data, which were found in Ref.~\cite{Holtmann96} and reproduced here as
$\Lambda_{\pi N} = \Lambda_{\rho N} = 1.56 \pm 0.07$~GeV and
$\Lambda_{\pi \Delta} = \Lambda_{\rho \Delta} = 1.39 \pm 0.07$~GeV,
the light-cone momentum distributions $f(z)$ are shown in
Fig.~\ref{fig:fpi-rho}.
The principal model uncertainty in these results comes from the
ultraviolet regulator $G_{MN}(s)$ used to truncate the $k_\perp$ integrations
in the distribution functions.  Various functional forms have been
advocated in the literature aside from the $s$-dependent exponential
form factor in Eq.~(\ref{eq:FF}), and we compare several of these,
including $s$- and $t$-dependent dipole forms) in Fig.~\ref{fig:FFdep}.
For the $s$- and $t$-dependent forms in particular, the differences
are noticeable mostly at small values of $z$, where the $t$-dependent
parametrization (of the form $G \sim 1/(t-\Lambda^2)^2$)
gives somewhat larger distributions that are peaked at smaller $z$
compared to the $s$-dependent form, which tend to be broader.

\begin{figure}[h]
\vspace*{1.5cm}
\includegraphics[height=7.5cm]{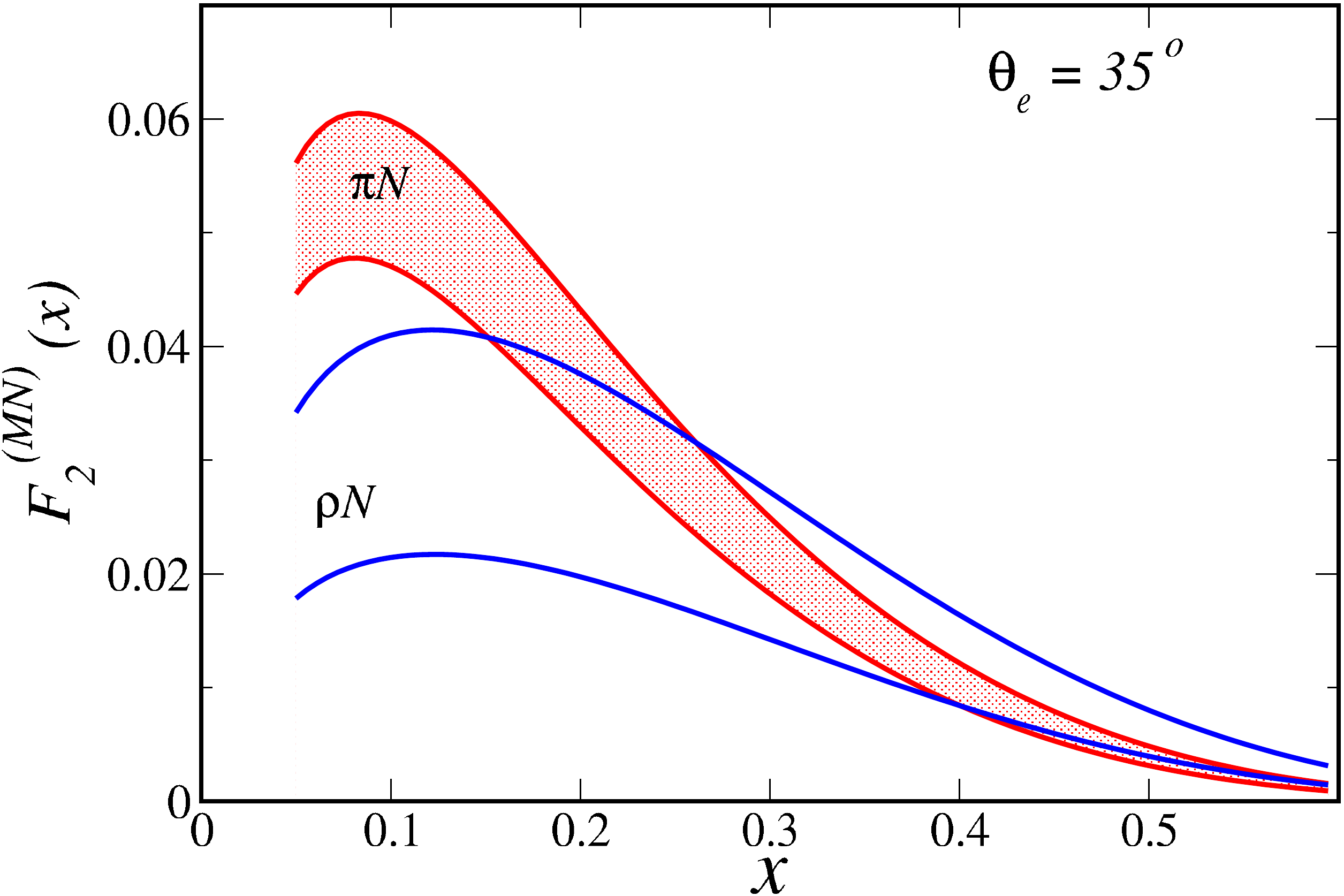}
\caption{Contributions from $\pi N$ and $\rho N$ intermediate
	states to the inclusive $F_2$ structure function of
	the proton for a typical fixed electron scattering angle.}
\label{fig:kT-int}
\end{figure}

Convoluting the light-cone distributions with the structure function
of the meson as in Eq.~(\ref{eq:conv}), the resulting contributions
from the $\pi N$ and $\rho N$ intermediate states to the inclusive
$F_2$ structure function of the proton are illustrated in
Fig.~\ref{fig:kT-int}.
For the meson structure function we use a parametrization
from GRV \cite{GRVpi}, and assume $F_{2\pi}(x) \approx F_{2\rho}(x)$.
The results are plotted for a plausible value of the scattering angle
of the final state electron $\theta_e \sim 35^\circ$, which determines the $Q^2$
dependence of the contribution at a given $x$ via a few simple
manipulations of the DIS definitions given in
Chap.~\ref{chap:ch-DIS}.\ref{sec:DIS}:
\begin{equation}
Q^2\ =\ 2xME\ \left\{ 1\ -\ \left({2E \over xM} \sin^2{{\theta_e \over 2}}\ +\ 1\right)^{-1} \right\}\ .
\end{equation}
We thus find that for typical lepton scattering angles at, \EG~JLab with
$E = 11$ GeV, the sensitivity to $Q^2$ is rather small at shallow $x \lesssim 0.2$
due to the mild scale dependence of the meson structure functions used.
For the fully integrated results of Fig.~\ref{fig:kT-int}, the model
uncertainties are greatest for the lowest accessible values of
$x \sim 0.05$; depending upon choice of the phenomenological
cutoff parameter $\Lambda$, the $\pi$ contribution to $F_2$ can either
be comparable to that of the $\rho$, or even larger at high values
of $x$ (where the calculation is also less reliable).

%
\section{Tagged structure functions}
\label{sec:TDISii}

While the inclusive reactions require integration of the pion
momentum over all possible values, detecting --- ``tagging'' --- the recoil proton
in the final state allows one to dissect the internal structure
with significantly more detail and increase the sensitivity to
the dynamics of the meson exchange reaction; this is attributable
to the fact that final state tagging effectively isolates contributions
of the type shown in Fig.~\ref{fig:feyn} in specific kinematical bins of
produced hadron momenta, angle, etc.
Our interests lie foremost in the relative contributions of
the semi-inclusive reaction with respect to the inclusive process.
In practice, the semi-inclusive structure function will be given
by the unintegrated product 
\begin{equation}
F_2^{(\pi N)}(x,z,k_\perp)
= f_{\pi N}(z,k_\perp)\ \cdot\ F_{2\pi}\Big(\frac{x}{z}\Big)\ ,
\label{eq:sidis}
\end{equation}
where the unintegrated distribution function $f_{\pi N}(z,k_\perp)$
is defined by
\begin{equation}
f_{\pi N}(z)
= {1 \over M^2} \int_0^{\infty} dk^2_\perp\ f_{\pi N}(z,k^2_\perp)\ .
\label{eq:incl}
\end{equation}

\begin{figure}[ht]       
\vspace*{1cm}
\includegraphics[width=9.0cm]{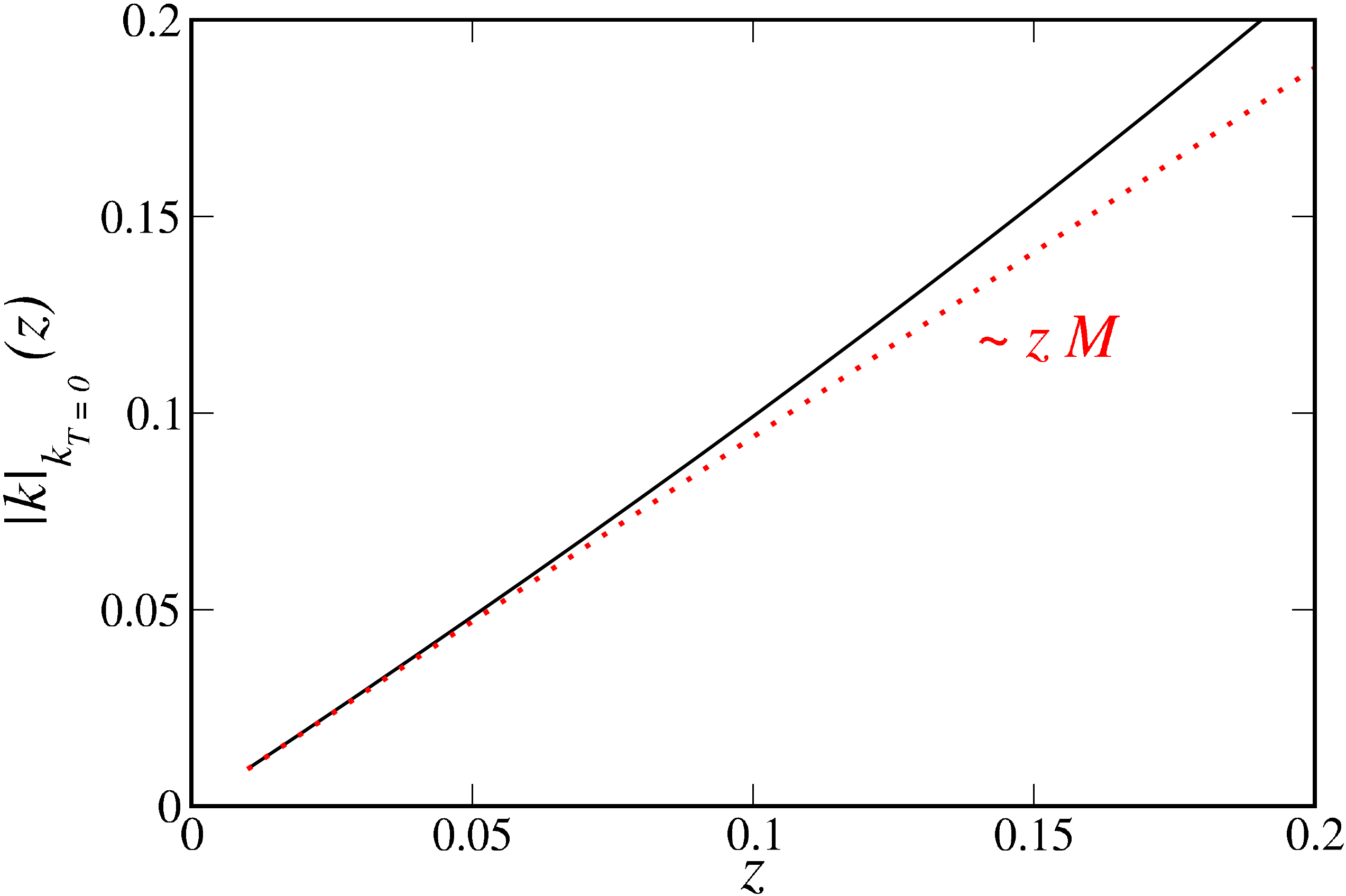}
\caption{Pion momentum $|\bm{k}|$ as a function of the light-cone
	fraction $z$ for $k_\perp = 0$ (black solid).  The linear
	approximation $\sim z M$ (red dotted) is shown for comparison.}
\label{fig:kMag_y}  
\end{figure}

The dependence of the tagged structure functions on the kinematical
variables that are measured experimentally can be studied by relating
the magnitude of the 3-momentum ${\bm k}$ of the exchanged pion in
the target rest frame to the pion's transverse momentum $k_\perp$
and light-cone fraction $z$,
\begin{align}
\bm{k}^2
&= k^2_\perp + k^2_\parallel \nonumber\\
&= k^2_\perp
+ \frac{\left[ k^2_\perp + (1 - [1-z]^2) M^2 \right]^2}
       {4 M^2 (1 - z)^2}\ .
\label{eq:kvec}
\end{align}
Experimentally, the quantities most readily measured are the
momentum of the produced proton, ${\bm p'}$, which in the
rest frame is ${\bm p'} = - \bm{k}$, and the scattering angle
$\theta_{p'}$ of the final state proton with respect to the virtual
photon direction.
%
%
In the limit $k^2_\perp = 0$, the magnitude of ${\bm k}$ becomes
\begin{equation}
|\bm{k}|_{k^2_\perp = 0}
= \frac{z M}{2} \left( \frac{2-z}{1-z} \right)\ ,
\label{eq:kvec_lim}
\end{equation}
which imposes the restriction $z \lesssim |\bm{k}| / M$.
This relation is illustrated in Fig.~\ref{fig:kMag_y} for
values of $z$ up to 0.2, and establishes the bound placed upon
the momentum fraction $z$ imposed by the soft limit proximate to
the $t = + m^2_\pi$ pole.

\begin{figure}[ht]
\vspace*{1cm}
\includegraphics[width=7.8cm]{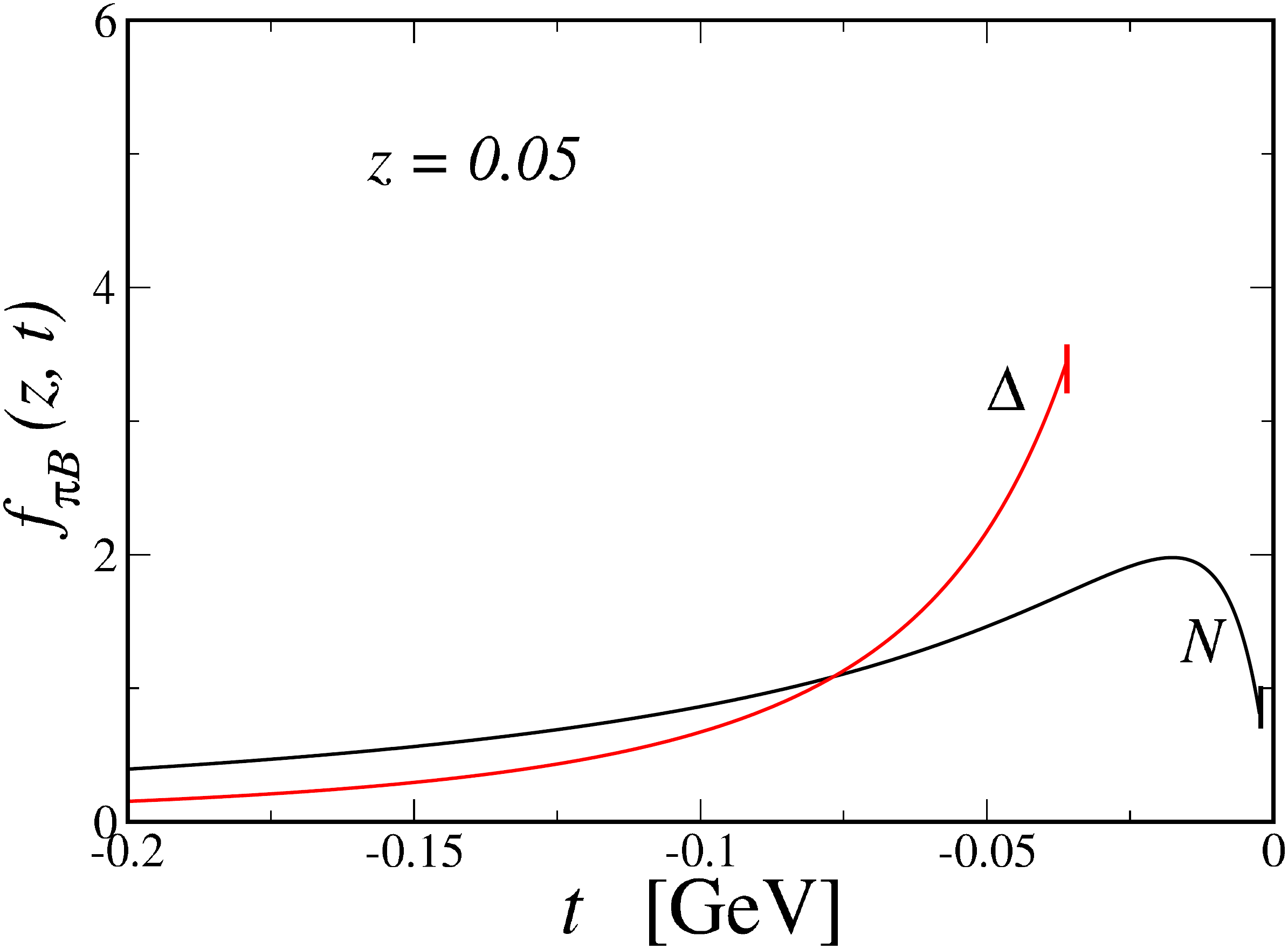} \ \ \ \ \ \
\includegraphics[width=7.8cm]{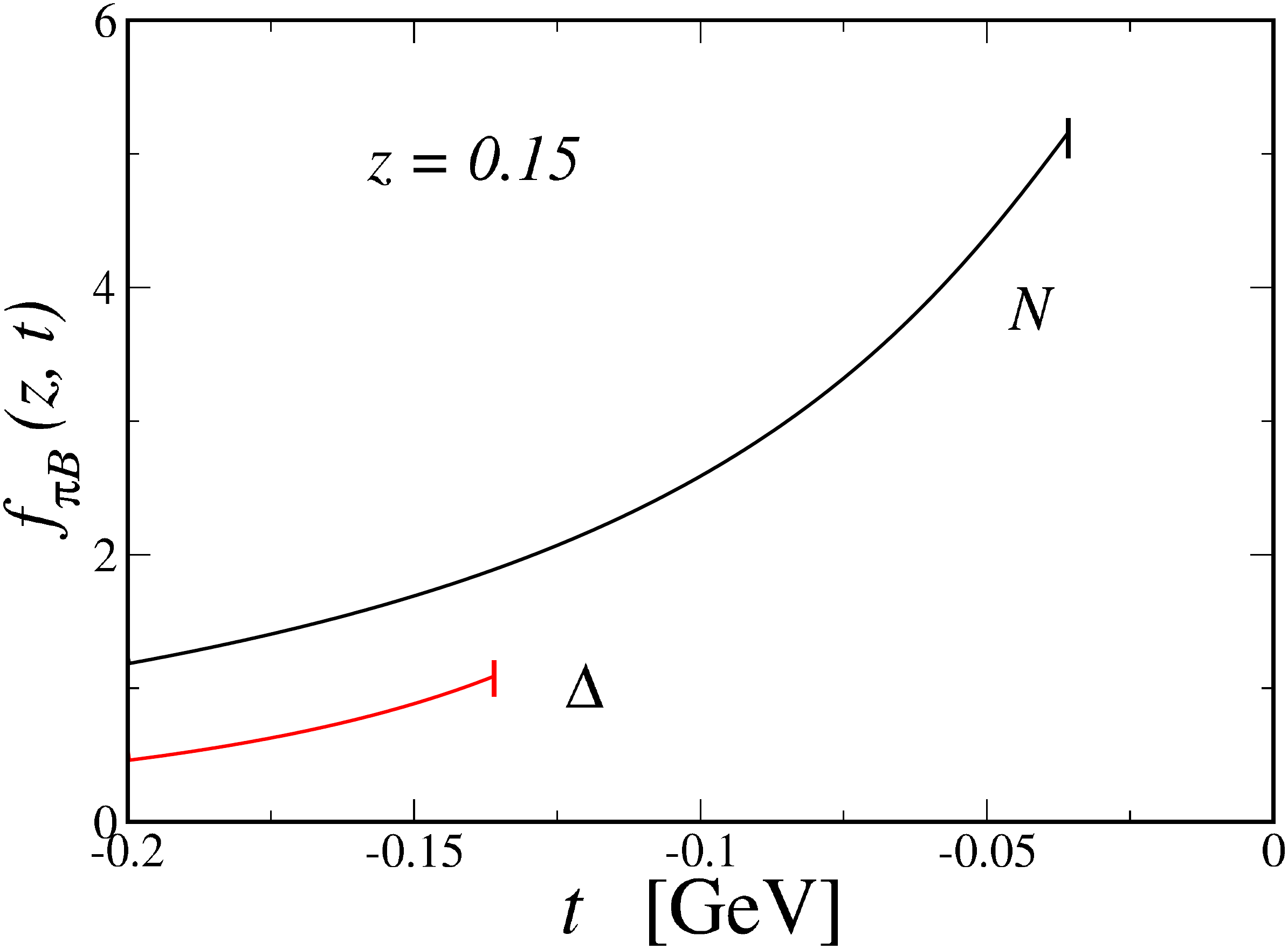}
\caption{Unintegrated light-cone distribution functions	for
	$\pi N$ (black solid) and $\pi \Delta$ (red solid)
	states as a function of $t$, for fixed values of
	$z=0.05$ (left)	and $z=0.15$ (right).}
\label{fig:t-dist}
\end{figure}

This kinematic restriction on $|\bm{k}|$ for a given $z$ can also be
illustrated by considering the unintegrated light-cone distribution
functions as a function of the variable $t$.
This is relevant since one way of identifying the pion exchange
mechanism is through its characteristic $t$ dependence, which
is pronounced near the pion pole.
The production of a physical proton (or $\Delta$ baryon)
in the final state restricts the maximum value of $t$, however
(corresponding to the minimum transverse momentum, $k_\perp=0$), to
\begin{equation}
t^{N}_{\rm max} = -{M^2 z^2 \over 1-z}\ ,\ \ \ \
t^{\Delta}_{\rm max}
= -{\left( M_\Delta^2 - (1-z) M^2 \right) z \over 1-z}\ ,
\label{eq:tlim}
\end{equation}
for final state nucleons $N$ and intermediate $\Delta$ states, respectively.
Implementing these limits, the $t$-dependence of the distributions
for $\pi$ exchange with a nucleon or $\Delta$ recoil is illustrated
in Fig.~\ref{fig:t-dist}.  Note that at the larger $z$ value there
is a considerable gap between the values of $t$ at which $\Delta$
production is possible compared with $N$ production.

The most judicious way of proceeding experimentally would then be to measure
the semi-inclusive cross section in specific bins of recoil proton momentum
$|\bm p'| = |\bm k|$ and scattering angle $\theta_{p'}$ (or equivalently $z$
and $k_\perp$). For this purpose, we hence define a partially integrated semi-inclusive
structure function $F_2^{(\pi N)}(x,\Delta z,\Delta k_\perp^2)$,
\begin{equation}
F_2^{(\pi N)}(x,\Delta z,\Delta k_\perp^2)
= {1\over M^2} \int_{\Delta z} dz \int_{\Delta k_\perp^2} dk^2_\perp\
  f_{\pi N}(z,k_\perp)\ \cdot\ F_{2\pi}\Big(\frac{x}{z}\Big)\ ,
\label{eq:sidis_int}
\end{equation}
integrated over the range $\Delta z = [z_{\rm min}, z_{\rm max}]$
and $\Delta k_\perp^2 = [k_{\perp \rm min}^2, k_{\perp \rm max}^2]$.
Of course, this is composed in the ``theory'' space spanned by $(z,k^2_\perp)$,
and lab frame expressions would be preferred; one can therefore translate to an
alternative expression for the semi-inclusive structure function integrated over
the target-frame variables $|\bm k|$ and $\theta_{p'}$.
With this, we shall evaluate, \EG
\begin{equation}
F_2^{(\pi N)}(x, \Delta |{\bm k}|, \Delta \theta_{p'})
= {1\over M^2} \int_{\Delta |{\bm k}|} d|{\bm k}| \int_{\Delta \theta_{p'}}d\theta_{p'}\
  J(x,|{\bm k}|,\theta_{p'})\ \cdot\
  f_{\pi N}(\tilde{z},\tilde{k}_\perp)\, F_{2\pi}\Big(\frac{x}{\tilde{z}}\Big)\ ,
\label{eq:sidis_int-vc}
\end{equation}
where now $\Delta |{\bm k}| = [|{\bm k}|_{\rm min}, |{\bm k}|_{\rm max}]$,
$\Delta \theta_{p'} = [\theta_{p'}^{\rm min}, \theta_{p'}^{\rm max}]$, and the
Jacobian appropriate to the lab-frame transformations
\begin{subequations}
\begin{align}
\tilde{z}(|{\bm k}|,\theta_{p'})\ &=\ {1 \over M} |{\bm k}| \cos{\theta_{p'}}\
+\ {1 \over M} \left( M - \sqrt{M^2 + |{\bm k}|^2} \right) \\
\tilde{k}_\perp (|{\bm k}|,\theta_{p'})\ &=\ |{\bm k}| \sin{\theta_{p'}}
\end{align}
\end{subequations}
can be put down as
\begin{align}
J(x,|{\bm k}|,\theta_{p'})\ &\defeq\ {\partial(x,\tilde{z},\tilde{k}^2_\perp) \over \partial(x,|{\bm k}|,\theta_{p'})} \nonumber\\
&=\ {2 \over M} |{\bm k}|^2 \sin{\theta_{p'}} \left(1 - \sin{\phi_k} \cdot \cos{\theta_{p'}} \right)\ ,
\end{align}
and we have defined the parameter $\phi_k \defeq \tan^{-1}(|{\bm k}|/M)$.
JLab experiments such as that proposed in \cite{TDIS} promise to probe the kinematical ranges
$0.05 \lesssim z \lesssim 0.2$ and $60 \lesssim |\bm k| \lesssim 400$~MeV,
and angles $30^\circ \lesssim \theta_{p'} \lesssim 160^\circ$, with $x$ in
the vicinity of $x \sim 0.05-0.2$, wherein we trace some of the more salient signals for DIS
from the nucleon's pion cloud using tagged structure functions such as Eq.~(\ref{eq:sidis_int-vc}).

\begin{figure}[h]
\vspace*{1.4cm}
\includegraphics[height=7.5cm]{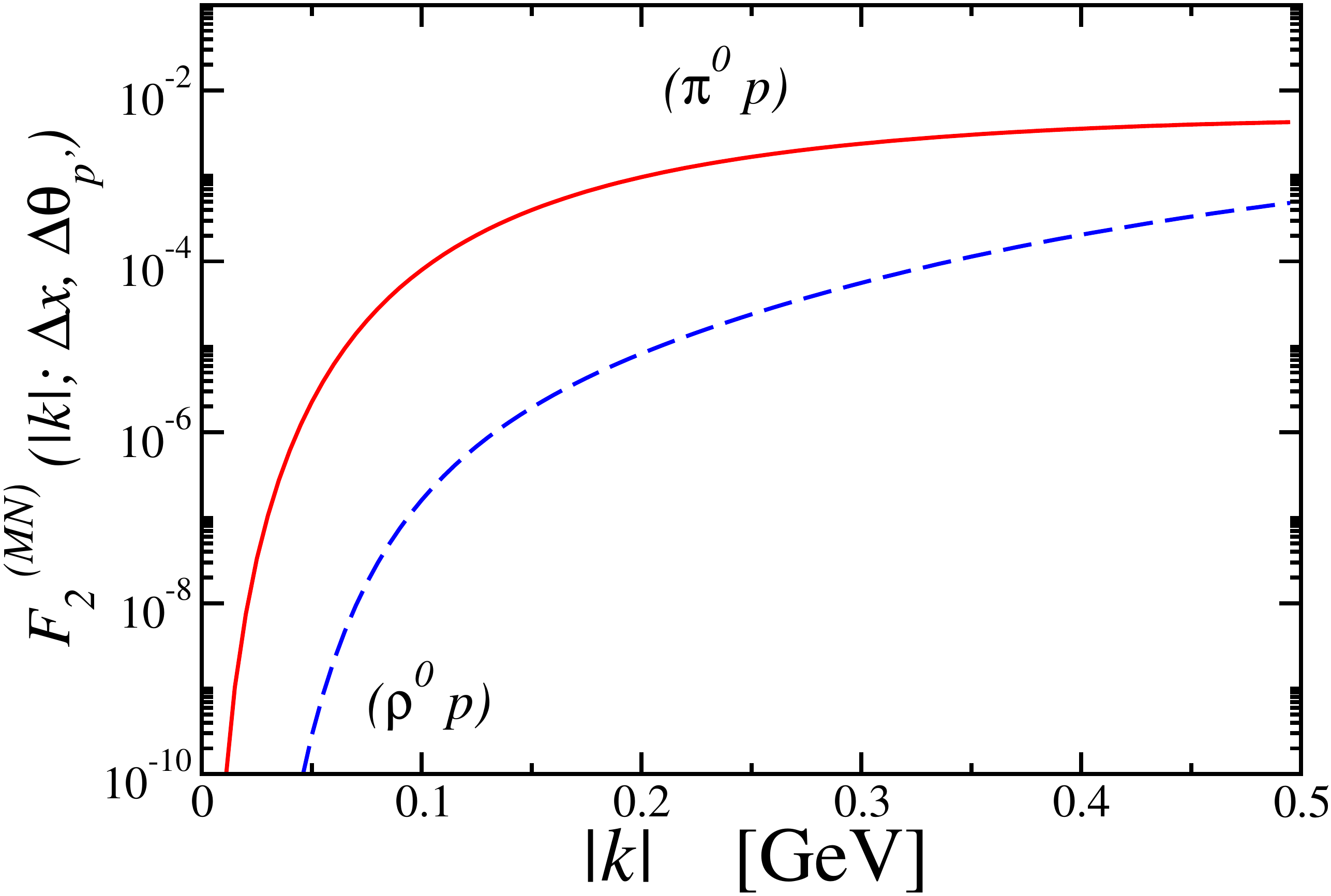}
\caption{
	Semi-inclusive structure functions
	$F_2^{(M N)}(|{\bm k}|;\Delta x,\Delta \theta_{p'})$
	over an illustrative range for
        $|{\bm k}|$ up to 0.5~GeV for the $p \to M\, p$ process,
	with $M=\pi^0$ (red solid) and $M=\rho^0$ (blue dashed),
	as a function of the recoil proton momentum $|{\bm k}|$,
	integrated over $\Delta x = [0,0.6]$ and all angles $\theta_{p'}$.}
\label{fig:F2MN_k}
\end{figure}

To that end, Fig.~\ref{fig:F2MN_k} shows the semi-inclusive structure functions\linebreak
$F_2^{(M N)}(|{\bm k}|;\Delta x,\Delta \theta_{p'})$ for the neutral-exchanges
$p \to \pi^0\, p$ and $p \to \rho^0\, p$, as a function of the
momentum $|{\bm k}|$, integrated over $x$ between 0 and 0.6,
and over all angles $\theta_{p'}$ from 0 to $\pi$. The $\rho$ contribution
is clearly suppressed relative to the pion contribution, with the structure functions
rising steadily at increasing $|{\bm k}|$ in the sampled region
$|{\bm k}| \lesssim 0.5$~GeV. At larger momenta, beyond the kinematic region plotted
in Fig.~\ref{fig:F2MN_k}, the effects of the meson--nucleon form factors become more
important, eventually suppressing the contributions from high-$|{\bm k}|$ tails of the distributions.
The peak in the $\pi$ distribution occurs at $|{\bm k}| \approx 0.6$~GeV, while the $\rho$
distribution peaks at higher momenta, $|{\bm k}| \approx 1.2$~GeV, and has a slower
fall-off with $|{\bm k}|$.

\begin{figure}[h]
\vspace*{1.4cm}
\includegraphics[width=8cm]{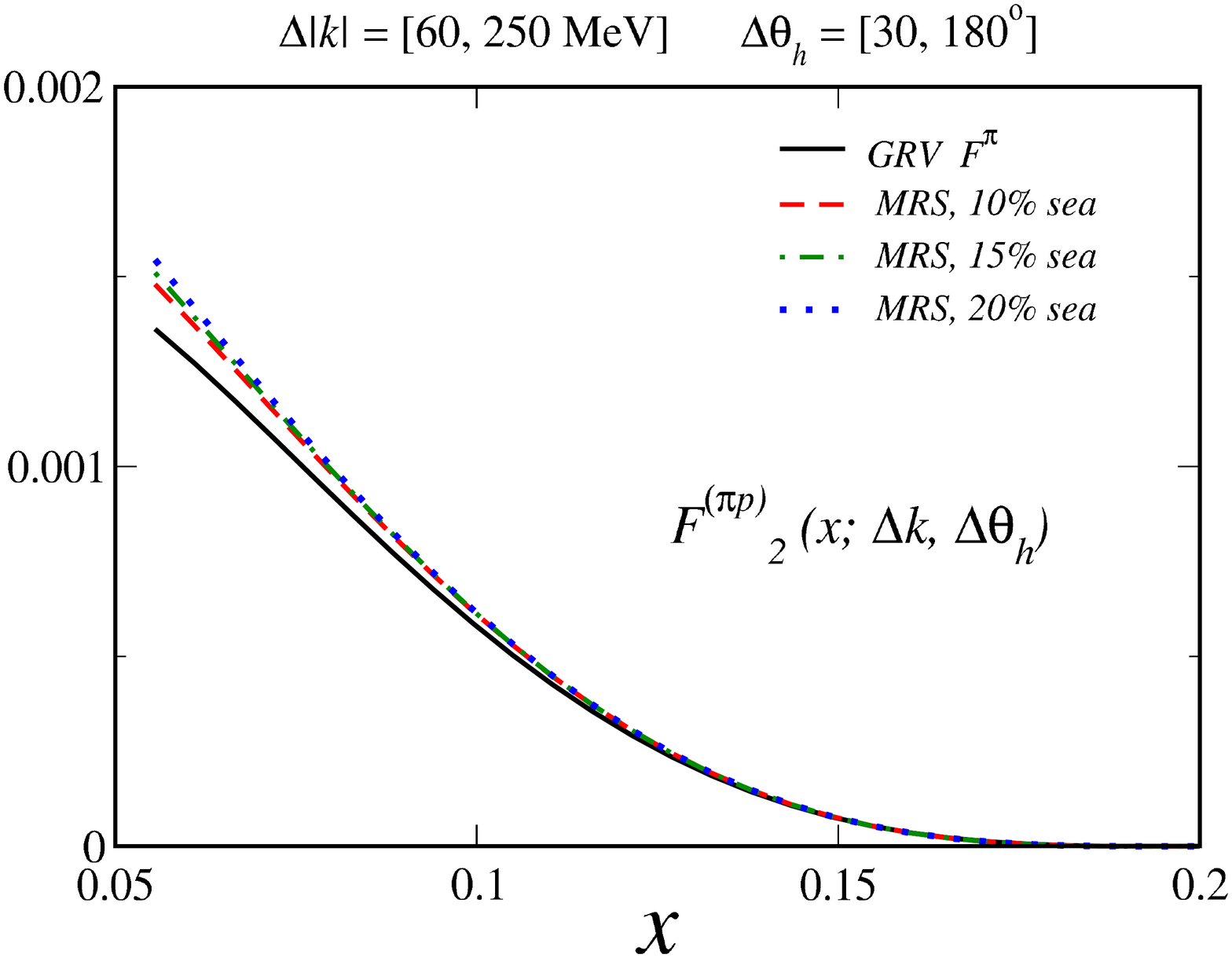} \ \ \
\includegraphics[width=8cm]{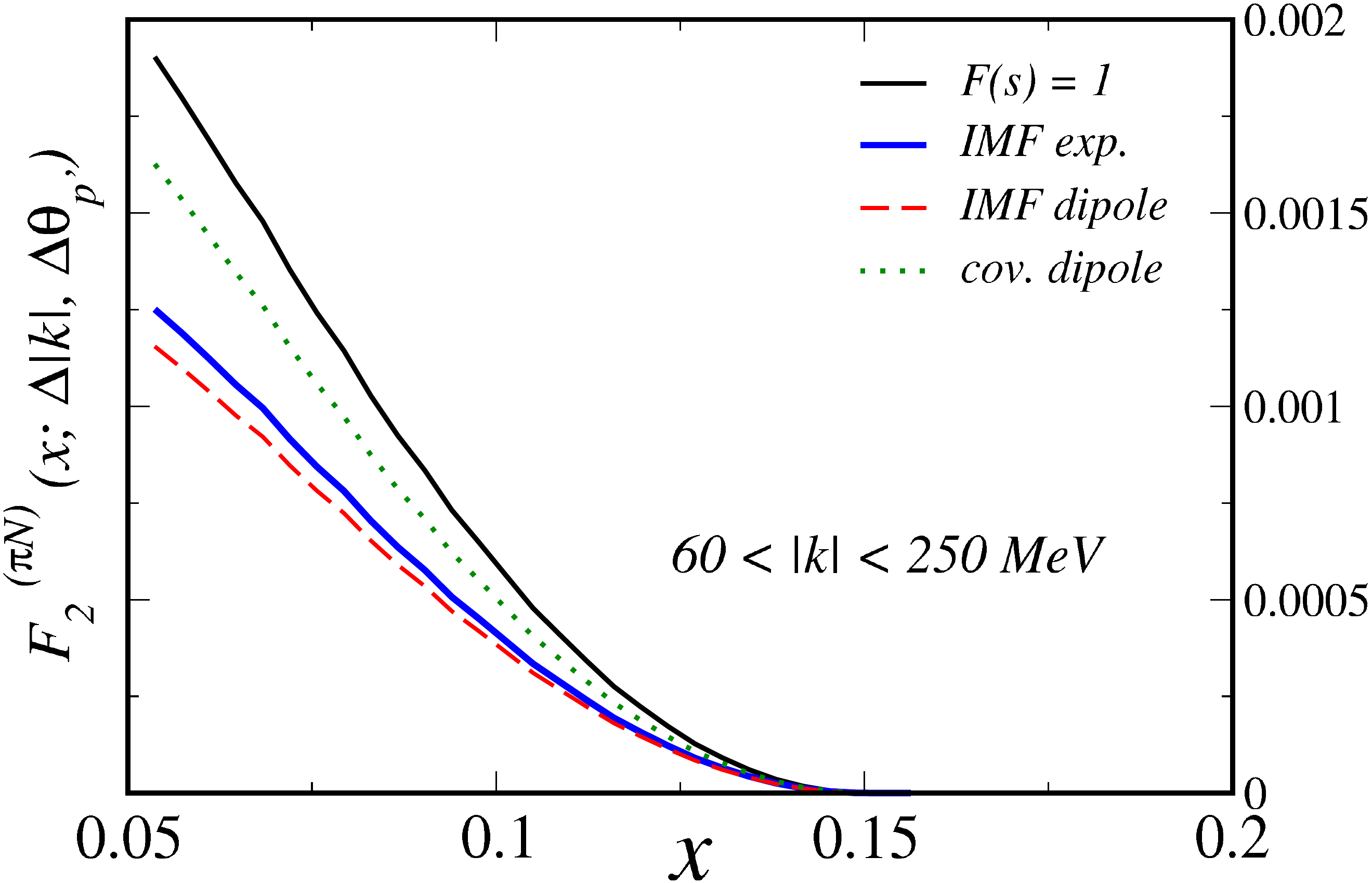}
\caption{(Left) Semi-inclusive structure function 
	$F_2^{(\pi N)}(x,\Delta |{\bm k}|,\Delta \theta_{p'})$
	for the $p \to \pi^0\, p$ process, integrated over
	the momentum range $\Delta |{\bm k}| = [60,250]$~MeV
	and angular range $\Delta \theta_{p'} = [30^\circ,180^\circ]$.
	The results with the GRV \cite{GRVpi} (black solid)
	parametrization of the pion structure function are
	compared with those using the MRS \cite{MRSpi} fit
	with different amounts of sea, 10\% (red dashed),
	15\% (green dot-dashed) and 20\% (blue dotted).
        (Right) The same quantity, but now examining the
        variations due to different choices for the vertex form factor
        $G_{\pi N}$, as explicitly indicated above.
}
\label{fig:F2pi_dep}
\end{figure}

The dependence of the semi-inclusive structure function
$F_2^{(\pi N)}(x,\Delta |{\bm k}|,\Delta \theta_{p'})$ on the
pion structure function parametrization is shown in the left of
Fig.~\ref{fig:F2pi_dep} as a function of $x$, integrated
over the proposed momentum range $\Delta |{\bm k}| = [60,250]$~MeV
and angular range $\Delta \theta_{p'} = [30^\circ,180^\circ]$.
The results with the GRV parametrization \cite{GRVpi} of the
pion parton distribution functions are compared with those
using the MRS parametrization \cite{MRSpi} with different
amounts of sea, ranging from 10\% to 20\%.
The pion structure function is relatively well constrained
from pion--nucleon Drell-Yan data at Fermilab at intermediate
and large values of $x$, but is not as well determined at
small $x$ values.

Nevertheless, the variation in the computed semi-inclusive
proton structure function from uncertainties in the pion
distribution functions is quite small, and considerably
smaller than the uncertainties from the pion--nucleon
vertex form factor dependence, which is illustrated in the right
panel of Fig.~\ref{fig:F2pi_dep}. Specifically, we exhibit the
semi-inclusive structure functions as a function of $x$ and
integrated over the momentum range $\Delta |\bm{k}|=[60,250]$~MeV;
we compare the $s$-dependent (IMF) exponential and dipole
form factors with the $t$-dependent dipole form factor, as well as
with a calculation without any form factor suppression.
In this case, the comparatively wide spread
in predictions for the $x$ dependence corresponding to different
scenarios for the $G_{\pi N}$ vertex suggests that
this element of the calculation is indeed the least under control,
and would benefit most from experimental guidance from direct measurement
as might come from \cite{TDIS}. For instance, while the computation of
$F^{(\pi N)}_2$ is apparently fairly insensitive to the specific functional
form employed for the vertex form factor in the IMF, the more general
choice of $s$ vs.~$t$-dependent parametrizations is quite significant,
being responsible for a $\sim$30\% systematic effect at lower $x$.

\begin{figure}[h]
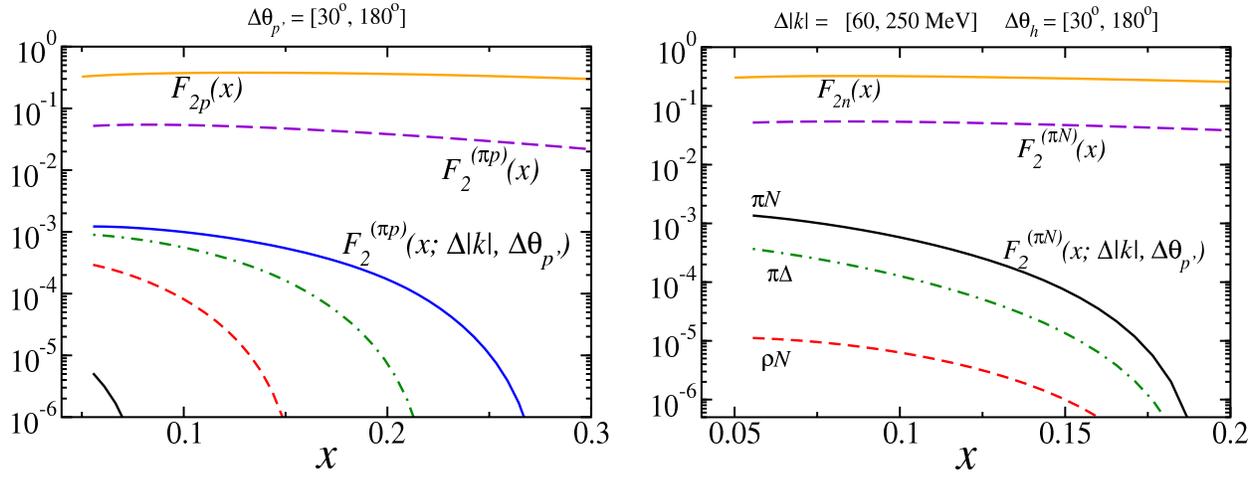

\vspace*{1cm}
\includegraphics[width=8cm]{F2pi_k-intervals_F2p_INT-theta-k.eps} \ \ \
\includegraphics[width=8cm]{F2-pi-rho-Delta_k-intervals_F2n.eps}
\caption{(Left) $x$ dependence of the semi-inclusive structure function
        $F_2^{(\pi p)}(x,\Delta |{\bm k}|,\Delta \theta_{p'})$.
	For comparison, the total integrated $\pi p$ contribution
	$F_2^{(\pi p)}$ to the inclusive proton structure function
	is shown (violet dashed), as is the total inclusive $F_{2p}$
	structure function (orange solid).The lower bands follow from
        varying the integration range $\Delta |{\bm k}|$ as described in
	the text.
	(Right) The corresponding quantity for charge-exchange in
	the $n \to \pi^-\, p$ process.
        The tagged semi-inclusive structure function for
        $(\pi^- p)$ (black, solid), $(\rho^- p)$ (red, dashed),
        and $(\pi^0 \Delta^0 + \pi^- \Delta^+)$ (green, dot-dashed)
        are compared with the inclusive structure function of the
        neutron $F_{2n}(x)$ (orange), and the fully-integrated
        $(\pi^- p)$ contribution $F_2^{\pi N}(x)$ (violet, dashed).
 }
\label{fig:F2pi_k-intervals_F2p_INT-theta-k}
\end{figure}

Lastly, in Fig.~\ref{fig:F2pi_k-intervals_F2p_INT-theta-k} we predict
the $x$ dependence for $F_2^{(\pi p)}(x,\Delta |{\bm k}|,\Delta \theta_{p'})$ in
several of the kinematical ranges we have proposed involving the charge-neutral
process. First, in the LHS of
Fig.~\ref{fig:F2pi_k-intervals_F2p_INT-theta-k} the colorful bands at bottom
follow from varying the integration range $\Delta |{\bm k}|$; namely, they correspond to
$\Delta |{\bm k}| = [60,100]$~MeV (black, solid),
$\Delta |{\bm k}| = [100,200]$~MeV (red, dashed),
$\Delta |{\bm k}| = [200,300]$~MeV (green, dot-dashed), and
$\Delta |{\bm k}| = [300,400]$~MeV (blue, solid). While these steps in the integration
range for $|{\bm k}|$ are fairly modest, the momentum dependence characterized in the
preceding discussion induces large, order-of-magnitude separations among the results for various
$|{\bm k}|$ intervals. At the same time, these observables are admittedly overshadowed by the much
larger inclusive structure functions, which are given for comparison, but current and future
facilities promise a high level of sensitivity. For instance, with a relatively
standard initial run, the JLab measurement proposed in \cite{TDIS} should be sensitive to
tagged structure functions down to $F_2^{(\pi p)} \sim 10^{-5}$.

That these measurements have the capacity to cleanly disentangle the role of scattering
from the pion cloud is made evident by the complementary information shown in the right panel of
Fig.~\ref{fig:F2pi_k-intervals_F2p_INT-theta-k}, which is plotted for the charge exchange process
$(n \rightarrow \pi^- p)$. As had already been hinted in Fig.~\ref{fig:F2MN_k}, $\pi$-exchange
is the dominant meson cloud process, eclipsing the corresponding mechanism involving the $\rho$
by $\sim 2$ orders of magnitude, even at the modest integration ranges shown, namely, 
$\Delta |{\bm k}| = [60,250]$~MeV and $\Delta \theta_{p'} = [30^\circ,180^\circ]$.
Contributions from scattering via channels that consist of intermediate $\Delta$ baryons are midway between
these two scenarios, being suppressed by an approximate order of magnitude relative to the simpler
reaction shown in the left diagram of Fig.~\ref{fig:feyn}. Of course, in the absence of direct coincidence
measurements of final state pions produced by the decay of the $\Delta$ (RHS of Fig.~\ref{fig:diag}),
these processes interfere, though the doubly-inclusive detection of a final state proton and pion may
allow the structure of the $\Delta$ to be better isolated.

These demonstrations emphasize the accessibility of the nucleon's pion cloud at a level which heretofore
had been unprecedented. Computing within the ambitious but achievable kinematics outlined
in this chapter, measurements have the novel ability to separate the contributions shown in Fig.~\ref{fig:feyn}
from various background processes, and better control their momentum dependence with sufficient resolution in
$|{\bm k}|$ (as established by the LHS of Fig.~\ref{fig:F2pi_k-intervals_F2p_INT-theta-k}). Moreover,
with these resolution enhancements, measurements of cloud contributions can in principle
be made very near the coveted $t = m^2_\pi$ pion pole as shown in Fig.~\ref{fig:t-dist}; much same, this
would also constrain the model dependence associated with the detailed behavior of the
pion-nucleon vertex shown in Fig.~\ref{fig:F2pi_dep} --- information that would significantly clarify
the pion cloud's role in the structure of the interacting nucleon.
%

%% file: the-conc.tex

This thesis has chiefly been concerned with hadronic structure as a
nonperturbative problem --- one that defies ready explanation in terms
of the standard QCD interactions introduced in Chap.~\ref{chap:ch-intro}
that have so successfully guided work at higher energies. In the
course of investigation, we have in fact seen these nonperturbative
properties to encompass an enormous range of phenomenology, and require
dedicated theoretical effort on many fronts to make even halting progress.
In spite of sustained attention, much remains fundamentally mysterious, and,
true to form, improved understanding if anything only leads to more and
deeper questions.

With the formal description in Chap.~\ref{chap:ch-DIS} of the DIS process
and operator structure of inclusive scattering, we were launched upon this
study by the dual recognition of the Compton amplitude's ability to provide
new information regarding the quark substructure of the nucleon
(Chap.~\ref{chap:ch-DIS}.\ref{sec:Compt}), and the problematic aspects
of performing DIS measurements at intermediate $Q^2$ (Chap.~\ref{chap:ch-Q2}).
In the latter case, we noticed that parity-violating DIS
is uniquely adapted as a tool to study novel aspects of the
partonic structure of the nucleon, including the flavor dependence of PDFs
in the region $x \sim 1$ or quark-level charge-symmetry violation, to say
nothing of more exotic physics beyond the Standard Model.

While the numerical suppression of the leptonic vector couplings $C_{2q}$
relative to the axial-vector couplings $C_{1q}$ is the main culprit in
the dominance of the hadronic-vector term $a_1$ inside the
parity-violating asymmetry $A^{\rm PV}$, we observed that the
axial-vector $a_3$ contribution can amount to as much as 20\% of the
total; the fact that this occurs for both proton and deuteron targets,
and in particular, introduces dependence on the still somewhat uncontrolled
electromagnetic ratio $R^\gamma$, implies that work remains in
quantitatively understanding the vector/axial-vector separation of
$A^{\rm PV}$.

A perhaps greater concern unique to the nonperturbative, low-$Q^2$ regime
of DIS comes from the interference ratio $R^{\gamma Z}$, which enters through
the larger $C_{1q}$-weighted vector term. Given its entirely unexplored phenomenology,
$R^{\gamma Z}$ may well contribute to $A^{\rm PV}_p$ if it differs significantly from
$R^{\gamma}$. We found that even modest deviations away from
$R^{\gamma Z} \approx R^\gamma$ could potentially obscure the main physics objectives
of DIS on the proton and deuteron --- namely, enhanced access to flavor information
(\EG~the $d/u$ PDF ratio) and quark-level CSV, respectively; this fact makes a more
thorough understanding of $R^{\gamma Z}$ and related finite-$Q^2$ effects a matter
of some urgency. Regarding the direct measurement of partonic CSV, we postulated
an alternative mechanism in Chap.~\ref{chap:ch-Q2}.\ref{sec:deut-pCSV} involving $ed$
scattering mediated by $W$-exchange; using a framework partly based on electromagnetically
induced CSV, we concluded a multi-percent signal could be observable at a TeV-scale
electron-ion collider --- yet another elusive physics objective such a machine might
advance.

Work on the possible extension of these considerations to the
domain of spin-polarized physics \cite{Bluemlein99,Detmold06,AM08,Leader10}
remains in its early stages, though we highlighted in
Chap.~\ref{chap:ch-Q2}.\ref{sec:pol} some suggestive ideas. The true push
would be to reduce the uncertainties on the spin-dependent quark distributions,
which remain quite egregious; for this purpose, new inputs from high precision
PVDIS spin asymmetry measurements would obviously be invaluable.

Prodded forward by the potential importance of the largely uncontrolled
ratio $R^{\gamma Z}$, we found a better understanding of the nonperturbative
contributions to that quantity necessary in order to be more
disposed in making definitive conclusions about its role in PVDIS.
Among the more accessible effects in this wise are the target mass
corrections, which we considered in both fully- and semi-inclusive
reactions. While the nature of the deuteron as an isoscalar target
insulates it from significant corrections from mass effects, this
is much less the case for the proton, which experiences a
multi-percent correction in both the asymmetry $A^{\rm PV}$ and
the phenomenological ratio $R^{\gamma Z}$. For the latter,
model dependence remains a serious issue, and a motivation for
additional calculation and measurement.

In the future, additional effects may need to be
considered at large $x$, not least of which are dynamical higher
twist corrections associated with nonperturbative multi-parton
correlations. These are of course very difficult to compute from first
principles, and until the present time, only rudimentary model estimates
have been available. Despite this limitation, our results on the
phenomenology of the target mass corrections should provide a benchmark
for future theoretical and experimental investigations of these additional
corrections.

As for the hadron mass corrections to semi-inclusive processes we
studied in Chap.~\ref{chap:ch-TMC}.\ref{sec:semi-intro}, their
most immediate use would be in leading twist analyses of SIDIS
cross sections, for which such corrections are an unavoidable
consideration before reliable extractions of parton distribution
and fragmentation functions can be made at large $x$ and $z_h$.
Clear applications of this work are also to be found in analyses
of semi-inclusive data in the nucleon resonance region, which has
been the recent focus of attention for the sake of understanding
quark-hadron duality \cite{MEK,ISGUR,CM}.

While the analysis of Chap.~\ref{chap:ch-TMC} was kept strictly to
leading order in $\alpha_s$, an extended formalism at NLO is becoming
more of a crying necessity; such a treatment is all the more imperative due to the
current lack of a more quantitative description of transverse mass
dependence of the produced hadrons, $p_{h\perp} \neq 0$. Such an
improvement would indubitably be a great service to
on-going studies of the Sivers, Boer-Mulders, and other effects
related to transverse momentum dependent parton distributions (TMDs),
in which nonzero parton transverse momentum, $k_\perp \neq 0$, is a
defining feature.

At the same time, unresolved challenges in the formalism itself remain.
This is clear given that the corrected SIDIS cross section
still exhibits the threshold problem which renders it nonzero as
$x \to x^{\rm max}$, much as we saw for inclusive DIS.
Solutions of this problem proposed in the literature for inclusive
structure functions \cite{Tung,Steffens,AQ} must be extended to SIDIS
in future work.

Throughout the analysis of Chaps.~\ref{chap:ch-Q2}--\ref{chap:ch-TMC}
we assumed (as is in fact the case) that the electroweak structure
functions are dominated in the QPM by their light $(u,d,s)$ degrees of
freedom. In this thesis' spirit of accounting for potential signals of
nonperturbative dynamics at large $x$ and moderate $Q^2$,
Chap.~\ref{chap:ch-charm} presented a comprehensive analysis of intrinsic
charm in the nucleon using a phenomenological model formulated in terms
of effective meson--baryon degrees of freedom; in doing so, we derived
couplings from $DN$ and $\bar{D} N$ Lippmann-Schwinger analyses
\cite{Hai07, Hai08, Hai11}, which permitted us to constrain model parameters
in the form of ultraviolet regulators using hadroproduction data \cite{Chauvat:1987kb}
and consistency checks \cite{Garcia:2001xj}.

In stark contrast to previous studies that neglected the spin structure \cite{Pum05}
and assumed dominance by the lowest mass state \cite{MT97, Nav96},
we included all low-lying hadrons of the $SU(4)$ spectrum in an ans\"atz that directly
connects the properties of the bound state spectrum to the multiplicity and momentum dependence
of the underlying charm quarks. Additionally, it was necessary to model internal distributions
for $c$ and $\bar c$ quarks within the charmed baryons and mesons, and this was accomplished by
means of a relativistic quark--spectator model, in which the momentum distributions of the quarks
are parametrized through phenomenological quark--spectator--nucleon vertex functions.
This procedure involved a number of assumptions, and we therefore compared several numerical approaches
to understand the systematic sources of model dependence. While this permitted a range of final
shapes for charm and anticharm momentum distributions, we turned to the technology of QCD global
analysis to further constrain the overall normalization of the intrinsic charm multiplicity.
Keeping a special eye to SLAC $ep$ and $ed$ data at $Q^2 \gtrsim 1$ GeV$^2$ and $W^2 \gtrsim 3.5$ GeV$^2$
\cite{Whitlow:1991uw}, the fits presented in Chap.~\ref{chap:ch-charm}.\ref{sec:GA} strongly constrained
the momentum carried by intrinsic charm; in the absence of high-$x$ $F^c_2$ measurements from EMC, we
found a quite restrictive upper bound, $\lan x \ran_{\rm IC} \le 0.1$\% at the $5\sigma$ level, though
fits that do include EMC data favor a rather small intrinsic charm component, $\lan x \ran_{\rm IC} = 0.13 \pm 0.04$\%.

In the end, however, direct charm production data at high $x$ and moderate
$Q^2$ would have the most immediate bearing on the intrinsic charm question,
and such measurements would advance this topic markedly. One might
proceed experimentally by searching for the characteristic asymmetries in
the $x$ dependence of the $c$ and $\bar c$ distributions (or analogously,
charge asymmetries in the hadroproduction of, \EG~$D^+/D^-$ or
$\Lambda_c/\bar{\Lambda}_c$), that are an inevitable consequence of MBMs such
as the one presented in Chap.~\ref{chap:ch-charm}. Slightly more indirect
channels might also be of use: for instance, the measurement of $W$ and $Z$
cross sections at the LHC \cite{Halzen13}, which receive significant
contributions from charm production, or photon plus charm jet production,
also at LHC kinematics \cite{Bednyakov13}.

Extension in a more detailed fashion to the strange $SU(3)$
sector (which possesses more plentiful data in specific meson and baryon
production channels), as well as confrontation with new charm production
data that might emerge from an eventual electron-ion collider would aid in
gauging the reliability of some of the model assumptions made in
Chap.~\ref{chap:ch-charm} --- particularly regarding the truncation of the
Fock state expansion of Eq.~(\ref{eq:Fock}).

Finally, in Chap.~\ref{chap:ch-TDIS} we provided further phenomenological
justification for the Fock state expansion of the nucleon used in
Chap.~\ref{chap:ch-charm}, exploiting the spontaneously broken chiral
symmetry of QCD as motivation for modeling the nucleon's ``pion cloud.''
This effect provides an avenue to precise intermediate $Q^2$ determinations of
both the much-sought pion form factor $F_\pi$, and, as we demonstrated for
a proposed JLab experiment \cite{TDIS}, the DIS structure function $F^\pi_2$. These
measurements stand to improve the connection between the pion
as a pseudo-Goldstone boson and its light quark substructure --- a
relationship that has been the subject of numerous modeling efforts, but
as always, requires experimental input to constrain the diverse range of
predictions.
%

%% file: the-pub.tex
\newpage

\vspace*{0.3cm}
{\bf Several of the author's scientific publications have contributed to this thesis with permission.}

\vspace*{0.3cm}

\begin{itemize}

\item
  T.~Hobbs and W.~Melnitchouk, \\
  \hspace*{0.5cm} Phys.\ Rev.\ D {\bf 77}, 114023 (2008)
  [arXiv:0801.4791 [hep-ph]]. \\
  \hspace*{0.5cm} {\it Finite-$Q^2$ corrections to parity-violating DIS.} \\
  \hspace*{0.5cm} {\bf \small Copyright 2008 by the American Physical Society.}

\vspace*{0.2cm}

\item
  A.~Accardi, T.~Hobbs and W.~Melnitchouk, \\
  \hspace*{0.5cm} JHEP {\bf 0911}, 084 (2009)
  [arXiv:0907.2395 [hep-ph]]. \\
  \hspace*{0.5cm} {\it Hadron mass corrections in semi-inclusive deep inelastic scattering.}

\vspace*{0.2cm}

\item
  T.~J.~Hobbs, J.~T.~Londergan, D.~P.~Murdock and A.~W.~Thomas, \\
  \hspace*{0.5cm} Phys.\ Lett.\ B {\bf 698}, 123 (2011)
  [arXiv:1101.3923 [hep-ph]]. \\
  \hspace*{0.5cm} {\it Testing Partonic Charge Symmetry at a High-Energy Electron Collider.} \\
  \hspace*{0.5cm} {\bf \small Copyright 2011 by Elsevier.}

\vspace*{0.2cm}

\item
  L.~T.~Brady, A.~Accardi, T.~J.~Hobbs and W.~Melnitchouk, \\
  \hspace*{0.5cm} Phys.\ Rev.\ D {\bf 84}, 074008 (2011)
  [arXiv:1108.4734 [hep-ph]]. \\
  \hspace*{0.5cm} {\it NLO analysis of target mass corrections to structure functions and asymmetries.} \\
  \hspace*{0.5cm} {\bf \small Copyright 2011 by the American Physical Society.}

\vspace*{0.2cm}

\item
  M.~Gorchtein, T.~Hobbs, J.~T.~Londergan and A.~P.~Szczepaniak, \\
  \hspace*{0.5cm} Phys.\ Rev.\ C {\bf 84}, 065202 (2011)
  [arXiv:1110.5982 [nucl-th]]. \\
  \hspace*{0.5cm} {\it Compton Scattering and Photo-absorption Sum Rules on Nuclei.} \\
  \hspace*{0.5cm} {\bf \small Copyright 2011 by the American Physical Society.}

\vspace*{0.2cm}

\pagebreak

\item
  T.~J.~Hobbs, J.~T.~Londergan and W.~Melnitchouk, \\
  \hspace*{0.5cm} Phys.\ Rev.\ D {\bf 89}, 074008 (2014)
  [arXiv:1311.1578 [hep-ph]]. \\
  \hspace*{0.5cm} {\it Phenomenology of nonperturbative charm in the nucleon.} \\
  \hspace*{0.5cm} {\bf \small Copyright 2014 by the American Physical Society.}

\vspace*{0.2cm}

\item
  P.~Jimenez-Delgado, T.~J.~Hobbs, J.~T.~Londergan and W.~Melnitchouk, \\
  \hspace*{0.5cm} arXiv:1408.1708 [hep-ph]. \\
  \hspace*{0.5cm} {\it New limits on intrinsic charm in the nucleon from global analysis
       of parton distributions.}

\vspace*{0.2cm}

\item
  A.~Camsonne, et al., \\
  \hspace*{0.5cm} Jefferson Lab Proposal PR12-14-010, (2014). \\
  \hspace*{0.5cm} {\it Measurement of Tagged Deep Inelastic Scattering (TDIS).}

\end{itemize}

%% file: thesis.bbl
\begin{thebibliography}{99} 

\bibitem{Heinz}
  H.~R.~Pagels,
  {\it The Cosmic Code: Quantum Physics as the Language of Nature}
  (Simon \& Schuster, New York, 1982).

\bibitem{Aristotle}
  Aristotle,
  {\it On Generation and Corruption}
  1.2 316a13--b16.

\bibitem{Dalton:1808}
  J.~Dalton, {\it A New System of Chemical Philosophy} (William Dawson \& Sons LTD., London, 1808).

\bibitem{Thomson:1897}
  J.~J.~Thomson, Phil.\ Mag.\ {\bf 44}, (1897) 293–316.

\bibitem{Rutherford:1911}
E.~Rutherford, Phil.\ Mag.\ {\bf 21}, (1911) 669–688.

\bibitem{Dirac:1928hu} 
  P.~A.~M.~Dirac,
  Proc.\ Roy.\ Soc.\ Lond.\ A {\bf 117}, 610 (1928).

\bibitem{Regge:1959mz} 
  T.~Regge,
  Nuovo Cim.\  {\bf 14}, 951 (1959).

\bibitem{Chew:1968fe} 
  G.~F.~Chew and A.~Pignotti,
  Phys.\ Rev.\  {\bf 176}, 2112 (1968).

\bibitem{Landau:1960}
  L.~D.~Landau,
  ``Fundamental Problems,'' in {\it Pauli Memorial Volume}, pg. 245, (Interscience, New York, 1960).

\bibitem{GellMann:1964nj} 
  M.~Gell-Mann,
  Phys.\ Lett.\  {\bf 8}, 214 (1964).

\bibitem{Zweig:1964jf} 
  G.~Zweig,
  Developments in the Quark Theory of Hadrons, Volume 1. Edited by D. Lichtenberg and S. Rosen. pp. 22-101

\bibitem{Hofstadter:1956qs} 
  R.~Hofstadter,
  Rev.\ Mod.\ Phys.\  {\bf 28}, 214 (1956).

\bibitem{Janssens:1965kd} 
  T.~Janssens, R.~Hofstadter, E.~B.~Hughes and M.~R.~Yearian,
  Phys.\ Rev.\  {\bf 142}, 922 (1966).

\bibitem{Bjorken:1968dy} 
  J.~D.~Bjorken,
  Phys.\ Rev.\  {\bf 179}, 1547 (1969).

\bibitem{'tHooft:1972fi} 
  G.~'t Hooft and M.~J.~G.~Veltman,
  Nucl.\ Phys.\ B {\bf 44}, 189 (1972).

\bibitem{Gross:1973id} 
  D.~J.~Gross and F.~Wilczek,
  Phys.\ Rev.\ Lett.\  {\bf 30}, 1343 (1973).

\bibitem{Allasia:1985hw} 
  D.~Allasia, C.~Angelini, A.~Baldini, L.~Bertanza, A.~Bigi, V.~Bisi, F.~Bobisut and T.~Bolognese {\it et al.},
  Z.\ Phys.\ C {\bf 28}, 321 (1985).

\bibitem{Allasia:1990nt} 
  D.~Allasia {\it et al.}  [New Muon Collaboration (NMC)],
  Phys.\ Lett.\ B {\bf 249}, 366 (1990).

\bibitem{Melnitchouk:1992eu} 
  W.~Melnitchouk and A.~W.~Thomas,
  Phys.\ Rev.\ D {\bf 47}, 3783 (1993)
  [nucl-th/9301016].

\bibitem{Feynman:1969ej} 
  R.~P.~Feynman,
  Phys.\ Rev.\ Lett.\  {\bf 23}, 1415 (1969).


\bibitem{Beringer:1900zz} 
  J.~Beringer {\it et al.}  [Particle Data Group Collaboration],
  Phys.\ Rev.\ D {\bf 86}, 010001 (2012).

\bibitem{Cutkosky:1960sp} 
  R.~E.~Cutkosky,
  J.\ Math.\ Phys.\  {\bf 1}, 429 (1960).

\bibitem{Bacchetta:2002xd} 
  A.~Bacchetta (PhD thesis),
  hep-ph/0212025.

\bibitem{Hobbs:2011} 
  M.~Gorchtein, T.~Hobbs, J.~T.~Londergan and A.~P.~Szczepaniak,
  Phys.\ Rev.\ C {\bf 84}, 065202 (2011)
  [arXiv:1110.5982 [nucl-th]].

\bibitem{toll56} J.~S.~Toll, Phys.\ Rev.\  {\bf 104}, 1760 (1956).

\bibitem{Levinger:1960} J.~S.~Levinger, Nuclear Photo-Disintegration,
  Oxford University Press, London, 1960.




    
\bibitem{Damashek:1969xj}
  M.~Damashek and F.~J.~Gilman,
  Phys.\ Rev.\  D {\bf 1}, 1319 (1970).

  
  
  

\bibitem{Brodsky:2008qu}
  S.~J.~Brodsky, F.~J.~Llanes-Estrada and A.~P.~Szczepaniak,
  Phys.\ Rev.\  D {\bf 79}, 033012 (2009). 
  

\bibitem{Dominguez:1970wu}
  C.~A.~Dominguez, C.~Ferro Fontan and R.~Suaya,
  Phys.\ Lett.\  B {\bf 31}, 365 (1970).

\bibitem{Shibasaki:1971}
I.~Shibasaki, T.~Minamikawa, T.~Watanabe, Prog.~Theor.~Phys. {\bf 46},
173 (1971).


\bibitem{Tait:1972}
N.~R.~S.~Tait, J.~N.~J.~White, Nucl.~Phys.~B {\bf 43}, 27 (1972).


\bibitem{Thomas:1925a} 
W. Thomas, Naturwissenschaften {\bf 13}, 627 (1925). 



\bibitem{Harvey:1964} R.~R.~Harvey, J.~T.~Caldwell, R.~L.~Bramblett,
 and S.~C.~Fultz, 
 Phys.~Rev.~{\bf 138}, B126 (1964).

\bibitem{Hesse:1970cy}
  W.~P.~Hesse, D.~O.~Caldwell, V.~B.~Elings, R.~J.~Morrison, F.~V.~Murphy, B.~W.~Worster and D.~E.~Yount,
  Phys.\ Rev.\ Lett.\  {\bf 25}, 613 (1970).

\bibitem{Caldwell:1973}
  D.~O.~Caldwell, V.~B.~Elings, W.~P.~Hesse, R.~J.~Morrison, F.~V.~Murphy, and D.~E.~Yount,
  Phys.\ Rev.\ D\ {\bf 7}, 1362 (1973).
 
\bibitem{Caldwell:1979}
  D.~O.~Caldwell {\it et al.}, 
  Phys.\ Rev.\ Lett.\ {\bf 42}, 553 (1979).
  
\bibitem{Bianchi}
N.~Bianchi {\it et al.}, Phys.~Rev.~C {\bf 54}, 1688 (1996).


\bibitem{Froissart:1961} M. Froissart, Phys. Rev. {\bf 123}, 1053 (1961). 

\bibitem{PDG}
  J.~Beringer {\it et al.}  [Particle Data Group Collaboration],
  Phys.\ Rev.\ D {\bf 86}, 010001 (2012).

\bibitem{Breitweg:1999}
J. Breitweg {\it et al.} [ZEUS Collaboration], Eur. Phys. J. C {\bf
  7}, 609 (1999);
B. Surrow, Eur. Phys. J. direct C {\bf 2}, 1 (1999).

\bibitem{Wilson:1969zs} 
  K.~G.~Wilson,
  Phys.\ Rev.\  {\bf 179}, 1499 (1969).

\bibitem{Symanzik:1970rt} 
  K.~Symanzik,
  Commun.\ Math.\ Phys.\  {\bf 18}, 227 (1970).

\bibitem{TMC} 
  I.~Schienbein, V.~A.~Radescu, G.~P.~Zeller, M.~E.~Christy, C.~E.~Keppel, K.~S.~McFarland, W.~Melnitchouk and F.~I.~Olness {\it et al.},
  J.\ Phys.\ G {\bf 35}, 053101 (2008)
  [arXiv:0709.1775 [hep-ph]].

\bibitem{Greiner}
  W.~Greiner, S.~Schramm, and E.~Stein,
  {\em Quantum Chromodynamics} 
  (Springer-Verlag, Berlin, 2002).

\bibitem{Ross:1978xk} 
  D.~A.~Ross and C.~T.~Sachrajda,
  Nucl.\ Phys.\ B {\bf 149}, 497 (1979).


\bibitem{Stephenson:2003dv} 
  E.~J.~Stephenson, A.~D.~Bacher, C.~E.~Allgower, A.~Gardestig, C.~Lavelle, G.~A.~Miller, H.~Nann and J.~Olmsted {\it et al.},
  Phys.\ Rev.\ Lett.\  {\bf 91}, 142302 (2003)
  [nucl-ex/0305032].

\bibitem{Prescott}
C.~Y.~Prescott {\it et al.},
Phys.\ Lett.\ B {\bf 77}, 347 (1978);
%
C.~Y.~Prescott {\it et al.},
Phys.\ Lett.\ B {\bf 84}, 524 (1979).

\bibitem{Cahn}
R.~N.~Cahn and F.~J.~Gilman,
Phys.\ Rev.\ D {\bf 17}, 1313 (1978).

\bibitem{JLab6}
Jefferson Lab experiment E-05-007,
R.~Michaels, P.~Reimer and X.~Zheng spokespersons;
%
Jefferson Lab experiment E12-07-102,
K.~Paschke, P.~Reimer and X.~Zheng spokespersons.

\bibitem{JLab12}
P.~Souder, talk given at the Workshop {\em Inclusive and Semi-Inclusive
Spin Physics with High Luminosity and Large Acceptance at 11 GeV},
Jefferson Lab, Dec. 13-14, 2006;
%
%
K.~S.~Kumar,
{\it 15th International Workshop on Deep-Inelastic Scattering and
Related Subjects (DIS2007)},
Munich, Germany, Apr. 16-20, 2007.


\bibitem{Kumar:2013yoa} 
  K.~S.~Kumar, S.~Mantry, W.~J.~Marciano and P.~A.~Souder,
  Ann.\ Rev.\ Nucl.\ Part.\ Sci.\  {\bf 63}, 237 (2013)
  [arXiv:1302.6263 [hep-ex]].

\bibitem{Glashow:1961tr} 
  S.~L.~Glashow,
  Nucl.\ Phys.\  {\bf 22}, 579 (1961).

\bibitem{Salam:1968rm} 
  A.~Salam,
  Conf.\ Proc.\ C {\bf 680519}, 367 (1968).

\bibitem{Weinberg:1967tq} 
  S.~Weinberg,
  Phys.\ Rev.\ Lett.\  {\bf 19}, 1264 (1967).

\bibitem{Cabibbo:1963yz} 
  N.~Cabibbo,
  Phys.\ Rev.\ Lett.\  {\bf 10}, 531 (1963).

\bibitem{Hob10} 
  T.~Hobbs and J.~L.~Rosner,
  Phys.\ Rev.\ D {\bf 82}, 013001 (2010)
  [arXiv:1005.0797 [hep-ph]].

\bibitem{Wang:2014bba} 
  D.~Wang {\it et al.}  [PVDIS Collaboration],
  Nature {\bf 506}, no. 7486, 67 (2014).

\bibitem{Deandrea:1997wk} 
  A.~Deandrea,
  Phys.\ Lett.\ B {\bf 409}, 277 (1997)
  [hep-ph/9705435].

\bibitem{Souder}
P.~A.~Souder,
AIP Conf.\ Proc.\ {\bf 747}, 199 (2005).

\bibitem{SLAC3}
SLAC proposal E-149 (1992),
P.~E.~Bosted spokesperson.

\bibitem{NP}
W.~Melnitchouk and A.~W.~Thomas,
Phys.\ Lett.\ B {\bf 377}, 11 (1996).

\bibitem{A3}
I.~R.~Afnan {\em et al.},
Phys.\ Lett.\ B {\bf 493}, 36 (2000);
%
I.~R.~Afnan {\it et al.},
Phys.\ Rev.\ C {\bf 68}, 035201 (2003).

\bibitem{BONUS}
L.~L.~Frankfurt and M.~I.~Strikman,
Phys.\ Rept.\ {\bf 76}, 215 (1981);
%
S.~Simula,
Phys.\ Lett.\ B {\bf 387}, 245 (1996);
%
W.~Melnitchouk, M.~Sargsian and M.~I.~Strikman,
Z.\ Phys.\ A {\bf 359}, 99 (1997);
%
Jefferson Lab experiment E03-012,
S.~Kuhn {\em et al.} spokespersons.

\bibitem{Comment}
W.~Melnitchouk, I.~R.~Afnan, F.~R.~P.~Bissey and A.~W.~Thomas,
Phys.\ Rev.\ Lett.\ {\bf 84}, 5455 (2000).

\bibitem{Hobbs08}
T.~Hobbs and W.~Melnitchouk,
Phys.\ Rev.\ D {\bf 77}, 114023 (2008).
Note that in Eqs.~(9) and (13) the mass-dependent correction
factor should have the energy $E$ replaced by $2E$.

\bibitem{Ans}
M.~Anselmino, P.~Gambino and J.~Kalinowski,
Z.\ Phys.\  C {\bf 64}, 267 (1994).
%
Note that the electroweak couplings used here differ by a factor
of two relative to those of Ref.~\cite{PDG},
$C_{iq} = 2 C_{iq}^{\rm Anselmino}$.

\bibitem{Hobbs:2011vy} 
  T.~J.~Hobbs, J.~T.~Londergan, D.~P.~Murdock and A.~W.~Thomas,
  Phys.\ Lett.\ B {\bf 698}, 123 (2011)
  [arXiv:1101.3923 [hep-ph]].

\bibitem{TWbook}
A.~W.~Thomas and W.~Weise,
{\em The Structure of the Nucleon}
(Wiley-VCH, Berlin, 2001).


\bibitem{HT}
S.~Fajfer and R.~J.~Oakes,
Phys.\ Rev.\ D {\bf 30}, 1585 (1984);
%
P.~Castorina and P.~J.~Mulders,
Phys.\ Rev.\ D {\bf 31}, 2760 (1985);
%
M.~Dasgupta and B.~R.~Webber,
Phys. Lett. B {\bf 382}, 273 (1996);
%
E.~Stein {\em et al.},
Phys. Lett. B {\bf 376}, 177 (1996);
%
A.~I.~Signal,
Nucl. Phys. {\bf B497}, 415 (1997);
%
E.~Stein {\em et al.},
Nucl. Phys. {\bf B536}, 318 (1998);
%
M.~Beneke,
Phys. Rep. {\bf 317}, 1 (1999).

\bibitem{CSVreview}
J.~T.~Londergan and A.~W.~Thomas,
Prog.\ Part.\ Nucl.\ Phys.\  {\bf 41}, 49 (1998);
%
J.~T.~Londergan and A.~W.~Thomas,
J.\ Phys.\ G {\bf 31}, 1151 (2005).

\bibitem{R1990}
L.~W.~Whitlow {\em et al.},
Phys.\ Lett.\ B {\bf 250}, 193 (1990).

\bibitem{R1998}
K.~Abe {\it et al.},
Phys.\ Lett.\  B {\bf 452}, 194 (1999).

\bibitem{RJLab}
  V.~Tvaskis, M.~E.~Christy, J.~Arrington, R.~Asaturyan, O.~K.~Baker, H.~P.~Blok, P.~Bosted and M.~Boswell {\it et al.},
  Phys.\ Rev.\ Lett.\  {\bf 98}, 142301 (2007)
  [nucl-ex/0611023].

\bibitem{CTEQ}
W.~K.~Tung {\em et al.},
JHEP {\bf 0702}, 053 (2007).

\bibitem{Peng}
W.~Melnitchouk and J.~C.~Peng,
Phys.\ Lett.\  B {\bf 400}, 220 (1997).

\bibitem{FJ}
G.~R.~Farrar and D.~R.~Jackson,
Phys.\ Rev.\ Lett.\  {\bf 35}, 1416 (1975).

\bibitem{RlowQ}
S.~A.~Kulagin and R.~Petti,
Phys.\ Rev.\  D {\bf 76}, 094023 (2007).

\bibitem{CSVmodels}
E.~Sather,
Phys.\ Lett.\  B {\bf 274}, 433 (1992);
%
%
J.~T.~Londergan {\em et al.},
Phys.\ Lett.\  B {\bf 340}, 115 (1994);
%
C.~Boros, F.~M.~Steffens, J.~T.~Londergan and A.~W.~Thomas,
Phys.\ Lett.\  B {\bf 468}, 161 (1999).

\bibitem{MRSTCSV}
A.~D.~Martin, R.~G.~Roberts, W.~J.~Stirling and R.~S.~Thorne,
Eur.\ Phys.\ J.\  C {\bf 35}, 325 (2004).

\bibitem{MRSTQED}
A.~D.~Martin, R.~G.~Roberts, W.~J.~Stirling and R.~S.~Thorne,
Eur.\ Phys.\ J.\  C {\bf 39}, 155 (2005).

\bibitem{GJRQED}
M.~Gluck, P.~Jimenez-Delgado and E.~Reya,
Phys.\ Rev.\ Lett.\  {\bf 95}, 022002 (2005).

\bibitem{KK}
K.~S.~Kumar,
private communication.

\bibitem{Bjorken:1966jh} 
  J.~D.~Bjorken,
  Phys.\ Rev.\  {\bf 148}, 1467 (1966).

\bibitem{BB}
J.~Bluemlein and H.~Bottcher,
Nucl.\ Phys.\  B {\bf 636}, 225 (2002).

\bibitem{AAC}
M.~Hirai, S.~Kumano and N.~Saito,
Phys.\ Rev.\  D {\bf 69}, 054021 (2004).

\bibitem{DNS}
D.~de Florian, G.~A.~Navarro and R.~Sassot,
Phys.\ Rev.\  D {\bf 71}, 094018 (2005).

\bibitem{LSS}
E.~Leader, A.~V.~Sidorov and D.~B.~Stamenov,
Phys.\ Rev.\  D {\bf 73}, 034023 (2006).


\bibitem{Henley:1979} E.M.~Henley and G.A.~Miller, \textit{Mesons in Nuclei},
     ed. M. Rho and D.H. Wilkinson, (North-Holland, Amsterdam, 1979),  
     p. 116. 

\bibitem{Miller:1990iz} G.A.~Miller, B.M.K.~Nefkens and I.~Slaus, Phys. Rept. 
{\bf 194}, 1 (1990).

\bibitem{Miller:2006tv} 
  G.~A.~Miller, A.~K.~Opper and E.~J.~Stephenson,
  Ann.\ Rev.\ Nucl.\ Part.\ Sci.\  {\bf 56}, 253 (2006)
  [nucl-ex/0602021].

%
\bibitem{Bentz:2009yy}
  W.~Bentz, I.~C.~Cloet, J.~T.~Londergan {\it et al.},
  Phys.\ Lett.\  {\bf B693}, 462-466 (2010).
  [arXiv:0908.3198 [nucl-th]].
%
\bibitem{Londergan:2003ij}
  J.~T.~Londergan, A.~W.~Thomas,
  Phys.\ Rev.\  {\bf D67}, 111901 (2003).
  [hep-ph/0303155].

\bibitem{MRST03} A.D. Martin, R.G. Roberts, W.J. Stirling and 
	R.S. Thorne, Eur.\ Phys.\ J.\ C{\bf 35}, 325 (2004).

\bibitem{Londergan:2009kj}
  J.~T.~Londergan, J.~C.~Peng, A.~W.~Thomas,
  Rev.\ Mod.\ Phys.\  {\bf 82}, 2009-2052 (2010).
  [arXiv:0907.2352 [hep-ph]].

\bibitem{Lo98a} J. T. Londergan and A. W. Thomas, 
	Prog.\ in Part.\ Nucl.\ Phys.\ {\bf 41}, 49 (1998).

\bibitem{LHeC} For details see:  http://www.lhec.org.uk

\bibitem{Accardi:2012qut} 
  A.~Accardi, J.~L.~Albacete, M.~Anselmino, N.~Armesto, E.~C.~Aschenauer, A.~Bacchetta, D.~Boer and W.~Brooks {\it et al.},
  arXiv:1212.1701 [nucl-ex].

\bibitem{Sather:1991je} E. Sather, Phys. Lett. B{\bf 274}, 433 (1992).  

\bibitem{Rodionov:1994cg} E.N. Rodionov, A.W. Thomas and J.T. Londergan, 
   Mod.\ Phys.\ Lett. A{\bf 9}, 1799 (1994).

\bibitem{Martin:2004dh} A.D. Martin, R.G. Roberts, W.J. Stirling and 
	R.S. Thorne, Eur.\ Phys.\ J.\ C{\bf 39}, 155 (2005).

\bibitem{Gluck:2005xh} M. Glueck, P. Jimenez-Delgado and E. Reya, Phys.\ Rev.\ 
   Lett. {\bf 95}, 022002 (2005).

\bibitem{Dokshitzer:1977sg} Y.L. Dokshitzer, Sov. Phys. JETP {\bf 46}, 
   641 (1977). 

\bibitem{Gribov:1972ri} V.N. Gribov and L.N. Lipatov, Sov. J. Nucl. Phys. 
   {\bf 15}, 438 (1972). 

\bibitem{Altarelli:1977zs} G. Altarelli and G. Parisi, Nucl. Phys. B{\bf 146}, 
   298 (1977). 

\bibitem{Chekanov:2004wr} S. Chekanov \EA~(ZEUS Collaboration), 
    Phys.~Lett.~B{\bf 595}, 86 (2004). 

\bibitem{Sjostrand:2000wi} T. Sjostrand \EA, Comput.~Phys.~Commun. {\bf 135}, 
    238 (2001). 

\bibitem{Marchesini:1991ch} G. Marchesini \EA, Comput.~Phys.~Commun. {\bf 67}, 
    465 (1992). 

\bibitem{Jaffe:1983hp} R.L. Jaffe, Nucl. Phys. B{\bf 229}, 205 (1983). 

\bibitem{Signal:1989yc} A.I. Signal and A.W. Thomas, Phys. Rev. D {\bf 40}, 
   2832 (1989). 


\bibitem{Bickerstaff:1989ch} R.P. Bickerstaff and A.W. Thomas, J. Phys. 
   G  {\bf 15}, 1523 (1989). 

  
\bibitem{Martin:2001es} A.D. Martin, R.G. Roberts, W.J. Stirling and 
	R.S. Thorne, Eur.\ Phys.\ J.\ C{\bf 23}, 73 (2002).



\bibitem{Signal:1987gz}
  A.~I.~Signal, A.~W.~Thomas,
  Phys.\ Lett.\  {\bf B191}, 205 (1987).
  
\bibitem{Thomas:2000ny}
  A.~W.~Thomas, W.~Melnitchouk, F.~M.~Steffens,
  Phys.\ Rev.\ Lett.\  {\bf 85}, 2892-2894 (2000).
  [hep-ph/0005043].

\bibitem{Bazarko:1994tt} A.O. Bazarko \EA~(CCFR Collaboration), 
   Z. Phys. C{\bf 65}, 189 (1995). 

\bibitem{Goncharov:2001qe} M. Goncharov \EA~(NuTeV Collaboration), 
   Phys. Rev. D{\bf 64}, 112006 (2001).  

\bibitem{Lai:2007dq} H.~L.~Lai, P.~M.~Nadolsky, J.~Pumplin, D.~Stump, W.~K.~Tung 
   and C.~P.~Yuan, JHEP {\bf 0704}, 089 (2007).

\bibitem{Mason:2007zz} D.~Mason {\it et al.}, Phys.\ Rev.\ Lett.\  {\bf 99}, 
   192001 (2007).

\bibitem{Ball:2009mk} R.~D.~Ball {\it et al.}  [NNPDF Collaboration],
   Nucl. Phys. B{\bf 823}, 195 (2009). 

\bibitem{Martin:2009iq} A.~D.~Martin, W.~J.~Stirling, R.~S.~Thorne and G.~Watt,
  Eur. Phys. J. C{\bf 63}, 189 (2009). 

\bibitem{Alekhin:2009mb} S. Alekhin, S. Kulagin and R. Petti, 
   Phys. Lett. B{\bf 675}, 433 (2009). 








\bibitem{CJ11}
A.~Accardi {\it et al.},
Phys. Rev. D {\bf 84}, 014008 (2011).


\bibitem{SLAC}
L.~W.~Whitlow {\it et al.},
Phys.\ Lett.\ B{\bf 282}, 475 (1992).


\bibitem{BONUS12}
Jefferson Lab Experiment E12-10-102 [BONUS12],
S.~B\"ultmann, M.~E.~Christy, H.~Fenker, K.~Griffioen, C.~E.~Keppel,
S.~Kuhn and W.~Melnitchouk, spokespersons.

\bibitem{MARATHON}
Jefferson Lab Experiment E12-10-103 [MARATHON],
G.~G.~Petratos, J.~Gomez, R.~J.~Holt and R.~D.~Ransome,
spokespersons.

\bibitem{SOLID}
Jefferson Lab Experiment E12-10-007 [SoLID],
P.~Souder, spokesperson.

\bibitem{Brady11}
L.~T.~Brady, A.~Accardi, W.~Melnitchouk and J.~F.~Owens,
arXiv:1110.5398 [hep-ph].

\bibitem{Kuhlmann00}
S. Kuhlmann {\it et al.},
Phys. Lett. B {\bf 476}, 291 (2000).

\bibitem{Brady:2011uy} 
  L.~T.~Brady, A.~Accardi, T.~J.~Hobbs and W.~Melnitchouk,
  Phys.\ Rev.\ D {\bf 84}, 074008 (2011)
  [Erratum-ibid.\ D {\bf 85}, 039902 (2012)]
  [arXiv:1108.4734 [hep-ph]].

\bibitem{Nachtmann73}
O.~Nachtmann,
Nucl.\ Phys.\  B {\bf 63}, 237 (1973).

\bibitem{GP76}
H.~Georgi and H.~D.~Politzer,
Phys.\ Rev.\  D {\bf 14}, 1829 (1976).


\bibitem{EFP83}
R.~K.~Ellis, W.~Furmanski and R.~Petronzio,
Nucl.\ Phys.\  B {\bf 212}, 29 (1983).

\bibitem{AOT94}
M.~A.~G.~Aivazis, F.~I.~Olness and W.~K.~Tung,
Phys.\ Rev.\  D {\bf 50}, 3085 (1994).

\bibitem{KR02}
S.~Kretzer and M.~H.~Reno,
Phys.\ Rev.\  D {\bf 66}, 113007 (2002).

\bibitem{AQ08}
A.~Accardi and J.~W.~Qiu,
JHEP {\bf 07}, 090 (2008).


\bibitem{Hobbs11}
T.~Hobbs,
AIP Conf. Proc. {\bf 1369}, 51 (2011).

\bibitem{Bj78}
J.~D.~Bjorken,
Phys.\ Rev.\ D {\bf 18}, 3239 (1978).






\bibitem{Greenberg71}
O.~W.~Greenberg and D.~Bhaumik,
Phys.\ Rev.\  D {\bf 4}, 2048 (1971).

\bibitem{Tung79}
K.~Bitar, P.~W.~Johnson and W.~K.~Tung,
Phys.\ Lett.\  B {\bf 83} (1979) 114;
%
P.~W.~Johnson and W.~K.~Tung,
Print-79-1018 (Illinois Tech), Contribution to Neutrino '79,
Bergen, Norway (1979).

\bibitem{Steffens06}
F.~M.~Steffens and W.~Melnitchouk.
Phys.\ Rev.\ C {\bf 73}, 055202 (2006).

\bibitem{KP06}
S.~A.~Kulagin and R.~Petti,
Nucl.\ Phys.\  A {\bf 765} (2006) 126.

\bibitem{AHM09}
A.~Accardi, T.~Hobbs and W.~Melnitchouk,
JHEP {\bf 11}, 084 (2009).

\bibitem{Qiu90}
J.~-W.~Qiu,
Phys.\ Rev.\  {\bf D42}, 30-44 (1990).

\bibitem{Altarelli78}
G.~Altarelli and G.~Martinelli,
Phys.\ Lett.\  B {\bf 76}, 89 (1978).

\bibitem{Bardeen78}
W.~A.~Bardeen, A.~J.~Buras, D.~W.~Duke and T.~Muta,
Phys.\ Rev.\  D {\bf 18}, 3998 (1978).

\bibitem{CSS88}
J.~C.~Collins, D.~E.~Soper and G.~F.~Sterman,
Adv.\ Ser.\ Direct.\ High Energy Phys.\  {\bf 5}, 1 (1988).



\bibitem{MEK05}
W.~Melnitchouk, R.~Ent and C.~E.~Keppel,
Phys. Rep. {\bf 406}, 127 (2005).


%



%
%


        



\bibitem{Bluemlein99}
J.~Bl\"umlein and A.~Tkabladze,
Nucl.\ Phys.\ {\bf B553}, 427 (1999).

\bibitem{Detmold06}
W.~Detmold,
Phys.\ Lett.\  B {\bf 632}, 261 (2006).

\bibitem{AM08}
A.~Accardi and W.~Melnitchouk,
Phys.\ Lett.\ B {\bf 670}, 114 (2008).

\bibitem{Leader10}
U.~D'Alesio, E.~Leader and F.~Murgia,
Phys.\ Rev.\  D {\bf 81}, 036010 (2010).

\bibitem{Schienbein}
I.~Schienbein {\it et al.},
J.\ Phys.\ G {\bf 35} (2008) 053101.

\bibitem{Nachtmann}
O.~Nachtmann,
Nucl.\ Phys.\  B {\bf 63} (1973) 237.


\bibitem{Tung}
K.~Bitar, P.~W.~Johnson and W.~K.~Tung,
Phys.\ Lett.\  B {\bf 83} (1979) 114;
%
P.~W.~Johnson and W.~K.~Tung,
Print-79-1018 (Illinois Tech) {\it Contribution to Neutrino '79},
Bergen, Norway, June 18-22, 1979.





\bibitem{Steffens}
F.~M.~Steffens and W.~Melnitchouk,
Phys.\ Rev.\  C {\bf 73} (2006) 055202.

\bibitem{EFP}
R.~K.~Ellis, W.~Furmanski and R.~Petronzio,
Nucl.\ Phys.\  B {\bf 212} (1983) 29.

\bibitem{Collins}
J.~C.~Collins, D.~E.~Soper and G.~Sterman,
Adv.\ Ser.\ Direct.\ High Energy Phys.\  {\bf 5} (1988) 1.

\bibitem{Collins2}
J.~C.~Collins and D.~E.~Soper,
Nucl.\ Phys.\  B {\bf 194} (1982) 445.



\bibitem{AQ}
A.~Accardi and J.~W.~Qiu,
JHEP {\bf 0807} (2008) 090.


\bibitem{Albino}
S.~Albino, B.~A.~Kniehl, G.~Kramer and C.~Sandoval,
Phys.\ Rev.\  D {\bf 75} (2007) 034018.

\bibitem{Mulders}
P.~J.~Mulders,
{\em ``Transverse momentum dependence in structure functions
in hard scattering processes''}, lecture notes,
\texttt{http://www.nikhef.nl/$\sim$pietm/COR-0.pdf}, 2001 (unpublished).

\bibitem{Greenberg}
O.~W.~Greenberg and D.~Bhaumik,
Phys.\ Rev.\  D {\bf 4} (1971) 2048.

\bibitem{TMD}
A.~Bacchetta, M.~Diehl, K.~Goeke, A.~Metz, P.~J.~Mulders and M.~Schlegel,
JHEP {\bf 0702} (2007) 093.

\bibitem{Drell:1969jm} 
  S.~D.~Drell, D.~J.~Levy and T.~-M.~Yan,
  Phys.\ Rev.\  {\bf 187}, 2159 (1969).


\bibitem{PDF}
J.~Pumplin, D.~R.~Stump, J.~Huston, H.~L.~Lai, P.~M.~Nadolsky and
W.~K.~Tung,
JHEP {\bf 0207} (2002) 012.

\bibitem{KKP}
B.~A.~Kniehl, G.~Kramer, B.~Potter,
Nucl.\ Phys.\  B {\bf 582} (2000) 514.

%

\bibitem{HallC}
T.~Navasardyan {\it et al.},   
Phys.\ Rev.\ Lett.\  {\bf 98} (2007) 022001.


\bibitem{EMC2}
J.~Ashman {\it et al.},
Z.\ Phys.\ C {\bf 52} (1991) 361.

\bibitem{MEK}
W.~Melnitchouk, R.~Ent and C.~Keppel,
Phys.\ Rept.\  {\bf 406} (2005) 127.

\bibitem{ISGUR}
F.~E.~Close and N.~Isgur,
Phys.\ Lett.\  B {\bf 509} (2001) 81.

\bibitem{CM}
F.~E.~Close and W.~Melnitchouk,
Phys.\ Rev.\ C {\bf 79} (2009) 055202.

\bibitem{Adloff}
C.~Adloff {\it et al.}, 
Nucl.\ Phys.\  B {\bf 504} (1997) 3.


\bibitem{SMT99} 
F.~M.~Steffens, W.~Melnitchouk and A.~W.~Thomas,
Eur. Phys. J. C {\bf 11}, 673 (1999).

\bibitem{Stavreva:2012bs} 
  T.~Stavreva, F.~I.~Olness, I.~Schienbein, T.~Jezo, A.~Kusina, K.~Kovarik and J.~Y.~Yu,
  Phys.\ Rev.\ D {\bf 85}, 114014 (2012)
  [arXiv:1203.0282 [hep-ph]].

\bibitem{Hobbs13} 
  T.~J.~Hobbs, J.~T.~Londergan and W.~Melnitchouk,
  Phys.\ Rev.\ D {\bf 89}, 074008 (2014)
  [arXiv:1311.1578 [hep-ph]].

\bibitem{Globe}
 P.~Jimenez-Delgado, T.~J.~Hobbs, J.~T.~Londergan and W.~Melnitchouk,
  arXiv:1408.1708 [hep-ph].

\bibitem{BHPS}
S.~J.~Brodsky, P.~Hoyer, C.~Peterson and N.~Sakai,
Phys. Lett. B {\bf 93}, 451 (1980).













\bibitem{MT97}
W.~Melnitchouk and A.~W.~Thomas,
Phys. Lett. B {\bf 414}, 134 (1997).

\bibitem{Nav96}
F.~S.~Navarra, M.~Nielsen, C.~A.~A.~Nunes and M.~Teixeira,
Phys. Rev. D {\bf 54}, 842 (1996).




\bibitem{Pum05}
J.~Pumplin,
Phys. Rev. D {\bf 73}, 114015 (2006).

\bibitem{Pum07}
J.~Pumplin, H.~L.~Lai and W.-K.~Tung,
Phys. Rev. D {\bf 75}, 054029 (2007).

\bibitem{Dulat13} 
S.~Dulat, T.-J.~Hou, J.~Gao, J.~Huston, J.~Pumplin, C.~Schmidt,
D.~Stump and C.-P.~Yuan,
arXiv:1309.0025 [hep-ph].

\bibitem{EMC3}
J.~J.~Aubert \EA,	
Nucl. Phys. {\bf B213}, 31 (1983);
Phys. Lett. B {\bf 94}, 96 (1980);
{\it ibid.} B {\bf 110}, 73 (1982).

\bibitem{H1}
A.~Aktas \EA,		
Eur. Phys. J. C {\bf 45}, 23 (2006).

\bibitem{ZEUS}
S.~Chekanov \EA,	
JHEP {\bf 0707}, 074 (2007).


%

%
%


\bibitem{Sib01}
A.~Sibirtsev, K.~Tsushima and A.~W.~Thomas,
Phys. Rev. C {\bf 63}, 044906 (2001).

\bibitem{Hai07}
J.~Haidenbauer, G.~Krein, U.-G.~Mei\ss ner and A.~Sibirtsev,
Eur. Phys. J. A {\bf 33}, 107 (2007).

\bibitem{Hai08}
J.~Haidenbauer, G.~Krein, U.-G.~Mei\ss ner and A.~Sibirtsev,
Eur. Phys. J. A {\bf 37}, 55 (2008).

\bibitem{Hai11}
J.~Haidenbauer, G.~Krein, U.-G.~Mei\ss ner and L.~Tolos,
Eur. Phys. J. A {\bf 47}, 18 (2011).

\bibitem{Weinberg:1966jm} 
  S.~Weinberg,
  Phys.\ Rev.\  {\bf 150}, 1313 (1966).

\bibitem{Chauvat:1987kb}
P.~Chauvat \EA,		
Phys. Lett. B {\bf 199}, 304 (1987).

\bibitem{Garcia:2001xj}
F.~G.~Garcia \EA,	
Phys. Lett. B {\bf 528}, 49 (2002).

\bibitem{Zoller92}
V.~R.~Zoller,
Z. Phys. C {\bf 53}, 443 (1992).

\bibitem{MT93} 
W.~Melnitchouk and A.~W.~Thomas,
Phys. Rev. D {\bf 47}, 3794 (1993).

\bibitem{Holtmann96}
H.~Holtmann, A.~Szczurek and J.~Speth,
Nucl. Phys. {\bf A596}, 631 (1996).



\bibitem{MST94} 
W.~Melnitchouk, A.~W.~Schreiber and A.~W.~Thomas,
Phys. Rev. D {\bf 49}, 1183 (1994).

\bibitem{Catani04}
Heavy quark asymmetries can be generated perturbatively,
but only at three-loop order, see
S.~Catani, D.~de Florian, G.~Rodrigo and W.~Vogelsang,
Phys. Rev. Lett. {\bf 93}, 152003 (2004).

\bibitem{Adler:2003qs} 
  S.~S.~Adler {\it et al.}  [PHENIX Collaboration],
  Phys.\ Rev.\ Lett.\  {\bf 92}, 051802 (2004)
  [hep-ex/0307019].

\bibitem{Bur13} 
M.~Burkardt, K.~S.~Hendricks, C.~-R.~Ji, W.~Melnitchouk and 
A.~W.~Thomas,
Phys. Rev. D {\bf 87}, 056009 (2013).

\bibitem{Machleidt87} 
R.~Machleidt, K.~Holinde and C.~Elster,
Phys. Rep. {\bf 149}, 1 (1987).

\bibitem{HDHPS95}
M.~Hoffman, J.~W.~Durso, K.~Holinde, B.~C.~Pearce and J.~Speth,
Nucl. Phys. {\bf A593}, 341 (1995).

\bibitem{Holzenkamp89}
B.~Holzenkamp, K.~Holinde and J.~Speth,
Nucl. Phys. {\bf A500}, 485 (1989).

\bibitem{Cazaroto13}
E.~R.~Cazaroto, V.~P.~Goncalves, F.~S.~Navarra and M.~Nielsen,
arXiv:1302.0035 [hep-ph].


\bibitem{Hohler75}
G.~Hohler and E.~Pietarinen,
Nucl. Phys. {\bf B95}, 210 (1975).






\bibitem{Close88}
F.~E.~Close and A.~W.~Thomas,
Phys. Lett. B {\bf 212}, 227 (1988).

%
\bibitem{Miyama96}
M.~Miyama and S.~Kumano,
Comput. Phys. Commun. {\bf 94}, 185 (1996).


\bibitem{Halzen13}
F.~Halzen, Y.~S.~Jeong and C.~S.~Kim,
arXiv:1304.0322.

\bibitem{Bednyakov13}
V.~A.~Bednyakov, M.~A.~Demichev, G.~I.~Lykasov, T.~Stavreva and M.~Stockton,
arXiv:1305.3548.  

\bibitem{DLY70} 
S.~D.~Drell, D.~J.~Levy and T.~-M.~Yan,
Phys. Rev. D {\bf 1}, 1035 (1970).

\bibitem{Mulders:1992za} 
P.~J.~Mulders, A.~W.~Schreiber and H.~Meyer,
Nucl.\ Phys.\ A {\bf 549}, 498 (1992).

\bibitem{Brock:1993sz} 
  R.~Brock {\it et al.}  [CTEQ Collaboration],
  Rev.\ Mod.\ Phys.\  {\bf 67}, 157 (1995).

\bibitem{Hoffmann:1983ah} 
  E.~Hoffmann and R.~Moore,
  Z.\ Phys.\ C {\bf 20}, 71 (1983).

\bibitem{Whitlow:1991uw} 
  L.~W.~Whitlow, E.~M.~Riordan, S.~Dasu, S.~Rock and A.~Bodek,
  Phys.\ Lett.\ B {\bf 282}, 475 (1992).

\bibitem{JR14} 
  P.~Jimenez-Delgado and E.~Reya,
  Phys.\ Rev.\ D {\bf 89}, 074049 (2014)
  arXiv:1403.1852 [hep-ph].

\bibitem{Brodsky:1995vr} 
  S.~J.~Brodsky, W.~-K.~Tang and P.~Hoyer,
  Phys.\ Rev.\ D {\bf 52}, 6285 (1995)
  [hep-ph/9506474].

\bibitem{Accardi:2011fa} 
  A.~Accardi, W.~Melnitchouk, J.~F.~Owens, M.~E.~Christy, C.~E.~Keppel, L.~Zhu and J.~G.~Morfin,
  Phys.\ Rev.\ D {\bf 84}, 014008 (2011)
  [arXiv:1102.3686 [hep-ph]].

\bibitem{deFlorian:2003qf} 
  D.~de Florian and R.~Sassot,
  Phys.\ Rev.\ D {\bf 69}, 074028 (2004)
  [hep-ph/0311227].

\bibitem{Tung:2006tb} 
  W.~K.~Tung, H.~L.~Lai, A.~Belyaev, J.~Pumplin, D.~Stump and C.~-P.~Yuan,
  JHEP {\bf 0702}, 053 (2007)
  [hep-ph/0611254].


\bibitem{Aubert:1982tt} 
  J.~J.~Aubert {\it et al.}  [European Muon Collaboration],
  Nucl.\ Phys.\ B {\bf 213}, 31 (1983); Phys.\ Lett.\ 
B{\bf 94}, 96 (1980); {\it ibid.} B{\bf 110}, 73 (1982).

\bibitem{Abramowicz:1900rp} 
  H.~Abramowicz {\it et al.}  [H1 and ZEUS Collaborations],
  Eur.\ Phys.\ J.\ C {\bf 73}, 2311 (2013)
  [arXiv:1211.1182 [hep-ex]].


\bibitem{Lattes:1947mw} 
  C.~M.~G.~Lattes, H.~Muirhead, G.~P.~S.~Occhialini and C.~F.~Powell,
  Nature {\bf 159}, 694 (1947).

\bibitem{Chang:2013nia} 
  L.~Chang, I.~C.~Cloët, C.~D.~Roberts, S.~M.~Schmidt and P.~C.~Tandy,
  Phys.\ Rev.\ Lett.\  {\bf 111}, 141802 (2013)
  [arXiv:1307.0026 [nucl-th]].

\bibitem{Huber:2008id} 
  G.~M.~Huber {\it et al.}  [Jefferson Lab $F_\pi$ Collaboration],
  Phys.\ Rev.\ C {\bf 78}, 045203 (2008)
  [arXiv:0809.3052 [nucl-ex]].

\bibitem{Melnitchouk:1998rv} 
  W.~Melnitchouk, J.~Speth and A.~W.~Thomas,
  Phys.\ Rev.\ D {\bf 59}, 014033 (1998)
  [hep-ph/9806255].

\bibitem{TDIS}
  A.~Camsonne, et al.,
  ``Measurement of Tagged Deep Inelastic Scattering (TDIS),''
  Jefferson Lab Proposal PR12-14-010, (2014).

\bibitem{Sullivan:1971kd} 
  J.~D.~Sullivan,
  Phys.\ Rev.\ D {\bf 5}, 1732 (1972).

\bibitem{Amaudruz:1991at} 
  P.~Amaudruz {\it et al.}  [New Muon Collaboration],
  Phys.\ Rev.\ Lett.\  {\bf 66}, 2712 (1991).

\bibitem{GellMann:1968rz} 
  M.~Gell-Mann, R.~J.~Oakes and B.~Renner,
  Phys.\ Rev.\  {\bf 175}, 2195 (1968).

\bibitem{Burkardt:2012hk} 
  M.~Burkardt, K.~S.~Hendricks, C.~-R.~Ji, W.~Melnitchouk and A.~W.~Thomas,
  Phys.\ Rev.\ D {\bf 87}, 056009 (2013)
  [arXiv:1211.5853 [hep-ph]].



\bibitem{GRVpi}
M.~Gl\"uck, E.~Reya, A.~Vogt,
Z. Phys. C {\bf 53}, 651 (1992).

\bibitem{MRSpi}
A.~D.~Martin, R.~G.~Roberts, W.~J.~Stirling and P.~J.~Sutton,
Phys. Rev. D {\bf 45}, 2349 (1992).

\end{thebibliography}
